\pdfoutput=1
\documentclass[12pt,a4paper,DIV=13,BCOR=15mm,chapterprefix=true]{scrbook}

\usepackage[english,german]{babel,varioref}   	  
\usepackage[latin1]{inputenc}											
\usepackage{graphicx}					
\usepackage[squaren]{SIunits}
\usepackage{amsmath,amssymb,array,setspace}    		
\usepackage{scrpage2}         										
\usepackage{color}
\usepackage[Sonny]{fncychap}											
\usepackage[pagebackref=true,colorlinks,linkcolor=black,citecolor=tumblaudunkel2,urlcolor=black]{hyperref} 
\usepackage[clearempty]{titlesec} 
\usepackage{epsfig}
\usepackage{longtable}
\usepackage{paralist}
\usepackage{subfig}
\usepackage{amssymb}
\usepackage{floatrow}
\usepackage{textcomp }
\usepackage{wasysym }
\usepackage{floatrow}
\usepackage{cite}
\usepackage{amsmath}

\usepackage{capt-of} 
\setkomafont{caption}{\itshape\normalfont}
\setkomafont{captionlabel}{\upshape\bfseries}

\usepackage{xcolor}			
\definecolor{tumblauTitel}{rgb}{0,.459,.737}	
\definecolor{tumgruen}{rgb}{.633,.676,0}	%
\definecolor{tumblaudunkel}{rgb}{0,.199,0.348}	%
\definecolor{tumblaudunkel2}{rgb}{0,.320,0.574}	%

\definecolor{tumblaudunkel}{cmyk}{1.0,0.43,.00,.0} 
\ChTitleVar{\LARGE\sf\color{tumblauTitel}}
\ChNameVar{\Huge\sf\color{tumblauTitel}}
\ChNameUpperCase
\ChNumVar{\Huge\sf}

\setcapindent{0.cm}

\setheadsepline[\textwidth]{0.6pt} 

\newcommand{\superscript}[1]{\ensuremath{^{\textrm{#1}}}} 
\newcommand{\subscript}[1]{\ensuremath{_{\textrm{#1}}}}   
\newcommand{\mathsym}[1]{{}}
\newcommand{\unicode}[1]{{}}

\renewcommand*{\backref}[1]{
}%
\renewcommand*{\backrefalt}[4]{%
	(Cited on %
	\ifnum#1=1 %
	page~%
	\else
	pages~%
	\fi #2.)
}%

\def\deadline{August 2016}
\def\title{Upgrade and Characterization of the SPIFFI/SINFONI optics}
\def\author{Dominik Gr\"aff}
\def\typeOfThesis{Master's Thesis}

\renewcommand\maketitle{
	\begin{titlepage}
		\begin{figure}[htbp]
			\begin{minipage}[t]{3.3cm}
				\vspace{0pt}
				\begin{flushleft}
					\includegraphics[width=3.3cm]{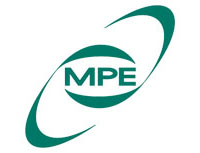}
				\end{flushleft}
			\end{minipage}
			\hfill
			\begin{minipage}[t]{5cm}
			\end{minipage}
			\hfill
			\begin{minipage}[t]{1cm}
				\vspace{3mm}
				\includegraphics[width=3.3cm]{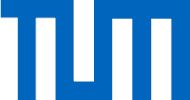}
			\end{minipage}
			\hfill
			\begin{minipage}[t]{2cm}
				\vspace{0pt}
				\begin{flushright}
					\includegraphics[width=2cm]{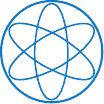}
				\end{flushright}
			\end{minipage}
		\end{figure}
		
		\begin{minipage}[t]{6cm}
			\begin{flushleft} \fontsize{12}{16}\selectfont\sffamily Max Planck Insitute for Extraterrestrial Physics
			\end{flushleft}
		\end{minipage}
		\hfill
		\begin{minipage}[t]{6cm}
			\begin{flushright} \fontsize{12}{16}\selectfont\sffamily Technical University of Munich Physics Department
			\end{flushright}
		\end{minipage}

		\vspace*{6cm}

		\begin{center}
			\noindent\fontsize{21}{26}\selectfont\sffamily\textbf{\textcolor{tumblauTitel}{\title}}

			\vspace{2.5cm}
			\noindent\fontsize{17}{22}\selectfont\textbf{\author}
			
			\vspace{0.5cm}
			\noindent\fontsize{14}{16}\selectfont\sffamily\typeOfThesis
			
			\vfill
			\noindent\deadline
		\end{center}
	\end{titlepage}
}
\begin{document}
\selectlanguage{english}
\pagenumbering{Roman} 

\maketitle

\thispagestyle{empty}
\newpage~

\vspace*{\fill}
\begin{flushleft} {First Reviewer (Supervisor): PD \ Dr.\ Frank Eisenhauer\\Second Reviewer: Prof.  Dr.\ Stefan Sch\"onert}
\end{flushleft}
\thispagestyle{empty}
\newpage

 \cleardoublepage
 \sloppy
\cleardoublepage
\thispagestyle{empty}
\selectlanguage{german}
\chapter*{Zusammenfassung}

\begin{sloppypar}
	SPIFFI ist ein Feldspektrograph am Very Large Telescope der Europ"aischen S"udsternwarte auf dem Berg Paranal in Chile. Momentan wird SPIFFI als Untereinheit von SINFONI betrieben, welches im nahen Infrarot Wellenl"angenbereich mit adaptiver Optik arbeitet um atmosph"arische St"orungen zu korrigieren. "Uber mehr als zehn Jahre hinweg hat sich SPIFFI als "au{\ss}erst produktiver Spektrograph in der Astronomie bew"ahrt. Im Januar 2016 wurden in SPIFFI vorzeitig einige optische Komponenten ausgetauscht um den Spektrographen f"ur die kommende Integration in das VLT Instrument ERIS vorzubereiten. Somit k"onnen schon bevor ERIS 2020 in Betrieb genommen wird neue Technologien genutzt werden um SPIFFI wettbewerbsf"ahig zu halten.
	
	Der Fokus dieser Masterarbeit liegt auf der Untersuchung und Beschreibung der Leistungsf"ahigkeit, Effizienz und G"ute SPIFFIs, welche durch das Upgrade gesteigert wurden. Insbesondere werden Messungen und Untersuchungen der ausgetauschten diamant-gedrehten Spektrometerspiegel dargestellt und analysiert. Hierbei wird sowohl die Oberfl"achenabweichung dieser, als auch die resultierende Wellenfront des SPIFFI Kollimators, dessen Einfluss auf die Leistung des Spektrographen und dessen Linienabbildungsfunktion untersucht. Genauer betrachtet werden die diversen nicht idealen Profile der Spektrallinien, die durch den Spektrometer hervorgerufen werden, ihre Variation und der Grund weshalb diese vorhanden sind sowie die Auswirkung des Upgrades auf diese. Mit Bezug hierauf wird das Aufl"osungsverm"ogen des Spektrographen diskutiert und wie sich dieses durch die ausgetauschten optischen Elemente ver"andert hat. Des Weiteren wird die relative Transmission des Instrumentes vor dem Upgrade mit danach verglichen und begr"undet. Durch eine Auswertung des integrierten Flusses innerhalb verschiedener Kreisfl"achen um das Zentrum der Punktabbildungsfunktion wird die Bildqualit"at festgestellt. Dies erm"oglicht es eine obere Schranke f"ur die Strehl-Zahl zu geben, welche mit SINFONI erreicht werden kann. Dar"uber hinaus werden Verbindungen zwischen den genannten Messgr"o{\ss}en und den ausgetauschten opto-mechanischen Komponenten gekn"upft um weitere Einfl"usse auf diese abzusch"atzen. Somit kann bestimmt werden, welche weiteren nicht ausgetauschten Komponenten einen Einfluss auf die Leistungsf"ahigkeit des Instruments haben k"onnen. Weiterhin werden Entscheidungen die innerhalb des Upgrades bez"uglich der Optik getroffen wurden zusammen mit deren Einfluss auf die Leistungsf"ahigkeit des Instrumentes diskutiert. Letztlich werden, basierend auf den vorausgegangenen Untersuchungen, Vorschl"age gemacht, inwiefern die Qualit"at und Leistungsf"ahigkeit SPIFFIs durch neue Optiken noch weiter verbessert werden kann.
\end{sloppypar}
\selectlanguage{english}
\chapter*{Abstract}

The SPIFFI integral field spectrometer is operated as a subunit of the adaptive optics instrument SINFONI at the Very Large Telescope of the European Southern Observatory at Paranal Observatory in Chile. It has been a very productive scientific instrument in astronomy over the last decade. To prepare it for the next generation VLT instrument ERIS, and thus keep it competitive, an early upgrade of several optical components was carried out in January 2016 to make use of new technological developments before ERIS will be commissioned in 2020.

In this thesis the focus lies on the determination and detailed description of the gain in performance due to the upgrade, with a particular focus on measurements and investigation of the diamond turned mirrors of the SPIFFI spectrometer that have been exchanged during the upgrade. An analysis of the surface deformation of these mirrors is done, followed by a determination of the resulting collimator wavefront and its influence on the instrument performance and the line spread function of the spectrometer. A careful analysis is done of the different shapes of the spectral line profiles, their variations, how they changed due to the upgrade, and the reasons for their existence. Related to this, the resolution of the spectrometer is discussed and how it was effected by the upgrade. Furthermore a relative throughput ratio between the pre- and post- upgrade instrument is determined and the causes for the actual behavior are described. The image quality of the spectrometer is evaluated via a determination of the encircled energy within different radii around a PSF center, and an upper limit of the Strehl ratio that can be reached with SINFONI is calculated. A connection between the observed quantities and the exchanged opto-mechanical components is made in order to detect and determine further influences on these quantities and link them to further critical components that were not exchanged in the upgrade, but from which the performance  of the instrument could suffer. Decisions made in the upgrade concerning to the optics are discussed together with their influence and effect on the instrument performance. Finally suggestions are made on optical components and their ability to even improve the quality of the instrument performance.

\newpage
\tableofcontents
\cleardoublepage
\mainmatter 
\section*{Organization of this Thesis}
This Master's thesis concerns the upgrade of the SPIFFI instrument in January 2016 and classifies its performance compared to the pre- upgrade performance.

First two chapters give an overview and background of the SPIFFI instrument and the modifications made during the upgrade. Chapter \ref{ch:chapter_introduction} provides the background of the SPIFFI instrument, the underlying principle of spectroscopy, as well as the future instrument in which SPIFFI will be integrated. In chapter \ref{ch:chapter_upgrade} the upgrade of the optically relevant components is described together with a their impact on the instrument performance. The third and fourth chapter contains my original personal contribution. In chapter \ref{ch:chapter_mirrors} the measurements of the collimator mirrors are described and an analysis of the mirror surfaces and the resulting wavefront is given together with a evaluation of the influence on the instrument performance. In chapter \ref{ch:chapter_performance} this instrument performance is then analyzed in detail - mainly as a pre- and post- upgrade comparison. Chapter \ref{ch:summary_and_outlook} gives a summary of the instrument performance and of the optical components limiting it. Finally an outlook on the future of SPIFFI is given with suggestions for a further performance improvement.

\chapter{Introduction}\label{ch:chapter_introduction}

\section{Near Infrared Integral Field Spectroscopy}\label{sec:ifs}

\begin{figure}
	\begin{center}
		\includegraphics[height=5cm]{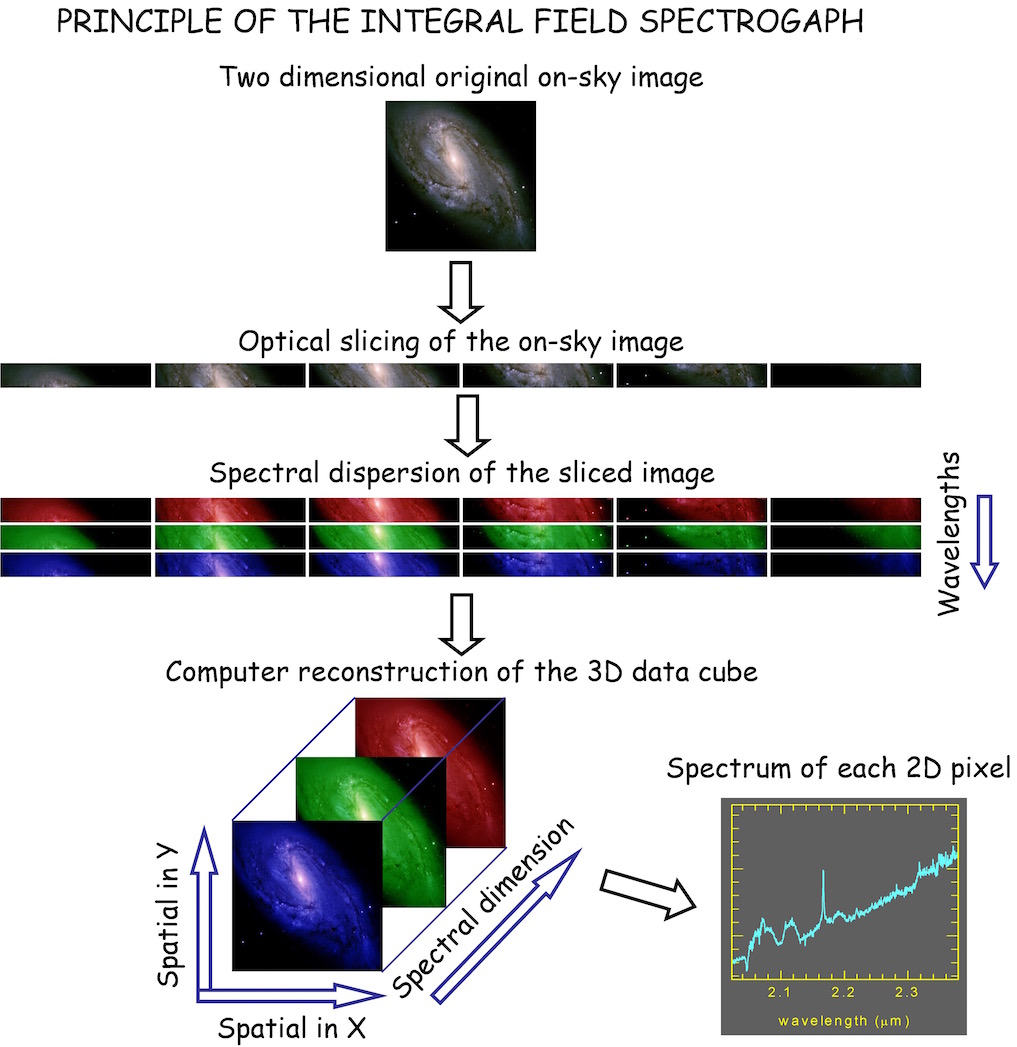}
		\caption[Principle of integral field spectroscopy]{The principle of integral field spectroscopy.\cite{esoifs}}
		\label{fig:schemeifs}
	\end{center}
\end{figure}

Many astronomical objects have a complex spatial structure. To characterize these objects, it is often useful to obtain the spectral information of every image point of a two dimensional field. The advantage of integral field spectroscopy is that it provides the spectra of every pixel in a two dimensional field in a single exposure (see figure \ref{fig:schemeifs}). This distinguishes integral field spectroscopy from techniques like slit scanning, which require several exposures for a single field.

Furthermore a long slit spectrograph does not provide a Point-Spread-Function (PSF), which is needed for deconvolution or even to treat the different contributions to the spectra of extended objects correctly. From Fabry Perot spectrometers and Fourier transform interferometers one gets information about the PSF, but the conditions are often not constant enough for the PSF to stay stable over a whole wavelength range scan. Integral field spectroscopy solves these problems by delivering the 2-d spatial information for the whole wavelength range in one integration.\cite{mengel00} Observations with integral field spectrometers are easier to correct for variations in the atmospheric transmission. This is especially important for the highly variable sky in the near infrared, where the atmosphere varies on the order of minutes. Together with the use of adaptive optics (AO), which provides images with a spatial resolution close to the diffraction limit of the telescope, the near infrared (NIR) is a good choice, since AO works better in IR than with visible light.

One scientific reasons why especially the NIR spectroscopy is of importance is for instance that many faint objects are redshifted. Due to the redshift many spectral lines which are used for the characterization of the object are shifted beyond 1 micron into the NIR. Another scientific motivation for spectroscopy in the NIR is that for instance nuclei of galaxies as well as star and planet forming regions are hidden behind dust, which is quite impenetrable for visible light but nearly transparent in the infrared.\cite{eisenhauer00}

How the SPIFFI integral field spectrograph and also the adaptive optics instrument MACAO is built up is described in the next section.

\section{The SINFONI Instrument at VLT}\label{spiffi}
SINFONI (\textbf{S}pectrograph for \textbf{IN}tegral \textbf{F}ield \textbf{O}bservations in the \textbf{N}ear \textbf{I}nfrared) is one of the four Cassegrain instruments at the \textbf{V}ery \textbf{L}arge \textbf{T}elescope (VLT) of the European Southern Observatory (ESO) at Paranal in Chile. SINFONI consists of the two subunits called SPIFFI (\textbf{SP}ectrometer for \textbf{I}nfrared \textbf{F}aint \textbf{F}ield \textbf{I}maging) and MACAO (\textbf{M}ultiple \textbf{A}pplication \textbf{C}urvature \textbf{A}daptive \textbf{O}ptics). SPIFFI is an integral field spectrograph for the near infrared. It provides a three dimensional (two spatial and one spectral) view of astronomical objects. MACAO is the adaptic optics system feeding SPIFFI with diffraction limited images. Together these form SINFONI which is mounted at the Cassegrain focus of the UT 4 (\textbf{U}nit \textbf{T}elescope 4), one of the four 8-meter class telescopes of the VLT. 

\begin{figure}[h]
	\begin{center}
		\includegraphics[width=7cm]{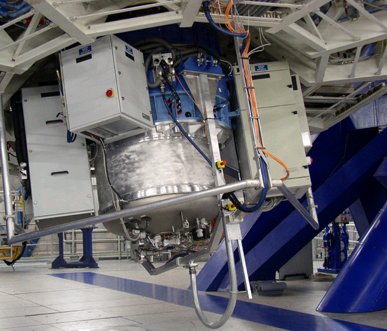}
		\caption[SINFONI at the Cassegrain focus of VLT UT4]{SINFONI at the Cassegrain focus of VLT UT4. The aluminum vessel is the cryostat of SPIFFI and the blue upper part is the housing of MACAO. The grey boxes are the electrical cabinets.}
		\label{fig:sinfoni}
	\end{center}
\end{figure}

The SINFONI field of view (FOV) is sliced by SPIFFI into 32 slitlets. The instrument allows for one of three different slitlet widths (25 milli-arcseconds (mas), 100 mas, or 250 mas) to be chosen. (The 25 mas pixelscale is in this thesis often called AO (adaptive optics) pixelscale, while the 100 mas and 250 mas pixelscale are often called seeing-limited pixelscales.) The resulting field-of-view is 0.8"x 0.8", 3.2"x 3.2" and 8"x 8" respectively. The length of each slitlet is imaged onto 64 pixels of the detector, a 2048 x 2048 pixel Hawaii 2RG. This results in spectra over 2048 pixels in a 32x64 pixel field-of-view in a wavelength range between $\mathrm{1.1 \ \mu m}$ to $\mathrm{2.45 \ \mu m}$ with a moderate resolving power of $\mathrm{R=1500}$ to $\mathrm{R=4500}$ depending on the observed wavelength. The SPIFFI spectrograph operates in the four NIR bands J, H, K and a combined H+K band.\cite{eisenhauer03} \cite{bonnet04}

In the following sections, the single optical components of SPIFFI and their modes of operation are described. Prior to this, there is a short section about the SINFONI adaptive optics module MACAO.

\subsection{MACAO - Optical Layout}\label{sec:macao}

The SINFONI adaptive optics module is an adapted clone of MACAO, which was initially developed for the VLT interferometer. MACAO can be operated with a natural guide star (NGS) or with the artificial sodium laser guide star (LGS) of UT 4. A bright wavefront reference near the line of sight of the science target is required for the adaptive optics (AO) corrections. In regions on sky where there is a low coverage of NGS the LGS is used as a wavefront reference.\cite{eisenhauer03}

MACAO is fed by the telescope input f/13.4 beam and it images the telescope pupil (VLT secondary mirror) onto a 60-element bi-morph deformable mirror (DM). There the beam is corrected for atmospheric distortions, reflected and re-imaged by further mirrors to the SPIFFI f/17.1 focus, which is located right after the infrared dichroic. The dichroic serves as the entrance window for the science beam ($\mathrm{1.05 \mu m}$ to $\mathrm{2.45 \ \mu m}$) into the SPIFFI cryostat. The visible light (450 to 950 nm) is reflected towards the wavefront sensor in MACAO. From the total 1' x 2' FOV the field selector picks up the reference star in an intermediate image plane and images it onto the membrane mirror, which is excited at a frequency of 2.1 kHz. The beam is collimated onto a lenslet array made of 60 lenses sitting in the pupil. From here the pupil, which is now divided into 60 sub-apertures, is fed to optical fibers, each ending at an avalanche photo diode. Through the vibration of the membrane mirror, the intra- and extra-pupil areas are imaged on the lenslet array. The curvature signal of the incoming wavefront is provided by the relative flux differences between the intra- and extra-pupil positions. This curvature signal is applied to the DM in real time with an update rate of 500 Hz.

For the calibration of MACAO and also of SPIFFI, a calibration unit is mounted in MACAO. It consists out of an integrating sphere with a halogen lamp and a set of spectral calibration lamps (Argon, Neon, Krypton, Xenon) to provide an uniformly illuminated field. A single mode fiber, used for sub-diffraction limited point source calibrations, is attached to the integrating sphere. For flat field calibrations, a folding mirror looking at the exit pupil of the integrating sphere is moved into the FOV and for point source calibrations the end of the fiber is moved into the center FOV.\cite{bonnet03} (For an schematic overview of the MACAO optical components see figure \ref{fig:scheme_macao}.)

After the science beam passed the MACAO unit in SINFONI it enters the SPIFFI instrument. The optically relevant parts are described in the next section.

\subsection{SPIFFI - Optical Layout}\label{sec:spiffi}
All the optical components of SPIFFI are located within an aluminum bath cryostat, which is cooled with liquid nitrogen to suppress the thermal background in the NIR. The liquid nitrogen tank is below the instrument plate.

\begin{figure}[htbp!]
	\center
	\subfloat[\label{fig:spiffi_pic}Optical components of SPIFFI. The camera in the image is the old one for the 1k x 1k detector.\cite{eisenhauer03}]
	{\includegraphics[height=6cm]{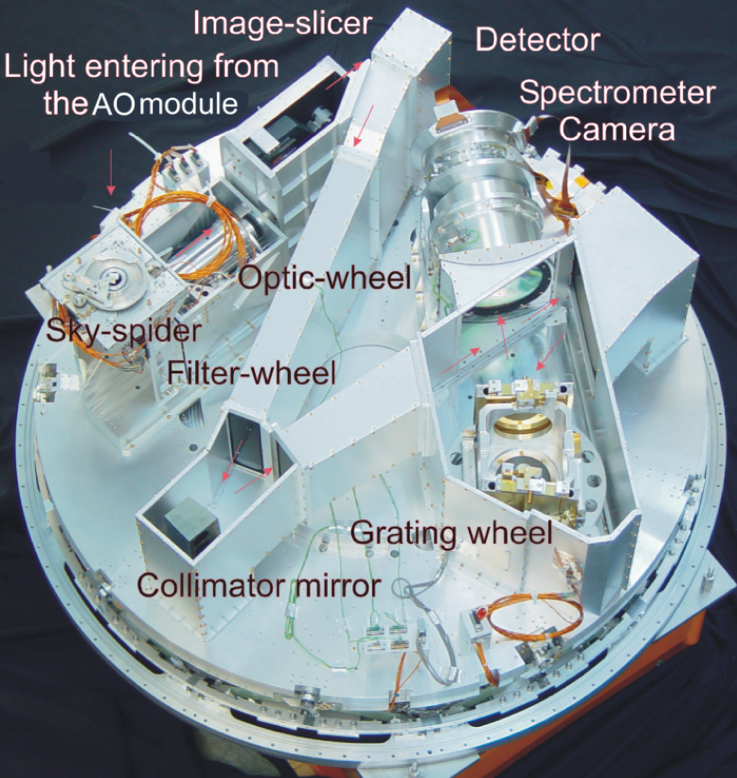}}
	\quad
	\subfloat[\label{fig:spiffi_scheme}Schematic drawing of the optical components in SPIFFI (top view). The light enters the instrument from the top (The sky spider is shown 90\degree \ rotated, the camera is the old one).\cite{eisenhauer00}]
	{\includegraphics[height=6cm]{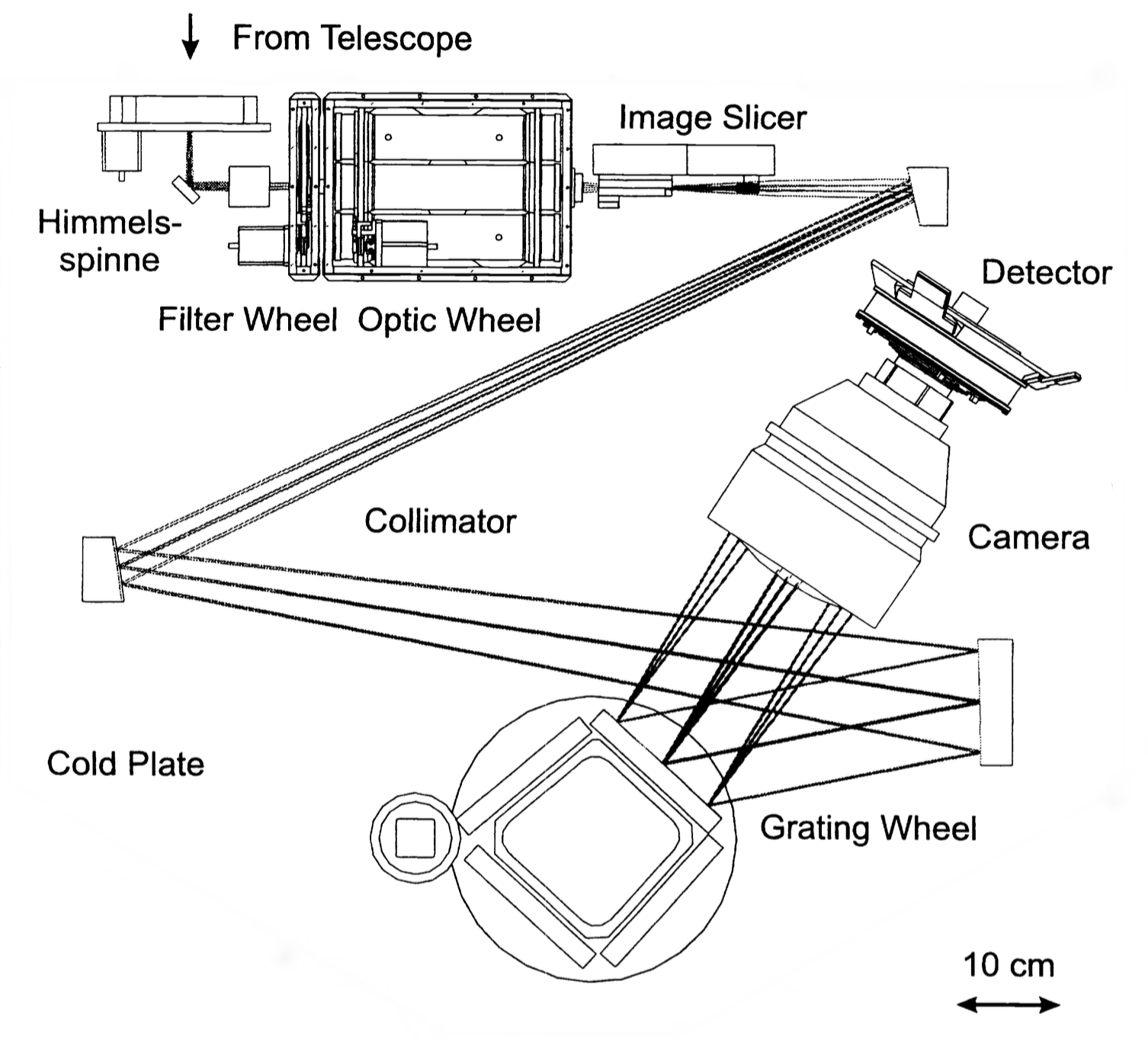}}
	\caption{Overview of the SPIFFI optics}
	\label{fig:overview}
\end{figure}

In the following the optical components of SPIFFI are described in detail. The order described is in the direction the light enters the instrument to the detector, which sits at the end of the beam.\cite{eisenhauer03} Figure \ref{fig:overview} shows on the left the instrument plate of SPIFFI while the instrument was assembled. Some mechanical parts such as the stiffening structure are missing as well as two gratings. The camera is the old one with the old detector that was exchanged in 2005. Likewise in the schematic overview on the right the old camera is drawn.

\subsubsection{Sky-Spider}
After the light from the telescope passes the AO unit of SINFONI, the NIR science beam enters the SPIFFI cryostat through the dichroic entrance window. It passes several cold baffles of the radiation shields and enters the first opto-mechanical element of SPIFFI, which is the sky-spider. The sky-spider is positioned in the entrance focal plane of SPIFFI and makes it possible to observe simultaneously a science target and the blank sky. One of three different fields with separations of 15", 30" and 45" from the SPIFFI field center may be chosen. The sky from the chosen field is redirected by two plane mirrors (the hypotenuses of prisms for an accurate alignment) onto a corner of the FOV, which is afterwards imaged on the image slicer.\cite{eisenhauer03}

Note: Due to the spatially varying and complex instrumental line profiles the sky-spider is not used for science observations.

\subsubsection{Pre-optics}

The purpose of the pre-optics is to re-image the focal plane after the adaptive optics onto the image slicer. It is made up of four major parts: a fixed collimator, a movable filter-wheel, a fixed cold stop, and a movable optics wheel. The design of the pre-optics is telecentric. In the pre-optics all lenses are spherical and made out of Barium-fluoride and IRG2.\cite{eisenhauer03}
\begin{itemize}
	\item \textbf{The Pre-optics Collimator} consists out of a double lens unit and a $45 \degree$ mirror, so that the light coming from the sky spider on top is afterwards parallel to the SPIFFI instrument plate. The pre-optics collimator reimages the entrance pupil onto a cold stop at the entrance of the optics wheel.
	\item \textbf{The Filter Wheel} is motorized and has 18 positions for filters and baffles. It houses the four band-pass filters for suppressing unwanted diffraction orders of the diffraction grating as well as open and closed positions. The filters cover the atmospheric transmission bands J ($1.1 - 1.4 \ \mu m$) , H ($1.45 - 1.85 \ \mu m$) and K ($1.95 - 2.45 \ \mu m$) and the combined bands H and K ($1.45 - 2.45 \ \mu m$). It is located directly in front of the cold stop at the pupil position.
	\item \textbf{The Cold Stop} is at the entrance of the optics wheel. It has a central obscuration for the suppression of thermal background radiation from outside of the telescope pupil. It is slightly undersized (smaller outer diameter, larger inner diameter) to account for tolerances in the f-number of the SPIFFI input beam, alignment, and the diffraction from the limited FOV.
	\item \textbf{The Optics Wheel} is also motorized and provides three different magnifications (17.8 x, 4.45 x, 1.78 x). The equivalent pixelscales are $\mathrm{25.0 \pm 1.3}$ mas, $\mathrm{100 \pm 5}$ mas and $\mathrm{250 \pm 13}$ mas per slitlet. For an accurate alignment of SPIFFI's optical axis with MACAO and the Telescope there is an additional pupil imaging objective. Each objective consists of two or three lenses in a tube. The different objectives image the field and the pupil respectively onto the image slicer. 
\end{itemize}

\subsubsection{Image Slicer}\label{sec:image_slicer}
\begin{figure}[htbp!]
	\center
	\subfloat[\label{fig:image_slicer}The large slicer in the background and the small slicer in the foreground]
	{\includegraphics[height=3.6cm]{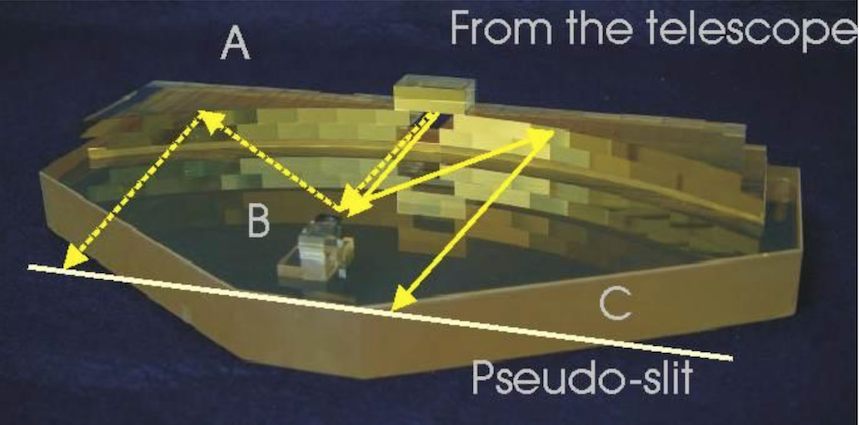}}
	\quad
	\subfloat[\label{fig:small_slicer}The small slicer]
	{\includegraphics[height=3.6cm]{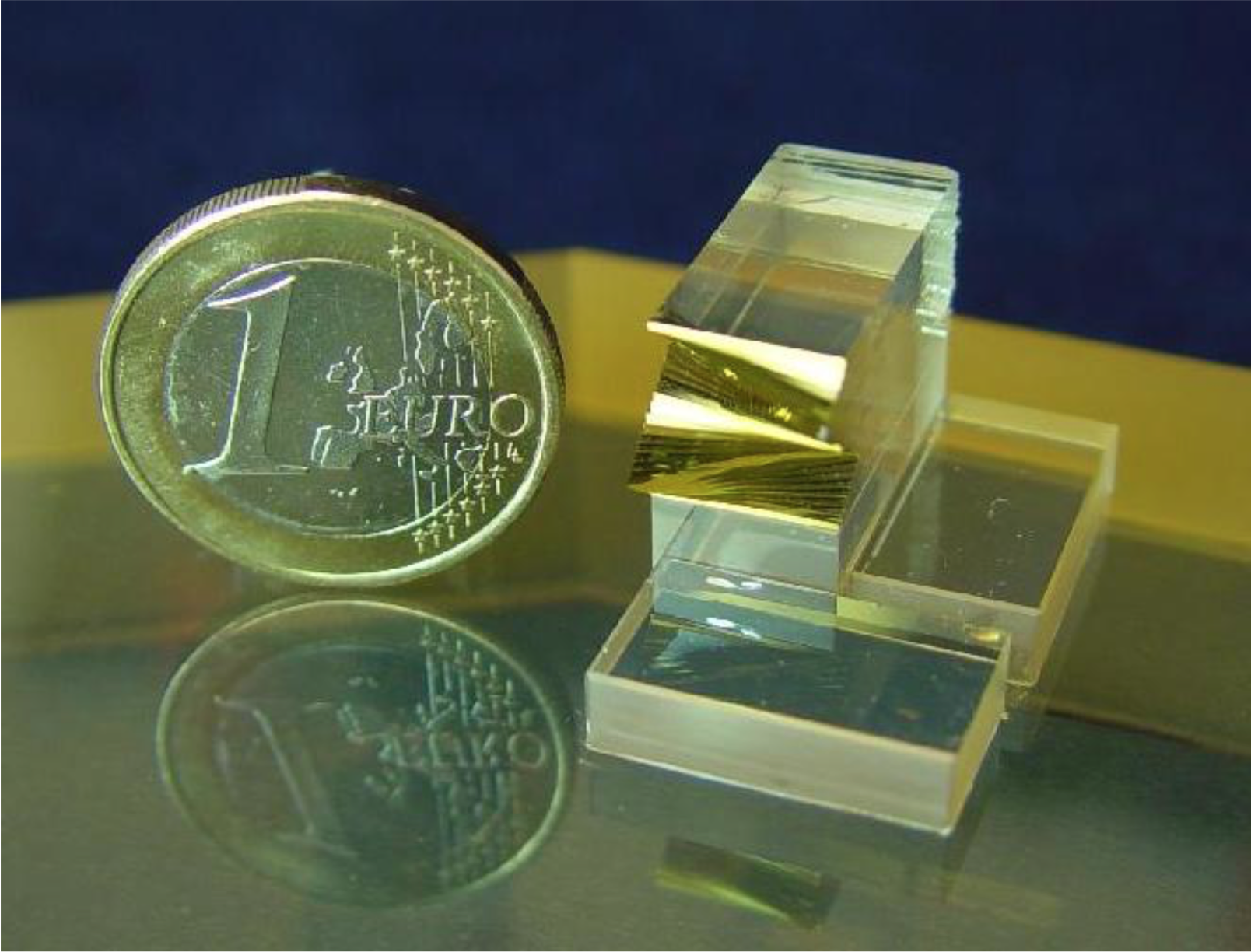}}
	\quad
	\subfloat[\label{fig:pattern}Illuminated pseudo slit]
	{\includegraphics[height=3.6cm]{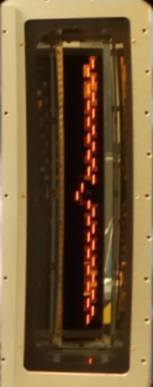}}
	\caption[Image slicer and pseudo long slit]{The parts of the image slicer in SPIFFI reproduced from \cite{iserlohe04} and the resulting brick wall pattern of the pseudo long slit.}
	\label{fig:slicer}
\end{figure}

The image slicer is the heart of the SPIFFI instrument. It consists of two optically relevant components, the so-called small slicer and the large slicer. Coming from the pre-optics the beam is focused on the small slicer. This is a stack of 32 plane mirrors that cuts the two-dimensional image into slitlets and redirects light from different field points into different directions. This separates the different slitlets from each other. At the location where the 32 slitlets are completely separated in the long direction, there is the large slicer. Because the beams diverge towards the large slicer, in the case that the slitlets would be aligned along a single layer, neighboring slitlets would slightly overlap at the edges. To avoid this, the mirrors of the large slicer are arranged in a brick wall pattern. As a result, the pseudo long slit consists of two dashed long slits. The large slicer realigns the pupils of the 32 different beams coming from the small slicer again, meaning, that the beams of the different slitlets are parallel at the exit of the image slicer and a pseudo long slit is formed. The whole slicer is manufactured out of monolithic Zerodur, while the mirrors are coated with a gold layer.\cite{eisenhauer00,tecza03} Figure \ref{fig:slicer} shows two images of the slicer and the resulting pseudo long slit.

\subsubsection{Spectrometer Collimator}
The SPIFFI spectrometer consists of three optical subunits. The collimator mirrors, the gratings, and the camera. A large part of the work for this thesis was focused on the collimator mirrors.

After the image has been sliced and arranged into a pseudo slit, it enters the spectrometer. A classical Three Mirror Anastigmat (TMA) collimates the light onto the diffraction gratings. The TMA consists of one concave spherical mirror (M1), an off-axis oblate elliptical convex mirror (M2) and an off-axis oblate elliptical concave mirror (M3). The three mirrors are diamond-turned from aluminum and gold coated for higher reflectivity. The benefit of this TMA is that it allows a large field of $\mathrm{\sim 6\degree}$ (corresponding to a pseudo slit length of 307 mm), while its dimensions stay quite compact. With this design the distortion along the slit is less than 0.2 \%, while the position of the exit pupil of the spectrometer varies for the different field points on the long slit by only 1 \%. The focal length of the collimator is 2886.5 mm.\cite{eisenhauer00} (For a more precise description of the mirrors see chapter \ref{ch:chapter_mirrors}.)

\subsubsection{Diffraction Gratings}
After the light passes the collimator mirrors it reaches the diffraction gratings. There are four plane gratings matched to the band pass filters mounted on a wheel, allowing the choice of one of the four gratings. The identical blanks of the gratings are 160 mm x 140 mm wide and 20 mm thick with seven 15mm diameter light-weighting holes drilled through the grating in each direction. They are made from 6061 aluminum and are electroless nickel plated with a high content of phosphorus to get an amorphous nickel layer. In order to get a flat surface, the two large faces of the grating blank were polished. Afterwards the blank was coated with a thick gold layer, in which the grating structure is ruled. All four gratings are direct-ruled and blazed to the center of the respective pass-band. The design resolution of the three gratings for J-, H-, and K-band is $\mathrm{\sim  4000}$ and $\mathrm{\sim 2000}$ for the H+K-band grating.\cite{eisenhauer00} The position sensitivity of the grating wheel is designed that the location of spectral lines on the detector is accurate to within 1/5 of a pixel.\cite{eisenhauer03} The J-, H-, and K-band gratings are operated in second order and the H+K-band grating in first order.

\subsubsection{Camera}
When SPIFFI was first commissioned it was equipped with a 1k x 1k pixel Hawaii detector and the associated camera.\cite{eisenhauer03} Already during the construction of the instrument a later upgrade to a 2k detector was forseen. The new camera, developed by ASTRON/NOVA\cite{kroes08}, has been installed in 2005 and consists of 5 lenses and a folding mirror. With its f/2.8 optics it is used to re-image the spectra on the 2048 x 2048 pixel Hawaii 2RG detector.

\subsection{Fields of Research}\label{fields_of_research}

The design of SPIFFI is optimized to minimize the background noise and to maximize the throughput.\cite{mengel00} Since SPIFFI can be operated in different observation modes (pixelscales) which show different sensitivities, the possible science applications are various. A short overview is given here. For an entire concrete and latest enumeration what kind of science is done with SINFONI see \lq ESO Telescope Bibliography\rq \ \cite{esotelbib}.

\subsubsection{Science Highlights with SPIFFI}\label{science}

\begin{itemize}
	
	\item \textbf{Galactic Center:} One of the reasons SPIFFI has been built was for observations and studies of the Galactic Center. Young B stars in close orbits around the central black hole $\mathrm{Sgr A^{*}}$, the so called \lq S-stars\rq \ have been discovered as well as infra-red flares of $\mathrm{Sgr A^{*}}$.\cite{eisenhauer05,gillessen06}. With the knowledge of the orbits of the S-stars from long term monitoring with SPIFFI, the mass of $\mathrm{Sgr A^{*}}$ and its distance could be determined.\cite{gillessen09} 
	
	\noindent The Galactic center with its supermassive black hole is an excellent target to test general relativity in the strong-field limit, which is currently done by the peri-center passage of the star S2.\cite{gillessen09a,genzel10} 
	
	\noindent Also by long-term monitoring with SINFONI, the gas cloud G2, which falls towards the black hole was discovered.\cite{gillessen13, gillessen13a,gillessen14,pfuhl15}
	
	\item \textbf{High-Redshift Galaxies:} With SINFONI a lot of research is done in the field of the formation and evolution of high-redshift galaxies.\cite{genzel06} In the SINS survey with SINFONI, a large sample of star forming galaxies at z $\mathrm{\sim 1-3}$ have been observed spatially resolved in order to get information about the gas kinematics, the star formation distribution, and the ionized gas properties of these galaxies.
	
	\noindent The MASSIV survey with SINFONI provided empirical evidence that the mass assembly of early massive galaxies with a majority of rotating, turbulent and gas rich disks is dominated by a smoother accretion via cold gas streams along the cosmic web and minor mergers.\cite{epinat09, epinat12}
	
	\noindent Metallicity gradient measurements of high redshift galaxies have also been done with SINFONI Some isolated galaxies showed an oxygen abundance towards the center. An explanation could be that the central gas has been diluted by inflowing primordial gas.\cite{cresci10}
	
	\item \textbf{Active Galactic Nuclei:} SINFONI is also used to study Active Galactic Nuclei (AGN). It was shown that in the nuclear regions of the observed objectes there were recent short lived starbursts in the last 10 -300 Myr that are no longer active. The intense starbursts seem to be Eddington limited, implying multiple short bursts. A time delay of 50-100 Myr between the onset of the star formation and the fueling of the central black hole was found.\cite{davies07}
	
	\noindent Furthermore with SINFONI investigations on AGN outflows are done. It was found that there is a correlation between the outflow velocity and the molecular gas mass. It indicates that the accumulation of gas around the AGN results in a more collimated outflow with higher velocities.\cite{muellersanchez11}
	
\end{itemize}

After almost 20 years of successful operation of SPIFFI the SINFONI instrument will be decommissioned in 2020. It will be replaced by the ERIS instrument, in which SPIFFI will be re-used. In the next section the ERIS instrument is explained briefly, and the scientific goals of ERIS are listed.

\section{ERIS - the 2\superscript{nd} Generation Instrument at the VLT}\label{sec:eris}
ERIS (\textbf{E}nhanced \textbf{R}esolution \textbf{I}mager and \textbf{S}pectrograph) is the next-generation instrument for the Cassegrain focus of VLT UT 4. It will be an NIR-imager ($\mathrm{1 - 5 \ \mu m}$) and spectrograph ($\mathrm{1 - 2.5 \ \mu m}$), taking advantage of the new deformable secondary mirror and the four-laser guide star facility. ERIS will consist of a calibration unit, the AO module, the NIR camera system NIX, and SPIFFI, which will be upgraded and integrated as the Integral Field Unit (IFU) subsystem SPIFFIER into ERIS. SPIFFIER will be on the main optical path of the telescope, while NIX will receive its light by a selector mirror. NIX will provide diffraction limited imaging, sparse aperture masking and pupil plane coronagraphy. It will be a cryogenic instrument with a FOV of 27" x 27" in J to Ks bands and 55" x 55 " in J to M bands. NIX with its 2k x 2k detector will be cooled to 40K.\cite{riccardi16}

\noindent The components of the upgrade SPIFFI has to undergo before it becomes SPIFFIER are the following:\cite{amico12}\cite{george16}
\begin{itemize}
	\item The cryostat lid will be changed. The dichroic of SINFONI will be replaced by a simple entrance window.
	\item The pre-optics collimator will be changed, because the incoming beam will be f/13.6 directly from the telescope instead of the f/17.1 focus inside SINFONI. The new collimator will be a three lens design instead of a two lens system.
	\item The J-, H-, and K-band filter will be replaced by ones with higher transmission. $\rightarrow$ Done in the upgrade covered in this thesis.
	\item All lenses in the optics wheel will be replaced by ones with better J-band transmission anti reflective coating. $\rightarrow$ Done in the upgrade covered in this thesis.
	\item All three spectrometer mirrors will be replaced by new ones. $\rightarrow$ Done in the upgrade covered in this thesis.
	\item Instead of the H+K grating, a high resolution dispersion grating (R$\mathrm{\sim 8000}$ for J - K) will be installed.
	\item The grating drive will be replaced by a new one.
	\item All motors (at filter wheel, optics wheel, grating wheel) will be replaced from Berger-Lahr motors to Pythron step motors.
	\item The detector will be exchanged by a new 2k x 2k Hawaii 2RG (HgCdTe) from Teledyne. The goal is to improve detector cosmetics, QE, and persistence performance.
	\item The electronics of SPIFFI will be refurbished.
\end{itemize}
This thesis deals with the early upgrade of SPIFFI in January 2016. The final upgrade to SPIFFIER will occur shortly before the integration into ERIS. First light of ERIS will by early 2020.

The main scientific goals of ERIS benefit from the close to diffraction limited (8m) imaging and spectroscopy in the NIR and IR. Here is a list of potential science topics for ERIS (from the ERIS science report):
\begin{itemize}
	\item \textbf{Disk Science} - Observations of circumstellar disks of young forming stars to place fundamental constraints on the initial conditions of planet formation as well as debris disks around mature main sequence stars to trace the dynamical evolution of planetary systems.
	\item \textbf{Exoplanets} - AO assisted observations in the infrared will provide a stable image quality with Strehl ratios close to 100\%. Furthermore spectral energy distributions can be taken to constrain the temperature of planetary-mass targets and the surface gravity. $\rightarrow$ low-resolution spectroscopy with SPIFFIER 
	\item \textbf{Starburst Clusters and the Initial Mass Function} - The initial mass function (IMF) is the main observable quantity in stellar populations - which can be resolved. The explanation of its origin is a key problem of modern astronomy. The correlation of observed variations of the IMF with the conditions in star forming regions is a main astronomy research topic. Ever more more extreme regimes of star formation are considered. From the densest stellar systems to the lowest masses permitted to fragment in the interstellar medium.
	\item \textbf{Solar System} - Observations of planets and their atmospheres, their satellites and small bodies in the solar system like asteroids provide clues on the formation and history of the solar system and the evolution of other planetary systems. $\rightarrow$ medium-resolution spectroscopy with SPIFFIER
	\item \textbf{Galactic Center} - It is of great interest because it houses the closest supermassive black hole, $Sgr A^{*}$. The determination of stellar orbits is needed for distance and mass measurements of the black hole. Just a few science cases for the Galactic Center include taking a closer look into young star formation in the central pc, measuring cluster structures at larger distances from $Sgr A^{*}$, and constraining the physics of the accretion flow, and measuring the proper motion of ISM features to understand the central outflow. $\rightarrow$ spectroscopy with SPIFFIER in H- and K-band with 25 mas pixelscale 
	\item \textbf{Centers of Nearby Galaxies} - Most of the energy of a galaxy is produced in its center where a massive object, like a supermassive black hole, is present. These massive objects should also be present in the nuclei of quiescent galaxies as remnants of a previous active galaxy nucleus (AGN).
	\item \textbf{Distant High Redshift Galaxies} - Questions on the nature of dark energy, dark matter, and the evolution and formation of galaxies and of black holes in AGNs can be tackled by studying the structure, contents, and kinematics of distant galaxies. The transformation of galaxies into the Hubble sequence of morphologies can be studied as well as the bulge and disk evolution. $\rightarrow$ spectroscopy with SPIFFIER in 25 mas pixelscale
\end{itemize}

Before SPIFFI will be re-used in the ERIS instrument in 2020, an early upgrade of the instrument was done in January 2016. This upgrade is the topic of the next chapter.

\chapter{SPIFFI Upgrade}\label{ch:chapter_upgrade}
In this section the upgrade of the SPIFFI instrument in January 2016 is described. This early upgrade belongs to the ERIS project described in section \ref{sec:eris}. It was completed such that ongoing science projects could take advantage of the advanced performance of the instrument until ERIS will be commissioned in the year 2020.

\section{Goals of the Upgrade}
There are two main reasons for an upgrade of the below listed optical and mechanical components in SPIFFI. The first reason pertains mostly to the pre-optics and as later discovered also to the grating wheel. It is simply a maintenance measure to reduce the probability of mechanical failures. This concerns the old worn toothed wheels, and old lenses for which there is a danger that the Anti-Reflection (AR) coatings degrade and peel off the lenses when the instrument is cryogenically cycled. This degradation induces a large amount of scattered light and the throughput will decrease dramatically from it which leads to the second reason for the upgrade.

Besides the maintenance there are also scientific reasons to upgrade the instrument: In order to get a increased throughput for observations of faint objects, sharper images to distinguish between close objects, and a higher spectral resolution relating to sharper spectral lines. These reasons and the possibility to improve SPIFFI with respect to these are tightly connected to the improvement in manufacturing processes since SPIFFI was originally built. These improvements are for example the ability to manufacture high transmission filters with hundreds of coating layers to get a constant high transmission over a band-pass, better anti-reflection coatings for lenses, and a higher accuracy in machine control for the manufacturing of reflective mirror surfaces. Taken together, these upgrades lead to a better and more reliably working instrument, and finally to a benefit for astronomers and science. In the next section the exchanged components are listed with the reason for the exchange and the problems that came up during the upgrade.

\section{Upgraded Components of SPIFFI}
This section describes in detail the SPIFFI components exchanged during the 2016 upgrade, including why they were exchanged, practical results of the exchange, and a brief statement about what the expected benefit of the new components is. A detailed analysis of the performance enhancement is presented in chapter \ref{ch:chapter_performance}

\subsection{Exchange of the Entrance Baffel}
At the entrance focal plane of SPIFFI there is a quadratic baffle positioned inside the sky-spider housing. This baffle is with 5.61 mm x 5.61 mm only $\mathrm{300 \ \mu m}$ oversized compared to the the size of the FOV in the largest pixelscale with 8" corresponding to a 5.3 mm x 5.3 mm field at the SPIFFI instrument focus. This baffle was shown to vignette the edges of the FOV in the largest pixelscale in some observations due to the tolerance on the repeatability of the optics wheel position. For this reason a new baffle with a larger aperture of 7.5 mm x 7.5 mm was installed. It is strongly oversized, so that it cannot vignette the outer parts of the small slicer anymore. A negative effect of oversizing this baffle was recognized while commissioning. It is now harder to find the correct centered position of the optics-wheel, since by moving the optics wheel the field can be moved but a sharp vignetting effect of the focal plane baffle cannot be seen anymore.

\subsection{Exchange of the Pre-optics Collimator}
During an intervention in September 2013, which was done because of several motor failures, it was found that one lens of the pre-optics wheel in the 25 mas pixelscale tube had degraded AR coatings. This led to a loss in transmission and an increase in stray light. The process of degradation happened when the instrument was warmed up at the beginning of the intervention. The coating of this single lens was then removed by polishing. In order to prevent the risk of the degradation of the coatings on all pre-optics BaF\subscript{2} lenses, a completely new collimator and optics wheel tubes equipped with new lenses and AR coatings were installed. The new AR coatings have a higher transmission in J-band compared to the old ones.

It was expected that no alignment of the pre-optics collimator would be necessary since the alignment of the entrance pupil can be done with the M5 mirror in MACAO. But the angle of the beam between SPIFFI and MACAO was misaligned by 6.7', too much to compensate it with M5. The result was a rotated and offset field. The old collimator was compared to the new one during a warm-phase using a test setup mimicking the optical input from MACAO with a grid laser. To correct for the mismatch, the sleeve around the alignment pin serving to fix the rotational position of the collimator was customized such that the angles of the new and old collimator matched.

\subsection{Exchange of the Filter Wheel}\label{sec:filters}
\begin{figure}[htbp!]
	\begin{center}
		\resizebox{1.0\textwidth}{!}{
			\includegraphics[trim={2cm 0 1.8cm 0},clip=true,width=1.0\textwidth]{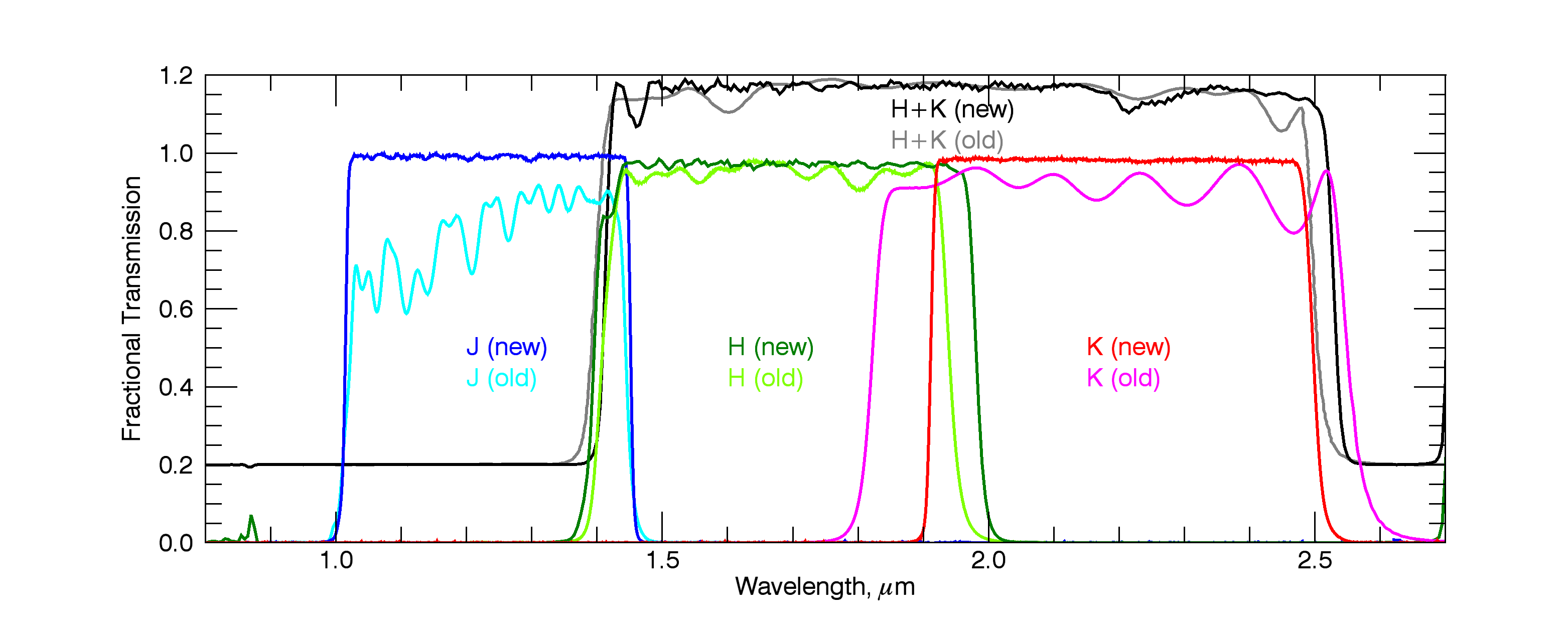}
		}
		\caption[Filter transmission curves]{Plot of the new filter transmission curves and the old filter curves. The H+K filters have been offset by +0.2 units for plot readability. Starting with the January 2016 upgrade, the default filters in use are J (new), H (new), K (new), and H+K (old).\cite{george16}}
		\label{fig:filters}
	\end{center}
\end{figure}

The band pass filters for J-, H- and K-band were exchanged by new filters with a higher throughput and also a more constant throughput function over wavelength for the individual band pass ranges. The new H+K filters were manufactured on the wrong substrate and do not show an improvement in throughput compared to the old filters and are now used as spares in the filter wheel.
The original plan was to replace the whole filter-wheel with its chassis and attach the old Berger Lahr motor at the new filter-wheel unit. After opening the filter wheel housing it was found, that the toothed wheels were  in a very good shape and that there are some differences in the positioning of the Marquardt switch compared to the drawings. So it was decided to leave the old housing and wheel and put additionally the new filters into the free positions of the filter wheel. The throughput curves of the old and new filters can be seen in figure \ref{fig:filters}.\cite{george16}

The new filters have many more layers, and hence much smaller (but higher frequency) oscillations in the transmission, as well as an overall higher throughput. The out-of-band suppression between 0.8 and 2.7 $\mu$m is quite similar to the old filters, while the edges of the new band-pass filters are way sharper.

\subsection{Exchange of the Pre-Optics Wheel}
The exchange of the optics-wheel with the pre-optics cameras in it was intended as a preventive maintenance measure to prevent a potential mechanical failure and the degradation of the AR coatings on the BaF\subscript{2} lenses. Furthermore, in the 25 mas pixelscale tube the loss in throughput and increase of stray light due to the uncoated concave side of the first lens (the coatings were removed in 2013 after they degraded) is eliminated with completely new lenses with AR coatings. As for the filter wheel the original update plan was to exchange the whole pre-optics wheel unit. When the old optics-wheel was inspected, there were a few differences found in comparison to the new manufactured unit. A different wheel back bearing mount was installed and the arrangement of the four optics tubes were different. Furthermore, the weight inset in the 250 mas pixelscale tube was from a different material (copper instead of aluminum) and had a different size. The copper weight had a mass of more than three times the mass of the aluminum weight. This different weight solved the problem of the imbalance of the new optics-wheel that was already noticed in the laboratory and that led to a resonance at $\mathrm{\sim 11\degree/s}$ when the wheel was turned. It was then decided to re-use the old pre-optics wheel and balancing weight with the new optics tubes inserted.

\subsection{Exchange of M1 - M3 Collimator Mirrors}\label{sec:mirror_exchange}
The decision to exchange the collimator mirrors was based on the old spare mirrors available in the laboratory. These mirrors had been installed into SPIFFI when it was a Guest Instrument (GI) at VLT UT1. They had been manufactured together with the ones that were installed in SPIFFI before the upgrade. From the original pre- upgrade collimator mirrors that were installed in SPIFFI until January 2016 there were no measurements available before the upgrade. The pre-upgrade spare mirrors available in the lab, however, showed strong residual marks from the diamond turning process. These diamond turning marks were thought to be the reason for the complex disturbed shape of the spectral line profiles of the SPIFFI spectrometer (see section \ref{sec:lineprofiles}). The assumption was that the mirrors that were installed in SPIFFI were virtually identical in mirror shape, surface quality, and coating to the spare mirrors, with only small improvements to the diamond turning marks from the post-polishing process that was performed on the installed mirrors.

Because of the huge progress in machine controlling in the last ten years, it was possible to produce new diamond turned mirrors without residual diamond turning marks. From those the spectrometer would benefit in sharper and less distorted spectral line profiles. (For a detailed description and characterization of all mirrors including the pre- upgrade SPIFFI mirrors, see chapter \ref{ch:chapter_mirrors}. Note again, that the surface shape and quality of the pre- upgrade collimator mirrors was not available before the upgrade.)

The mirrors (and also for M2 and M3 their sockets) are aligned with pins to the SPIFFI instrument plate. In order to not to change the alignment of the mirrors (therefore resulting in a misalignment of the new mirrors), a position verification setup was built onto the SPIFFI instrument plate for the exchange procedure. This consisted of an integrating sphere supplying the spectrometer with an incoherent uniform illumination. (A laser with 632.8 nm was found to produce too many diffraction effects at the small image slicer.) The exit pupil of the integrating sphere was imaged 1:1 onto the small slicer by the use of an $\mathrm{f = 200 \ mm}$ lens. The resulting slitlet pattern was then reflected out of the spectrometer optical path after the light passed M3 and imaged by a further $\mathrm{f = 100 \ mm}$ lens onto a CMOS sensor. The optical design of this was chosen in a way, such that the image on the CMOS has the same size in pixels as on the SPIFFI Hawaii 2k detector. When a movement of the image on the CMOS appeared by exchanging a mirror, it is immediately clear that the mirror surface position must have changed.
When M2 was replaced, the pseudo-slit position in the test setup moved horizontally by 15 pixel. This corresponds to a movement of $\mathrm{\sim 1.25}$ mrad of M2 which was measured by shimming the mirror in a way that the image on the pseudo-slit is again on its nominal position on the CMOS sensor. The reason for this was the left alignment pin on the old M2 socket when looking from the backside of M2. It was offset by around 80 microns from its nominal position. As a result, the footprint on M3 moves by around 2 mm horizontally. Due to the significant surface deviation of M3 (see section \ref{sec:M3}) and the the rotation applied to it in order to correct for it (see section \ref{sec:fullcoll}) the image was shifted horizontally by 225 pixels which is compensated by new default positions of the grating wheel and vertically by 22 pixels for which the detector was adjusted laterally in its mount.

\subsection{Exchange of the Grating Wheel Drive}
Originally there was no plan to change anything on the grating wheel. But after the baffles covering the grating section were removed during the intervention for an inspection, lots of metallic wear debris and a cage of a broken ball bearing of the grating gear were found. So a spare grating gear drive and the spare toothed wheel on which the gratings sit were installed. For this the entire grating carousel had to be removed. To preserve the optical alignment during the exchange, an optical verification setup was installed in front of the grating wheel. A laser beam is reflected from one of the gratings, and the position of reflected spot is recorded before and after the exchange, preserving the angle of the grating carousel. The angle of the grating wheel with respect to the cryostat did not change within an uncertainty of 0.05 mrad.

\subsection{Detector Adjustment}\label{sec:detector_adjustment}
The SPIFFI detector was not exchanged in this upgrade. It will be replaced by a new one with better performance in the SPIFFIER upgrade. Despite this, the detector position had to be adjusted for three reasons. First, the dead spot close to the center of the detector array should be shifted from slitlet 15 which is right in the middle of the FOV into slitlet 1 at the edge of the FOV. Connected to this shift, slitlet 24 at the upper edge of the detector fell off by 11 to 12 pixels. Second, due to the turning center offset of M3 the image was shifted upwards (as seen from the instrument plate) by 22 pixels and the focus changed slightly (see section \ref{sec:M3}). Third, because of the disassembling and assembling of the stiffening structure supporting the instrument plate and the exchange of the complete pre-optics, the instrument plate may have a different flexure from before and thus the image position could have changed. To correct for the shift induced by M3 and to shift slitlet 24 again onto the detector, the detector was shifted laterally upwards (with respect to the instrument plate) by 600 $\mathrm{\mu m}$. The size of a detector pixel is 18 $\mu m$. Given the pixel shifts measured, the corresponding numbers are a 400 $\mathrm{\mu m}$ shift to correct for M3, and an additional 200 $\mathrm{\mu m}$ shift in order to move the hotspot and shift the slitlets, so that slitlet 24 does not fall off the detector anymore. On top of that the focus position of the spectrometer camera changed due to the different radius of M3. To correct for this the detector has to be shifted along the beam. This is very sensitive, because of the f/2.8 beam of the camera. It results in a depth of focus of $\Delta f = \pm 2 \lambda N^2$ with the F-number N and the wavelength $\lambda$. This is the amount of defocus introducing a $\mathrm{\lambda /4}$ wavefront error. \cite{mclean06} It results in a few tens of micron for the depth of focus in SPIFFI. To measure it, a procedure was carried out in which the filter wheel is moved to two distinct positions located symmetrically on either side of the default position to vignette a part of the pupil. If the detector were not in focus, an image on the detector would move depending on the vignetting position. If the detector is in focus, an image would appear in the same position regardless of the pupil vignetting.\cite{iserlohe05} After the first cooldown of SPIFFI after the upgrade the new focus position was by chance slightly better than before the upgrade (by around 10 micron), so the detector did not have to be adjusted in beam direction. In section \ref{sec:resolution} the effect of the better focus on the spectrometer resolution is discussed.\\

Since from the exchange of the collimator mirrors the largest change in the instrument performance on the spectral line profiles was expected, the next chapter describes the measurements that were done with the collimator mirrors in the laboratory.

\chapter{The Collimator Mirrors}\label{ch:chapter_mirrors}
The SPIFFI spectrometer collimator is a Three Mirror Anastigmat (TMA). It collimates the light from the pseudo slit ($\mathrm{\sim 6 \degree}$ field) onto the grating, where it is dispersed. The surface of the collimator mirrors have been subjected to investigations since the original construction of SPIFFI. The first set of single-point diamond turned mirrors showed significant residual turning marks despite the fact that they were post-polished after manufacturing. Figure 6 of Eisenhauer et al 2003\cite{eisenhauer03} shows these turning marks on the collimator mirrors with a characteristic spatial frequency of 1 per mm and a depth of 200 nm PV. These more or less regular turning marks diffracted the NIR light in the instrument and caused a broadening and distortion of the instrumental spectral line profiles which limited the performance of SPIFFI. Since these measurements in 2003, a new set of mirrors were installed with a more aggressive post-polishing. The amplitude of the residual diamond turning marks was reduced significantly by the new post-polishing process, but the resulting surface showed additional deformations from the polishing process. This set of mirrors remained in the SPIFFI instrument from 2004 until 2016. Unfortunately from this set of mirrors no measurements were available, so there was no possibility to compare those mirrors to the ones that were manufactured in 2015 and which were installed in the upgrade 2016. It was known, that two off the collimator mirrors had been post-polished, while one of them was re-diamond turned.\cite{iserlohe04} Which one of the mirrors was re-machined is unclear; presumably due to the convex surface M2. In the surface plots presented later in this chapter, none of the mirrors appear to have been diamond turned without polishing, so it is possible that all three of the mirrors were post-polished. The second set of mirrors could only be measured after they were removed from SPIFFI during the upgrade in 2016. It was only then possible to compare their surfaces to the new mirrors that have been installed during the upgrade in 2016 (which had been measured before the upgrade). From the exchange of the collimator mirrors it was expected that the spectral line profiles of the instrument, which are non-ideal with shoulders, distinct side-peaks and show effects of out-smearing, will improve and become sharper and closer to the ideal shape.

Before the SPIFFI collimator mirrors themselves are investigated, in the next section the design of a three mirror anastigmat is briefly explained as well as the special case of the SPIFFI TMA.

\section{Three Mirror Anastigmat}
While two mirror imaging configurations can only practically correct for a maximum of two aberrations (usually spherical aberration and coma), the design of a TMA allows correction of four third-order aberrations, namely spherical aberration, coma, astigmatism and field curvature.\cite{korsch72} As a result, TMAs are used where a wide field is needed. The TMA in SPIFFI is used off-axis both in aperture and field. The advantages of a TMA used off-axis are twofold: the system is unobscured in comparison to an on-axis TMA, and the design is compact without auxiliary folding mirrors. Compared to an on-axis use, the shielding from stray light is improved as well as the accessibility of the focal plane and pupil. Additionally, an off-axis TMA can handle faster beams.\cite{cook79} Because there are eight degrees of freedom available in the TMA design, the system is capable of a high degree of correction. These degrees of freedom are the three curvatures and conic constants of the mirrors and the two separation distances of the three mirrors. The constraints are the system focal length, the location of the image surface and the four previously named third-order aberrations. So two degrees of freedom are left for higher order residuals.\cite{robb78}

\begin{figure}[htbp!]
	\begin{center}
		\includegraphics[width=0.8\textwidth]{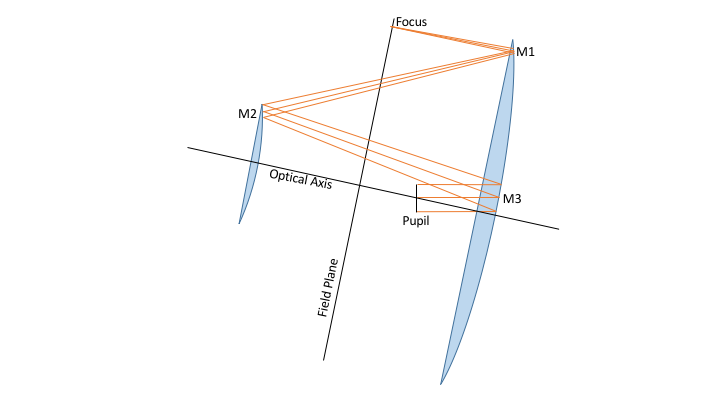}
		\caption{Scheme of the SPIFFI TMA}
		\label{fig:TMAscheme}
	\end{center}
\end{figure}

The SPIFFI collimator TMA (compare with figure \ref{fig:TMAscheme}) is designed symmetrically, which means that the minor axis of the oblate elliptical surfaces of M2 and M3 describes the optical axis of the system. The distance between M1 and M2 is identical to the distance between M2 and M3. The field is planar and perpendicular to the optical axis. The pupil as flat, but tilted with respect to the optical axis. This is an effect of operating the TMA off-axis. The smaller the angles in the TMA, the more perpendicular to the optical axis the pupil will be. A property that distinguishes the SPIFFI spectrometer TMA from traditional telescope TMAs, is that the field imaged by the SPIFFI TMA is the pseudo-slit, which has strongly differing dimensions in width and height. The field along the pseudo slit is with $\mathrm{\sim 6 \degree}$ in one dimension quite large, while perpendicular to the pseudo slit the field is more than one order of magnitude smaller. This especially constrains the design of the first mirror (M1) of the TMA. 

For a correct working TMA the surface of the mirrors is a crucial point. The next section shows therefore the technical aspects of the SPIFFI collimator mirrors.

\section{Technical Aspects}
The blanks of the three collimator mirrors are made of light-weighted thermally treated 5083 aluminum alloy. The same material is used for the sockets of the mirrors and for the coldplate in SPIFFI, resulting in an athermal design. The old collimator mirrors were manufactured in the years 2001 to 2003. The new mirrors were manufactured in 2015. In the decade between, there was a rapid improvement in machining control. The mirror surfaces are single point diamond turned. With this manufacturing method rotationally symmetric surface forms can be turned. Due to the improvement in machining control, the surface of the mirrors can be turned nowadays way more precise than over ten years ago.
The first way that the new mirrors differ from the ones that were in SPIFFI until 2016 are the coating layers. While on the old aluminum mirror blanks an electroless plated nickel-phosphor layer was deposited, on which afterwards the precise mirror surface was single point diamond turned, for the new mirrors the diamond turning is directly done on the aluminum blank. Since in this nickel layer a crystallization of the metal is not desired, the amount of phosphor is around $\mathrm{\sim 10\%}$ in order to get a thick amorphous thick layer. The nickel layer on the old mirrors made it possible to polish them after the diamond turning process. Since a post-polishing of the mirrors is not necessary anymore - because the machine control improved and so the specifications can be now fulfilled without post-polishing, on the new mirrors there is no nickel layer, which also eliminates the potential risk of bimetallic bending stresses when the mirrors are cooled to 80K. For a higher reflectivity the mirrors are coated with a gold layer. For the old mirrors on the polished nickel layer and for the new mirrors directly on the aluminum, there is a thin chromium binding layer that connects the final $\mathrm{2 - 3 \ \mu m}$ thick gold coating with the material below. The gold coating effects a high reflectance in the NIR and IR. Only on the new mirrors, on top of the gold coating there is a SiO\subscript{2} protection layer. This protection layer decreases the reflectivity and thus the throughput of the instrument slightly at short wavelengths. Bare gold shows a reflectivity of 98.5\% over all SPIFFI bandpasses, while the protected gold has a reflectivity of 96.5\% in J-band rising to 98.5\% in K-band where the reflectivity is identical to the unprotected gold.

The reflective surfaces of the aluminum mirrors should be manufactured with an average roughness of less than $\mathrm{5 nm}$ on spatial scales of $\mathrm{1 \mu m}$ to 1 mm. The centering accuracy is 1', and the cosmetic surface quality specification permits up to three surface imperfections with a maximum size of 0.16 mm. The surface form tolerance of all mirrors is identical for a circular region of 25 mm diameter. Namely a irregularities of 0.2 fringes corresponding to $\mathrm{0.1 \lambda}$ at $\mathrm{\lambda = 546.07 nm}$ resulting in $\mathrm{\sim 50 \ nm}$.

\begin{figure}[htbp!]
	\begin{center}
		\includegraphics[width=0.5\textwidth]{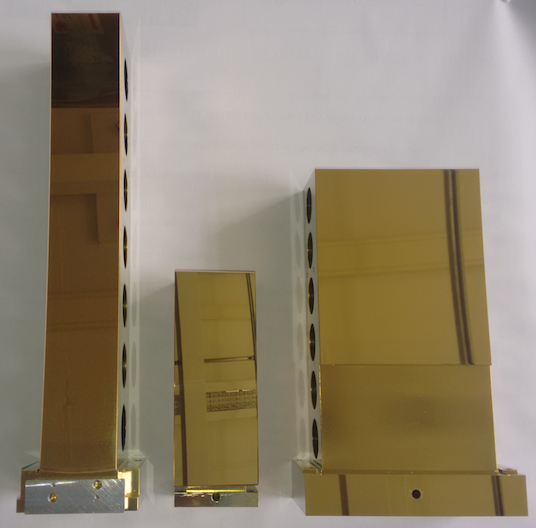}
		\caption[The three collimator mirrors]{The three mirrors of the SPIFFI spectrometer collimator. From left to right: M1, M2, M3}
		\label{fig:mirrors}
	\end{center}
\end{figure}

For the whole respective mirror, the maximum peak-to-valley surface deviation (sagitta error) should be within $\mathrm{2 \lambda}$ for M1, $\mathrm{1 \lambda}$ for M2 and $\mathrm{0.5 \lambda}$ for M3. Irregularities are tolerated on M2 and M3 up to $\mathrm{0.5 \lambda}$ and $\mathrm{1 \lambda}$ for M1.
Figure \ref{fig:mirrors} shows an image of the three collimator mirrors.

How the surface measurements on these mirrors were done is explained in the next section.

\section{Single Mirror Measurements}
For a properly working TMA two aspects are important. First, the the mirrors must be aligned accurately with respect to each other and second, the surfaces of the mirrors must be correct. The main challenge here is the alignment, done by alignment pins in the instrument plate, and the precise manufacturing of the mirror surfaces. The surface roughness has to be as small as possible to avoid stray light and the complex overall surface form has to fit to the constraints on it in order to remove all third order aberrations.

The goal of the surface measurements carried out for this thesis is to determine the overall surface form to an accuracy of less then a tenth of a micron and the surface roughness to approximately nanometer precession. To reach this, all surface measurements are done with an Twyman-Green interferometer that consists of both a CCD camera and a phase-shifter supplied by a He-Ne laser at 632.8 nm wavelength. The phase-shifting measurement is done via a movement of the reference surface in the interferometer by half a wavelength in total along the optical axis separated into five steps. The advantage of the five step method is that the first and fifth measurements should be identical, thus the method is self-checking. After a phase unwrapping one obtains a wavefront from the the measurement.

The single mirror measurements were carried out on an floating optical bench since the interferometer is very sensitive to vibrations. The optics table is located in a clean-tent in the laboratory to prevent dust from settling on the mirror surfaces. For all three single mirror measurements, lenses with the longest focal lengths and largest f-number beams possible were used in order to get a slow beam for an easier alignment with as low spherical aberrations as possible, while still covering the areas of the mirror which are illuminated in SPIFFI. In all single mirror measurements the collimated 6 mm diameter laser beam from the interferometer was sent through different planoconvex lenses to adapt the beam for the individual mirrors. To correct for aberrations induced by the lens a calibration exposure was taken for each mirror measurement. For this, a small spherical mirror was put into the beam after the focus of the lens, such that the focus of the beam was co-located with the center of the radius of curvature of the spherical mirror, retro-reflecting the beam back to the interferometer. The mirror used for this calibration has a maximum surface irregularity of $\mathrm{\lambda/4}$ at 633 nm. Also in all single mirror setups there is a large mirror used, for M2 a spherical mirror for correcting the induced astigmatism and for M1 and M3 a flat folding mirror. The folding mirror used is a the calibration mirror of the interferometer and the spherical mirror is a f = 1212.6 mm mirror with 20 cm diameter and $\mathrm{\lambda / 4}$ surface quality. Both mirrors are not calibrated, but the surface irregularities should not play a role for them, because in all three setups they are fare away from focus. The radius of the large spherical mirror was measured in order not to induce errors on the M2 surface measurement originating from the sagitta error of the spherical mirror. 

Four sets of collimator mirrors have been tested. These are the first mirrors that were in SPIFFI with strong residual diamond turning marks (called: old-spare), the mirrors - that have been in SPIFFI until 2016 (called: old-in-SPIFFI) and two sets of the new manufactured mirrors from 2015 (called: name of mirror + serial number). Since the decision to exchange the collimator mirrors was made on the surface quality of the old-spare mirror set, the surfaces of these mirrors are also shown here. In all interferograms the defocus and tilt was subtracted since both are degrees of freedom that cannot be verified in the single mirror measurement. In principle one could measure the relative defocus and tilt between the different new and old mirrors, but in practice the measurements of the single mirror surfaces are extremely sensitive. Even with a very sturdy setup with alignment pins, the fringes move over time due to thermal expansions and contraction. Furthermore, the tilt of a mirror is not reproducible, meaning the following: When one mirror was placed on its socket, then the folding mirror was aligned in order to minimize the number of fringes on the interferogram. The mirror was then taken off its socket and then replaced in its nominal position on the socket. The result was a wavefront that was tilted by many fringes. The same effect was measured, when screwing the mirror down. The tilt of the mirrors is easily changed by the force of the mirror screws when they are tightened. In all single mirror measurements the tilt and the defocus are degrees of freedom and cannot be measured. Also the relative tilt and defocus of the old mirrors compared to the new ones cannot be measured easily because the setup is so sensitive. Due to this in the interferograms the tilt and defocus is subtracted. However, in the full collimator test setup tilts and defocus can be measured relative to the other sets of mirrors, since the focus will move lateral when relative tilts are induced by the exchange of one mirror and along the optical axis for a change in focal length. The tilt can also be measured by the mirror replacement setup described in section \ref{sec:mirror_exchange} where the image of the pseudo slit will move horizontally and vertically induced by tilts of the exchanged mirrors.

In the following sections, the setups of the three single mirror measurements are described. The results of the measurements are shown and discussed.

\subsection{M1}
\begin{figure}[htbp!]
	\begin{center}
		\resizebox{0.6\textwidth}{!}{
			\includegraphics[width=1\textwidth]{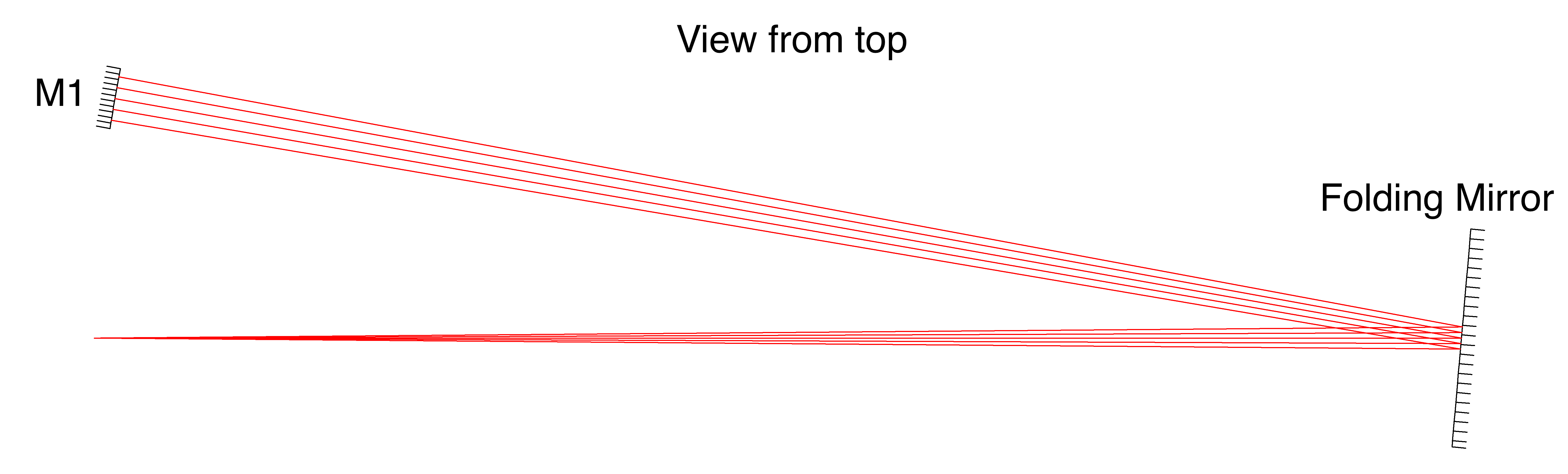}
		}
		\resizebox{0.6\textwidth}{!}{
			\includegraphics[width=1\textwidth]{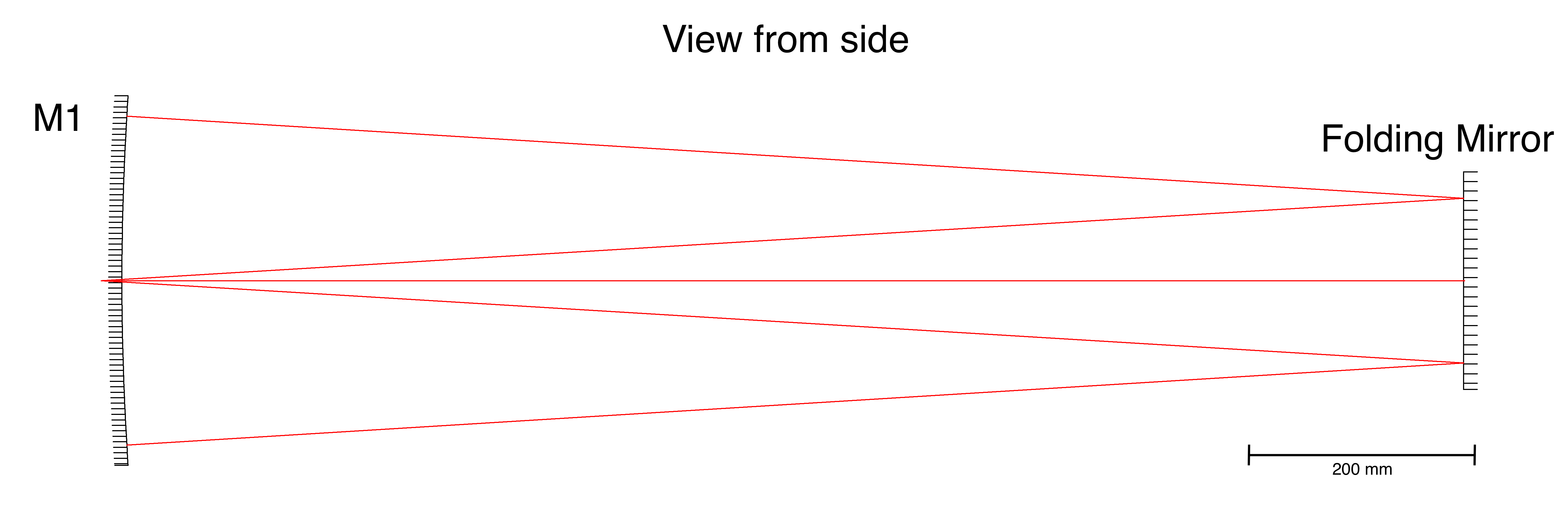}
		}
		\caption[M1 test setup]{The optical setup of the M1 single mirror test between focus position and M1. The interferometer is located left of the focus position. The collimated 6mm beam from the interferometer is bundled by a lens to the focus.}
		\label{fig:M1setup}
	\end{center}
\end{figure}

{\label{sec:M1}}
The surface of M1 is spherical concave, which makes it the easiest mirror surface to measure. The radius of the mirror is $\mathrm{R = 2599.58 \ mm}$. The reflective surface is 54 mm wide and 344 mm high. The height of M1 is correlated to the dimensions of the small slicer. Since it slices a quadratic field, the width of a slitlet must be the 32\superscript{nd} part of the height of one slitlet. Since there are technical and optical constraints on the minimum size (width) of one slitlet mirror, the width was chosen to be $\mathrm{300 \ \mu m}$. \cite{eisenhauer00} So the resulting slitlet height is 9.6 mm, corresponding to a resampled pseudo slit of 307.2 mm, which has to fit on M1.
Because M1 is a spherical mirror, the surface measurement was done in-line. The collimated 6 mm laser beam from the interferometer is sent through a planoconvex lens with $\mathrm{f = 50 \ mm}$ focal length resulting in an $f/8.3$ beam. M1 is placed so that its focus is co-located with the focus of the lens. Since the radius of M1 is so large, in the measurement a flat folding mirror in between the lens and M1 has been used, because the optical bench available was not large enough. The optical setup from the focus position to M1 is shown in figure \ref{fig:M1setup}. 

\begin{figure}[htbp!]
	\begin{center}
		\subfloat[old-spare M1]
		{\includegraphics[width=0.25\textwidth, trim={0.cm 0 0cm 0.1cm}, clip=true]{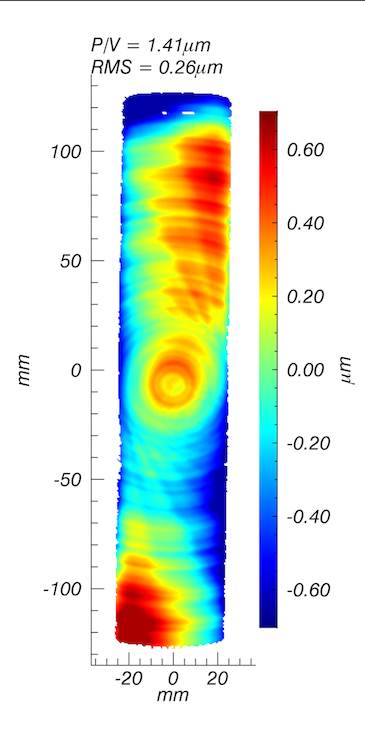}}
		\subfloat[pre- upgrade M1]
		{\includegraphics[width=0.25\textwidth, trim={0.cm 0 0cm 0.1cm}, clip=true]{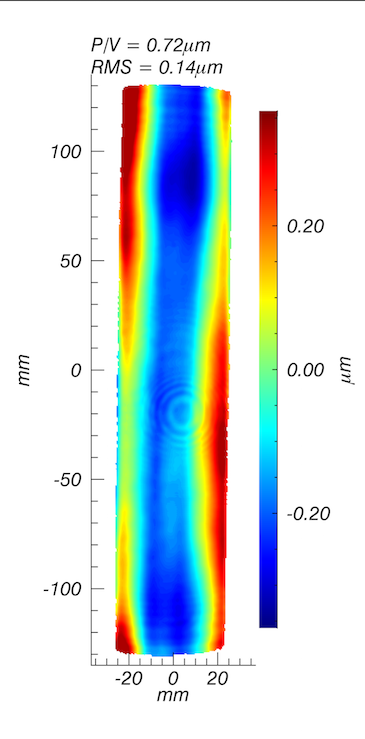}}
		\subfloat[new M1 S01]
		{\includegraphics[width=0.25\textwidth, trim={0.cm 0 0cm 0.1cm}, clip=true]{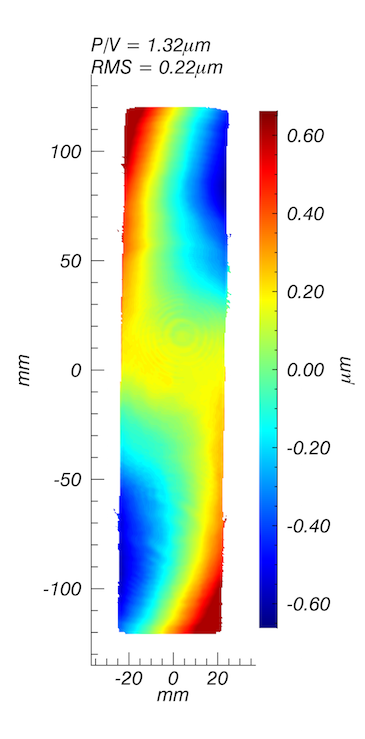}}
		\subfloat[new M1 S02]
		{\includegraphics[width=0.25\textwidth, trim={0.cm 0 0cm 0.1cm}, clip=true]{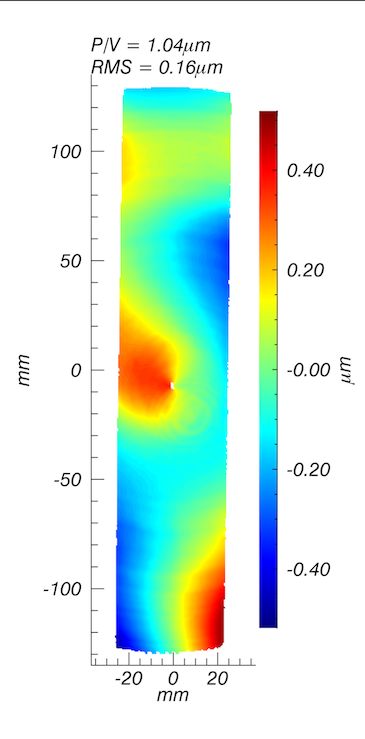}}
		\caption[M1 surface deviation maps]{The surface deviation maps of all M1 mirrors. Note the different surface scales!}
		\label{fig:M1surface}
	\end{center}
\end{figure}

While the position of the folding mirror does not matter for the measurement, the distance between M1 and the focus has to be identical to its radius in order to retro-reflect the laser light. The fine adjustment of the distance was done directly by looking at the fringes on the Real Time Display (RTD) of the interferometer software and adjusting the micrometers on the translation stages of the folding mirror. In order to get a sharp image of the surface of M1 also the position of the lens has to be adjusted until the edges of the mirror appear sharp on the RTD. This is simply to fulfill the Newtonian projection equation. Since the distance to M1 is large in comparison to the focal length of the lens used to adapt the interferometer beam, the distance between the interferometer and the lens is roughly the focal length of it. Since the surface of M1 is spherical, the measured wavefront divided by a factor of two corresponds directly to the surface form deviation from an ideal sphere. Figure \ref{fig:M1surface} shows this surface deviation for all four M1 collimator mirrors.

The footprint on M1 is $\mathrm{\sim 250 \ mm}$ in diameter, covering around 75 percent of the mirror surface. The upper and lower 12.5\% of the reflecting surface are missing.
While on the old-spare M1 the diamond turning marks are very obvious, during the post-polishing process of the old-in-SPIFFI M1, all turning marks were smoothed out so that none of them are visible anymore. The concentric rings close to the center of that mirror are reflections within the interferometer. The post-polishing of the spherical M1 surface was excellent resulting in a smooth surface with no turning marks and a smaller P/V surface deviation than reached with the new mirrors (though of course this was not measured until after the mirror had been replaced). The new mirrors are mostly affected by a point symmetric wavefront deviation which indicates a torsional instability in the tall mirror during machining. This torsional effect can also be seen in the wavefront of the old-spare M1 though with opposite sign, meaning the valleys and peaks are reversed. For the old-in-SPIFFI mirror this torsion was removed by polishing. The concentric rings in the surface plot of the new M1 S01 are again internal reflections inside the interferometer. When zooming in on the new mirrors, very small amplitude diamond turning marks can be recognized with an amplitude less than $\mathrm{10 \ nm}$ and close to the surface roughness. Another high frequency surface deviation the polishing removed is the volcano-shaped turning center on M1. This turning center shows up in the new M1 mirrors as a cusp that is strongest in series 02. However in series 01 it also exists with a much smaller amplitude of $\mathrm{\sim 0.10 \ \mu m}$ in comparison to $\mathrm{\sim 0.60 \ \mu m}$ for series 2 (see appendix figure \ref{fig:cusp}). Since slitlet 1 and 32 (located at the center of the pseudo-slit) are offset from the center of M1, the cusp does not play a big role in the wavefront quality of the reflected light. Since the cusp is much smaller an M1 S01, this mirror was chosen to be built into SPIFFI for the upgrade. Even if the torsion effect is larger, the footprint of any individual slitlet is very small on this mirror, so the large-scale variations do not matter as much. The surface roughness of the new mirrors is with $\mathrm{\sim 5 \ nm}$ in the specifications. The sagitta error is close to the specifications, but the strong cusp on M1 S02 deviates strongly from the tolerances fo the irregularities since the it extents over more than 0.16 mm x 0.16 mm.

For the upgrade it was chosen to implement M1 series 01 because of the smaller cusp at the turning center.

\subsection{M2}{\label{sec:M2}}
M2 is the smallest of the three collimator mirrors and sits in the middle of the optical path between M1 and M3. Its surface form is an off-axis convex oblate ellipse with a central radius of $\mathrm{R = 1368.89 \ mm}$ and a conic constant of $\mathrm{k = 2.82}$ corresponding to a numerical eccentricity of $\mathrm{\epsilon ^2 = \left(1-a^2/b^2\right)}$ with the major axis a and the minor axis b. The reflective surface is 62 mm wide and 148 mm high. During the diamond turning process of the new M2 mirrors the reflective surface got scratched. It was re-machined to remove the scratches. So the surface of the new mirrors is 0.5 mm lower with respect to the backside of the mirror. Since the backside of M2 is pushed against alignment pins in SPIFFI, on the backside of M2 a 0.5 mm aluminum shim was attached to correct for the surface offset.

The measurement of the M2 surface is the most complicated one of the three mirrors, since it is convex. For this reason a large 20 cm diameter spherical mirror was used to converge the expanded beam on M2 and reduce the astigmatism in the measurement of the mirror. The measurement setup was designed and then optimized in Zeemax using the constraint to have a minimum spot size at focus. In this setup third order aberrations are removed, but a small amount of higher order aberrations remain. To remove these remaining aberrations from the surface measurement of M2, the aberrations have to be known. This is achieved by simulating the experimental setup in an optics program to obtain the wavefront distortion for a perfectly manufactured M2. For this reason the experimental setup in the laboratory has to be as close as possible to the setup in the optics program. To achieve this with a sub-millimeter accuracy, the optical components in the laboratory were aligned with the help of a measurement arm to the correct position calculated by the optics program. (This alignment procedure is very delicate. For a description of the alignment method see appendix section \ref{sec:M2alignment} and figure \ref{fig:M2setupnumbers}.) The fine adjustment for tip and tilt was done with the large spherical mirror in order to not change the angle and height of the optical axis of the test setup, which had been pre-aligned to be level with the optical table. Like for the M1 and also the M3 measurement initial lens in the setup (here: $\mathrm{f = 30 \ mm}$) had to be at the correct distance from the interferometer to get a sharp image of the M2 surface. The optical relevant setup from the focus position to M2 is shown in figure \ref{fig:M2setup} and more accurately in figure \ref{fig:M2setupnumbers} of the appendix. The interferometer and the lens to adjust the beam is not shown. These two elements are located left of the focus position.

\begin{figure}[htbp!]
	\begin{center}
		\resizebox{0.6\textwidth}{!}{
			\includegraphics[width=1\textwidth]{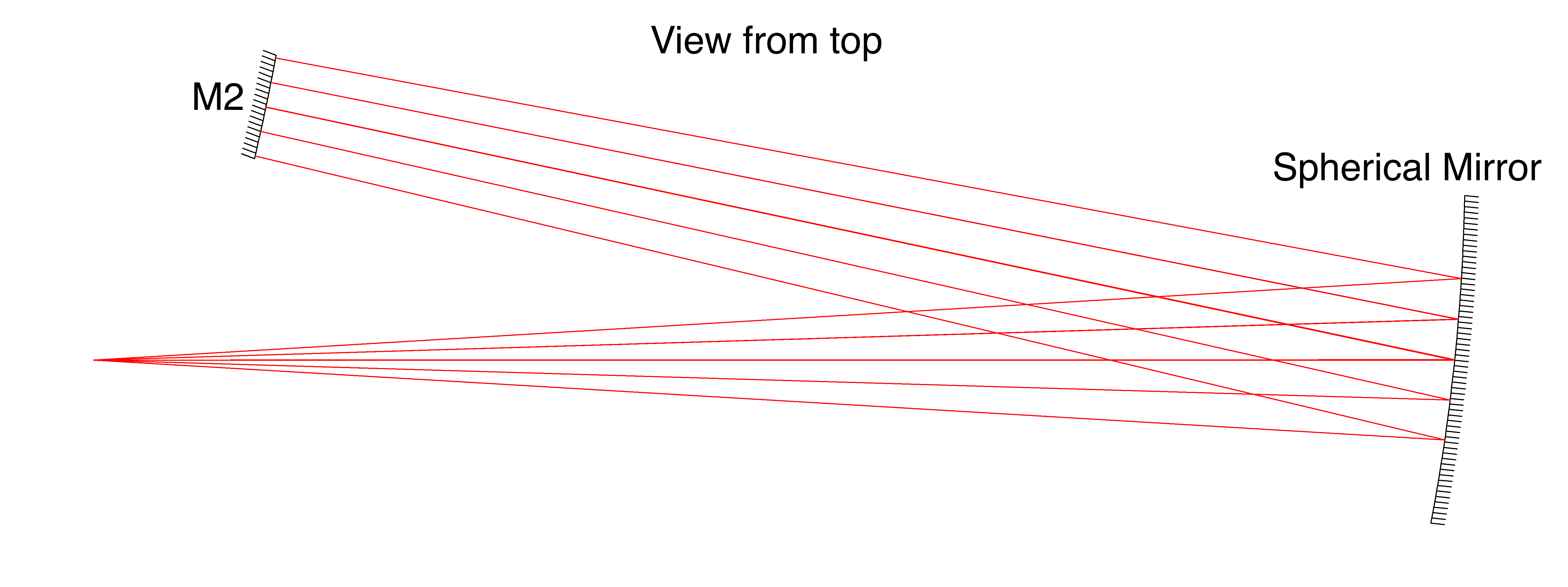}
		}
		\resizebox{0.6\textwidth}{!}{
			\includegraphics[width=1\textwidth]{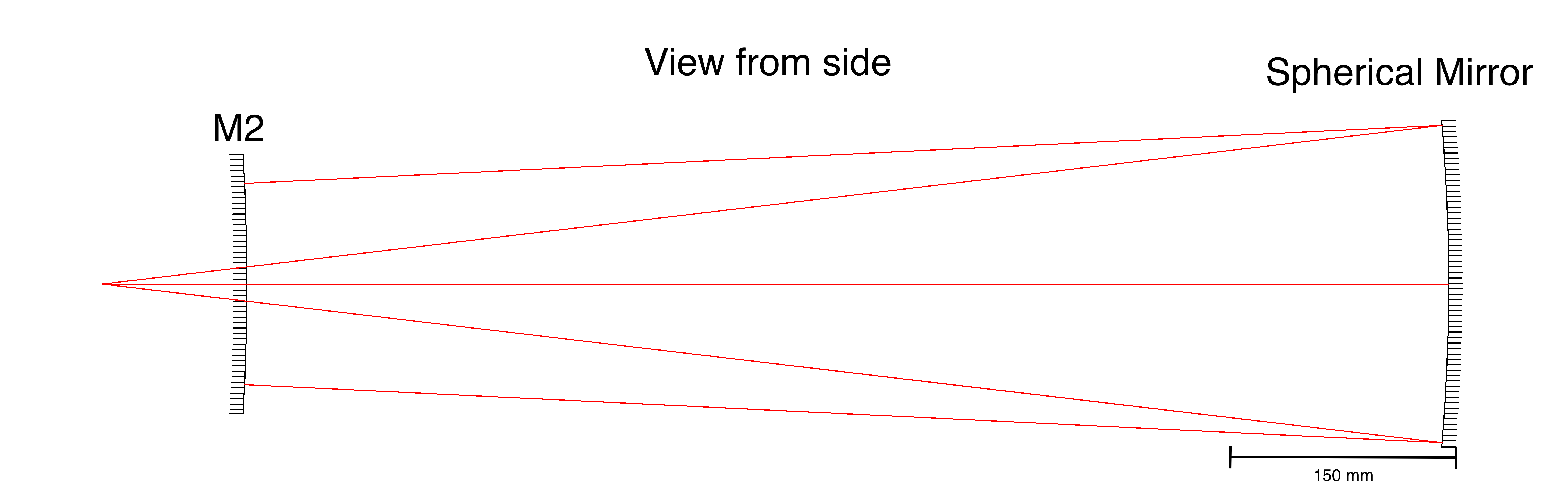}
		}
		\caption[M2 test setup]{The optical setup of the M2 single mirror test between focus position and M2. The interferometer is located left of the focus position. The collimated 6mm beam from the interferometer is bundled by a lens to the focus.}
		\label{fig:M2setup}
	\end{center}
\end{figure}

\begin{figure}[h!]
	\begin{center}
		\subfloat[default surface deviation]
		{\includegraphics[width=0.4\textwidth, trim={0.cm 0 0cm 0.1cm}, clip=true]{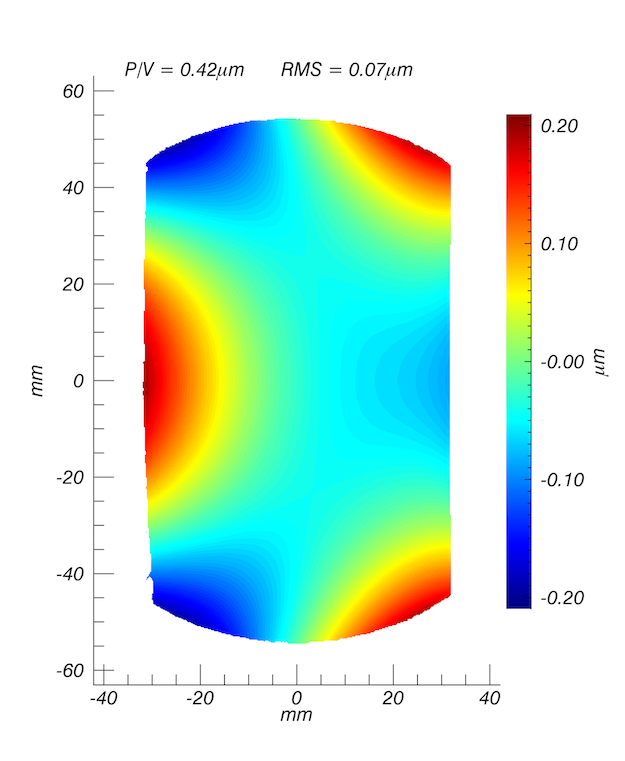}}
		\subfloat[Measured surface of the old-spare M3 without subtraction of the default surface deviation from the left.]
		{\includegraphics[width=0.385\textwidth, trim={0.cm 0 0cm 0.1cm}, clip=true]{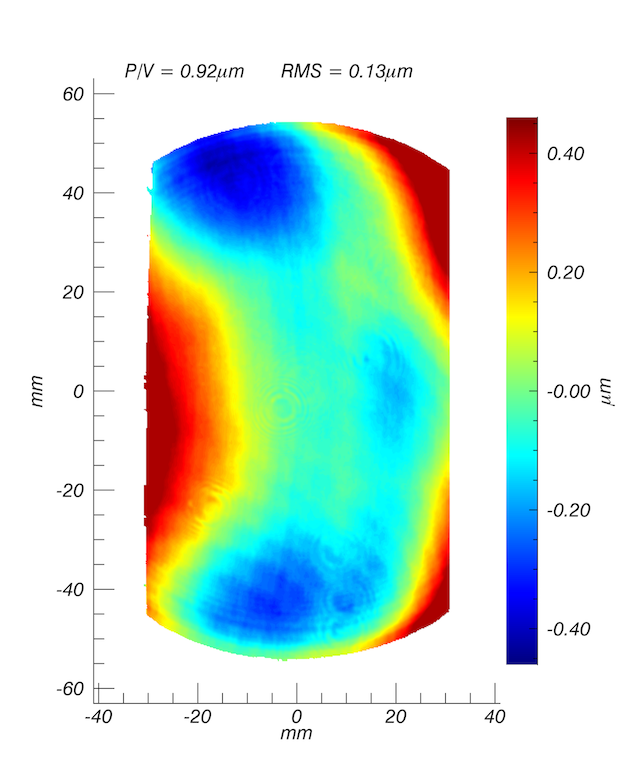}}
		\caption[M2 default surface deviation and raw measurement]{On the left the surface deviations induced by the M2 test setup are shown, while on the right side the raw surface deviation measurement is shown. To get the true surface deviation which is shown in figure \ref{fig:M2surface}, one simply has to subtract the default deviation from the measured one.}
		\label{fig:M2default}
	\end{center}
\end{figure}

The aberrations induced by this setup are shown in figure \ref{fig:M2default} in units of micron surface form error. This default surface form is then subtracted from the measured surface of M2. The residuals are the surface deformations from the ideal M2 surface. For reasons of consistency it is shown here in surface deviation and not in wavefront deviation.

\begin{figure}[htbp!]
	\begin{center}
		\subfloat[old-spare M2]
		{\includegraphics[width=0.5\textwidth, trim={0.cm 0 0cm 0.1cm}, clip=true]{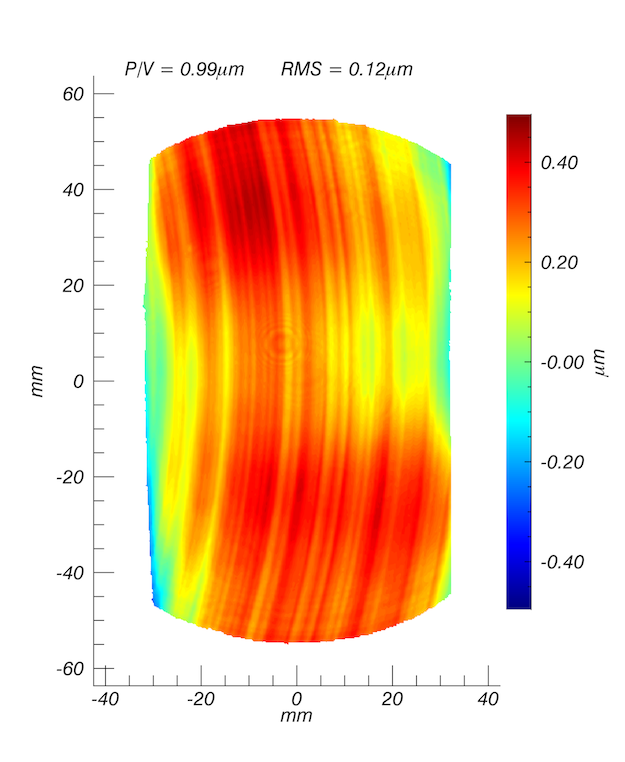}}
		\subfloat[pre- upgrade M2]
		{\includegraphics[width=0.5\textwidth, trim={0.cm 0 0cm 0.1cm}, clip=true]{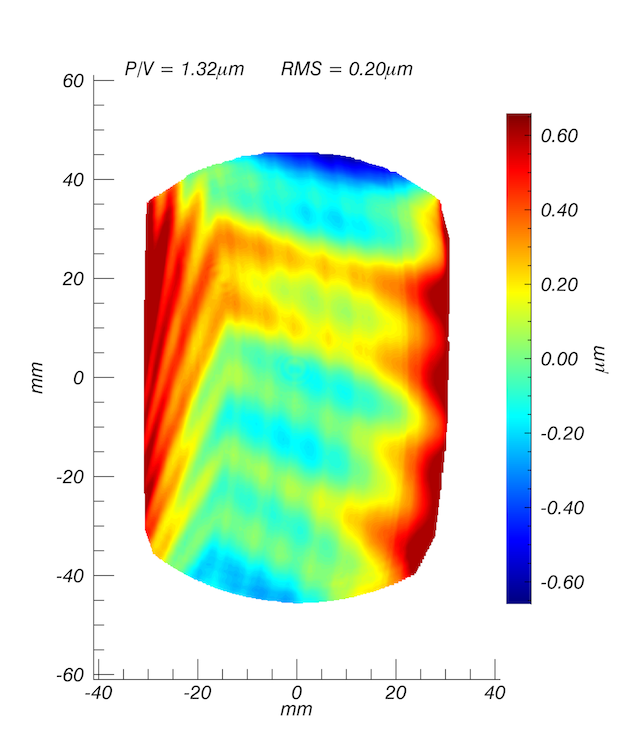}}
		\quad
		\subfloat[new M2 S01]
		{\includegraphics[width=0.5\textwidth, trim={0.cm 0 0cm 0.1cm}, clip=true]{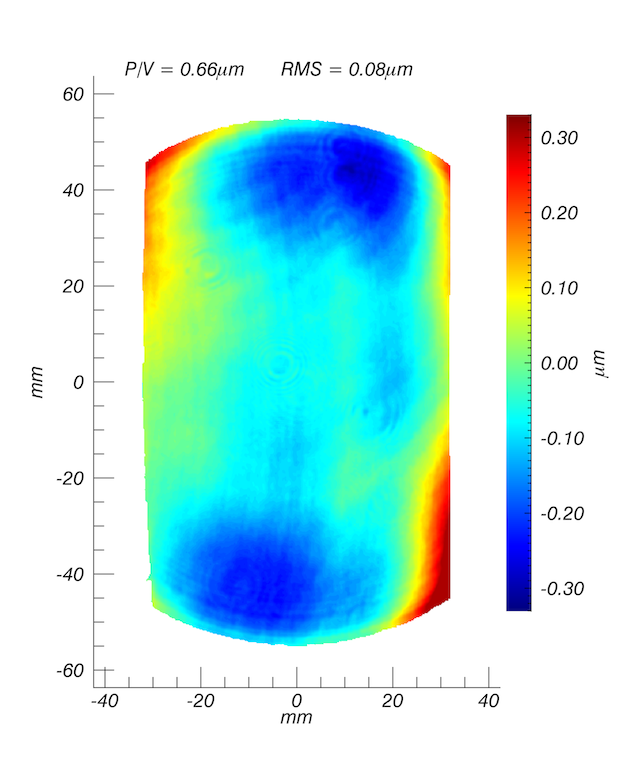}}
		\subfloat[new M2 S02]
		{\includegraphics[width=0.5\textwidth, trim={0.cm 0 0cm 0.1cm}, clip=true]{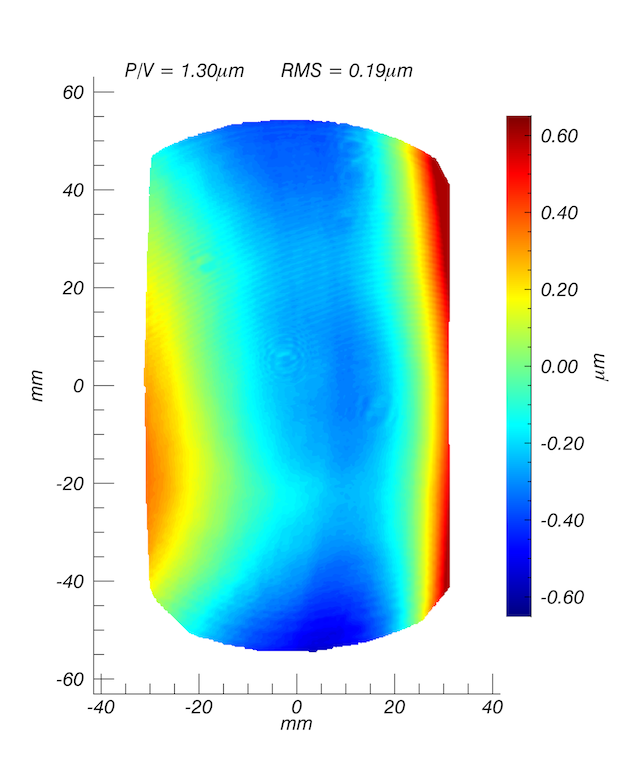}}
		\caption[M2 surface deviation map]{The surface deviation maps of all M2 mirrors. Note: The footprint of the old-in-SPIFFI M2 mirror, that was in SPIFFI is slightly smaller than the others, because for the measurement a lens with a slightly larger focal length was accidentally used. The different mirror surfaces have different color scales.}
		\label{fig:M2surface}
	\end{center}
\end{figure}

The setup corrects for third order aberrations, so the default surface form error introduced by the setup is mainly trefoil with an amplitude of less than half a micron and higher order aberrations as shown in the left plot of figure \ref{fig:M2default}. The footprint measured with the interferometer on M2 is $\mathrm{\sim 110 \ mm}$ in diameter covering $\mathrm{\sim 70\%}$ of the reflective surface of M2. It is centered, so 15\% of the upper and lower parts of the mirror respectively are missing. Despite this, the covered area is nearly identical to the full area of the SPIFFI beam on M2.
Figure \ref{fig:M2surface} shows the measured surfaces of all M2 mirrors.

Starting with the old-spare M2 one can clearly see the diamond turning marks on spatial scales of the order of millimeters. Beside the turning marks, the overall surface is quite close to the ideal surface. In the center regions it only differs by $\mathrm{\sim 0.2 \ \mu m}$. The old-in-SPIFFI M2 has different surface features. The first thing one recognizes is the wavy structure in the vertical direction along the mirror surface with a spatial frequency of around 1.5 cm and an amplitude 0.2 to 0.4 $\mathrm{\mu m}$. The turning marks are still visible with a spatial frequency of about 5 mm. It is likely that this mirror was post-polished by hand due to convex surface. This mirror is a good example of how the nickel layer behaves when it is polished. Especially at the left edge, the structure reveals that the nickel behaves like a highly viscous liquid when it is polished. In the polishing process, the nickel was literally pushed from the inner areas of the mirror to the sides. This surface form error is mainly responsible for the overall distortion of the wavefront in the SPIFFI collimator that was installed until 2016.

The surface of the new M2 series 01 looks quite good. Again the concentric rings are internal reflections in the interferometer. The large concentric rings at the edges of the surface are caused by the imperfectly-calibrated spherical aberration in the lens used in the test setup. Very faint diamond turning marks ($\mathrm{P/V < 10 nm}$) in the direction like on the old-spare M2 can be recognized. The radial wavy structure is likely to be not real and caused by internal reflections in the interferometer or more probable by the imperfectly-calibrated lens. The surface of series 02 has some residual astigmatism of about one micron resulting in a valley on the surface. The horizontal grooves must be caused by a different optical element in the setup like for series 01. The spatial frequency of these is way higher and they are slightly tilted with respect to series 01. This can be caused by the slight adjustment of the large spherical mirror in order to remove tilts in the interferogram. However the origin of these groves is likely to be reflections or spherical aberration of the lens. They are too regular and in an untypical direction with a varying radius, which cannot be an effect of the manufacturing process. The sagitta error is for M2 S01 again very close to the specifications and since it has by a factor of two the smaller surface deviation, this serial number was chosen to be implemented into SPIFFI in the upgrade. The performance of the collimator will be effected by the exchange of M2 the most, since the old-in-SPIFFI M2 distorted the wavefront by a large amount. (see also figure \ref{fig:M2exchanged})

For the upgrade it was chosen to implement M2 series 01 because the surface deviations are smaller than in series 02.

\newpage
\subsection{M3}{\label{sec:M3}}
The third collimator mirror, M3, is positioned close to the pupil position of the SPIFFI spectrograph. The surface form is an off-axis concave oblate ellipse. Its radius at the vertex is $\mathrm{R = 2648.10 \ mm}$ and the conic constant is 6.237.

\begin{figure}[htbp!]
	\begin{center}
		\resizebox{0.6\textwidth}{!}{
			\includegraphics[width=1\textwidth]{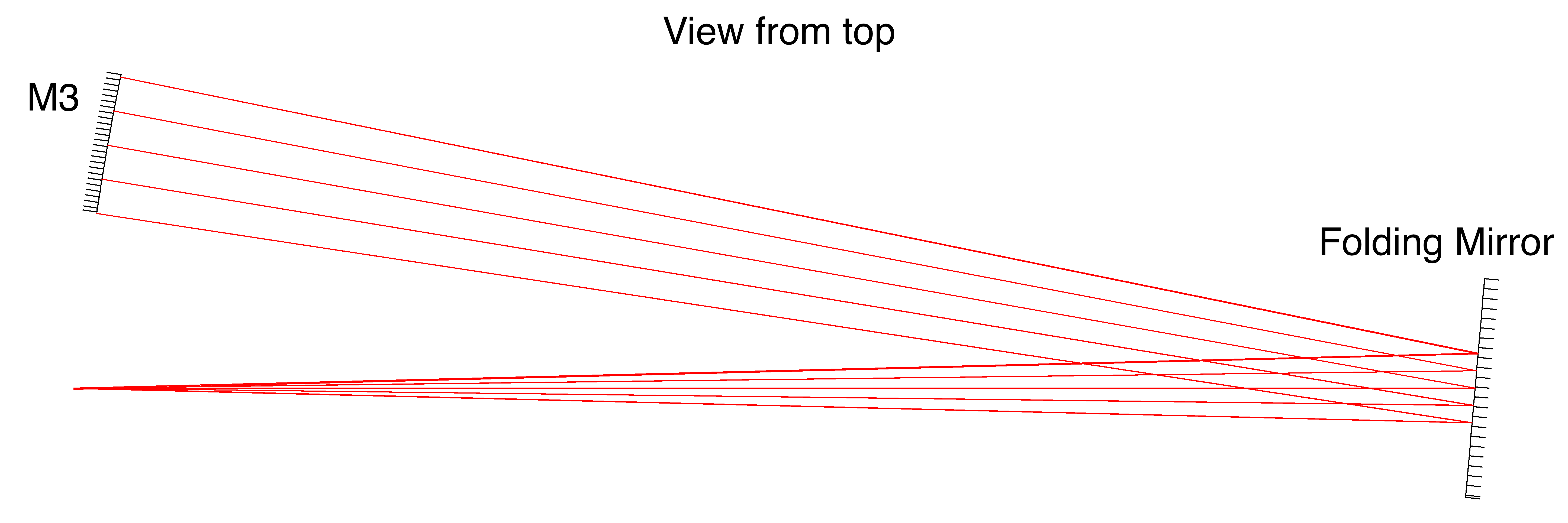}
		}
		\resizebox{0.6\textwidth}{!}{
			\includegraphics[width=1\textwidth]{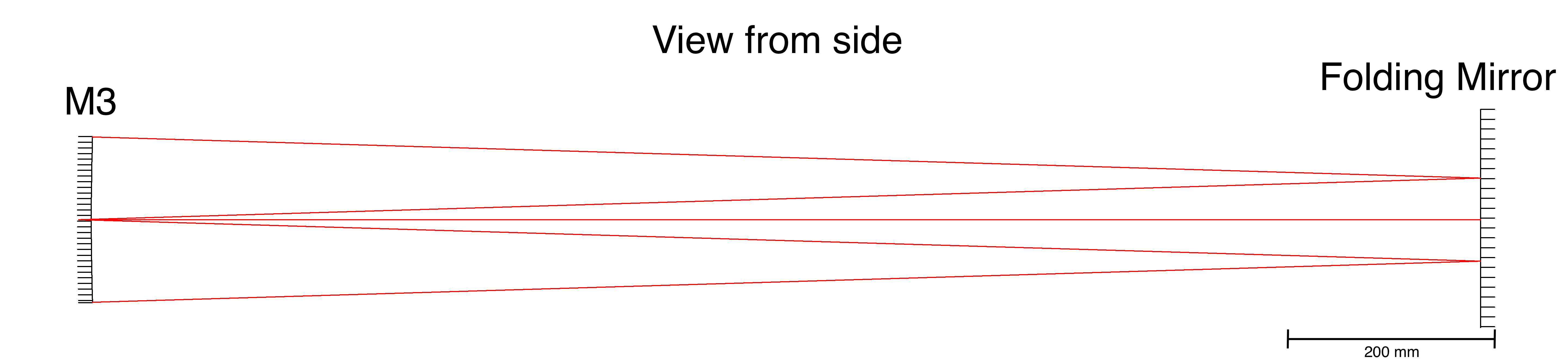}
		}
		\caption[M3 test setup]{The optical setup of the M3 single mirror test between focus position and M3. The interferometer is located left of the focus position. The collimated 6mm beam from the interferometer is bundled by a lens to the focus.}
		\label{fig:M3setup}
	\end{center}
\end{figure}

The interferometric surface measurement is implemented in-line as for M1 with a flat folding mirror. In this setup astigmatism is induced due to the elliptical surface form. The planoconvex lens used is a lens with $\mathrm{f = 80 \ mm}$ focal length resulting in an $f/13$ beam. The distance between the focus and the mirror is roughly the radius at the vertex with a small correction, because the measurement is not done in the paraxial approximation. As it was done in the M1 measurement setup, the focus and tilt adjustments were done with the flat folding mirror in order to make the wavefront as flat as possible. The optical setup from the focus position to M3 is shown in figure \ref{fig:M3setup}. It is quite similar to the M1 setup since both measurements are done in an in-line setup.

\begin{figure}[htbp!]
	\begin{center}
		\subfloat[default surface deviation]
		{\includegraphics[width=0.4\textwidth, trim={0.cm 0 0cm 0.1cm}, clip=true]{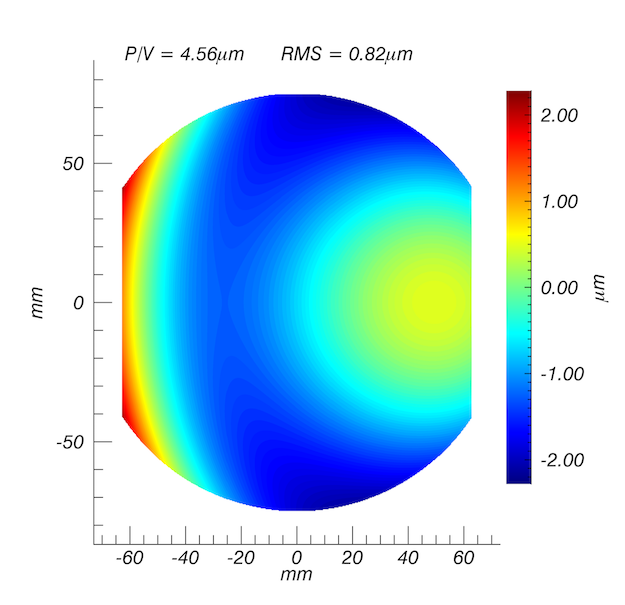}}
		\subfloat[Measured surface of the old-in-SPIFFI M3 without subtraction of the default surface deviation from the left.]
		{\includegraphics[width=0.365\textwidth, trim={0.cm 0 0cm 0.1cm}, clip=true]{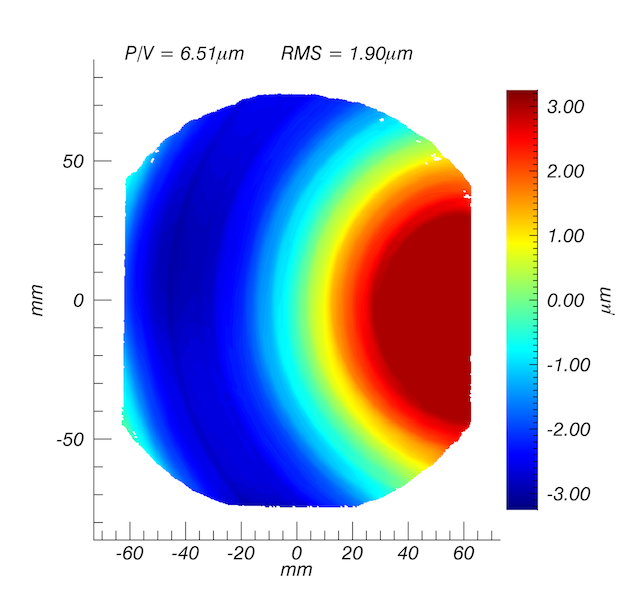}}
		\caption[M3 default surface deviation and raw measurement]{The figure on the left shows the surface deviation induced by the in-line measurement setup. One would also measure it for a perfect M3. On the right side the raw surface deviation measurement is shown. To get the true surface deviation one simply has to subtract the default deviation from the measured one.}
		\label{fig:M3default}
	\end{center}
\end{figure}

Figure \ref{fig:M3default} shows the error that is induced by the in-line measurement of M3. It is mainly astigmatism. For reasons of consistency it is shown here in surface deviation and not in wavefront deviation. The optical model is optimized for minimal spot size. This has to be done, because M3 is not a sphere, but an oblate ellipse. Thus in an in-line setup astigmatism is induced. Furthermore the distance to the focus is not exactly the radius of the ellipse at the vertex, because the measurement is not done in the paraxial approximation. Free parameters are, as on the measurement in the laboratory, the distance to the focus and the tilt of M3. From the measurements the defocus and tilt is subtracted. This leads to a setup that is close to the optimized optical model. The default surface deviation was subtracted afterwards from the measured one in order to get the real deviation from a perfect M3.

\begin{figure}[htbp!]
	\begin{center}
		\subfloat[old-spare M3]
		{\includegraphics[width=0.5\textwidth, trim={0.cm 0 0cm 0.1cm}, clip=true]{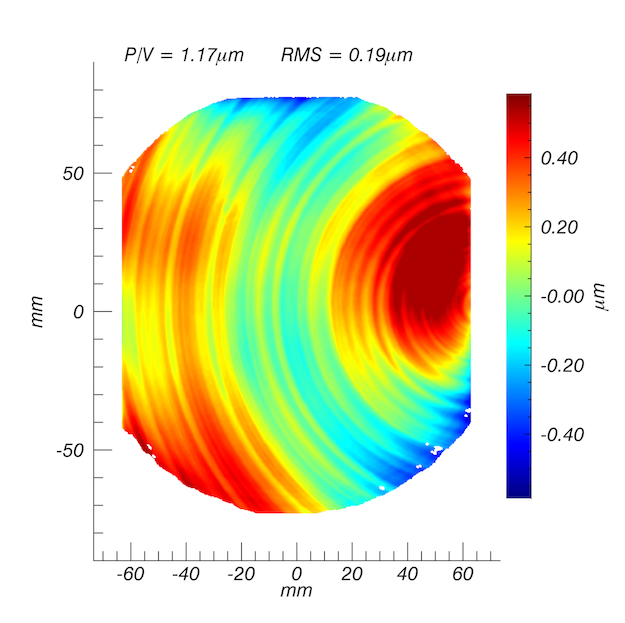}}
		\subfloat[pre- upgrade M3]
		{\includegraphics[width=0.5\textwidth, trim={0.cm 0 0cm 0.1cm}, clip=true]{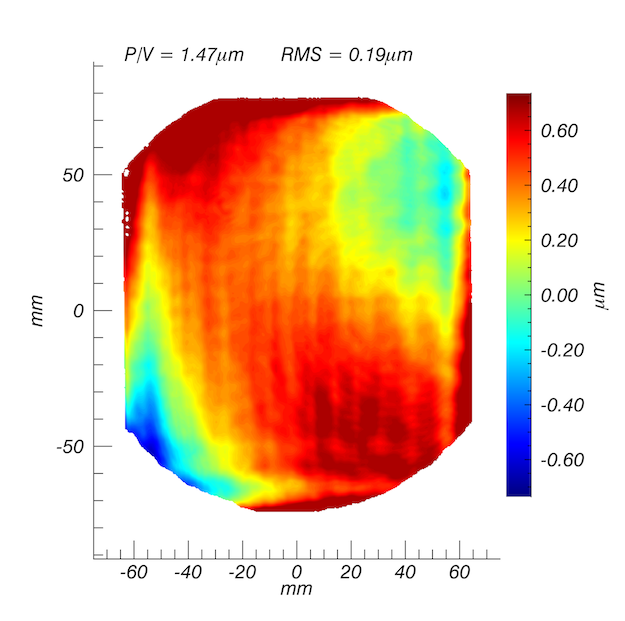}}
		\quad
		\subfloat[new M3 S01]
		{\includegraphics[width=0.5\textwidth, trim={0.cm 0 0cm 0.1cm}, clip=true]{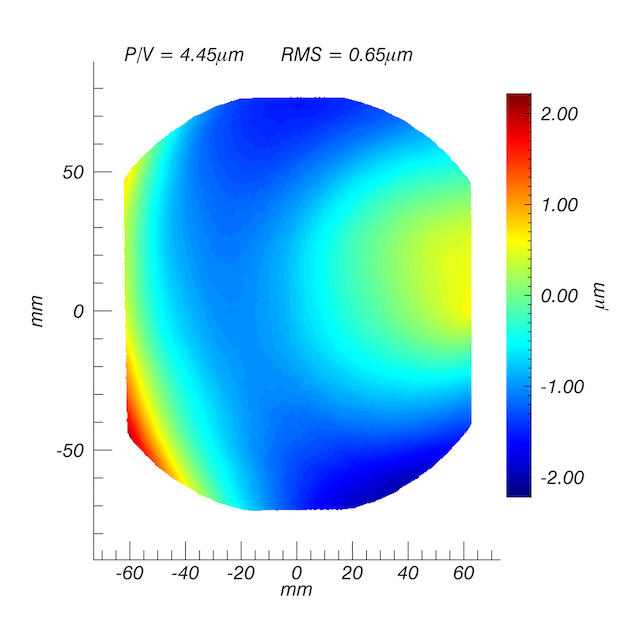}}
		\subfloat[new M3 S02]
		{\includegraphics[width=0.5\textwidth, trim={0.cm 0 0cm 0.1cm}, clip=true]{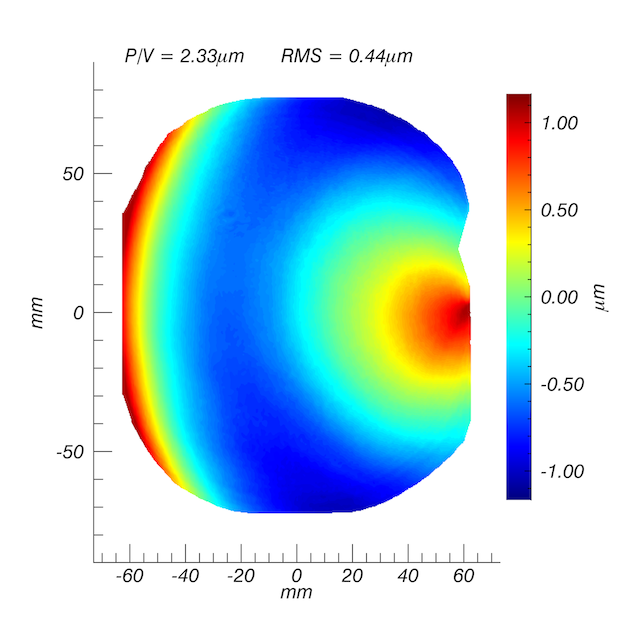}}
		\caption[M3 surface deviation maps]{The surface deviation maps of all M3 mirrors. Note the different surface scales!}
		\label{fig:M3surface}
	\end{center}
\end{figure}

The footprint measured with the interferometer on M3 is $\mathrm{\sim 190 \ mm}$ in diameter covering nearly the full reflective surface of M3 other than the corners of the mirror. The footprint covers the full area of the SPIFFI beam on M3. Figure \ref{fig:M3surface} shows the measured surface deviations of all M3 mirrors after the subtraction of the default surface form. At the edges in some interferograms the measurement did not work perfectly, so that for example in the plot of M3 S02 a piece at the right edge is missing.

Starting with the old-spare M3 surface deviation map it is again easy to see the diamond turning marks like on all old-spare mirrors. The overall surface form of this mirror is again surprisingly good. Off course for a mirror so close to the pupil like M3, the turning marks with a frequency of a few millimeters act like a diffraction grating resulting in non-Gaussian shaped spectral line profiles. The center of turning which is close to the edge of M3 (1.2 mm of the edge) can be seen very well. The old-in-SPIFFI M3 however was post-polished on a machine. As one can see, this produced some kind of a grid structure on the mirror surface. This structure is able to act like a diffraction grating especially because M3 is positioned close to the pupil of the collimator. It is possible that this mirror had an influence on the shape of the spectral line profiles (see section \ref{sec:lineprofiles}). It seems like the polishing machine was polishing in lines on the surface with distances of around 1 cm and then afterwards perpendicular to these lines resulting in a grid structure. The polishing of M3 nearly removed all diamond turning marks. Only very close to the turning center, a residual concentric structure can be seen but with a very low amplitude of $\mathrm{\sim 0.1 \ \mu m}$.

Looking at both of the new M3 mirrors, it is quite obvious that the wavefronts show lots of astigmatism. Second, one can see especially in the surface deviation of series 01 that the center of turning is offset. The measurement and the P/V value is in this case very sensitive, since it rises at the edges of the mirrors. So modulo the differences in the illuminated portions of the edges and corners of the mirrors (S01 shows a bit more of the upper and lower left corner in comparison to S02) the P/V surface deviations are approximately the same for both new mirrors. The turning center of series 01 is shifted $\mathrm{\sim 7.5 \ mm}$ upwards, while on series 02 it is shifted $\mathrm{\sim  2.5 \ mm}$ downwards. But this alone would not induce such a strong amount of astigmatism. On top of this the curvature of the mirror is wrong, meaning that either the radius at the vertex or the conic constant is wrong. Furthermore on series 02 one can see a small cusp close to the turning center of $\mathrm{\sim 0.5 \ \mu m}$ height.

In the full collimator measurement (see section \ref{sec:fullcoll}) it was found that the astigmatism in the collimator wavefront can be corrected by rotating M3 series 02 by 23' towards the center of the instrument plate. For series 02 rotations corrected the astigmatism much better than for series 01 where also a tilt of the mirror would have been required to remove the large 45\degree astigmatism. For that reason and because it was possible to nearly remove all the astigmatism for M3 series 02 it was decided to implement this mirror in the collimator.

A new M3 and a spare one will be manufactured soon. It will be installed into SPIFFI in 2019 in the SPIFFIER upgrade.

For the performance of the SPIFFI spectrometer, the collimator wavefront is of interest. In the next section the measurements of the wavefront are described and results are shown.

\section{Full Collimator Measurements}
\begin{figure}[htbp!]
	\begin{center}
		\subfloat[\label{fig:fullmirror}The full collimator test setup. The interferometer sits close to the pupil position (the location of the diffraction grating). A collimated beam passes through the collimator in reverse (M3-M2-M1), goes through focus, and then is retro-reflected by a return sphere.]
		{\includegraphics[height=0.4\textwidth]{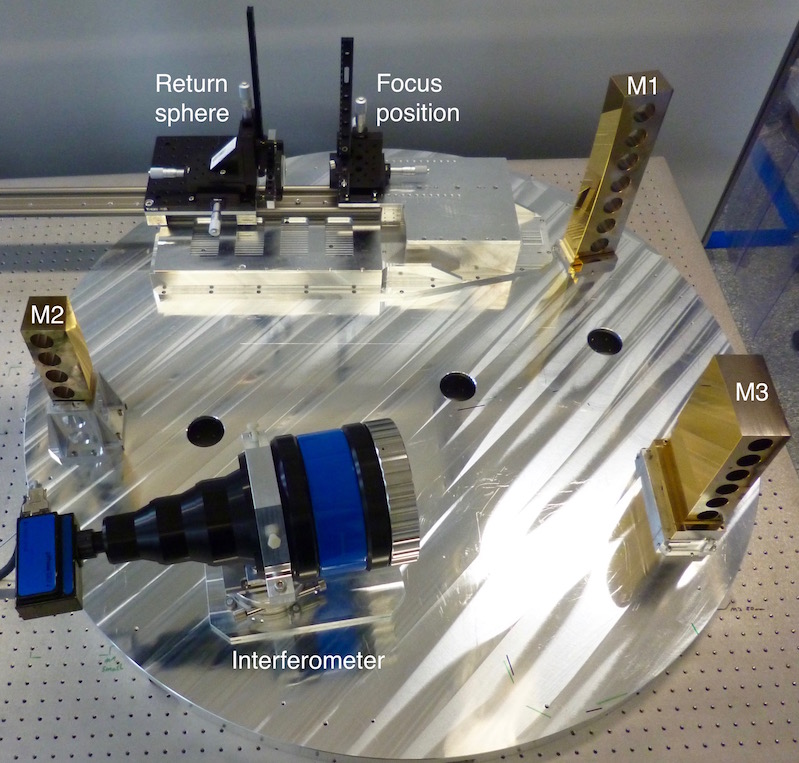}}
		\quad
		\subfloat[\label{fig:uncorrected}The nominal wavefront of the collimator using M1 S01, M2 S01 and M3 S02 without correcting for the strong surface deviation of M3.]
		{\includegraphics[height=0.4\textwidth, trim={0cm 0.2cm 0cm 0.5cm}, clip=true]{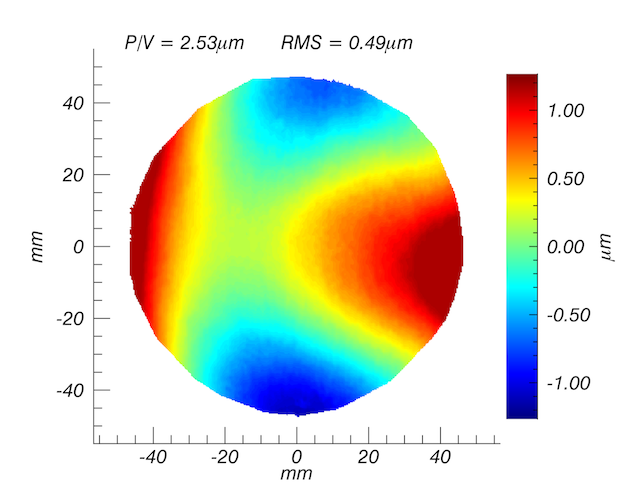}}
		\caption[Full collimator test setup and uncorrected nominal wavefront]{The test setup for the full collimator measurement and the uncorrected nominal wavefront.}
		\label{fig:fullcoll}
	\end{center}
\end{figure}

{\label{sec:fullcoll}}
The critical performance parameter of the collimator is the behavior of its wavefront. Because of this all plots shown in this section are wavefront deviations and not surface deviations as for the single mirror measurements. The design goal of a TMA is to minimize the wavefront distortion at the pupil position. Because the wavefront is very sensitive to the position of the mirrors, a duplicate of the SPIFFI instrument plate was used to align the mirrors to the correct positions, which was simply done by pushing them against the alignment pins in the plate. The position of those alignment pins is accurate to within 0.01 mm while the mirror surface reference point position tolerance is 0.05 mm. On the instrument plate, the interferometer with a beam expander that widens the beam from 6 mm to 150 mm was mounted close to the pupil position of the spectrometer collimator. The interferometric wavefront measurement was done in double pass. A pupil stop with a diameter of 95 mm was installed. The light from the interferometer passes M3, M2 and finally M1. After M1, the beam passes through focus and is retro-reflected by a spherical mirror back through the collimator and finally into the interferometer. The return mirror used has surface irregularities up to $\mathrm{\lambda / 4}$ at 633 nm. At the focus position a field stop was used in order to easily detect shifts of the focus, and therefore the image in SPIFFI. With this method one obtains the wavefront for one field direction, corresponding to a single position within a single slitlet. Due to the large beam footprint separations of the different slitlets on M1, the wavefront of only one slitlet is measured in this setup. Measurements were carried out for different field positions. For the new collimator the differences were negligible, while for the old collimator there were significant changes. These are also shown in figure \ref{fig:lowhighfootprint} A picture from the setup in the laboratory can be seen in figure \ref{fig:fullmirror}. As for the single mirror measurements the setup was placed on a floating optics table to reduce vibrations that would make the interferometric measurement impossible.

Already from the changing focus position between the old and the new collimator mirrors the effect of the large surface form error of M3 could be seen. In comparison to the old-spare mirrors, the focus position moved for both new M3 mirrors towards the center of the instrument plate and upwards (from the instrument plate) by around 1 mm. A wavefront of the old-spare mirrors can be seen in Eisenhauer et al. 2003 in figure 6. It is assembled out of several interferograms of different parts of the pupil. The wavefront of the old-spare mirrors is distorted so strongly by the diamond turning marks in the mirror surfaces it is not possible to measure a single interferogram of the full pupil. For this reason there is no comparison of the wavefront of the full collimator made up of the old-spare mirrors done in this thesis.

\begin{figure}[htbp!]
	\begin{center}
		\subfloat[pre- upgrade collimator wavefront]
		{\includegraphics[width=0.5\textwidth]{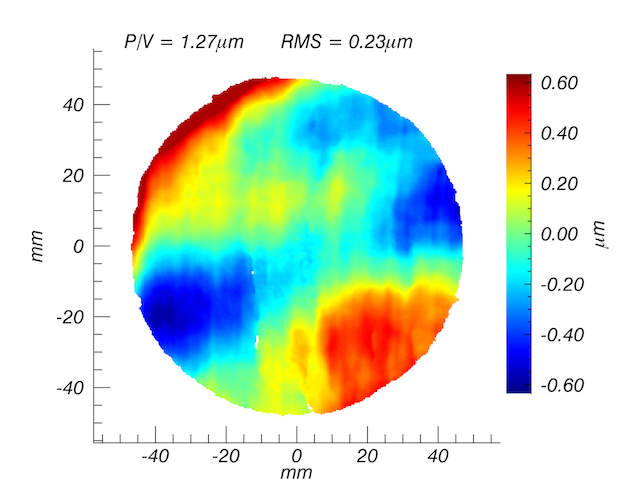}}
		\subfloat[post- upgrade collimator wavefront]
		{\includegraphics[width=0.5\textwidth]{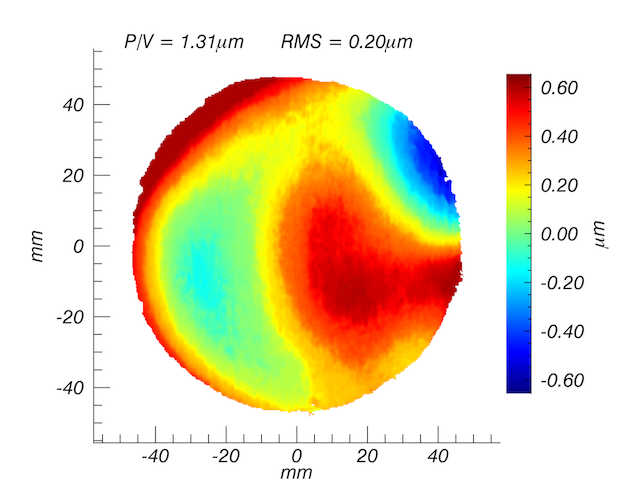}}
		\caption[Pre- and post- upgrade collimator wavefront]{The wavefront of the collimator in single path pre- and post- upgrade.}
		\label{fig:oldnewwavefront}
	\end{center}
\end{figure}

In figure \ref{fig:oldnewwavefront} the full collimator wavefronts using both the pre- upgrade SPIFFI mirrors and the post- upgrade SPIFFI mirrors are shown. Both wavefronts were measured in double pass but are shown here in single pass, since this is the configuration used in the instrument. The left image shows the wavefront of the collimator as it looked before the upgrade, while the right one shows the current SPIFFI collimator wavefront (always assuming that the mock instrument plate the wavefront of the collimator was measured on is identical to the instrument plate in SPIFFI, which should be true within the manufacturing tolerances.). The wavefront is displayed as one would see it when looking from M3 through the pupil position towards the grating.

\begin{sloppypar}
	
	The pre- upgrade wavefront shows a peak-to-valley surface deviation of $\mathrm{\sim 1.3 \ \mu m}$ and a RMS surface deviation of $\mathrm{\sim 0.23 \ \mu m}$. These two values were not affected a lot by the upgrade. Initially, the new mirrors with the nominal alignment showed $\mathrm{\sim 2}$ waves of 90\degree \ astigmatism in the wavefront at 633 nm wavelength (see figure \ref{fig:uncorrected}), due to the wrong surface form of M3. To correct for the wrong surface of M3, the mirror was rotated by 23' towards the the center of the instrument plate by implementing a 1.0 mm shim in between the alignment pin in the instrument plate and the M3 socket. This position was found to reduce most of the third order aberrations and especially nearly all 90\degree astigmatism. Due to this rotation a horizontal movement of the slit image is induced, which can be corrected by turning the grating wheel by the same amount.
	
	The new mirrors installed into SPIFFI during the upgrade were the following: M1 S01, because it does not show such a large cusp as M1 S02; M2 S01, because the surface deviations are a factor of 2 smaller than M2 S02; and M3 S02, because the decentering of the turning center is a factor of 3 less than in M3 S01.
	
\begin{figure}[htbp!]
	\begin{center}
		\subfloat[\label{fig:oldcollpsf}pre- upgrade collimator PSF]
		{\includegraphics[width=0.5\textwidth]{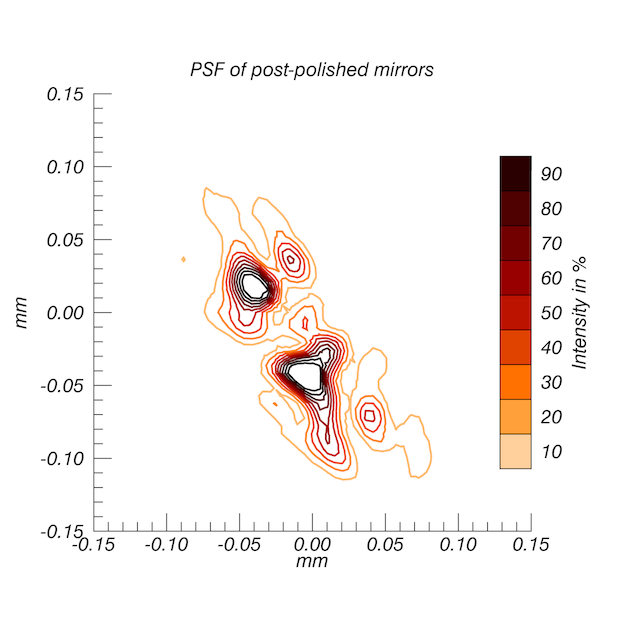}}
		\subfloat[\label{fig:newcollpsf}post- upgrade collimator PSF]
		{\includegraphics[width=0.5\textwidth]{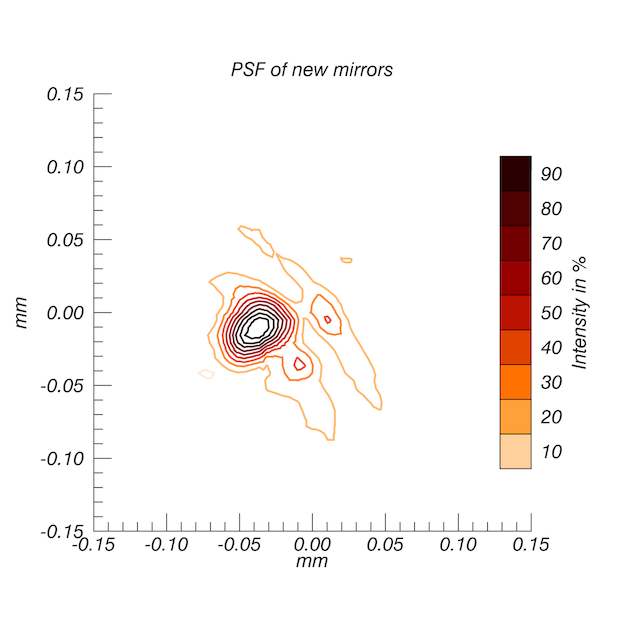}}
		\caption[Pre- and post- upgrade collimator PSF]{The best focus PSF of the full collimator mirror test setup at the position of the image slicer.}
		\label{fig:oldnewpsf}
	\end{center}
\end{figure}

\begin{figure}[htbp!]
	\begin{center}
		\subfloat[\label{fig:highfootprint}pre- upgrade collimator PSF (high footprint)]
		{\includegraphics[width=0.31\textwidth, trim={0.1cm 0 0.8cm 0.cm}, clip=true]{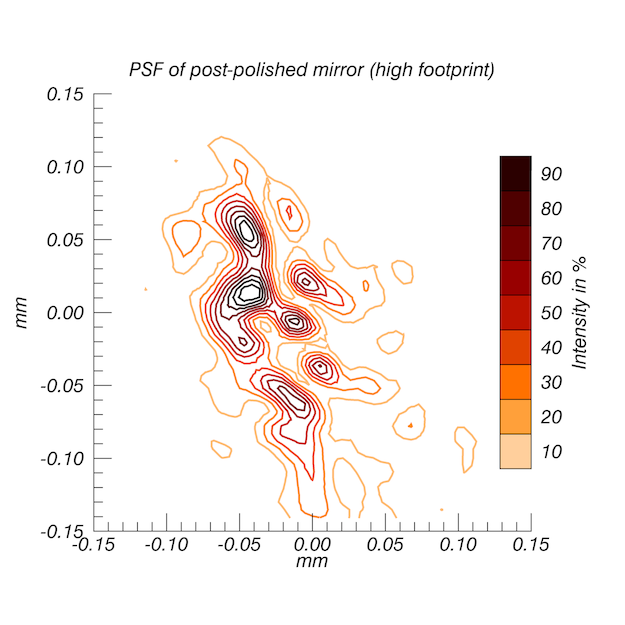}}
		\quad
		\subfloat[\label{fig:lowfootprint}pre- upgrade collimator PSF (low footprint)]
		{\includegraphics[width=0.31\textwidth, trim={0.1cm 0 0.8cm 0.cm}, clip=true]{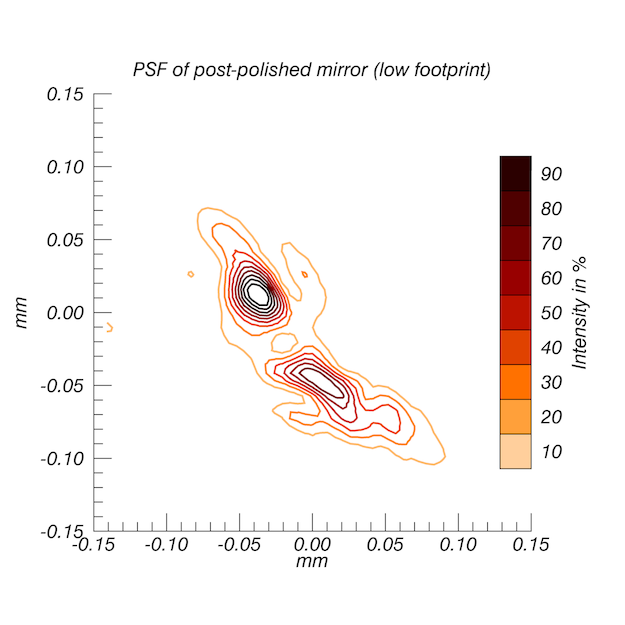}}
		\quad
		\subfloat[\label{fig:M2exchanged}pre- upgrade collimator PSF (center footprint) with M2 S02 implemented (compare with figure \ref{fig:oldcollpsf})]
		{\includegraphics[width=0.31\textwidth, trim={0.1cm 0 0.8cm 0.cm}, clip=true]{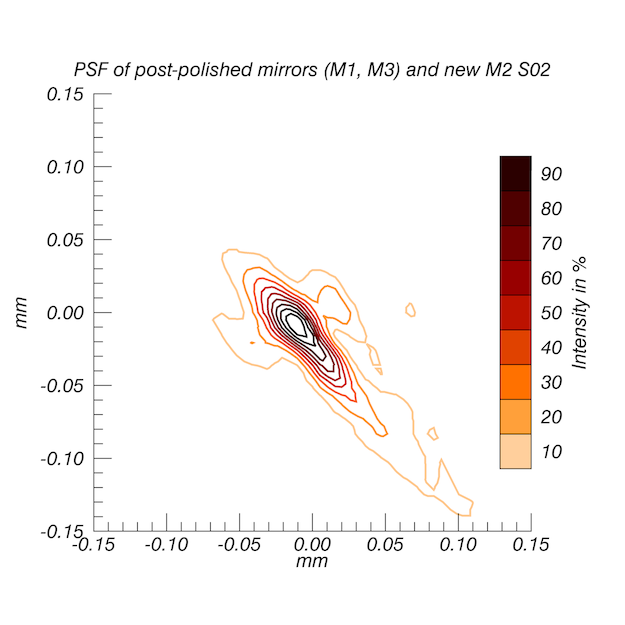}}
		\caption[Pre- upgrade collimator PSF with high and low footprint and with exchanged M2]{The PSF of the full collimator mirror test setup with a high field on M1 (left) and a low field on M1 (center). The right plot shows the PSF of the old collimator but with the new M2 S02 implemented.}
		\label{fig:lowhighfootprint}
	\end{center}
\end{figure}

	The post- upgrade wavefront shows a peak-to-valley surface deviation of $\mathrm{\sim 1.3 \ \mu m}$ and a RMS surface deviation of $\mathrm{\sim 0.20 \ \mu m}$. What significantly changed compared to the pre- upgrade wavefront is the variability and roughness of the wavefront. The pre- upgrade wavefront shows a complex non-symmetrical structure with valleys and humps irregularly distributed over the pupil. Furthermore residual diamond turning marks can be seen crossing vertically the pupil. They are mainly caused by the turning marks on M2. The grid structure from the polishing process of M3 has also part in the irregular appearance of the wavefront. On top of the small scale structure one can also recognize some 45\degree \ astigmatism of about half a micron in the distorted wavefront. The new wavefront however is much smoother than the pre- upgrade wavefront. The small scale structures are on much smaller scales than before - close to the spatial resolution limit of the interferometer. Clearly visible are the third order aberrations in the interferogram, which is mainly coma with an amplitude of $\mathrm{0.3 \ \mu m}$, and also some residual 45\degree \ astigmatism. The coma can also be seen nicely in the PSF of the new collimator in figure \ref{fig:oldnewpsf}. These aberrations are introduced by the surface form error of M3. With the described shifting of M3 by 23' the 90\degree \ astigmatism could be completely eliminated and the 45\degree astigmatism reduced to around a quarter of a micron in wavefront error. The amount of coma was nearly not effected by this rotation of M3. With the new manufactured M3 it is expected for the wavefront to have an RMS wavefront error in the orders of $\mathrm{\lambda/14}$ in J-band by design corresponding to less than $\mathrm{\sim 0.1 \ \mu m}$. 
	
\end{sloppypar}

This behavior of the collimator wavefronts was also seen in a measurement of the collimator PSF. For this, in the focus position of the full collimator mirror test setup a CMOS detector with a pixel size of $\mathrm{5.2 \mu m / px}$ was placed. To find the best focus position a through-focus scan was done. The results are shown in figure \ref{fig:oldnewpsf} and \ref{fig:lowhighfootprint}. The four peaks in the PSF of the pre- upgrade mirrors correspond to the four regions of the wavefront (two valleys and two humps). The dispersion direction of the grating is in the left-right direction. The width of one slitlet of the small image slicer is $\mathrm{300 \ \mu m}$. One can see that both the pre- and the post- upgrade collimator PSF is smaller than one slitlet. Because of this fact the collimator PSF is cannot be responsible for the double or triple peaks in the spectral line profiles, even if the collimator PSF is double-peaked. The effect of this will be an out-smearing of the line profiles and maximal the appearance of shoulders at the wings of the spectral line profiles. Distinct side-peaks are not possible to get by convolving the slitlet with the old collimator PSF. The collimator PSF double peak shape was changed significantly due to the upgrade. The residual coma is seen on the right of the new collimator PSF in figure \ref{fig:newcollpsf}. Furthermore, as one can see from the PSFs in figure \ref{fig:lowhighfootprint}, the PSF of the old collimator varied strongly in its shape over the field, resulting in varying spectral line profiles over the detector. For this measurement a high and a low footprint on M1 was taken in order to see the PSF variation over the field. As a result of the change in the collimator PSF it is expected that the spectral line profiles are less smeared out after the upgrade and shoulders (not side-peaks) at the spectral lines are less distinct.

The PSF varies strongly with the field position. While a low footprint on M1 looks quite similar to the center field PSF, the high footprint looks more disturbed. This variation was not caused by M1, as one could imagine because different field positions correspond to different spots on M1, instead it was the result of the M2 mirror. The post-polished M1 surface is nearly ideal, the post-polished M2 surface however has a wavy structure in vertical direction. The beam footprints of the different slitlets are slightly overlapping on M2, but still separated. Each slitlet is reflected by a different surface area of M2, resulting in a variation of the spectral line profiles along the pseudo-slit. That it was indeed the post-polished M2 which caused this variation of the PSF can be seen in figure \ref{fig:M2exchanged}. There the post-polished M2 was exchanged by the new M2 S01. The double peak behavior of the PSF is gone, however the PSF is still elongated, which is caused by the old M3 also inducing the faint shape around the elongated peak with two very slight side bumps. For the new collimator, this behavior of a variation of the PSF over the field is not seen. The PSF stays quite the same over all the field points along the pseudo-slit. As a result of the change in the collimator PSF it is expected that the spectral line profiles are less smeared out after the upgrade and shoulders (not side-peaks) at the spectral lines are less distinct.

\section{Interpretation and Effect on the Spectrometer Performance}
With the new collimator mirrors the effect of the now very tiny residual diamond turning marks is negligible. It is not expected that they have an influence on the LSF of the spectrograph. The collimator PSF was improved by a lot due to the exchange of the mirrors, mainly because of the exchange of M2 and M3. The now less disturbed but asymmetric wavefront leads as already before the upgrade to asymmetric spectral line profiles, which means that the shoulders of the line profiles will be reinforced on the one side of the line compared to the other side (see section \ref{sec:lineprofiles}). The main effect of the old collimator wavefront respectively PSF, was a smearing of the spectral line profiles, which is now expected to be much smaller. This leads to sharper line profiles, meaning that features that were smeared out before are visible now. Also the smoother surfaces of M2 and M3 result in a smaller variation in the line profiles over the pseudo-slit. This makes it possible to distinguish more detailed features in emission and absorption lines (see section \ref{sec:resolution}). 

With the upgrade to SPIFFIER and the implementation of the new manufactured M3, shoulders in the line profiles will even get less asymmetric and thus the LSF will be even be sharper by a small amount.

Due to the fact that the collimator mirrors have been exchanged and the new mirrors do not show diamond turning marks, it is clear that the collimator mirrors are not the source of the non-ideal spectral line profiles in the SPIFFI spectrometer. This is a major result of the upgrade and it is supported by the results of the next chapter, in which the performance of the SPIFFI instrument is under investigation.

\chapter{Performance Analysis}\label{ch:chapter_performance}
To determine how the SPIFFI upgrade affected the performance of the instrument (and therefore the science performance), an analysis of the throughput, the image quality, and the spectral line profiles is done in this chapter. This analysis connects the changes in performance to the opto-mechanical parts that have been exchanged in the upgrade. For the remaining non-idealities in the performance that were expected to improve due to the upgrade but did not, other optical components are taken into account in order to explain these effects.

\section{Relative Throughput}\label{sec:throughput}
Of strong interest for the operation of the telescope is the absolute throughput, meaning what fraction of photons entering the telescope are actually detected on the SPIFFI detector. In this thesis only a relative throughput to the pre- upgraded instrument is given. In order to determine the relative throughput, exposures of three standard stars were taken before and right after the upgrade in all bands and all pixelscales under clear conditions. These stars were HD38921 (A0V, 7.5 mag in H-band), HD49798 (O6, 9.0 mag in H-band), and HD75223 (A1V, 7.3 mag in H-band). Only the first two stars were used for the throughput measurement, because by the time the third star was measured on January 30\superscript{th}, the seeing increased to over 2". The raw data was reduced with the SINFONI \lq spred-pipeline\rq \ of MPE (see \cite{abuter06} and \cite{schreiber04}). Since in the upgrade, the J-, H- and K-band filters have been exchanged, the throughput as a function of wavelength changed (see section \ref{sec:filters}). Due to this, the images were not flat-fielded in order to see the relative filter characteristics on top the transmission of the new pre-optics lenses and the collimator mirrors. This has the effect that the spatial varying response of the detector is not taken into account, introducing the first photometric error. By trying to place the star in similar regions on the detector for the pre- and post- upgrade observations, an attempt is made to minimize this error. The result of the data reduction is a SINFONI data cube with two poorly resolved (60x64 pixel) spatial directions and one highly resolved spectral direction.

Depending on the seeing and the AO performance, the PSF of the observed star is widened. In the AO pixelscale, where the PSF is over the Nyquist sampling level, a not negligible fraction of the flux falls of the detector, depending on the position of the PSF on the detector. In the two larger pixelscales only a tiny fraction of the flux will fall of the detector. One important step to do reliable photometry is to define a region over which the flux for each wavelength is integrated. This is done here as following: To get the highest possible signal to noise ratio and to reduce the noise in the background the central 1000 spectral pixels are averaged for each spatial pixel. This is only done for the central 1000 pixel to get rid of all effects atmospheric absorption features that show up at the edges of the band pass filters. This wavelength averaged PSF is used to define the center of the circle for the integration and to estimate the fluctuation of the sky background as well as the offset of the background. The estimation is done by taking the mean flux in the corners of the image plus the mean absolute deviation. This estimated sky is now subtracted from the averaged image. The integrated encircled energy for radii reaching from 1 pixel to the maximum distance from the center of the PSF to the edge of the image frame in pixels is calculated. The radius over which it is integrated later is then determined by the radius at which the encircled energy function is flat. More accurate speaking, the radius at which the slope is less than 1\permil per pixel, exclusively. The rest of the PSF flux is then in the noise. The PSF for each spectral pixel of the detector is integrated in the spatial directions out to the radius determined in the previous step. This method of determining the integration radius compensates for the variations in seeing and AO performance.

From this one obtains spectra of the stars for all bands in the two larger pixelscales weighted by the total flux at a certain wavelength. These are now corrected for the airmass using the Beer-Lambert law \cite{horvath93} as a model for the attenuation of light by the atmosphere. This is done by using a simple decreasing exponential function with the extinction coefficient per magnitude for the corresponding wavelength multiplied by the airmass in the exponent (see table \ref{tab:extinction}).

\begin{table}[htbp!]
	\begin{center}
		\begin{tabular}{c|c}
			\hline
			Band&Extinction Coefficient [$mag / airmass$]\\
			\hline
			\hline
			J	&	$0.072 \pm 0.040$\\
			H 	&	$0.034 \pm 0.015$\\
			K	&	$0.043 \pm 0.013$\\
			H+K	&	$0.039 \pm 0.015$\\
			\hline
		\end{tabular}
	\end{center}
	\caption[Atmospheric extinction coefficients at Paranal]{Atmospheric extinction coefficients at Paranal used for airmass correction. \cite{lombardi11}. For H+K-band the average of H- and K-band is taken while the uncertainty is estimated to be as large as the maximum error of H- and K-band.}
	\label{tab:extinction}
\end{table}

\begin{figure}[htbp!]
	\begin{center}
		\resizebox{0.9\textwidth}{!}{
			\includegraphics[width=1.\textwidth]{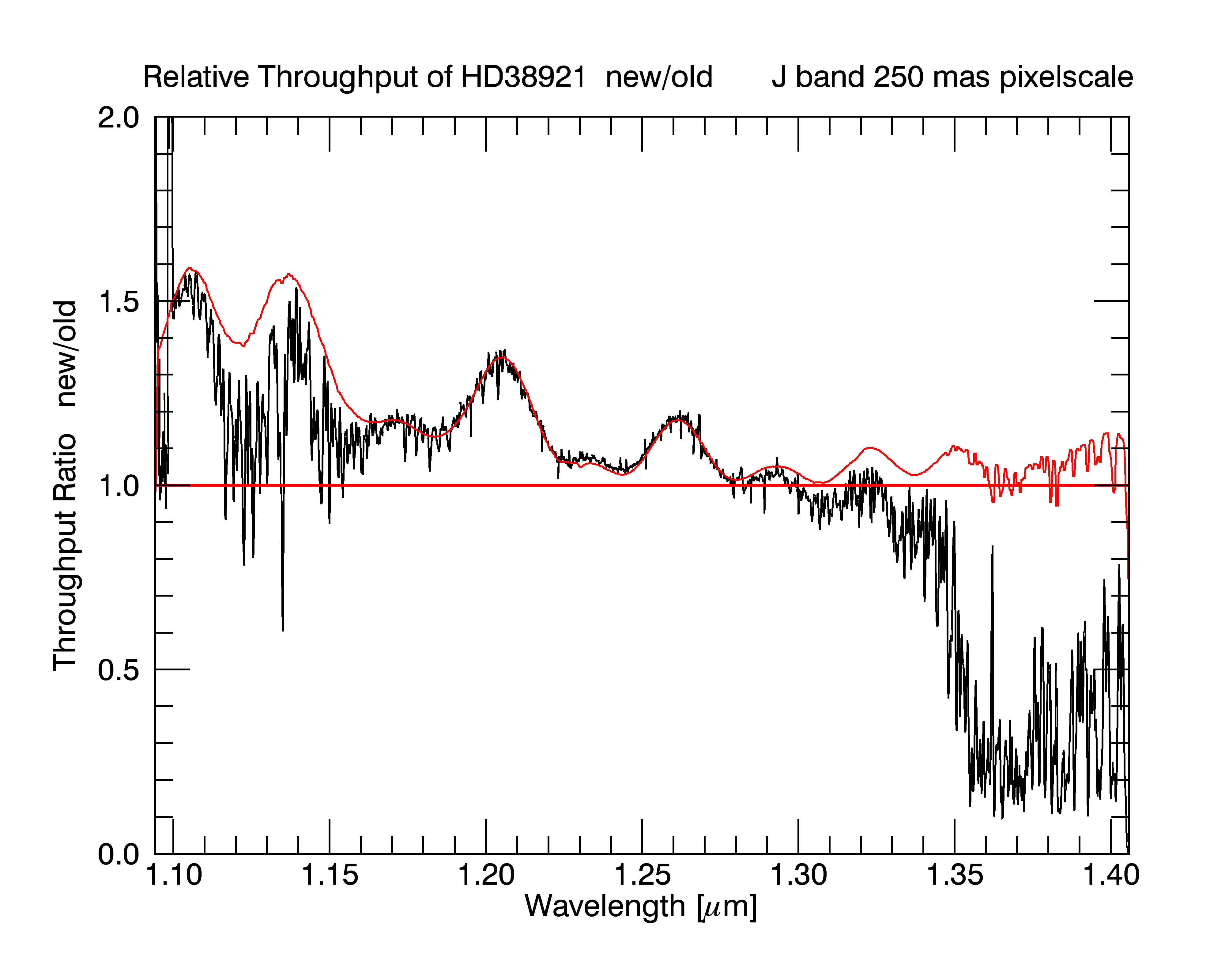}
			\includegraphics[width=1.\textwidth]{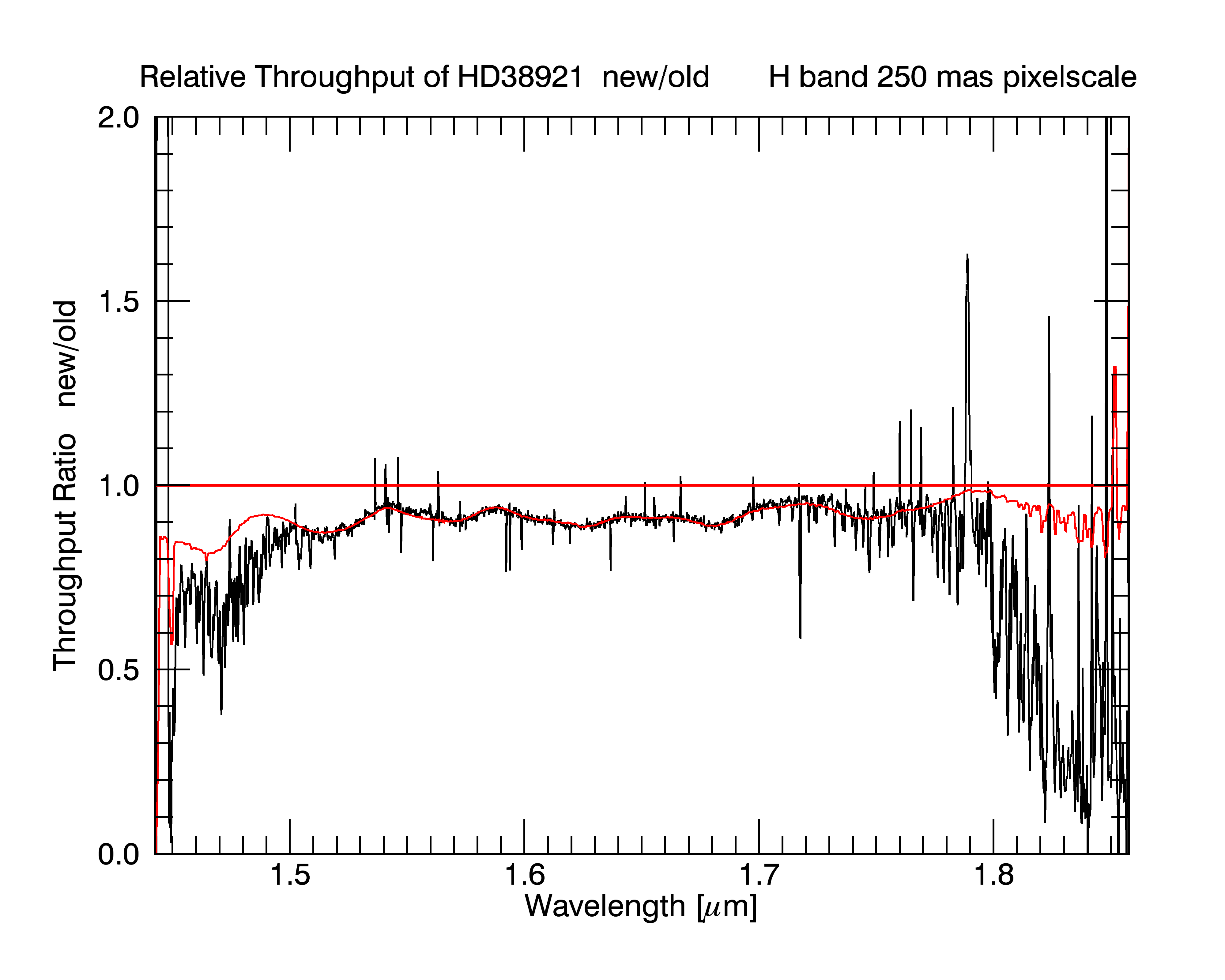}
		}
		\resizebox{0.9\textwidth}{!}{
			\includegraphics[width=1.\textwidth]{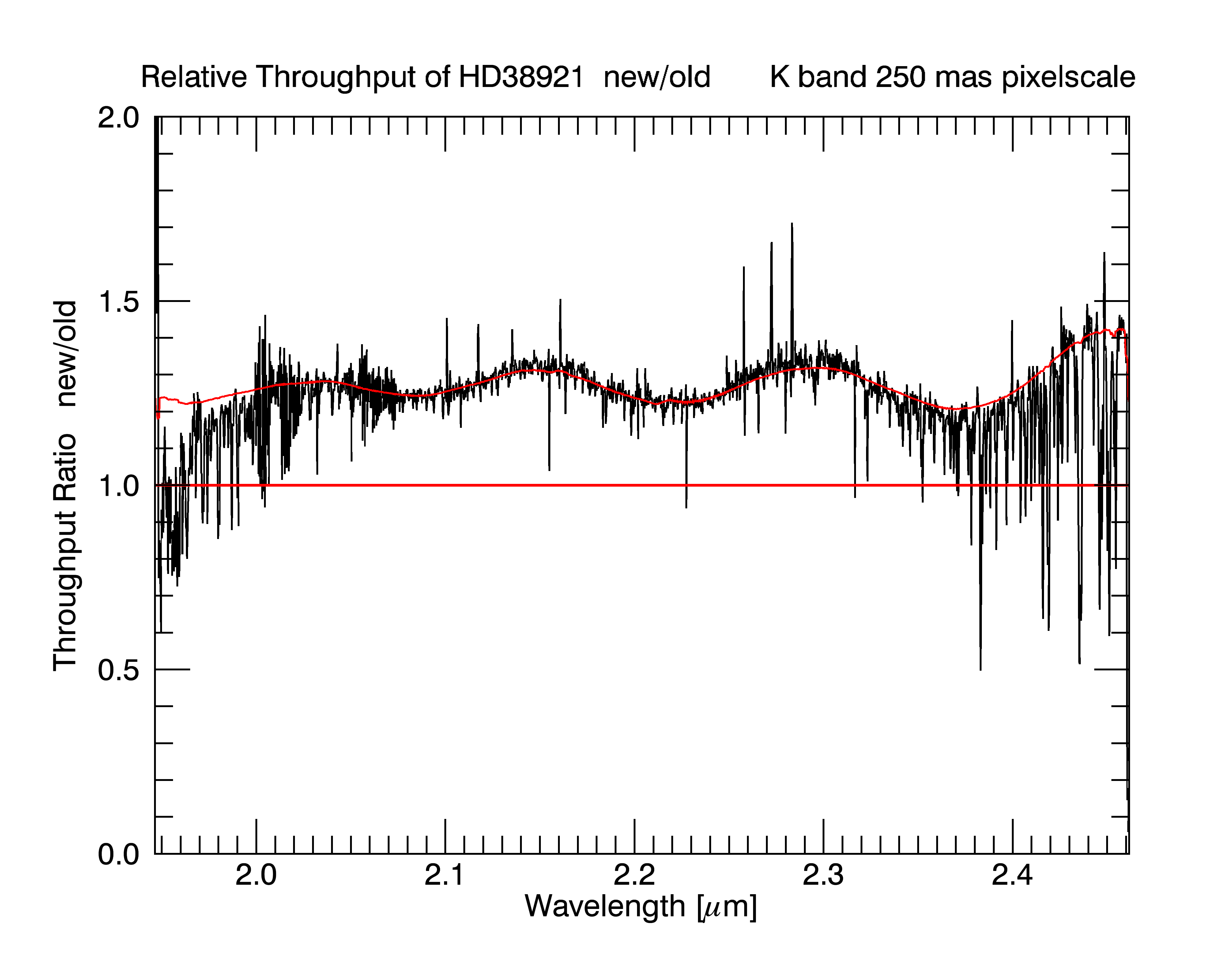}
			\includegraphics[width=1.\textwidth]{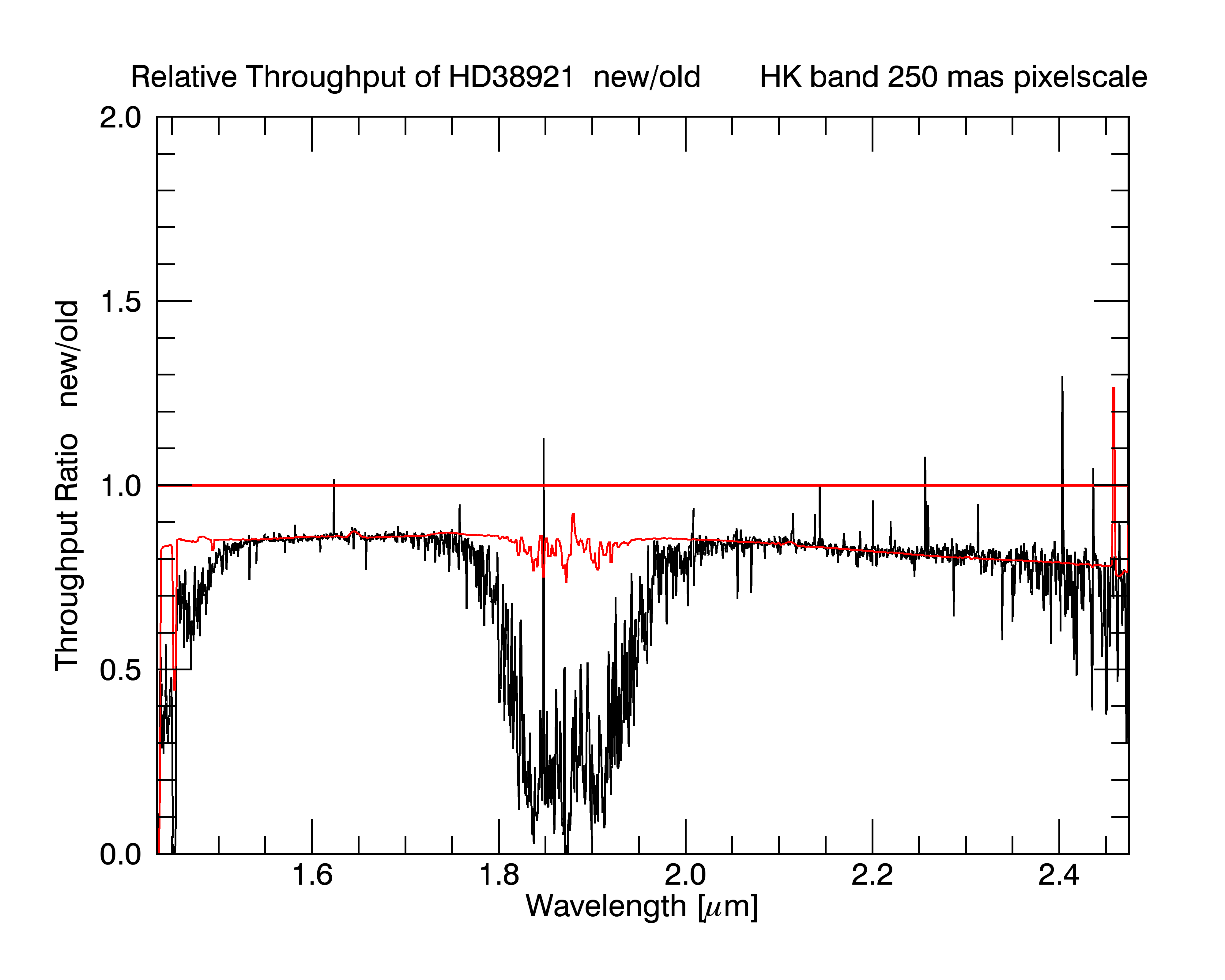}
		}
		\caption[Ratio between spectra for throughput]{The airmass corrected ratio between the spectra of HD38921 in the 250 mas pixelscale with the fitted flat field throughput ratio. Only the slope and the offset are fitted.}
		\label{fig:flatfit}
	\end{center}
\end{figure}

In these spectra there are atmospheric absorption lines that vary in depth depending on the atmospheric conditions at which the exposure was taken. The first approach here was to remove them by using the ESO atmosphere modeling tool \lq molecfit \rq \ \cite{smette14}\cite{kausch15}. This worked quite well in the center regions of the atmospheric bands, but it could not correct for the strong $\mathrm{H_2O}$ absorption lines at the edges of the bands. In order to get also a reliable relative pre- post- upgrade throughput at the edges of the atmospheric bands, flat field exposures taken with the calibration unit were used. Since the halogen and spectral lamps of the calibration unit in MACAO were exchanged in the upgrade and since the photon flux of the halogen lamp depends on its temperature, the relative throughput measured with these lamps is not a good metric. But beside the slope and the offset caused by the different temperatures, the comparison of the flats shows the wavelength dependant transmittance of SINFONI. From this the shape of the relative throughput (new flux/old flux) is obtained. This curve is now fitted to the well shaped inner band curves of the relative throughput measurements of the two standard stars HD38921 and HD49798 in the two larger pixelscales in order to correct for the slope and the offset (see figure \ref{fig:flatfit}). These adapted throughput ratios of the flats are used for further analysis. They show a undisturbed shape, because they are not affected by atmospheric absorption lines. In the central region of H+K-band one can also see noise in the throughput ratio curves from the flats. This noise is caused by the absorption of water vapor in MACAO.

\begin{figure}[htbp!]
	\begin{center}
		\includegraphics[width=1.0\textwidth]{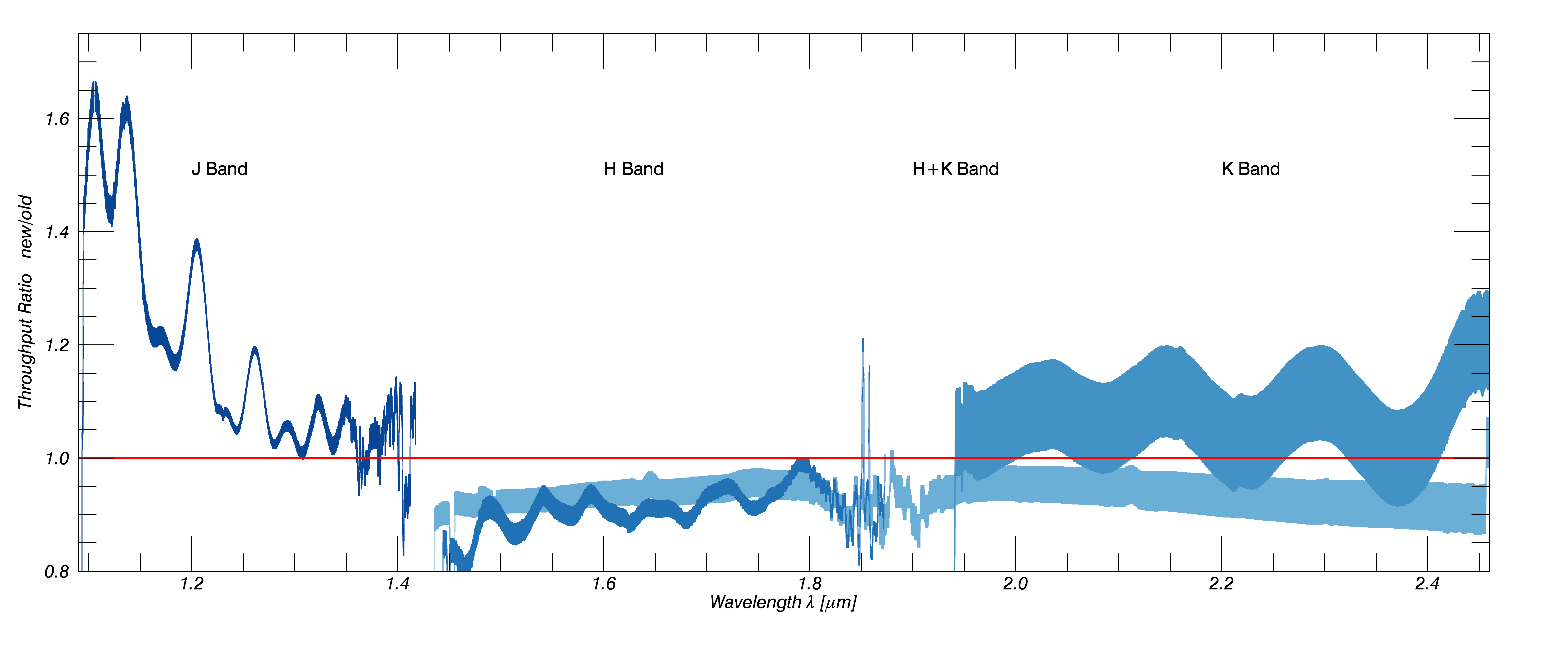}
		\caption[Relative throughput ratio pre- and post- upgrade]{The throughput ratio (new/old) over the wavelength range of the instrument. Each band is plotted separately, with the shaded regions indicating the standard deviation of the 4 ratios measured in each band. The shaded region does not include additional sources of systematic errors which are estimated to be of order $\mathrm{\sim 10\%}$.}
		\label{fig:throughput}
	\end{center}
\end{figure}

To reach a more reliable result, the average of both stars and both larger pixelscales was taken for each of the four bands. However, there is a significant offset between the measurements of the two stars, which can only be caused by the atmosphere, since for both stars the same method is applied. The result can be seen in figure \ref{fig:throughput}. The shaded region indicates the 1$\sigma$ error of the four averaged measurements. Systematic errors could be induced by pupil misalignments between SPIFFI and MACAO or dust buildup on the primary and secondary mirror of the telescope during the two month periode between the measurements. Also the result depends slightly on the radius chosen for the integration of the flux. These systematic errors are estimated to be $\mathrm{\sim 10\%}$.

That there is an systematic offset in the plot, one can see from the H+K-band throughput curve. Since no filter was exchanged in H+K-band and the transmittance of the pre-optics lenses should be higher, one would expect in general, that the relative throughput of SINFONI is a straight line around 1 with a tendency to a higher throughput.

In J-, H- and K-band the intrinsic wavelength dependence of the old filter (the new filters are flat over wavelength, see figure \ref{sec:filters}) can be seen quite well in the peak valley shape of the throughput ratio. In J- and K- band the throughput has essentially improved, while in H-band it remained the same or even slightly decreased. Since in H+K-band neither a large offset nor a variance dependent on the wavelength is expected, from both factors an systematic error of $\mathrm{\sim 10\%}$ can be estimated. There is one aspect of the throughput curves that indicates that the shape of H+K-band throughput ratio is real. This is that the H-band curve as well as the K-band curve follow the same behavior as the H+K-band curve. The H-band ratio is increasing with wavelength, while K-band ratio is falling. This bandpass filter independent but wavelength dependent behavior could be caused by the new coatings of the pre-optics lenses or the SiO\subscript{2} protection layer on the collimator mirrors. The reflectivity of the protected gold coating on the collimator mirrors is indeed by a few percent lower than the reflectivity of unprotected gold at edge of the H-band and increases to K-band in the same way as the relative throughput curve in H+K-band does. In general the relative position of the throughput ratios are consistent with the improvement of the filter throughput. The largest improvement by the filters is expected in J-band, followed by K-band and only a very small improvement in H-band.

The large throughput gain in J-band will make it easier to study the galaxy formation and evolution using the H$\alpha$ line in the redshift range around $\mathrm{z \sim 1}$. Furthermore, the more constant throughput over the wavelength range will result in a more constant flux over the spectra, better statistics, and easier flat fielding.

Beside the throughput, the image quality of the instrument is of interest. In order to do so the Strehl ratio and the encircled energy per radius is measured in the next section.

\section{Image Quality}\label{sec:psf}
The quality of the image on the image slicer is mainly influenced by the pre-optics and the AO performance. The spectrometer and camera wavefront distortion effect therefore mostly the quality of the spectra. As a metric to measure the image quality, the Strehl Ratio (SR) is used. In order to compare the imaging quality of SPIFFI before and after the upgrade and not be influenced by the performance of the AO and atmospheric conditions, the Strehl measurements are done by using the calibration source of SINFONI. This sets an upper limit for the on-sky performance. The AO is used in closed-loop operation for this measurement. The single mode fiber of the calibration unit in MACAO is moved into the VLT focus providing a point source. The fiber is a standard $\mathrm{9 \ \mu m}$ single mode fiber with a Mode Field Diameter (MFD) from $\mathrm{\sim 9 \ \mu m}$ in J-band to $\mathrm{\sim 13 \ \mu m}$ in K-band (the diameter at which the power density is reduced to $\mathrm{1/e^2}$ of the central power density). Because there are only 32 image slicer slitlets and thus the reconstructed image has dimensions of only 32 x 60 pixels with pixels being twice as long as wide (The detector has 2048 x 2048 pixels so in an ideal case each slitlet gets 64 pixels for spatial imaging and 2048 in dispersion direction. Because of the distortion not all slitlets get the full pixel range of 64 pixels, so the reconstructed image is cut to 60 pixels within a slitlet). This poorly sampled image plane is the reason why a valid statement about the image quality can only be stated in the 25 mas pixelscale, where the diffraction limited PSF is sampled relatively well in comparison to the two larger pixelscales.

\subsection{Method for Strehl Ratio Determination}
The diffraction limiting optical component of SINFONI is the lyot cold stop between the pre-optics filter-wheel and the optics-wheel. For the simulation of the diffraction limited SPIFFI PSF only the diameter of the free pupil stop of $\mathrm{d_{free} = 6.436 \ mm}$ and the diameter of the central obscuration $\mathrm{d_{obscured} = 1.004 \ mm}$ is used. The spider arms holding the central obscuration are not taken into account, nor is the resulting Airy pattern convolved with the MFD of the calibration fiber. This is because the FWHM of the Gaussian intensity distribution from the fiber is small in comparison to the diffraction limited PSF. With a magnification of the fiber image on the slicer of

\begin{equation}
	M = \frac{F_{SPIFFI}}{F_{Cassegrain}}\frac{f_{Camera}}{f_{Collimator}} = 23.6
\end{equation}

\noindent with the F-number $F_{Cassegrain} / 14.1$ of the telescope, $F_{SPIFFI} / 17.1$ of the SPIFFI input beam and the focal length of the 25 mas pixelscale tube of the pre-optics camera $f_{Camera} = 2087.8 \ mm$ as well as the focal length of the pre-optics Collimator $f_{Collimator} = 112.7 \ mm$. The FWHM of the fiber intensity profile results thus in\\
\noindent$FWHM_{fiber} = \sqrt{\left(2 \ ln \ 2\right)} \cdot MFD \cdot M \approx 250 \ \mu m$    for J-band and\\
\noindent $FWHM_{fiber} \approx 350 \ \mu m$    for K-band on the small slicer, which is close to the width of one slitlet ($\mathrm{300 \ \mu m}$).

The radial part of the resulting Airy intensity distribution pattern at the paraxial focal plane of an diffraction-limited optical system with an annular circular aperture is given by\cite{rivolta86} 

\begin{equation}
	I(r) = \frac{1}{\left(1-\epsilon^2\right)^2} \left[\frac{2J_1(x)}{x}-\frac{2\epsilon J_1(\epsilon x)}{x} \right]^2
	\label{eq:bessel}
\end{equation}

\noindent which is the Fraunhofer diffraction pattern with the radial distance $r$ in the focal plane from the optical axis, the annular aperture obscuration ratio $\epsilon = d_{obscured} / d_{free} = 0.156$ and the Bessel function of the first kind in first order $J_1(x)$ with its argument $x = \frac{2 \pi r}{\lambda} sin(\alpha)$ the wavelength $\lambda$ and the angle $\alpha$ between optical axis and the line between aperture center and observation point.

\begin{figure}[htbp!]
	\begin{center}
		\resizebox{1.0\textwidth}{!}{
			\includegraphics[width=0.333\textwidth, trim={2.cm 0 2cm 0cm}, clip=true]{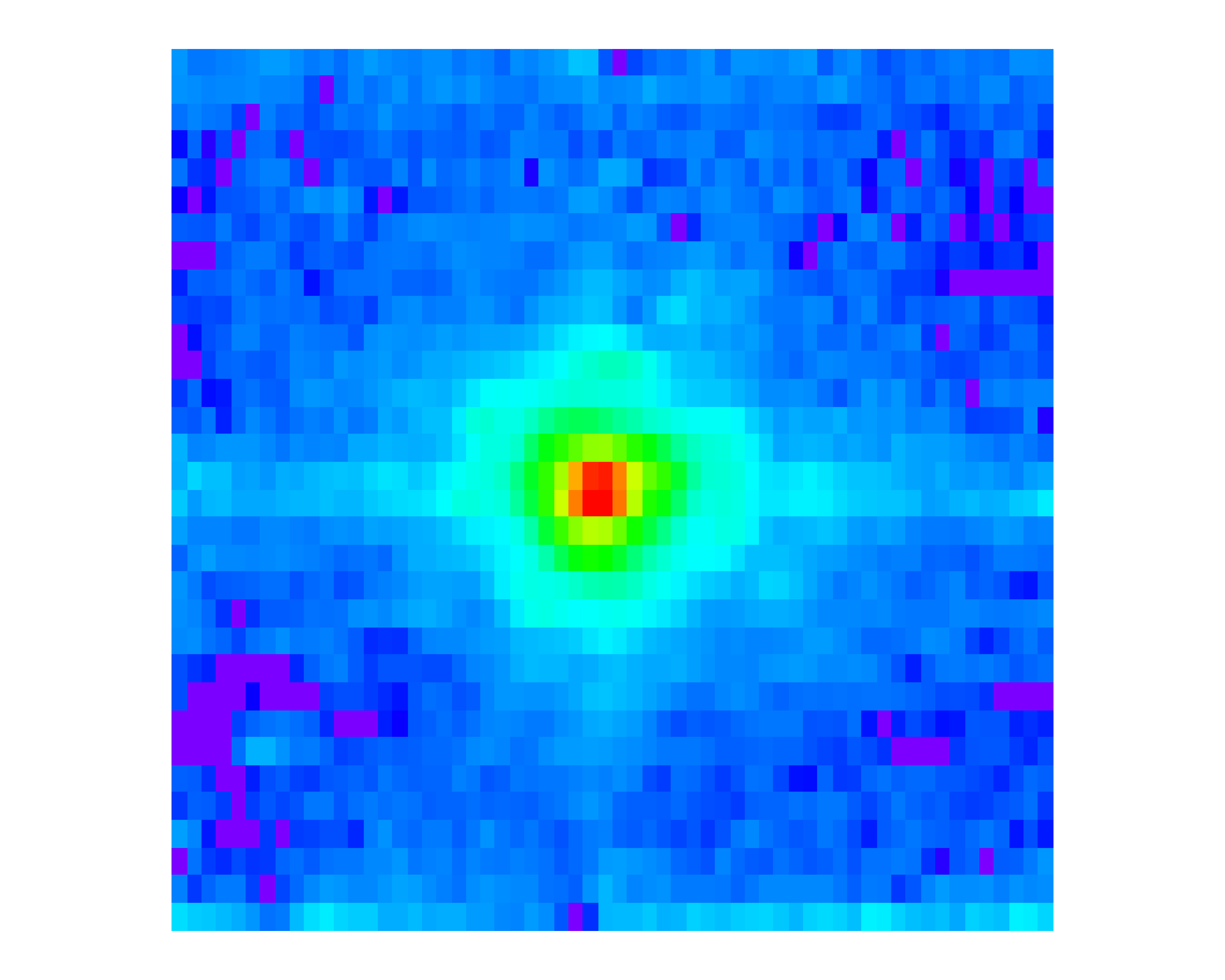}
			\quad
			\includegraphics[width=0.333\textwidth, trim={2.cm 0 2cm 0cm}, clip=true]{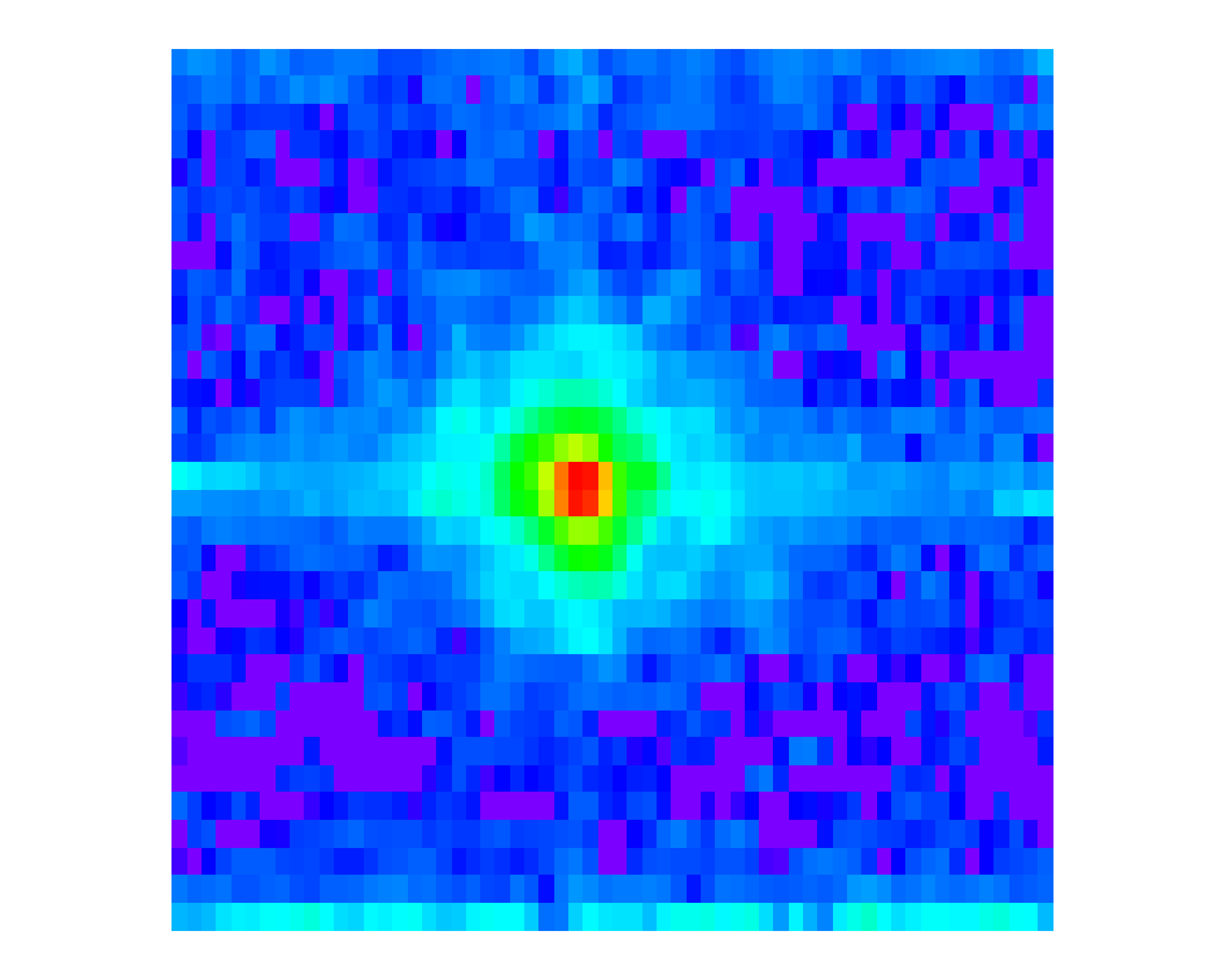}
			\quad
			\includegraphics[width=0.333\textwidth, trim={2.cm 0 2cm 0cm}, clip=true]{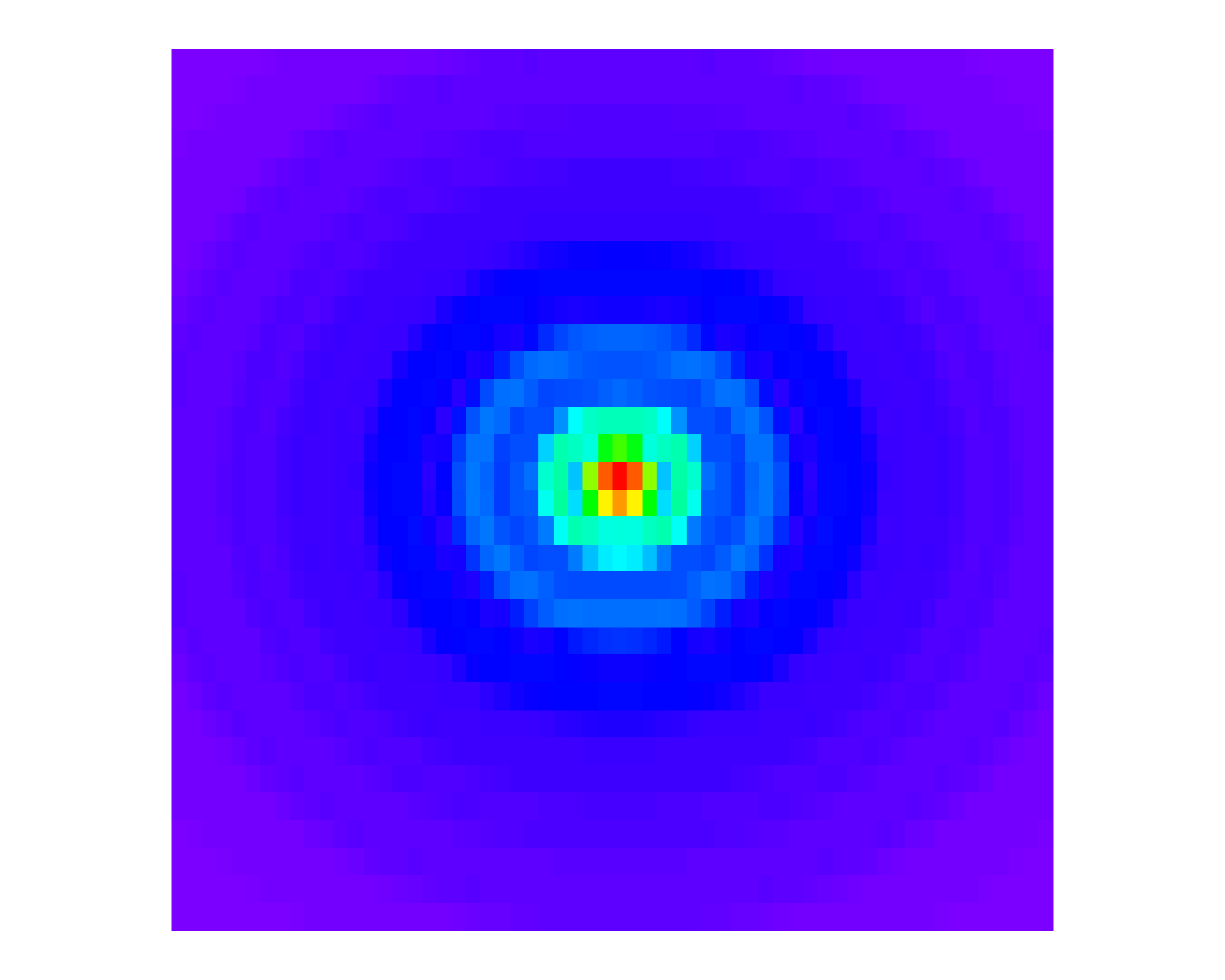}
		}
		\resizebox{1.0\textwidth}{!}{
			\includegraphics[width=0.333\textwidth, trim={2.cm 0 2cm 0cm}, clip=true]{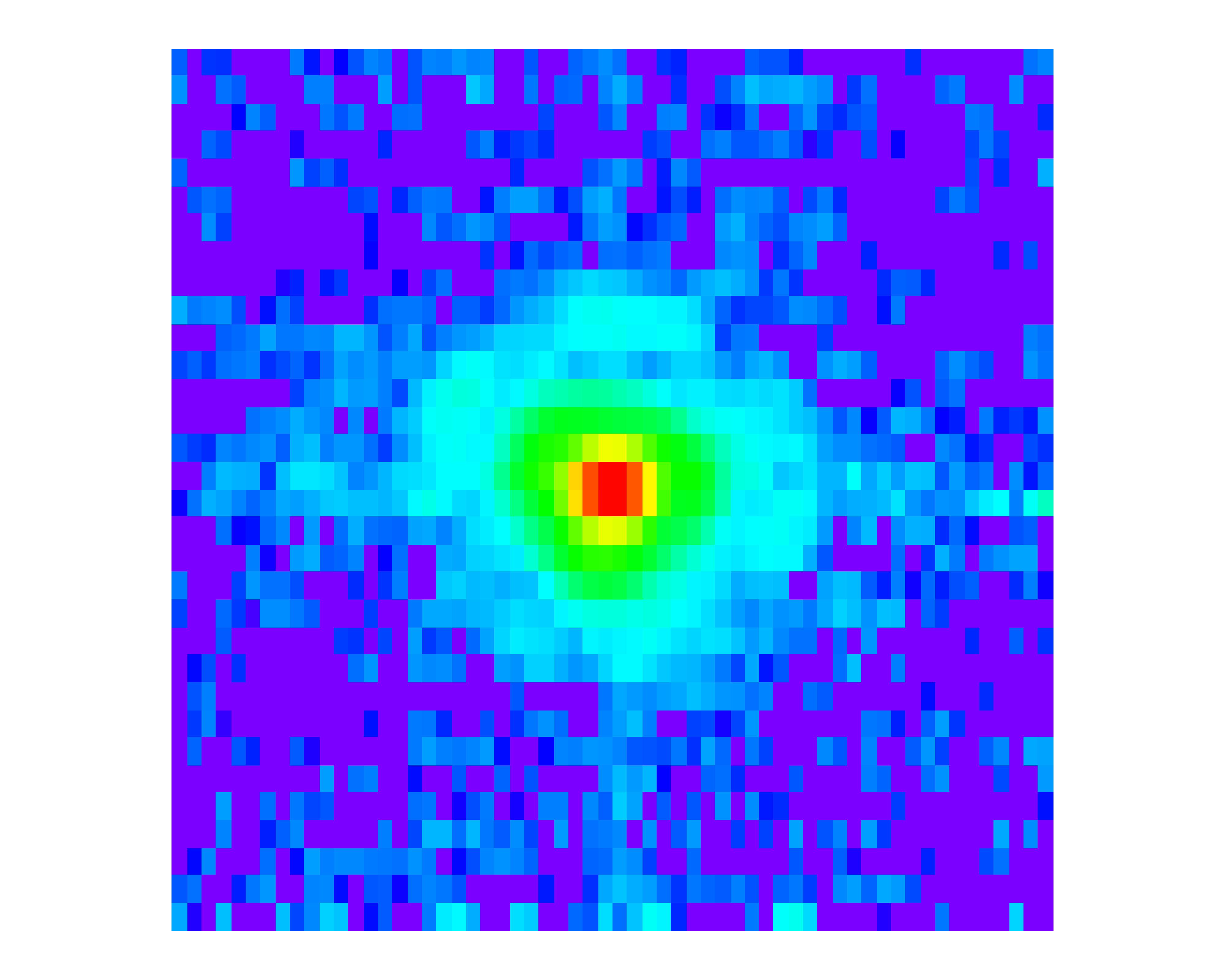}
			\quad
			\includegraphics[width=0.333\textwidth, trim={2.cm 0 2cm 0cm}, clip=true]{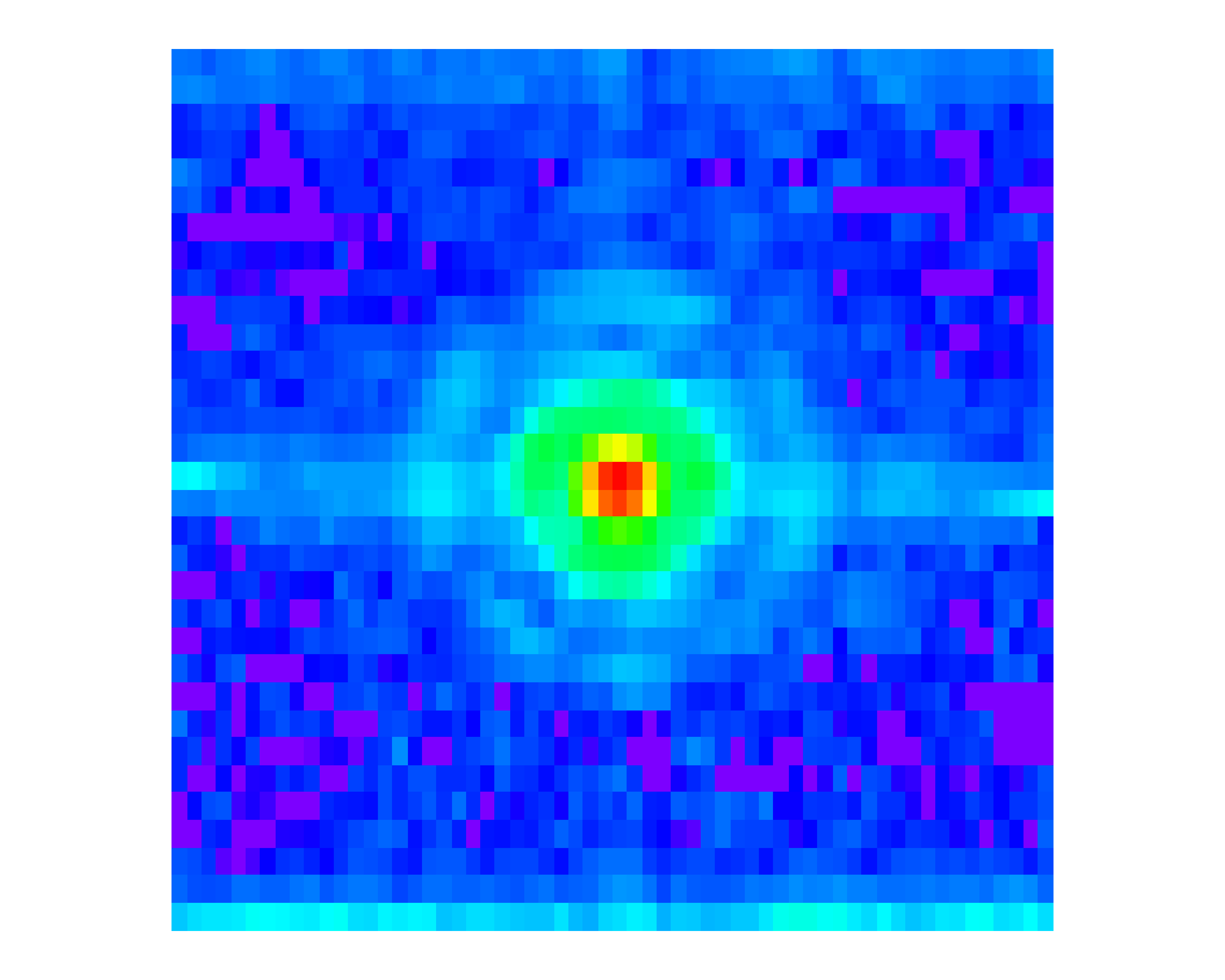}
			\quad
			\includegraphics[width=0.333\textwidth, trim={2.cm 0 2cm 0cm}, clip=true]{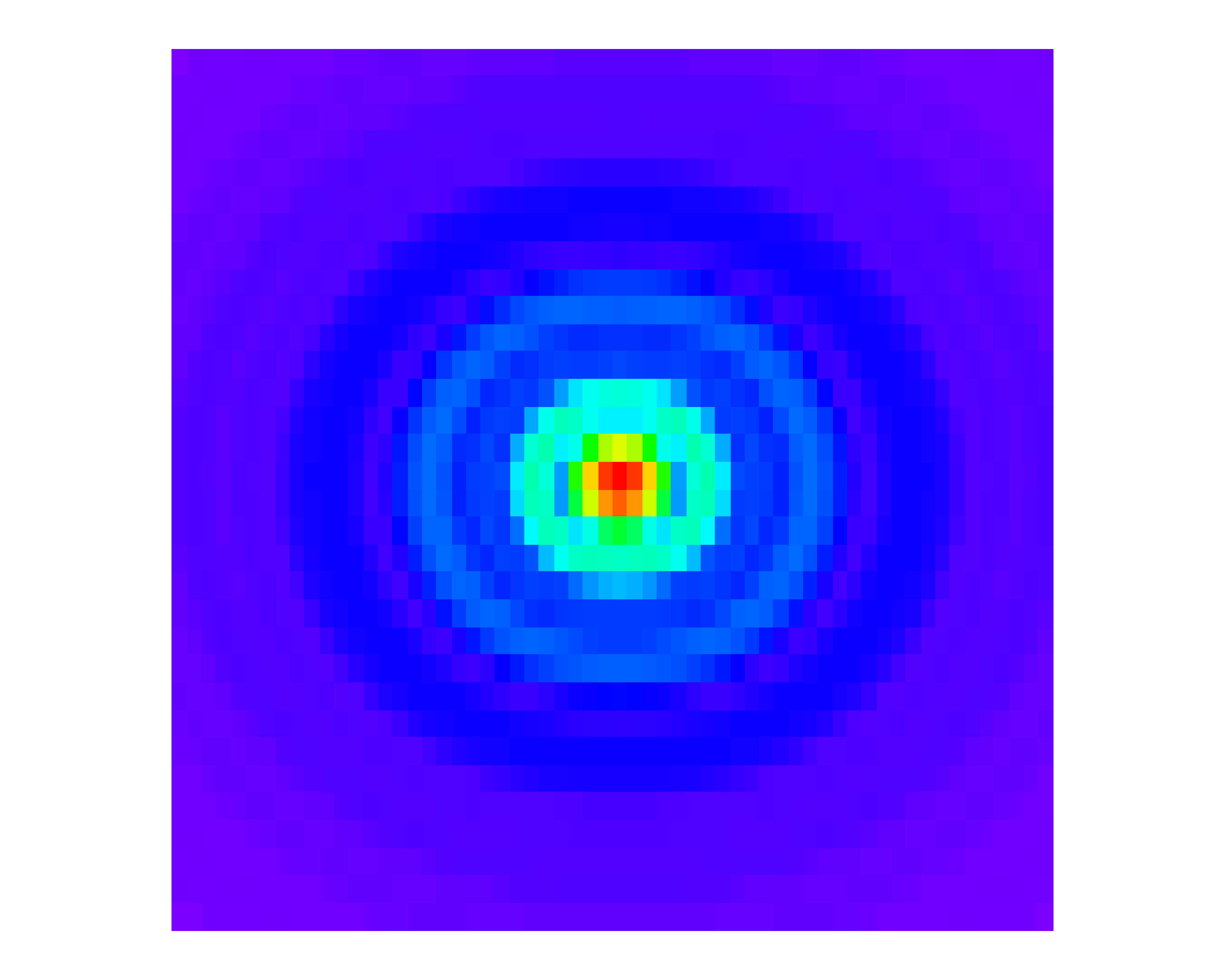}
		}
		\resizebox{1.0\textwidth}{!}{
			\includegraphics[width=0.333\textwidth, trim={2.cm 0 2cm 0cm}, clip=true]{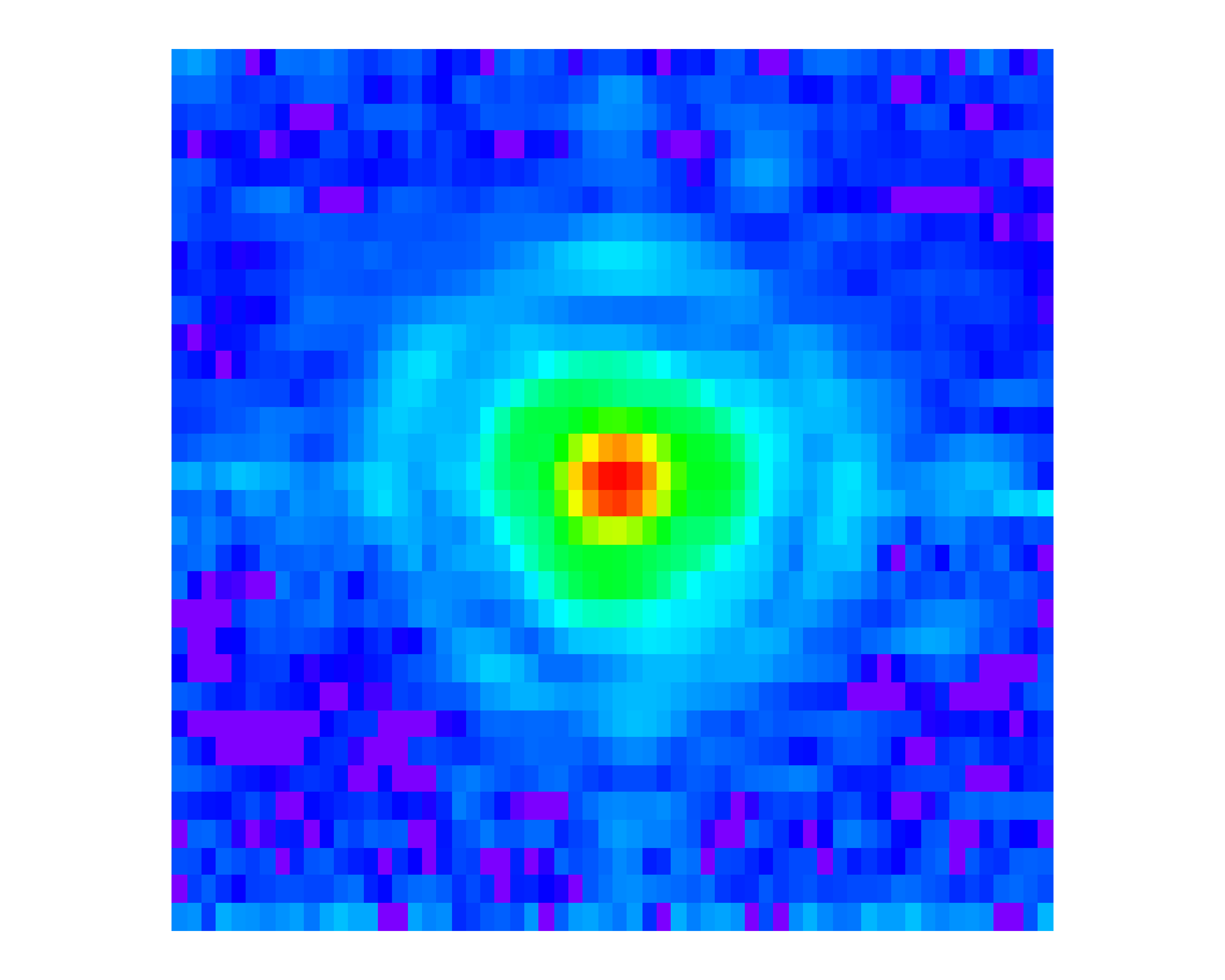}
			\quad
			\includegraphics[width=0.333\textwidth, trim={2.cm 0 2cm 0cm}, clip=true]{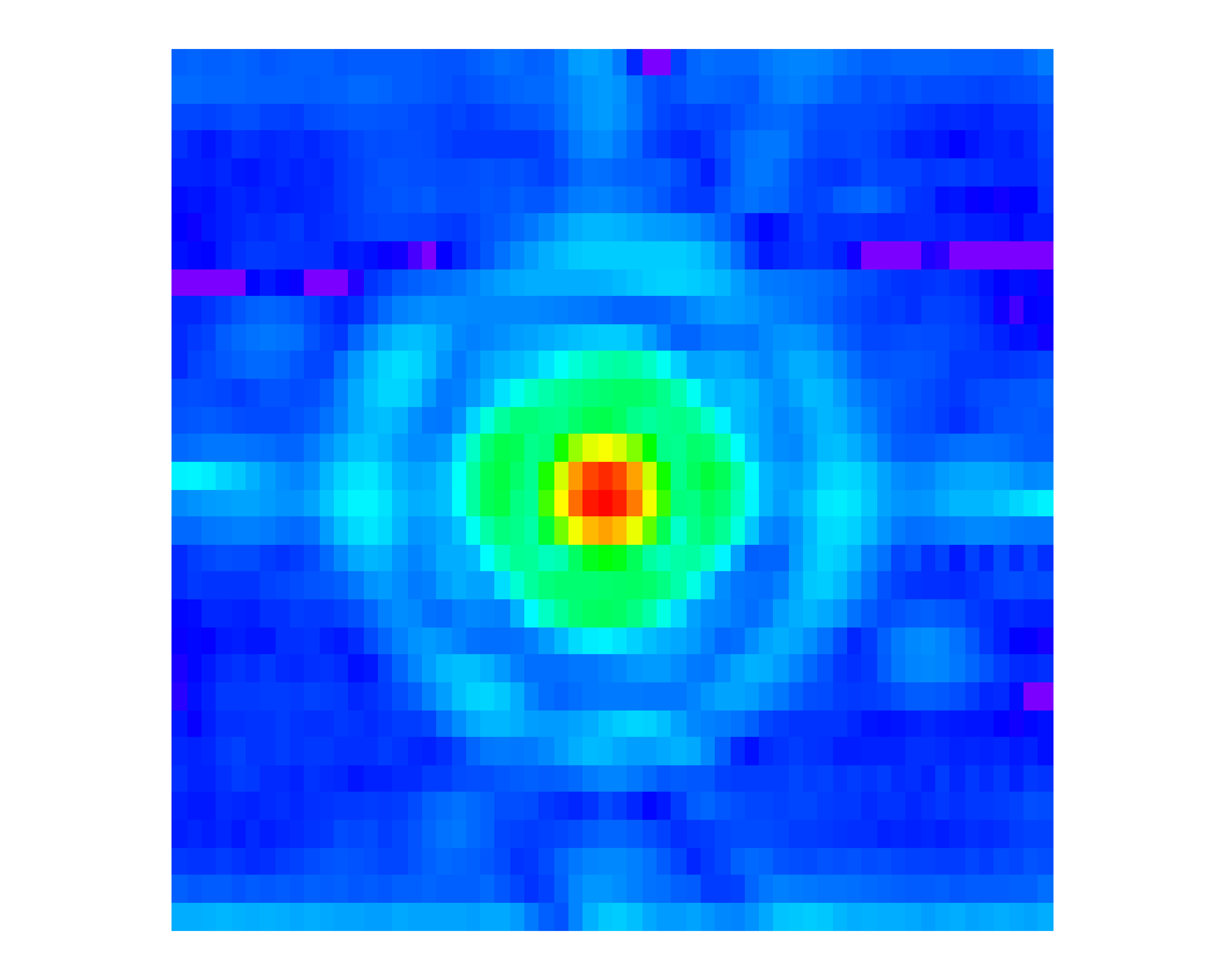}
			\quad
			\includegraphics[width=0.333\textwidth, trim={2.cm 0 2cm 0cm}, clip=true]{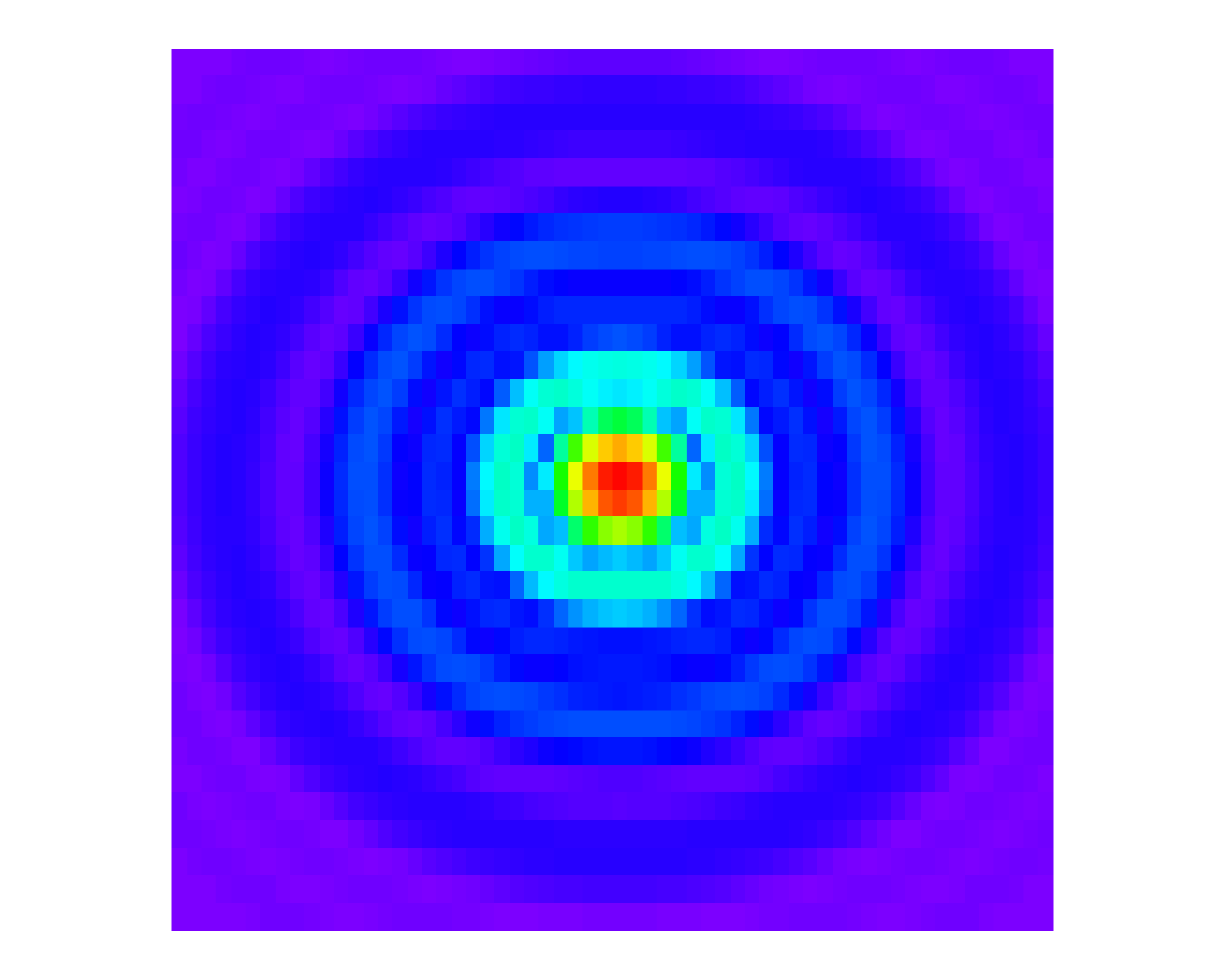}
		}
		\resizebox{1.0\textwidth}{!}{
			\includegraphics[width=0.3\textwidth, trim={2.cm 0 2cm 0cm}, clip=true]{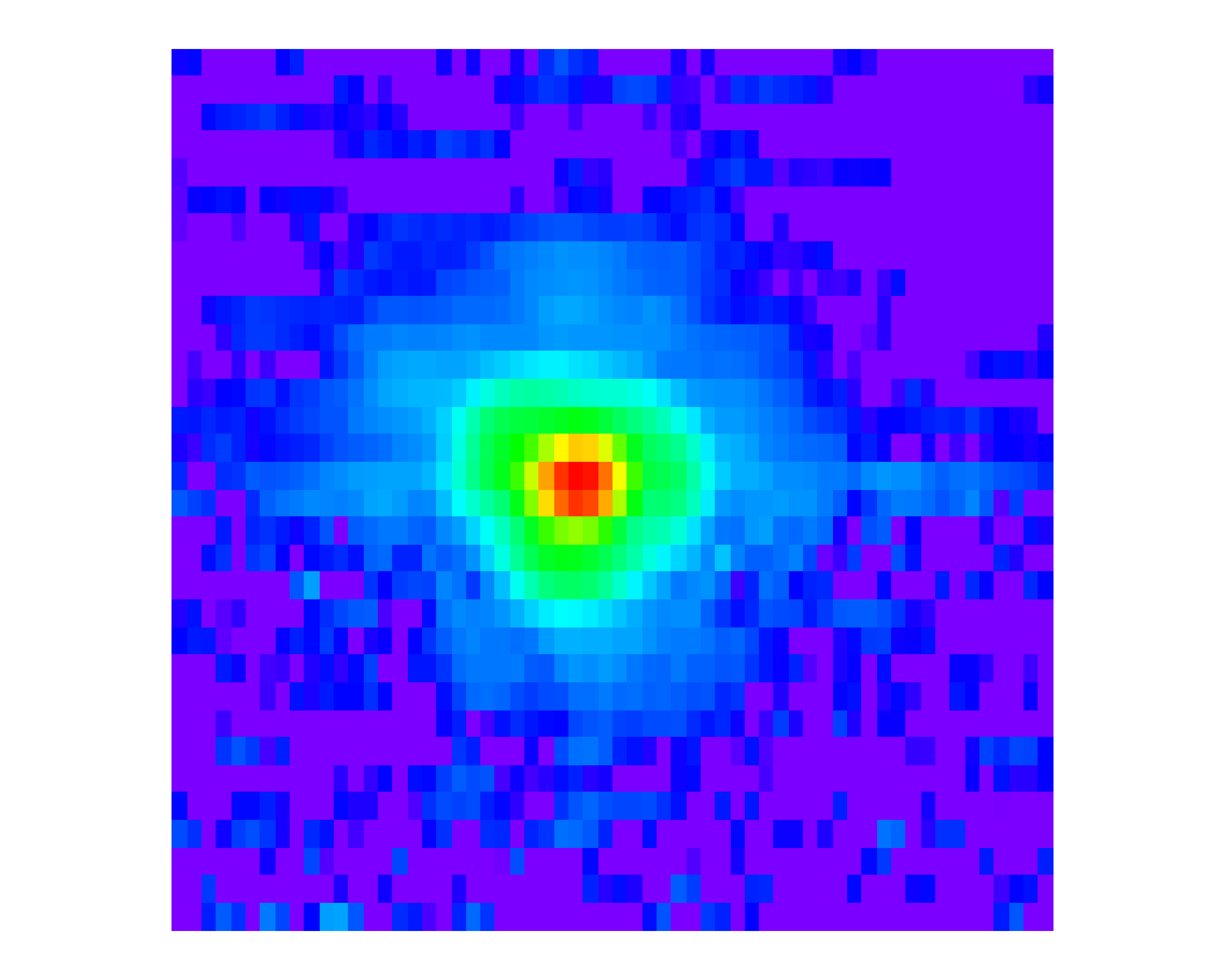}
			\quad
			\includegraphics[width=0.3\textwidth, trim={2.cm 0 2cm 0cm}, clip=true]{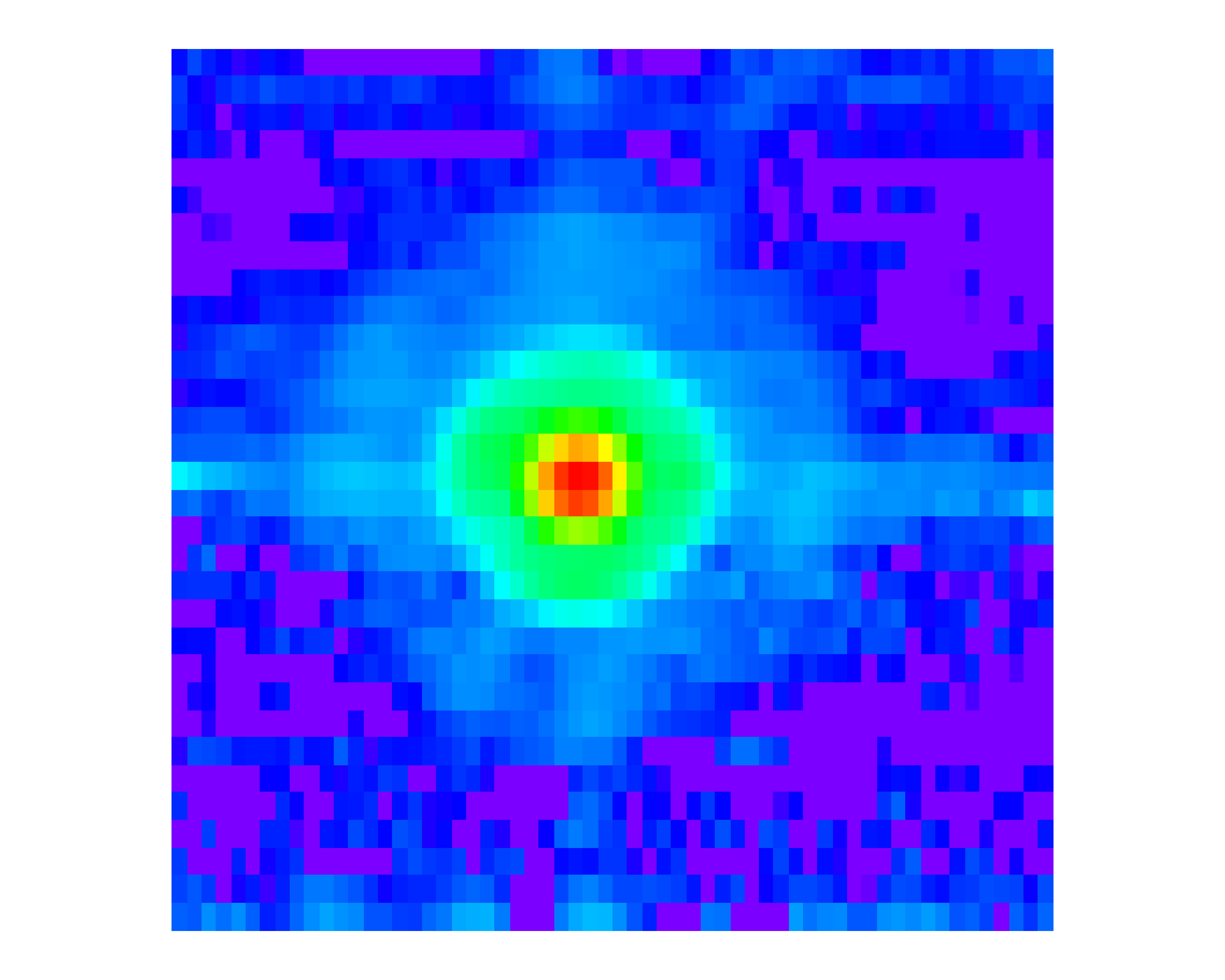}
			\quad
			\includegraphics[width=0.3\textwidth, trim={2.cm 0 2cm 0cm}, clip=true]{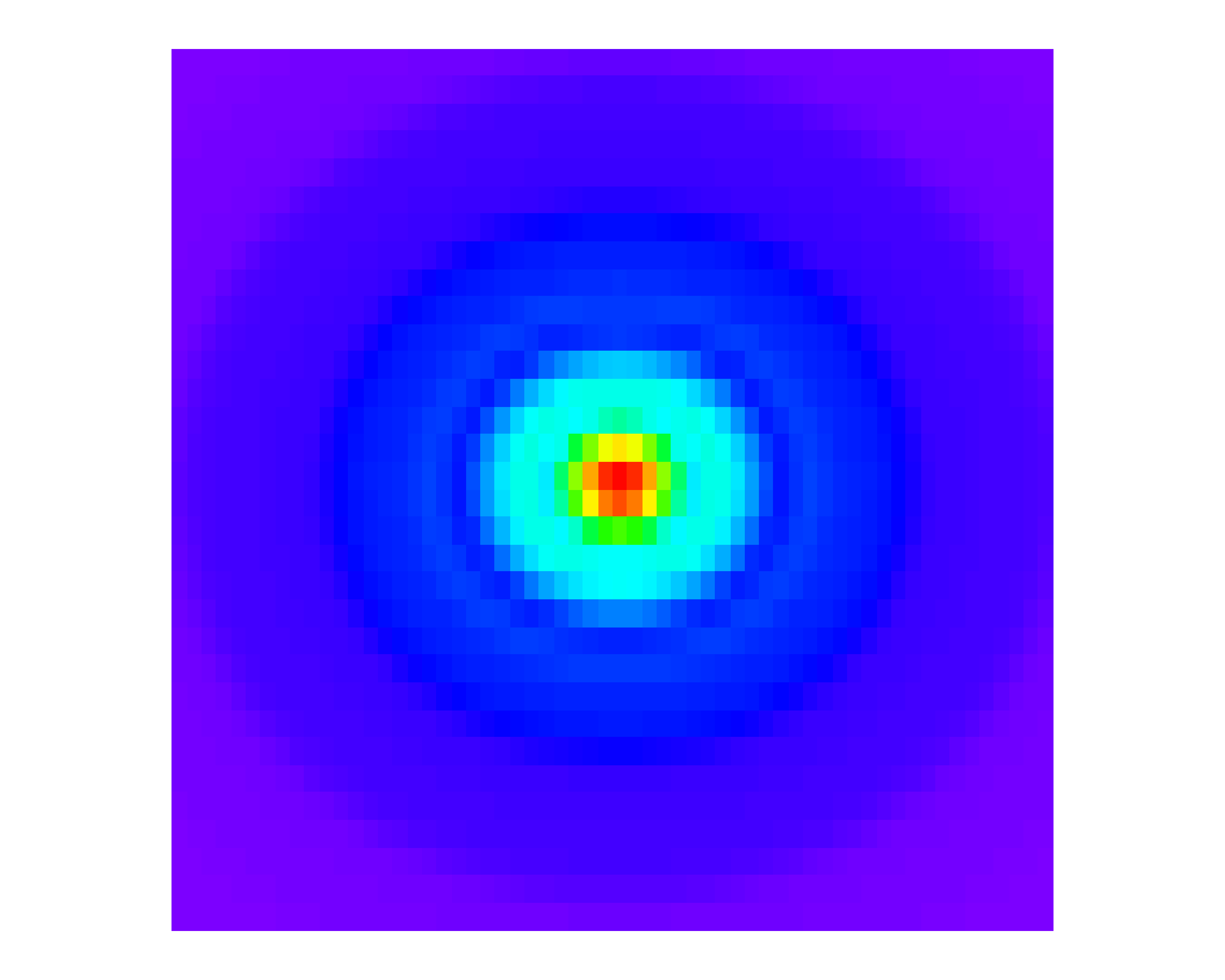}
		}
		\subfloat[measured pre- upgrade PSF]
		{\includegraphics[width=0.3\textwidth, trim={2.cm 0 2cm 16cm}, clip=true]{PSFtheory_hk_withoutfiber_new_color_rainbow.png}}
		\quad
		\subfloat[measured post- upgrade PSF]
		{\includegraphics[width=0.3\textwidth, trim={2.cm 0 2cm 16cm}, clip=true]{PSFtheory_hk_withoutfiber_new_color_rainbow.png}}
		\quad
		\subfloat[simulated PSF]
		{\includegraphics[width=0.3\textwidth, trim={2.cm 0 2cm 16cm}, clip=true]{PSFtheory_hk_withoutfiber_new_color_rainbow.png}}
		\caption[PSF in 25 mas pixelscale]{The measured PSFs in the 25 mas pixelscale. Rows from top to bottom: J-band, H-band, K-band, H+K-band. The cross in the images is a result of the spider arms supporting the central obscuration of the cold stop. They were not simulated and thus not visible in the simulated PSF}
		\label{fig:psfs}
	\end{center}	
\end{figure}

To simulate a PSF for a certain wavelength range, from equation \ref{eq:bessel} a theoretical PSF was calculated for 10000 values within the wavelength range of the band. Those were stacked and normalized afterwards. Since in the measured data the data quality in at the edges of the band passes were not so good, only the following wavelengths were taken into account: J ($\mathrm{1.1 - 1.4 \ \mu m}$) , H ($\mathrm{1.47 - 1.83 \ \mu m}$), K ($\mathrm{2.0 - 2.4 \ \mu m}$) and H+K ($\mathrm{1.5 - 2.4 \ \mu m}$). Here it is assumed that the spectrum is a constant function of wavelength and that the quantum efficiency of the detector is also constant over wavelength. The measured PSFs are taken from the data of the North-South test, a calibration template for distortion which is taken weekly at the telescope. The data was reduced with the SINFONI \lq spred-pipeline\rq \ of MPE (see \cite{abuter06} and \cite{schreiber04}). The simulated PSFs as well as the reconstructed PSFs of the calibration fiber for the individual bands with pre- and post- upgrade data can be seen in figure \ref{fig:psfs}. The measured images are scaled to $\mathrm{I^{0.2}}$ while the theoretical PSFs are scaled to $\mathrm{I^{0.1}}$ for better visibility.

The performance of the new pre-optics that was installed during the upgrade is nearly optimal, with SRs measured in the laboratory of nearly 100\%. \cite{george16}. This results, as one can see in the images of figure \ref{fig:psfs} in a sharper PSF, and therefore a higher SR.
The Strehl Ratio, defined by Strehl \cite{strehl02} as the ratio of the peak intensity of a measured PSF to the peak intensity of a perfect diffraction-limited PSF for the same optical system is given as 

\begin{equation}
	S = \frac{I(\vec{x} = 0)}{P(\vec{x} = 0)} = \frac{I(\vec{x} = 0)}{\int I(\vec{x}) \ d\vec{x}}\frac{\int P(\vec{x}) \ d\vec{x}}{P(\vec{x} = 0}
\end{equation}

with the 2D position image space vector $\vec{x}$, the maximum of the measured PSF $I(\vec{(x)} = 0)$ and the maximum of the diffraction limited PSF $P(\vec{(x)} = 0)$.
In order to determine the Strehl Ratio the peak flux of the measured and simulated PSF was normalized by the integrated flux within a radius corresponding to an angle of $\alpha = \mathrm{5 \lambda_c / d_{free}}$ with the central wavelengths of each band J: $\mathrm{\lambda_c = 1.25 \ \mu m}$, H: $\mathrm{\lambda_c = 1.65 \ \mu m}$, K: $\mathrm{\lambda_c = 2.22 \ \mu m}$, H+K: $\mathrm{\lambda_c = 1.95 \ \mu m}$ and the free diameter of the cold stop $d_{free} = 6.436 \ \mu m$. Normally one would integrate the flux up to $\mathrm{\alpha = 7 \lambda_c / d_{free}}$, but since in K-band this would correspond to a radius of 33 pixels (half slitlets), while the detector is only 60 pixels wide. The radius in pixel is obtained dividing by the field angle of one pixel ($\mathrm{sin(\alpha) = x_{pixel} / f_{camera}}$) with the size of half a slitlet $x_{pixel}$ and the focal length of the 25 mas pixelscale pre-optics camera $f_{camera}$. Anyway in these regions the signal to noise ratio is too low to get a reliable result. For reasons of comparison in all bands the flux is then integrated up to $\mathrm{\alpha = 5 \lambda_c / d_{free}}$ corresponding to a radius of $\mathrm{r = 14 \ pixel}$ in J-band and $\mathrm{r = 24 \ pixel}$ in K-band. In order to get rid of the background of the detector, the average of the pixel values outside this integration radius is subtracted from the integrated region and of course also from the peak value. To get the best signal to noise ratio the average image over the given wavelength range was used.

The largest error that affects the SR is the low spatial sampling of the reconstructed SPIFFI image. This easily results in errors up to 10 \% \cite{roberts04}. The photometric error is much smaller - on the order of parts per thousand - in this measurement, because there is a large number of photons measured, and the statistical error only grows with the square root of the number of photons. Assuming that there is no photometric uncertainty in the simulated PSF, the measured SR is influenced by the photometric error as 

\begin{equation}
	S = \frac{I(\vec{x} = 0)}{P(\vec{x} = 0)} = \frac{I(\vec{x} = 0)}{\int I(\vec{x}) \ d\vec{x} + \epsilon}\frac{\int P(\vec{x}) \ d\vec{x}}{P(\vec{x} = 0} = S \left( 1 \pm \frac{\epsilon}{\int I(\vec{x}) \ d\vec{x}}\right)
\end{equation}

\noindent meaning that the fractional error in the measurement is equal to the fractional photometric error

\begin{equation}
	\frac{\epsilon}{\int I(\vec{x}) \ d\vec{x}} = \frac{1}{\int I(\vec{x}) \ d\vec{x}} \left[ \frac{1}{\sqrt{\int I(\vec{x}) \ d\vec{x}}} + \sqrt{nB} = n \sqrt{\frac{B}{n_{sky}}} \right]
\end{equation}

\noindent where the first term is the photon noise contribution with intensity measured in photons, the second therm is the sky background error with the mean background intensity B in photons/pixel and the third term which is the uncertainty of the mean background growing with the square root of the number of pixels in the measured sky background \cite{roberts04}. Furthermore systematic errors in flat fielding or a non-uniform sky background will likely increase the uncertainty in sky background subtraction. To estimate the error introduced by the low sampling, the simulated PSF is moved in sub-pixel steps within the image frame. For each of the resulting images the Strehl Ratio of the measured PSFs is calculated. From the change in SR the error is estimated. The error of Nyquist sampling is much larger than the photometry error. The error made by assuming a constant detector quantum efficiency and a constant light source spectrum over wavelength is $\mathrm{\sim 1\%}$ for J-, H-, K-band and $\mathrm{\sim 4\%}$ for H+K with a halogen lamp color temperature of 5000 K, but is like the photometry error, is not relevant for the relative SR comparison.

All in all the total error is summed up from the uncertainty due to the low sampling of image, the photometry error, the uncertainty rising from the lamp temperature and the quantum efficiency of the detector as well as systematic errors in the flat fielding and background subtraction.

\subsection{Strehl Ratio}
A summary of the Strehl Ratios measured with pre- and post- upgrade data is shown in table \ref{tab:strehl}. The FWHM of the PSF is given in pixels corresponding to half a slitlet at the small slicer ($\mathrm{150 \ \mu m}$). It is calculated as the average of the FWHM in x-direction and y-direction. The change in Strehl Ratio is mostly affected by the new pre-optics, with its better AR coatings reduced stray light. Furthermore the slightly better focus position of the detector could have a positive effect on the image quality. The smoother collimator mirror surfaces cause less stray light than the old mirrors and thus improve the SR. The SR was only measured in the smallest pixelscale, because this is the only pixelscale, where the PSF is sampled over the Nyquist level. In this pixelscale the collimator has a negligible effect on the image quality, because the PSF is sliced. In the 250 mas pixelscale, where the PSF fits on one slitlet the collimator mirrors will have an affect on the image quality and the SR. Due to the imperfectness of M3 the induced wavefront RMS error of $\mathrm{0.2 \ \mu m}$ will reduce the SR in the largest pixelscale. For spherical aberrations the loss in SR is connected with the RMS wavefront error by\cite{mahajan83}

\begin{equation}
	\Delta S = 1 - exp\left(-\left(2\pi\sigma\right)^2\right)
\end{equation}

\noindent with the RMS wavefront error $\sigma$ normalized by the central wavelength of a band. For coma and astigmatism this is a close approximation. The formula is indeed valid for the seeing limited pixelscales of SPIFFI, but not for the AO scale. 

The pre- and post- upgrade Strehl ratios are shown in table \ref{tab:strehl}.

\begin{table}[htbp!]
	\begin{tabular}{c|cc|cc|c}
		
		& \multicolumn{2}{|c|}{\textbf{pre - upgrade}} & \multicolumn{2}{|c|}{\textbf{post - upgrade}} &\\
		\hline
		{\small Band}&{\small FWHM}&{\small Strehl Ratio}&{\small FWHM [px]}&{\small Strehl Ratio}&{\small Improvement}\\
		&[pixels]&[\%]&[pixels]&[\%]&[\%]\\
		\hline
		\hline
		J &  3.6	&	$56 \pm 16$	&	3.5	&	$58 \pm 17$	&	2\\
		H&	4.1		&	$71 \pm 9$	&	3.7 &	$83 \pm 10$ &	12 \\
		K& 4.8		&	$85 \pm 5$	&	4.6	&	$95 \pm 6$	& 	10\\
		H+K& 4.4	&	$80 \pm 7$ &	4.5 &	$78 \pm 7$	&	-2\\
	\end{tabular}
	\caption[FWHM and Strehl ratio pre- and post- upgrade]{Comparison of FWHM of the PSFs and the Strehl ratios pre- and post- upgrade}
	\label{tab:strehl}
\end{table}

Since the Strehl Ratios are measured with the calibration fiber the values from table \ref{tab:strehl} are an upper limit for the performance on sky. The upgrade did not effect the SR in J- and H+K-band, but improved the SR in H- and K-band. The effect is mainly caused by the performance of the new pre-optics.

With the increased SR the detection of faint objects will now be improved because it directly results in a higher signal to noise ratio. Furthermore in fields of the sky that are crowded by many luminous objects, the distinction between the different objects will be improved.

\subsection{Encircled Energy}
If one wants to know how close the measured PSF is to a diffraction limited theoretical PSF of a certain optical system, a close look at the encircled energy is helpful. In this measurement the encircled energy within a certain radius from the PSF center is of interest. The theory encircled energy gives an upper limit for the energy deposited within a radius for an optical system. In the largest pixelscale most of the flux of a point source falls into one slitlet. For this case the encircled energy provides a fractional throughput measurement which is a more accurate measure of the instrument performance than the Strehl ratio for the large pixelscales. Different from the pre-optics test in the laboratory, where at the position of the image slicer a detector that delivers high-resolution images of a point source is positioned, in the SPIFFI cryostat the PSF is only sampled by 60 x 64 pixel. Thus a measurement of the encircled energy in the two larger pixelscales would not be helpful, since most of the energy is deposited in one pixel. Due to this the measurement of the encircled energy is only shown for the 25 mas pixelscale in this thesis. From this measurement information is obtained in which way the flux is shifted either to the center or the edges of the PSF.

\begin{figure}[htbp!]
	\begin{center}
		\resizebox{1.0\textwidth}{!}{
			\includegraphics[width=1\textwidth, trim={1.cm 0 1cm 0cm}, clip=true]{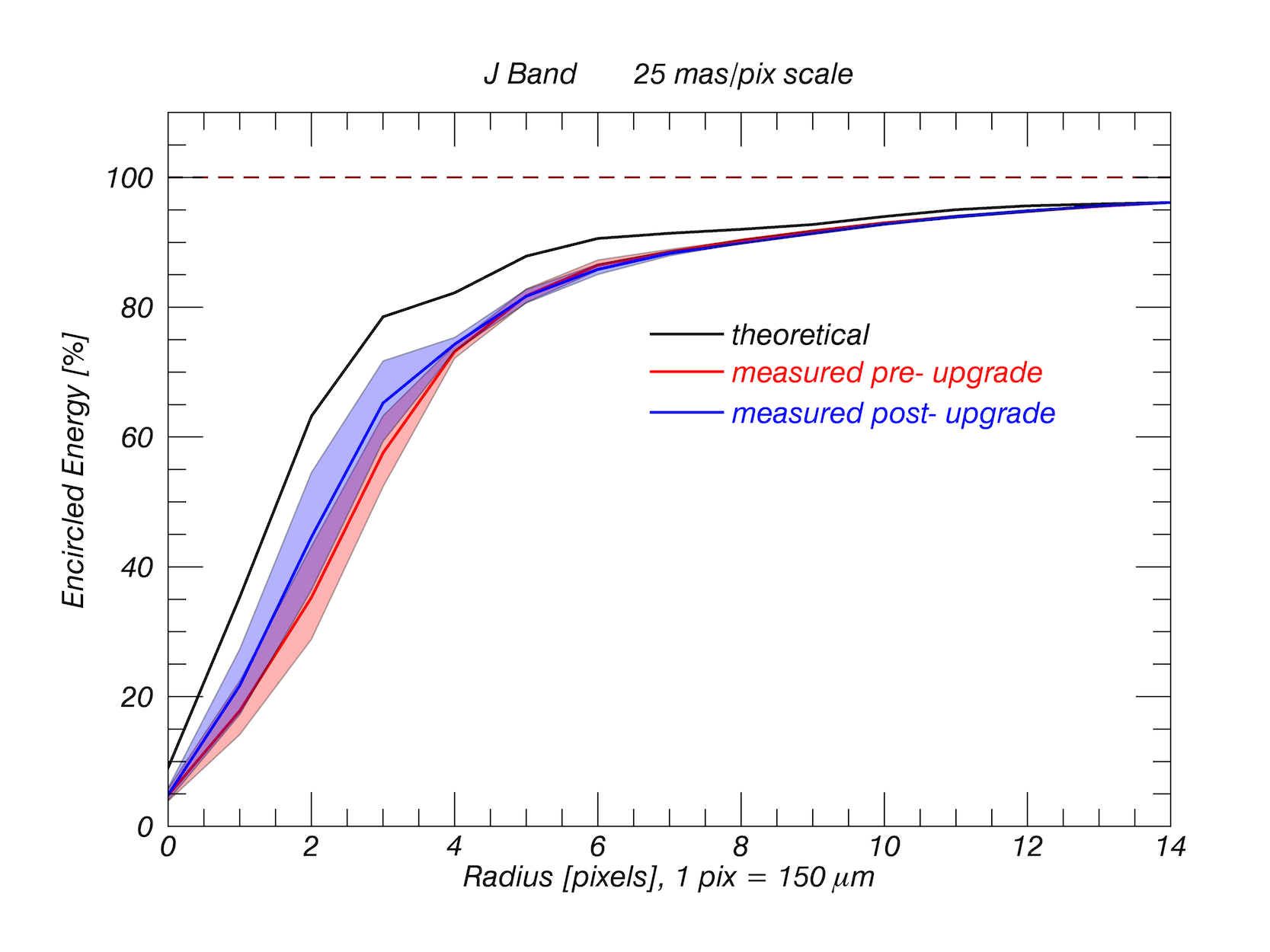}
			\includegraphics[width=1\textwidth, trim={1.cm 0 1cm 0cm}, clip=true]{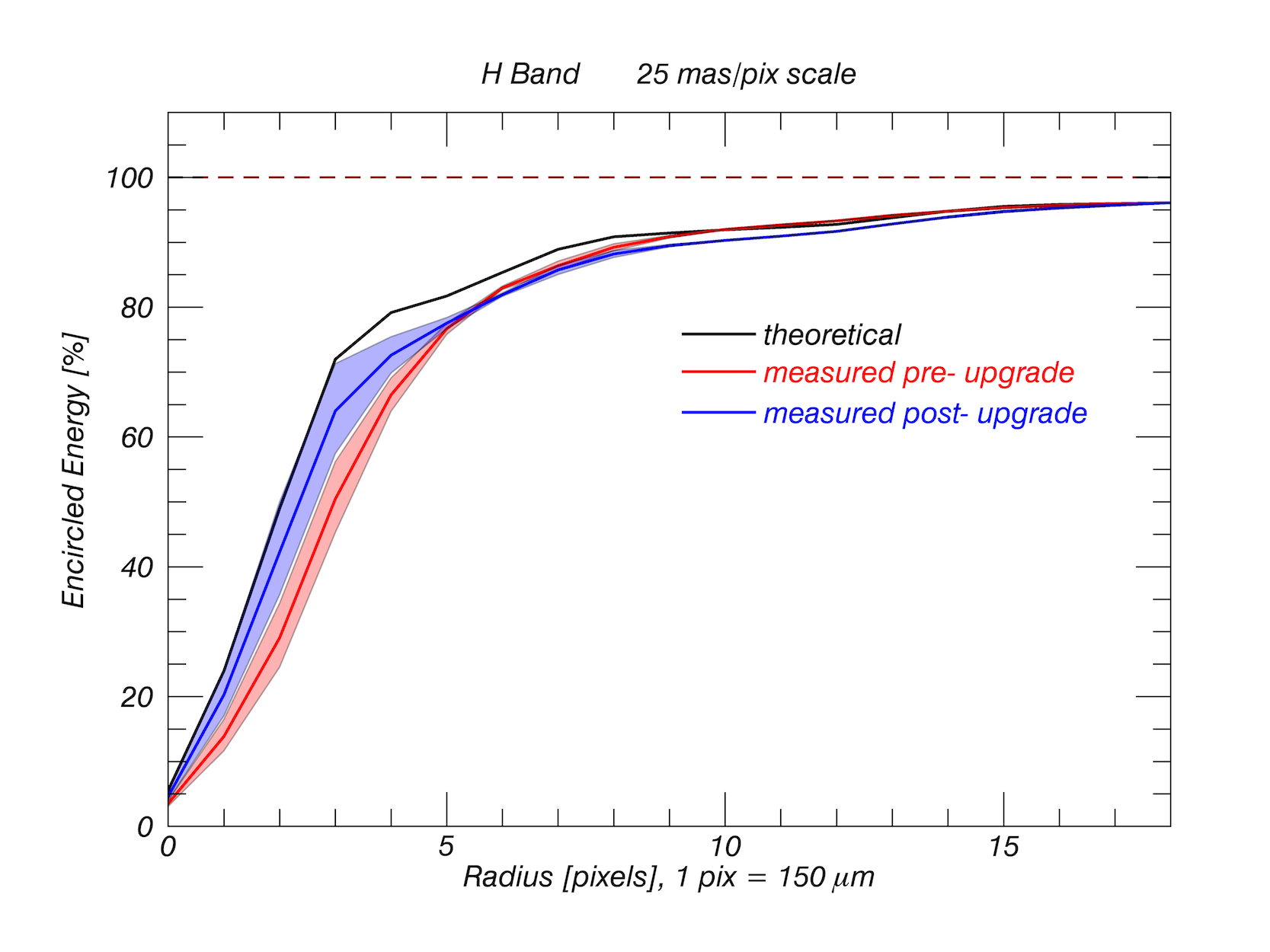}
		}
		\resizebox{1.0\textwidth}{!}{
			\includegraphics[width=1\textwidth, trim={1.cm 0 1cm 0cm}, clip=true]{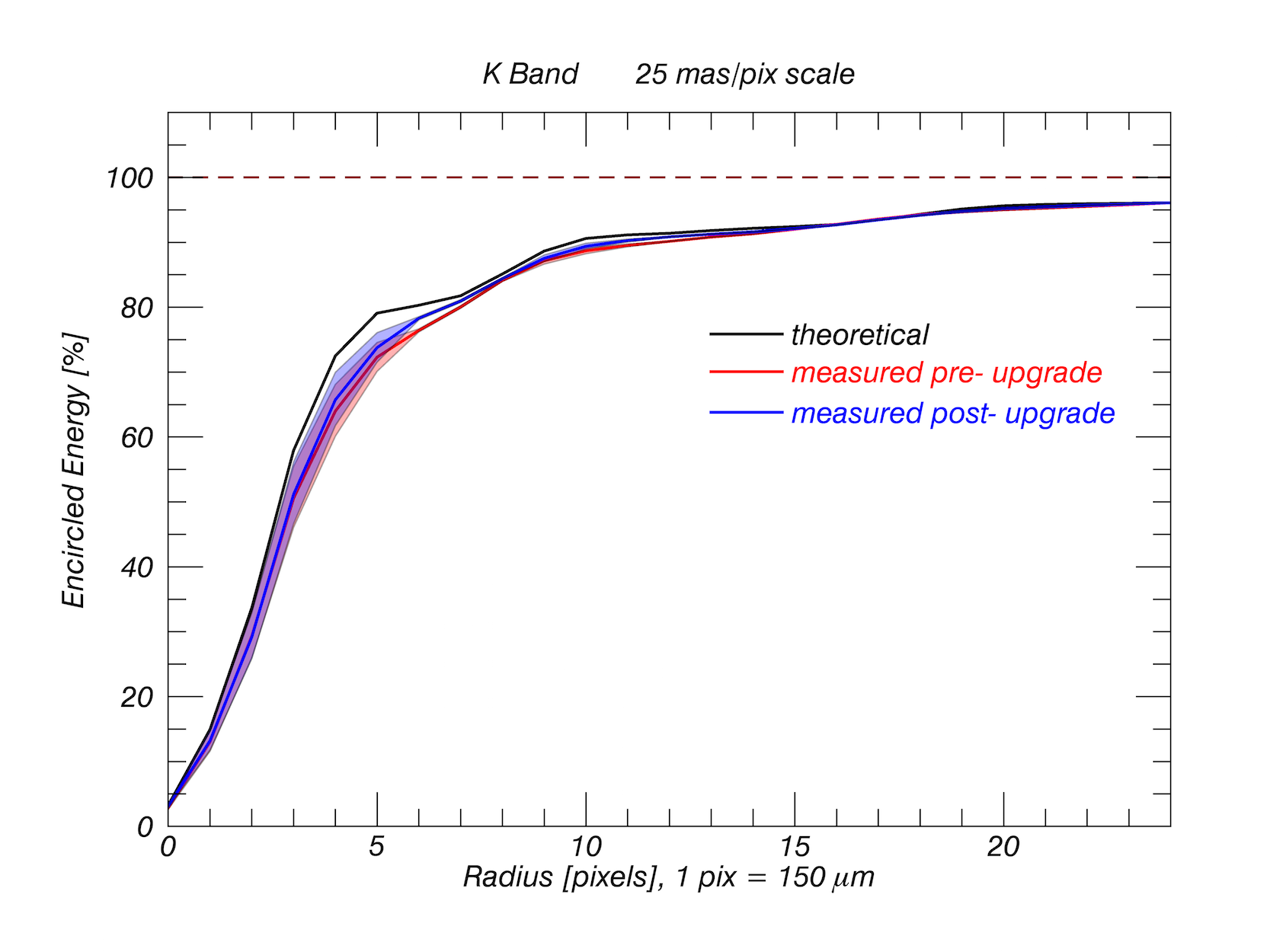}
			\includegraphics[width=1\textwidth, trim={1.cm 0 1cm 0cm}, clip=true]{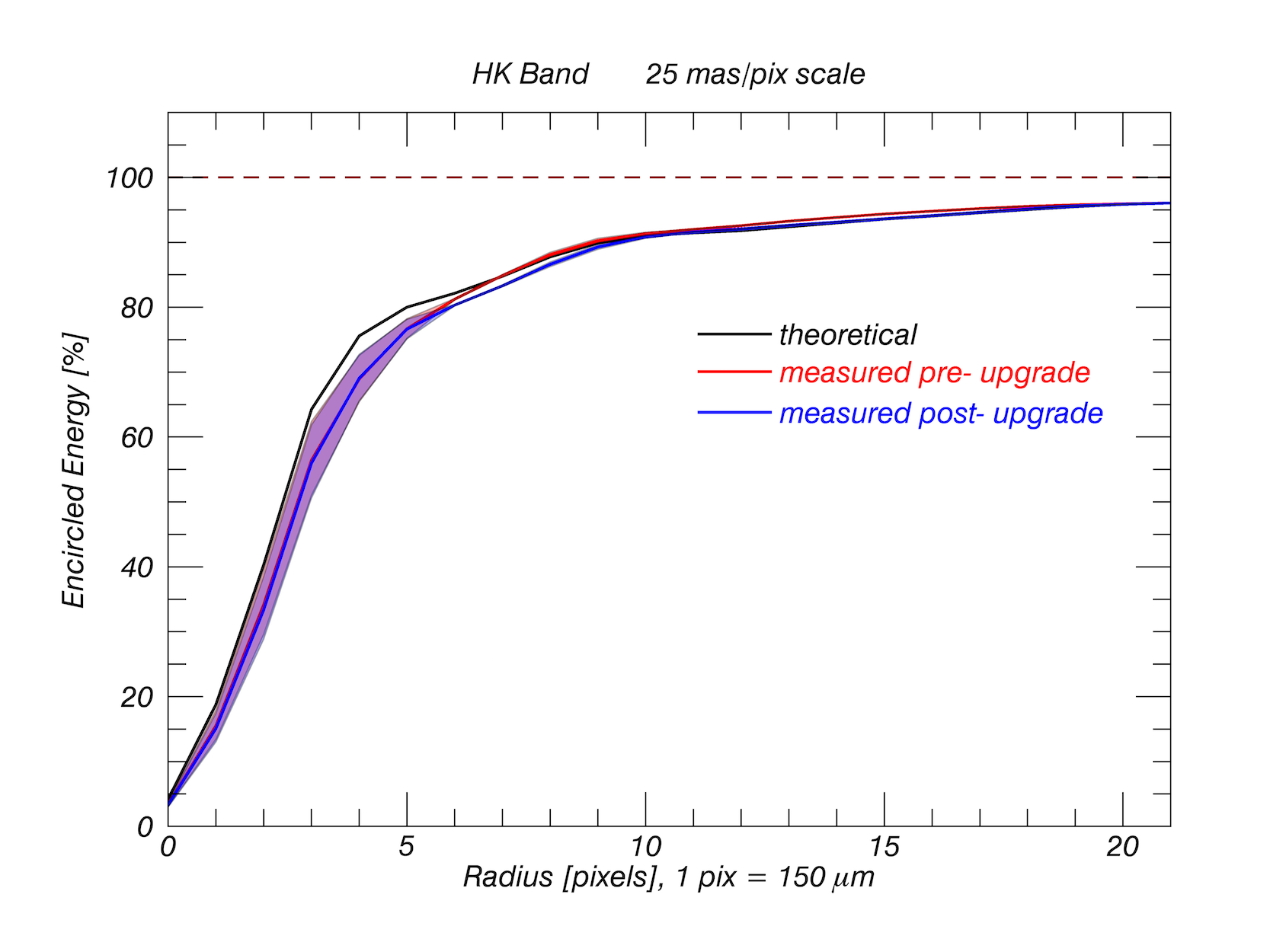}
		}
		\caption[Encircled energy in 25 mas pixelscale]{The encircled energy as a function of the radius with $\alpha = \mathrm{5 \lambda_c / d_{free}}$ corresponding to 100 \%. One pixel equals half a slitlet width on the small slicer.}
		\label{fig:encircled_energy}
	\end{center}
\end{figure}

Figure \ref{fig:encircled_energy} shows the trend of the encircled energy as a function of the radius. It builds on the assumption that all energy which can be theoretically deposited within a radius of $r = \mathrm{5 \lambda_c / d_{free}}$ is encircled in that radius. This is simply done because the encircled energy has to be normalized in some way. The so resulting plot is not related to the throughput, but shows only the fraction of the energy of a PSF within a certain radius. The theoretical upper limit of the fraction of the encircled energy within a certain radius is given by\cite{stamnes82}

\begin{equation}
	\begin{aligned}
		\frac{\int_0^r I(r') dr'}{\int_0^{\infty} I(r') dr'}= \frac{1}{1-\epsilon^2} \bigg( 1 - \left(J_0(x)\right)^2 - \left(J_1(x)\right)^2 +\epsilon^2 \left[\left(J_0(\epsilon x)\right)^2 - \left(J_1(\epsilon x)\right)^2\right]\\
		\left.- 4 \epsilon \int_0^x \frac{\left(J_1(x')\right)\left(J_1(\epsilon x')\right)}{x'} dx' \right)
	\end{aligned}
\end{equation}

\noindent with the argument $x = \frac{2 \pi r}{\lambda} sin(\alpha)$ and the ratio between the diameter of the central obscuration and the free diameter of the pupil stop $\epsilon$. For a radius corresponding to $\alpha = \mathrm{5 \lambda_c / d_{free}}$ the maximal total energy is 95\%. The measured encircled energy is normalized to this value. This normalization introduces maximal an error between 1\% and 2\%.
As in the Strehl Ratio measurement, the photometric error is small here. The uncertainty is dominated by the low sampling. To get an error estimation for the encircled energy, the simulated PSF is shifted in sub-pixel steps within a rectangle. Half the difference between the maximum and minimum encircled energies per radius obtained by this procedure gives a relative uncertainty, which is then applied to the measured data. Here it is assumed that statistically the measured curve lies in the middle of the maximum and minimum encircled energy.

From the plots in figure \ref{fig:encircled_energy} it can be seen that especially in J- and H-band the encircled energy at small radii changed between the measurements before and after the upgrade. The fraction of encircled energy at any given radius at small radii is systematically higher. In K-band the function of the encircled energy per pixel did not change by a lot. In the inner regions of the H+K-band PSF there is nearly no change. For larger radii, the deposited energy is slightly lower after the upgrade, which means that flux got shifted outward towards a radius of $\mathrm{5 \lambda_c / d_{free}}$. Numbers for the change of the encircled energy within a slitlet are not given here. Because one slitlet is only sampled by two pixels this number strongly depends where the center of the PSF is located (in the middle of a pixel, close to the edge or between two pixels). This results in errors of this measure with the available sampling around 30\% to 50\%. Thus a comparison of the pre- and post- upgrade encircled energy within the central slitlet does not provide helpful information.

Of scientific interest is that the higher encircled energy within a certain area also corresponds to a better spatial resolution that can be reached after the upgrade. Further an increase in the relative encircled energy within small radii results in higher sensitivities, which is important for faint objects. Due to the low sampling in the central parts of the PSF it is not helpful to give number for the encircled energy within one slitlet, because the sampling error would be far too high.

After looking at the throughput and the image quality a third quantity is of interest, which is the spectral line profiles of the instrument. Investigations on this are shown and discussed in the next section.

\section{Spectral Line Profiles}\label{sec:lineprofiles}
The resolution limiting factor of the SPIFFI spectrometer are the spectral line profiles. Many investigations have been done in the past to determine the cause of the non-Gaussian line profiles. The line profiles have an asymmetric distinct shape with shoulders, deviating from the roughly Gaussian profile. For an ideal spectrograph a Gaussian line profile is a good estimation for the intrinsic Voigt function profile of the spectral line . Since SPIFFI was built with the first 1k detector there have been lots of investigations to determine the root cause of the line profiles. In the early days of the instrument, the different widths of the lines in H and K bands was thought to be a result of a chromatic axial defocus of the spectrometer camera.\cite{iserlohe04} The deformity of the line profiles in all wavelengths was assumed to be caused by the collimator mirrors. The mirrors were replaced by post-polished ones, which were installed from 2004 to 2016 (see chapter \ref{ch:chapter_mirrors}), but there was only minimal improvement. After the 1k detector in SPIFFI was exchanged with the 2k detector the line profiles in J-band were blamed on the surface quality of the J-band grating.\cite{iserlohe04} 

For science done with the spectra of SPIFFI it is very important to have spectral lines that are as ideal shaped as possible. For example the sky subtraction with the sky-spider is not used because the line profiles vary too much over the detector, making a reliable subtraction of the sky lines not possible by only using a fraction of the FOV. Furthermore due to the broadened spectral lines information gets lost, because small features can not be resolved. On top of this for example velocity dispersions of galaxies are measured by the broadening of the spectral lines. If these lines are intrinsically broadened by the spectrometer performance and non-Gaussian shaped the determination of the velocity dispersion cannot be done as accurate as with not broadened line profiles. 

\subsection{The Grating Babysteps}
The width of the spectral lines on the detector is around two to four pixels depending on wavelength and pixelscale chosen. This means that the shape of the spectral lines is undersampled on the detector. In order to provide hypersampled line profiles a technique from \cite{thatte12} is used. Originally this method originated from the idea to perform an accurate sky subtraction without the need for dedicated blank sky exposures. It relies on precise measurements of the line spread function of the spectrograph at all wavelengths corresponding to night-sky emission lines for all spatial elements (spaxels) of the reconstructed SPIFFI FOV. From these measurements a database is formed, giving the LSF for each slitlet in each spectral line. Afterwards it is identified which part of the FOV can be used as sky. The spectral lines from this region are then fitted to the database and afterwards subtracted from the whole FOV.

For this thesis, the Thatte et al. method was adapted to get information about the shape of the spectral line profile. In order to obtain these hypersampled spectral line profiles, a series of IFU calibration exposures with SINFONI was taken for each band (J, H, K, H+K) and each pixelscale (25 mas/px, 100 mas/px, 250 mas/px). The calibration unit in MACAO was used to take a series of arc-lamp exposures. Following lamps were used: J-band: Argon, H-band: Xenon \& Argon, K-band: Neon \& Argon, H+K-band: Xenon. For a series of exposures of a particular band and pixelscale, the grating wheel was turned by a few encoder positions between each exposure, which corresponds to a shift of the central wavelength on the detector by approximately 0.1 pixels. Between 10 and 26 shifted exposures per band and pixelscale were obtained. These slightly shifted exposures are here called \lq babysteps\rq. The exposures were reduced with the \lq spred-pipeline\rq \ of MPE for SINFONI data reduction (see \cite{abuter06} and \cite{schreiber04}). The standard SINFONI waveflats were used in order to get a bad pixel map as well as the flat fields and the distortion measurements.

Each of the exposures was then wavelength-calibrated separately. In order to find the right position of the spectral lines, the measured spectrum was cross-correlated with a synthetic spectrum. Afterwards the emission lines were fitted using a Gaussian to get the position. From these positions the dispersion relation is determined for every detector column and smoothed within the slitlet. These distortion-, bad pixel- and flat field-corrected, and wavelength calibrated images were used for further analysis of the line profiles, as well as the next section about the spectrometer resolution.

\begin{figure}[htbp!]
	\begin{center}
		\resizebox{1.0\textwidth}{!}{
			\includegraphics[width=0.8\textwidth]{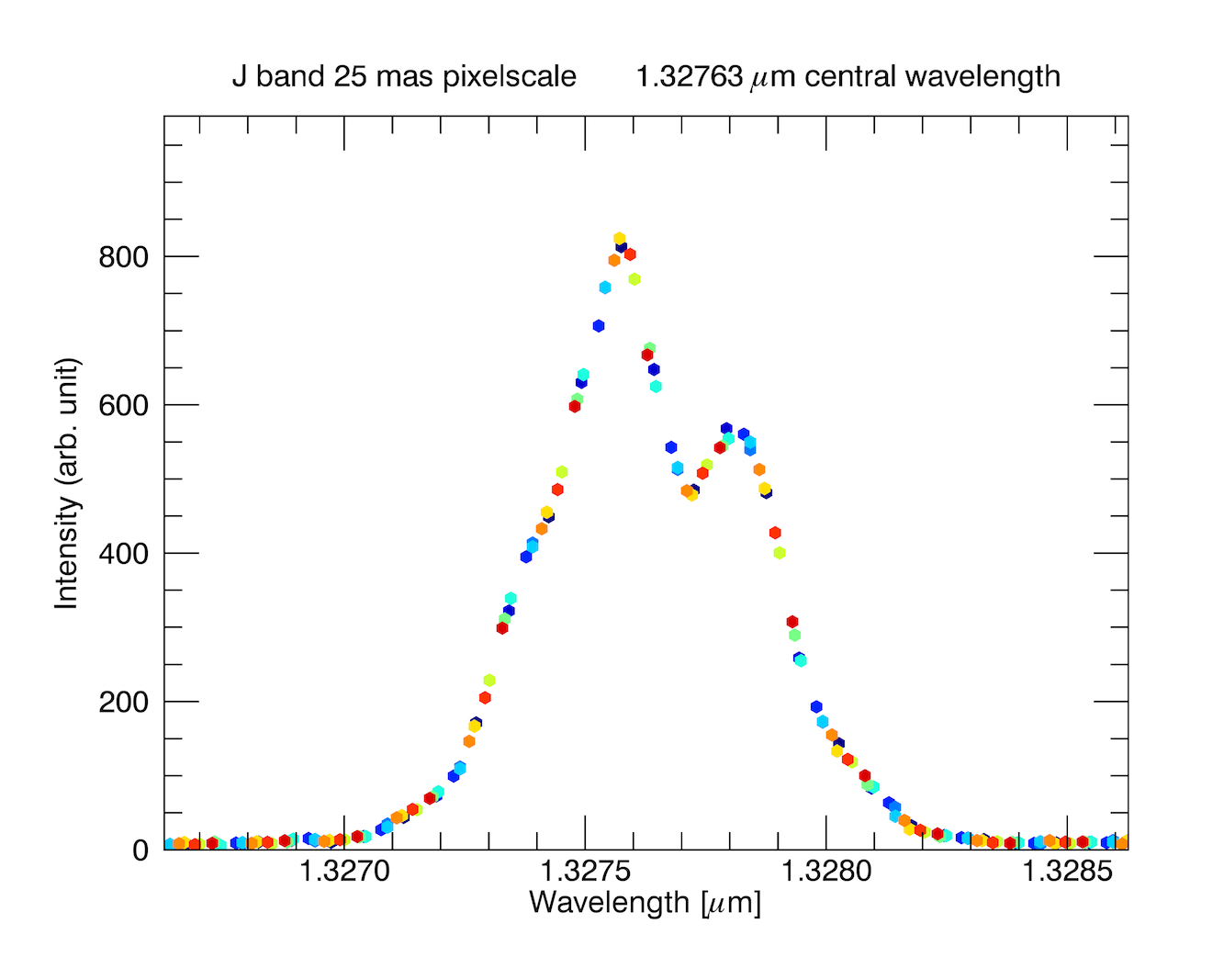}
			\includegraphics[width=0.8\textwidth]{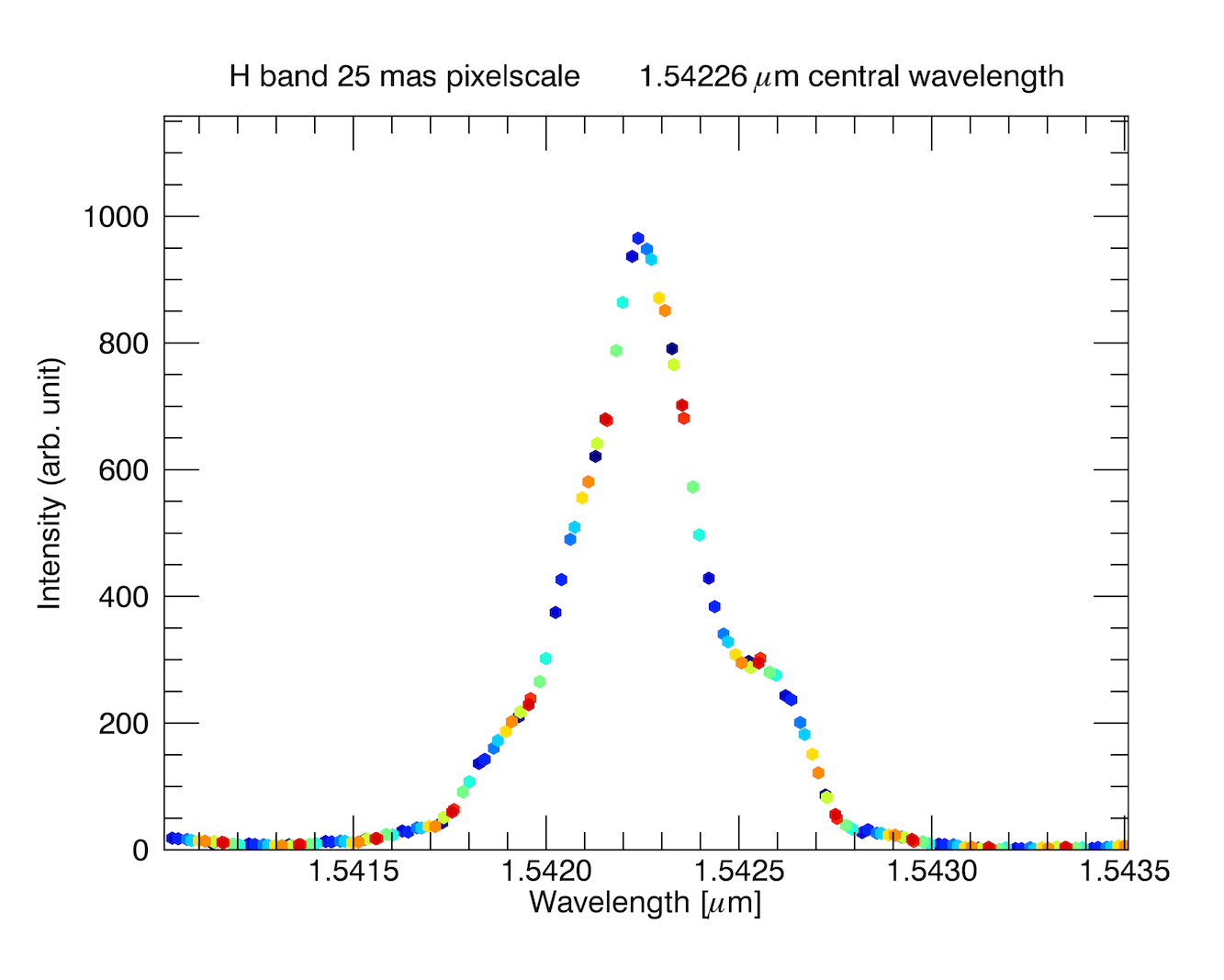}}
		\resizebox{1.0\textwidth}{!}{
			\includegraphics[width=0.8\textwidth]{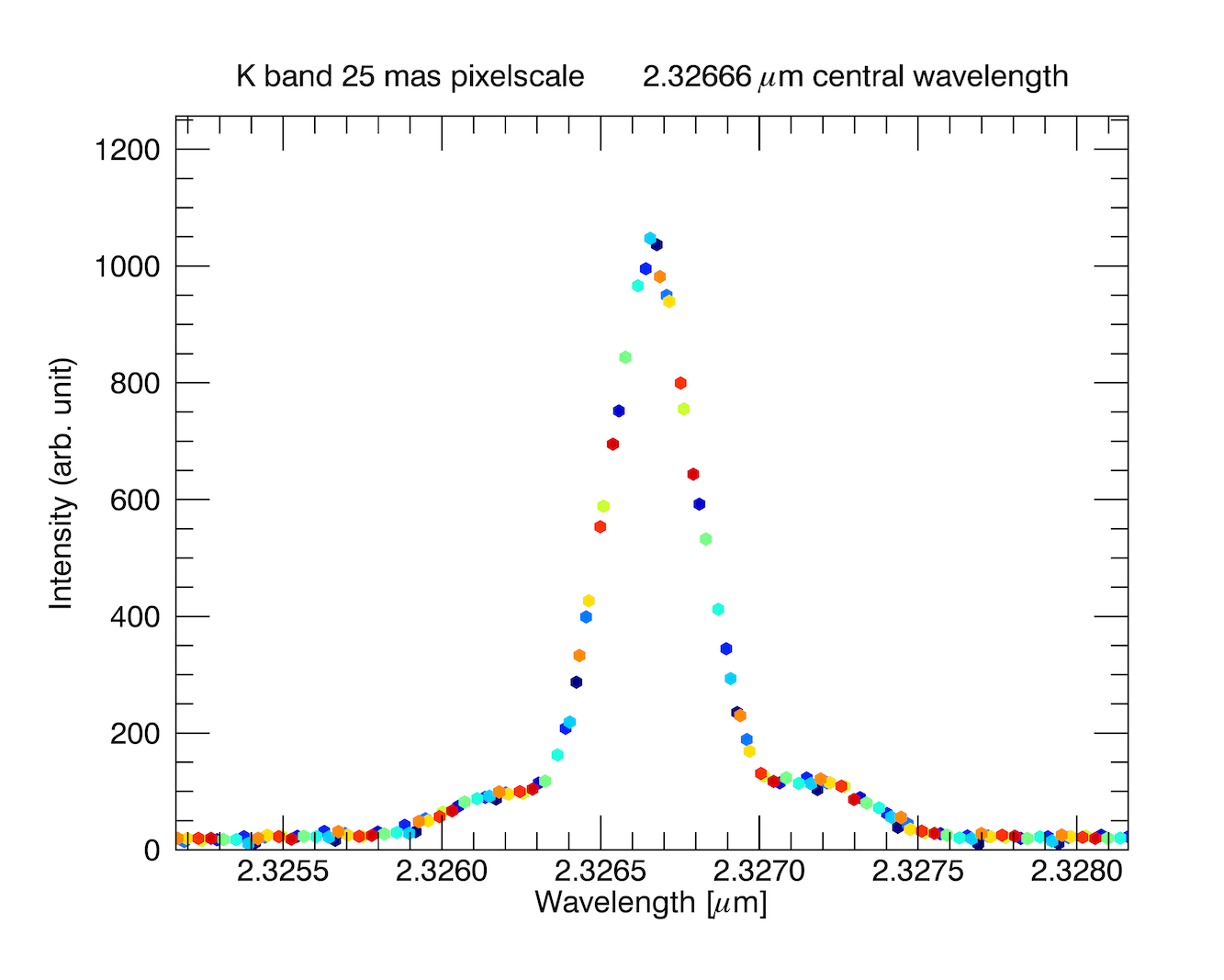}
			\includegraphics[width=0.8\textwidth]{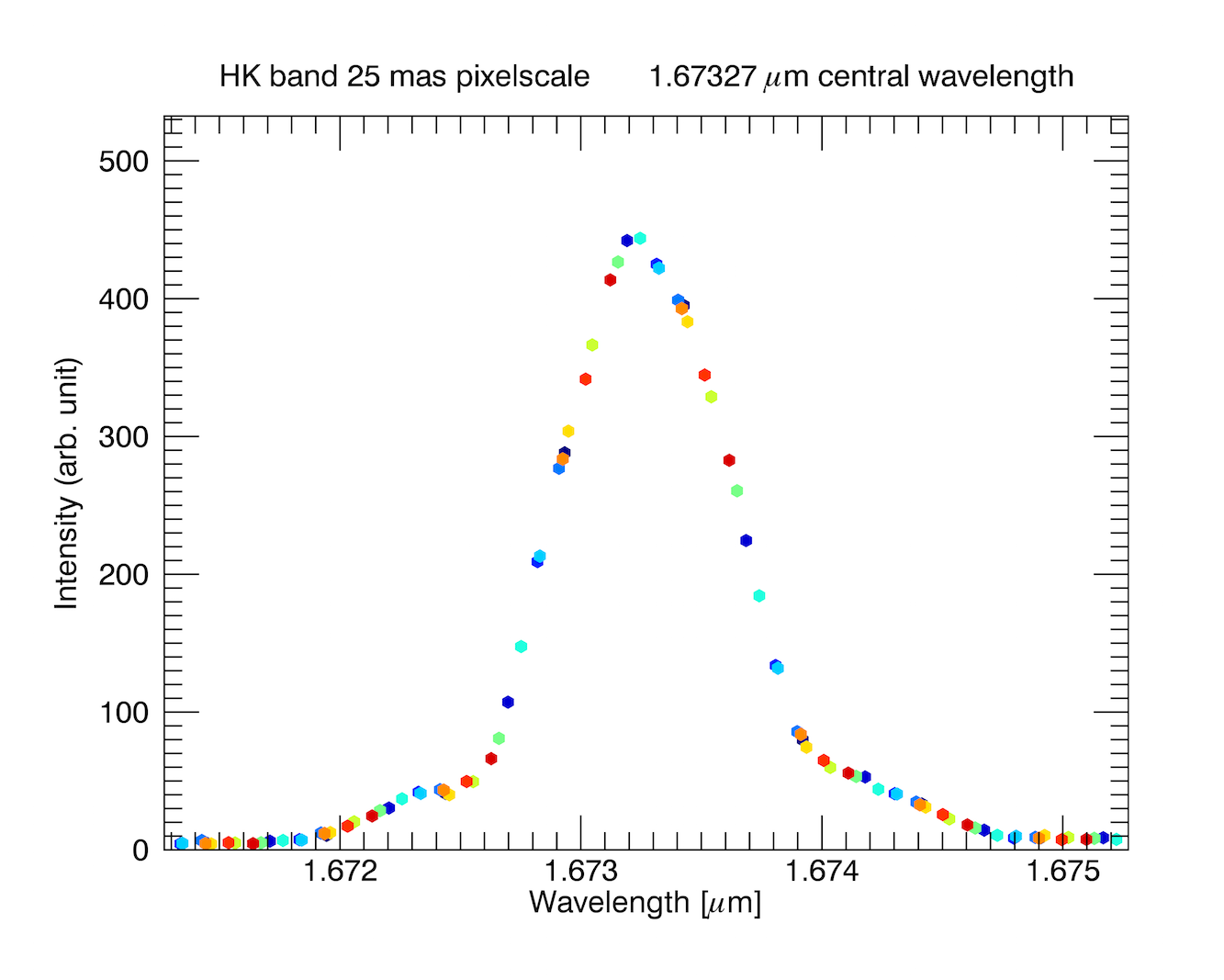}}
		\caption[Assembling of individual babysteps]{The assembling of the individual babysteps exposures on a single wavelength scale. Each color represents a certain exposure. The displayed line is at a wavelength close to the center of each band. The chosen column is column 790 in the distortion corrected image, which is in the center of slitlet 15.}
		\label{fig:accuracy}
	\end{center}
\end{figure}

Because the brightness of the spectral calibration lamps vary slightly over the time in which all babysteps were taken, it is not sufficient to place all of the exposures for each band and pixelscale on a common wavelength scale. It is also important that the exposures are placed on a common flux scale. In order to do so and to correct for variations in the line flux between the single exposures, the lines are normalized to the flux. This is done by fitting a Gaussian to each undersampled line in each column of the detector and integrating to get the total flux in the line. This Gaussian fit was found to produce the smoothest line profiles compared to using the total or peak line fluxes in accordance with \cite{thatte12}.The normalized exposures were then placed on a common wavelength scale. The accuracy of this process can be seen in figure \ref{fig:accuracy} where the different colored dots refer to the individual babysteps exposures (each color for a certain grating position).

\subsection{Shape of the Spectral Emission Lines}
\begin{figure}[htbp!]
	\begin{center}
		\resizebox{1.0\textwidth}{!}{
			\includegraphics[width=1.0\textwidth, trim={1.0cm 0 1.0cm 0cm}, clip=true]{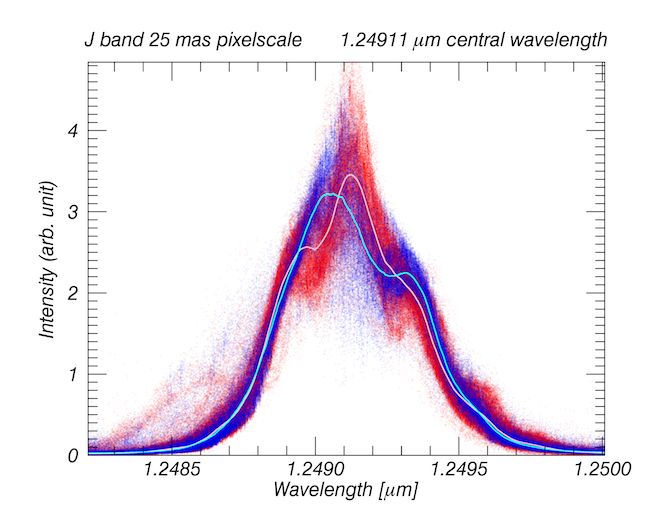}
			\includegraphics[width=1.0\textwidth, trim={1.3cm 0 0.7cm 0cm}, clip=true]{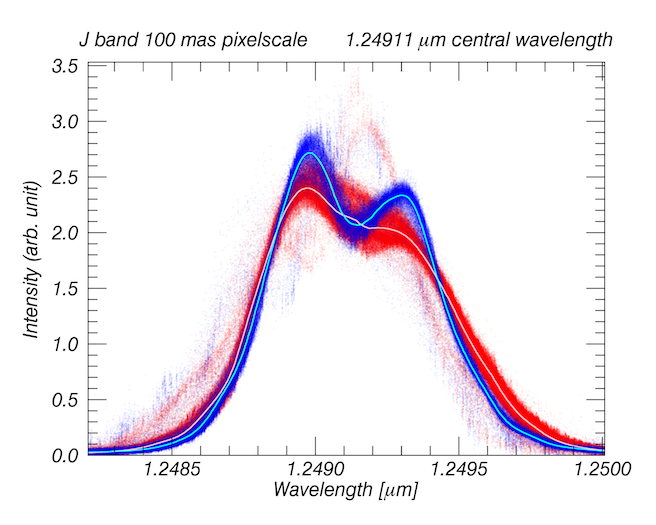}
			\includegraphics[width=1.0\textwidth, trim={1.3cm 0 0.7cm 0cm}, clip=true]{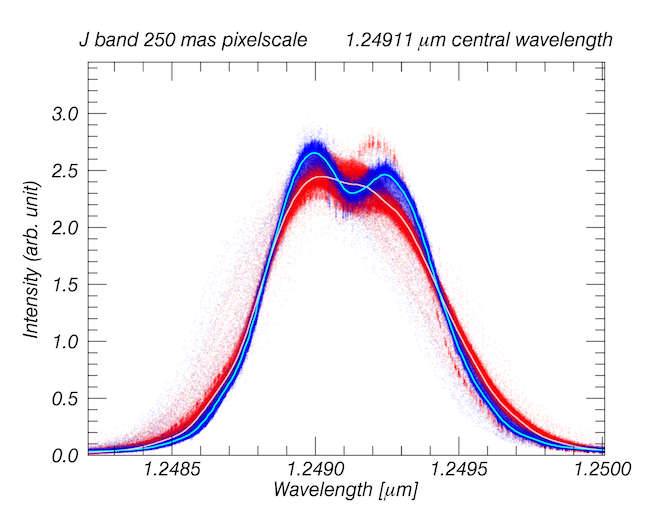}
		}
		\resizebox{1.0\textwidth}{!}{
			\includegraphics[width=1.0\textwidth, trim={1.0cm 0 1.0cm 0cm}, clip=true]{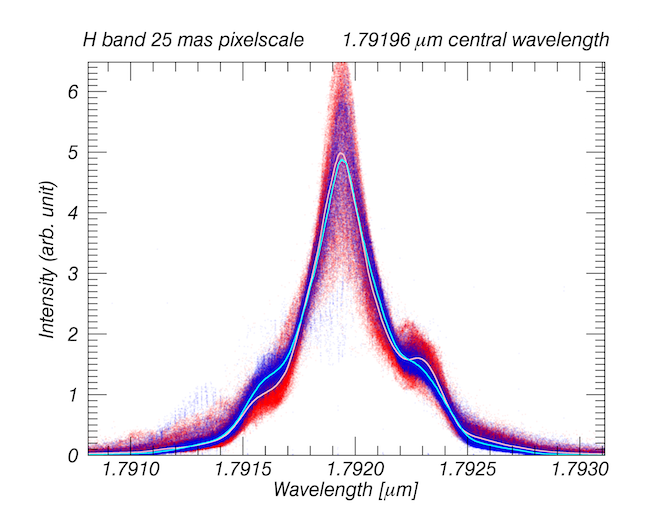}
			\includegraphics[width=1.0\textwidth, trim={1.5cm 0 0.5cm 0cm}, clip=true]{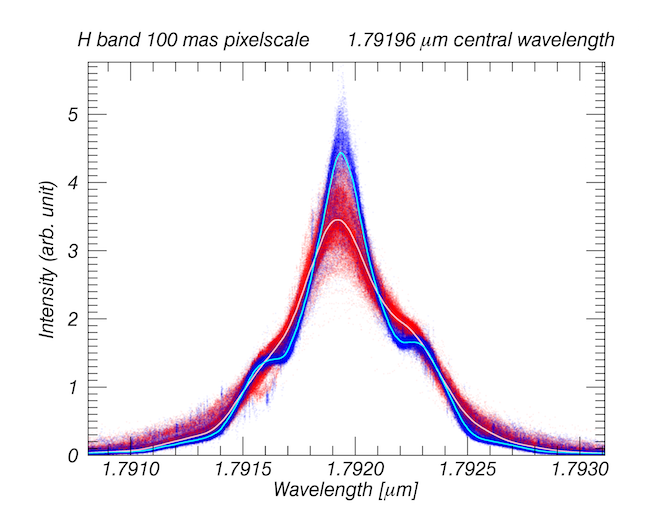}
			\includegraphics[width=1.0\textwidth, trim={1.5cm 0 0.5cm 0cm}, clip=true]{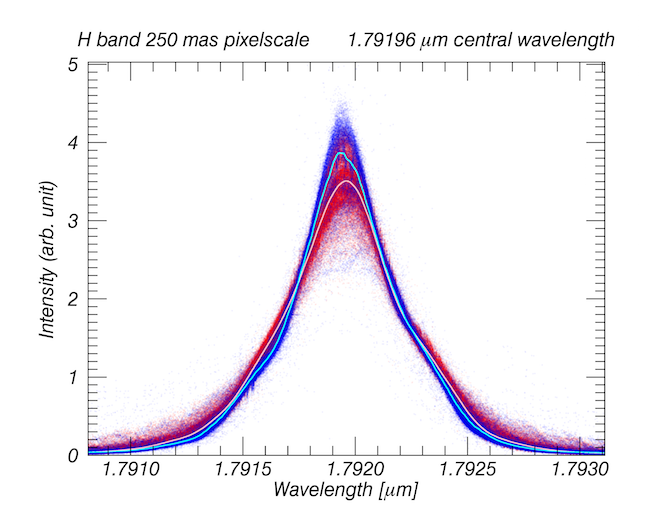}
		}
		\resizebox{1.0\textwidth}{!}{
			\includegraphics[width=1.0\textwidth, trim={1.0cm 0 1.0cm 0cm}, clip=true]{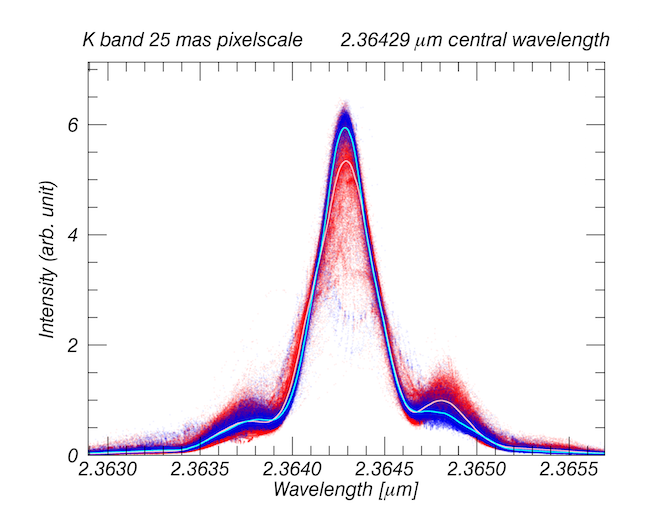}
			\includegraphics[width=1.0\textwidth, trim={1.5cm 0 0.5cm 0cm}, clip=true]{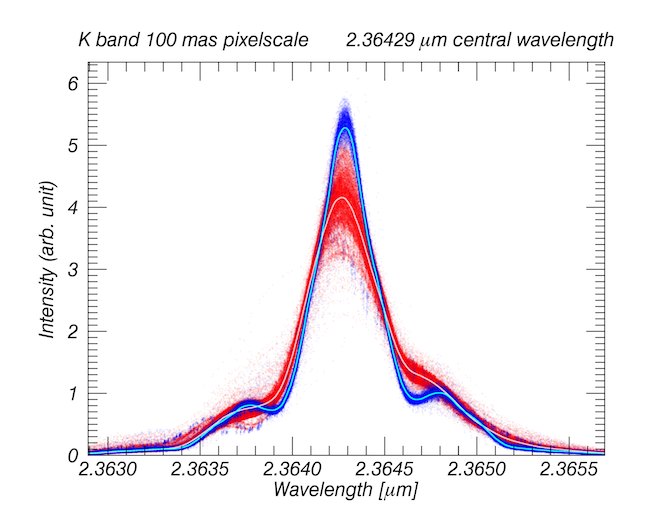}
			\includegraphics[width=1.0\textwidth, trim={1.5cm 0 0.5cm 0cm}, clip=true]{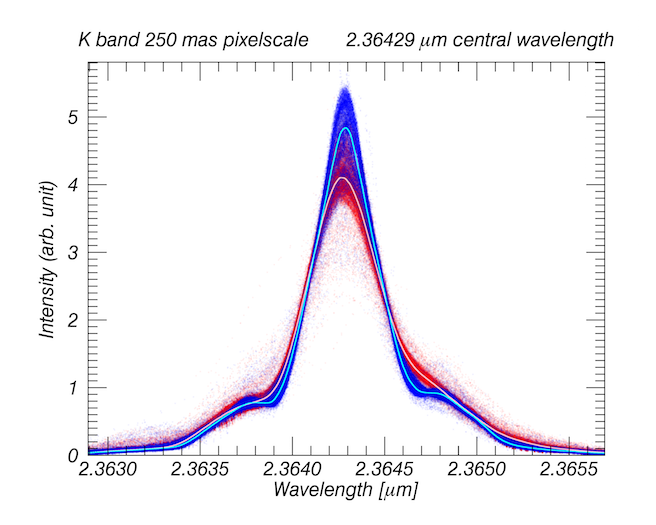}
		}
		\resizebox{1.0\textwidth}{!}{
			\includegraphics[width=1.0\textwidth, trim={1.0cm 0 1.0cm 0cm}, clip=true]{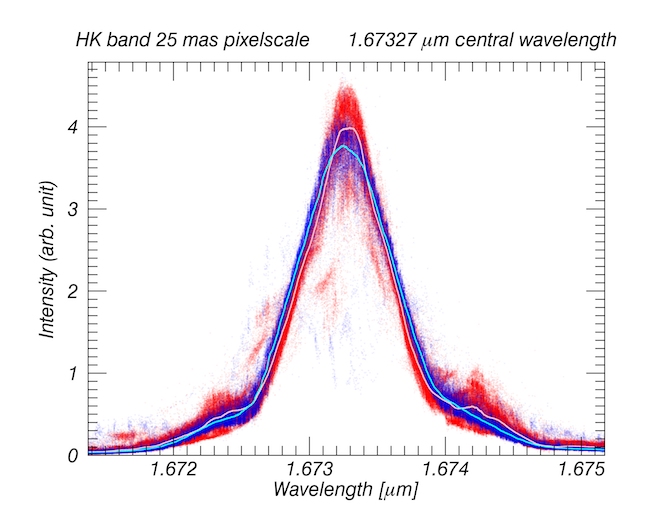}
			\includegraphics[width=1.0\textwidth, trim={1.5cm 0 0.5cm 0cm}, clip=true]{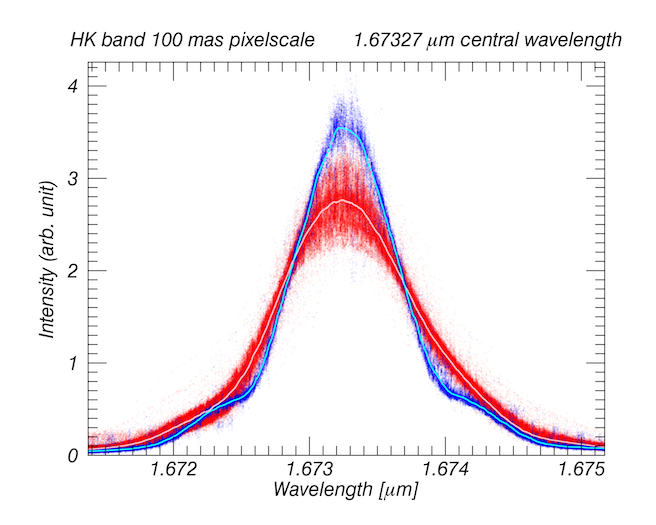}
			\includegraphics[width=1.0\textwidth, trim={1.5cm 0 0.5cm 0cm}, clip=true]{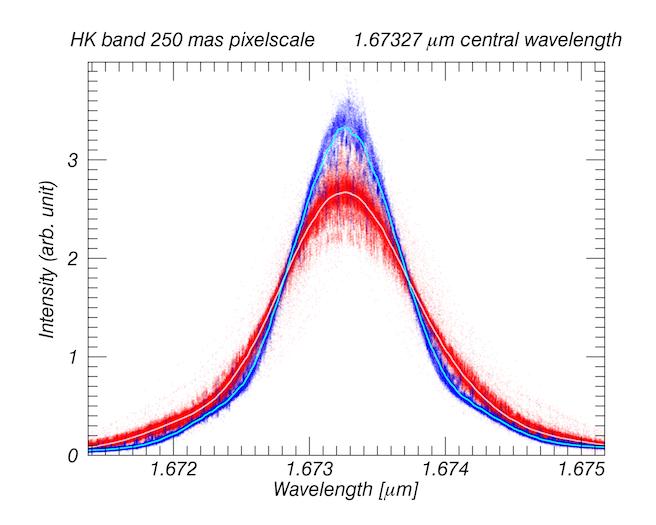}
		}
		\caption[Accumulated line profile along a single wavelength]{Accumulated line profiles at a single wavelength over the whole detector in each band and pixelscale. Rows from top to bottom: J, H, K, H+K gratings. Columns from left to right: 25, 100, 250 mas pixelscales. The cloud of points represents all data points of the supersampled lines from all locations on the detector, while the solid lines are the medians of the clouds. {\it Blue}: Post-upgrade {\it Red}: Pre-upgrade.}
		\label{fig:lineprofiles}
	\end{center}
\end{figure}

To learn about the average emission line over the detector as opposed to at a single location on the detector, we create a hypersampled line profile for each emission line in each detector column. This makes the statements more statistically reliable and does not only focus on selected parts of the detector. Figure \ref{fig:lineprofiles} shows the data points of the accumulated spectral line profiles along one emission line in each band and pixelscale. Blue dots are from the post- upgrade measurements, while red dots are from pre- upgrade measurements. The solid line shows the median that was calculated within a box filter over 40 equal fractions of the wavelength range of each image to get a fairly smooth curve without smoothing out too many details. Emission lines close to the center of band, and therefore detector, were chosen. In order to compare the old and new LSF, the line profiles are normalized to the median flux. The y-scale is in arbitrary units and should not be compared between the individual plots. From the cloud of data points the variance of the emission line shape can be determined. Instead of being Gaussian, the hypersampled line profiles have a distinct shape varying with wavelength and pixelscale.

The most striking feature of the line profiles is the presence of shoulders on the main peak. The trend for the shape of the profiles as a function of wavelength is to have higher shoulders at shorter wavelengths. When moving from longer to shorter wavelengths the shoulders which are quite symmetrical at longer wavelengths (K-band) gradually rise up, and become asymmetric. Finally in J-band (top row of figure \ref{fig:lineprofiles}) the shoulders are nearly as high as the central peak and cause the whole line to be broadened. A different case is present in the H+K-band (lowermost row of figure \ref{fig:lineprofiles}). There the shoulders are less distinct, so that they don't form defined peaks, but show a kink at the lower parts of the profiles, which is mainly an effect of the lower resolution of $\mathrm{\sim 2000}$ in H+K-band. Looking at the trend in pixelscale one finds that in the 25 mas pixelscale the shoulders are easy distinguishable (see left column of figure \ref{fig:lineprofiles}, while in the larger pixelscales (middle and right columns of figure \ref{fig:lineprofiles}) the shoulders get washed out, leading to a broadened line profile.

For more details about the variation of the line profiles, including at multiple wavelengths within a single band, see appendix \ref{fig:lineprofilesj} to \ref{fig:lineprofileshk}.

It is easiest to see the differences in the line profiles before and after the SPIFFI upgrade in Figure \ref{fig:lineprofiles}. There the accumulated line profiles of one emission line across the detector are shown. Red dots correspond to the pre- upgrade and blue to the post- upgrade data. In all bands and pixelscales it can be seen, that the side peaks or shoulders became more distinct after the upgrade. Also the distribution of the cloud of data points became sharper. In the seeing limited pixelscales one can see that the central peak has a higher amplitude after the upgrade (all plots are normalized by the flux under the median curve). At the base the broadened LSFs got slightly narrower for most bands and pixelscales or stayed the same. Furthermore the shoulders got slightly shifted towards the center. In the 25 mas pixelscale the changes are not that obvious as in the larger pixelscales. The side-peaks and shoulders became slightly more distinct and sharper due to the upgrade, while the amplitude of the central peak stayed quite constant. A special case shows up in J-band in the AO pixelscale, where the left shoulder vanished after the upgrade. What makes the J-band line profile furthermore special, is that in the two larger pixelscales two bumps show up after the upgrade very clearly that can only be recognized very vague in the pre- upgrade data. This effect, though not as strong can also be recognized in H- and K-band mainly in the 100 mas pixelscale, where two distinct shoulders or side-peaks are present after the upgrade, while in the pre- upgrade data there was only one or even no peak and maximal a slight shoulder.

The new collimator mirrors and their very tiny residual diamond turning marks can be ruled out to produce such complex line profiles as in figure \ref{fig:lineprofiles}. The building up effect of new side-peaks and more distinct shoulders after the upgrade can be explained by the old collimator wavefront, respectively the old collimator mirror surfaces. Due to the residual diamond turning marks, and the surface deformation on M2 and M3 caused by the post-polishing process the wavefront of the collimator was disturbed with high and low frequency perturbations. This disturbed wavefront washed out the small features of the line profiles especially in the larger pixelscales, since there the footprint on especially M3 is large enough to be affected by many deformation features of the mirror surface and thus a superposition of those. For the new collimator mirrors and thus the collimator wavefront this is not the case anymore. As a result shoulders, side-peaks or even as in J-band distinct bumps which have been washed out before, show up. This is in consonance with the observation that in the 25 mas pixelscale the side peaks and shoulders vanished after the upgrade as in J-band or H-band in the appendix in figure \ref{fig:lineprofilesh}. Since in the small pixelscale the footprint of the beam on M2 and especially on M3 is much smaller only surface deformations with a high spatial frequency affect the collimator wavefront and thus the spectral line profiles. This leads to a diffraction effect that built up side-peaks in H-band and especially in J-band, where the wavelength is closest to the surface deviations of the mirrors and the wavefront errors of the collimator wavefront. An effect of washing out cannot be observed in the smallest pixelscale as strong as in the larger pixelscales, since not so many wavefront deformation features participate due to the smaller footprint, neither before nor after the upgrade.

On top of that there could be another additional factor that influenced the change of the line profiles due to the upgrade. This factor is the different collimator pupil spot on the gratings. Because M3 was rotated by 23' in order to correct for the astigmatism induced by the wrong surface, the default grating carousel position also have to be different by 23'. The result is that the reflection angle stays the same as before and thus the image on the detector is at the same position. However the another result is, that the grating positions changed with respect to the pupil position. This has two consequences. The first one is that the \lq footprint\rq \ of the beam on the gratings moved by around 2 mm and second that now different locations of the gratings intersect with the pupil plane. If one assumes a non ideal surface of the grating, like it is done in section \ref{sec:discussion_lineprofiles}, then slightly different regions of the grating deformations influence now the beam than before the upgrade. This has an effect on the spectral line profiles and can change them slightly, but this is not expected to be large enough to vanish the side left side-peak in the 25 mas pixelscale in J-band or H-band, while it might contribute by a small amount to the enforcement of the shoulders in the larger pixelscales in H- and K-band.

The effect of the coma in the collimator wavefront makes the line profiles asymmetric, because it reinforces one side of the spectral line by giving it more power while the other side gets attenuated. One can see this asymmetrical behavior best in the 100 mas pixelscale in H- and K-band. While the bumps in the larger pixelscales of J-band may be affected by the coma in the wavefront of the collimator, they are definitely not caused by it. This can be seen in figure \ref{fig:differentorders} in section \ref{sec:discussion_lineprofiles}. For all three plots a J-band emission line is used and the J-, H-, and K-band grating in different orders. The bumps are only when the J-band grating is used. For the other gratings these bumps disappear, indicating that it is a grating intrinsic feature and not caused by the coma of the collimator wavefront.

In the next section the focus lies on the variation of the lineprofiles within a slitlet and across the detector. This section shows in what way the variance of the line profiles got sharper after upgrade what can be already seen in the plots of figure \ref{fig:lineprofiles}, where the cloud of blue data points is sharper as the red one.

\subsection{Variation of the Line Profiles}\label{sec:var_line}
\begin{figure}[htbp!]
	\begin{center}
		\includegraphics[width=0.47\textwidth, trim={0.3cm 0 0.8cm 0cm}, clip=true]{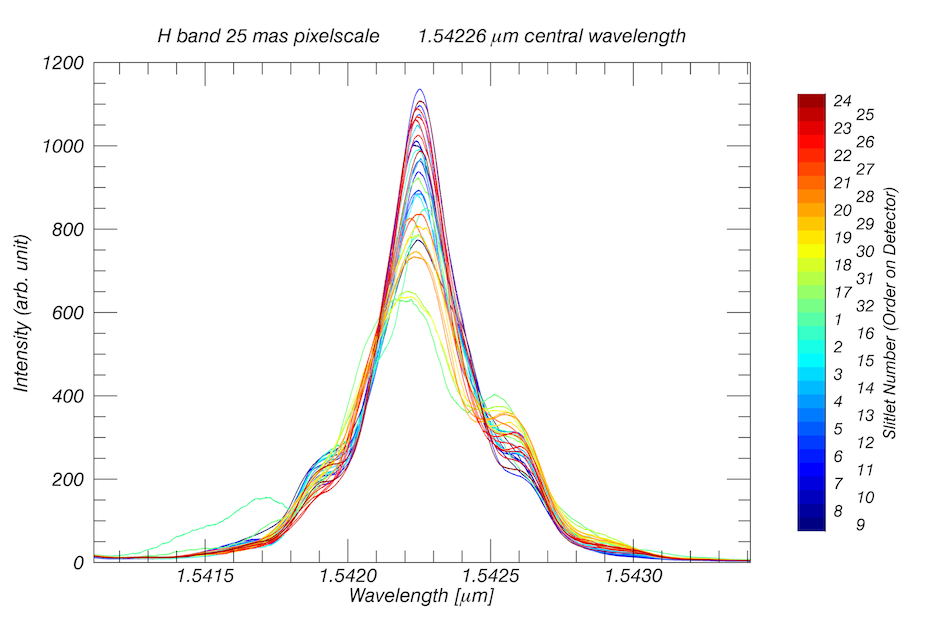}
		\includegraphics[width=0.52\textwidth, trim={0.8cm 0.99cm 0.8cm 0cm}, clip=true]{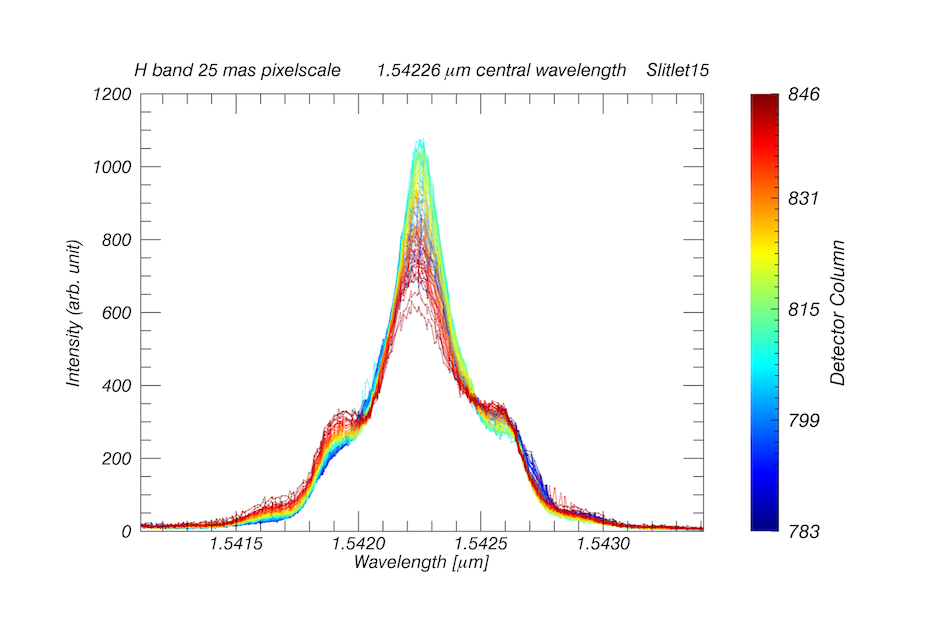}
		
		\subfloat[\label{fig:detector_vatiation}Median LSF of individual slitlets across the detector along the pseudo slit.]
		{\includegraphics[width=0.47\textwidth, trim={0.3cm 0 0.8cm 0cm}, clip=true]{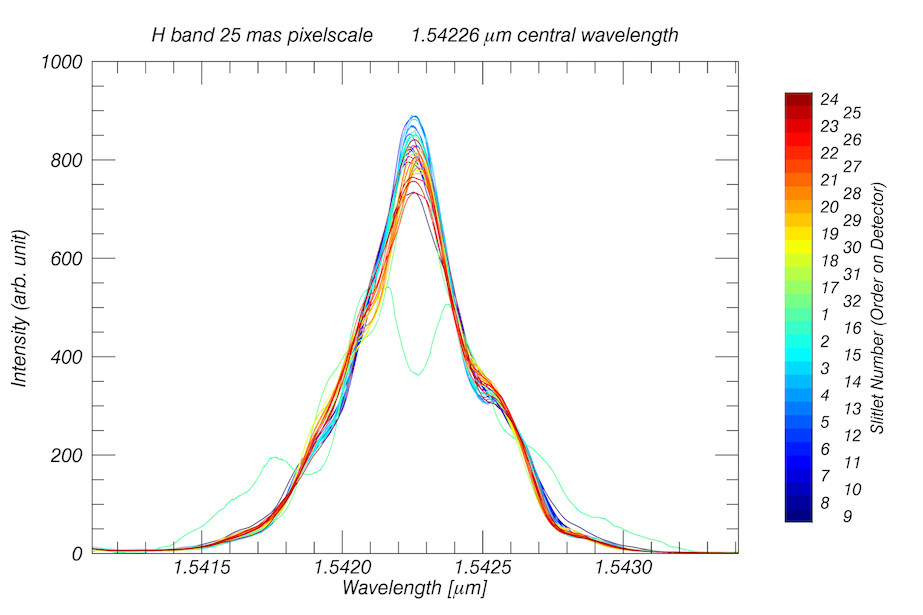}}
		\subfloat[\label{fig:slitlet_vatiation}Variation of the LSF within slitlet 15.]
		{\includegraphics[width=0.52\textwidth, trim={0.8cm 0.99cm 0.8cm 0cm}, clip=true]{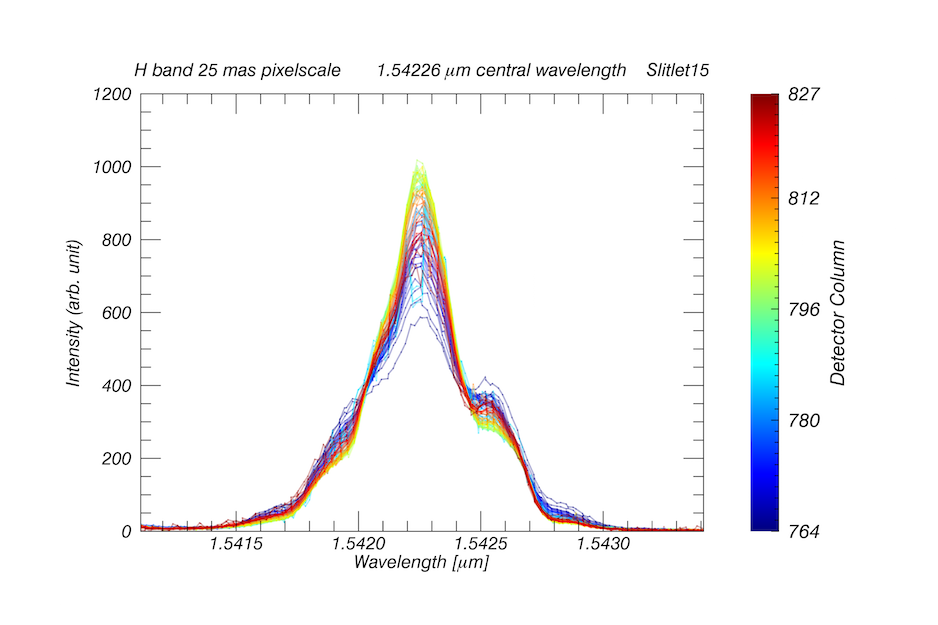}}
		\caption[Variations of the line profiles in H-band 25 mas pixelscale]{Variations of the line profiles in H-band 25 mas pixelscale. (upper row: pre- upgrade, lower row: post- upgrade)}
		\label{fig:variations}
	\end{center}
\end{figure}

To find out the origin of the non Gaussian line profiles, it is important to take a closer look at the variation of the LSF across the detector in individual emission lines. In figure \ref{fig:detector_vatiation} the variation of the hypersampled line profile along one emission line in H-band in the 25 mas pixelscale is shown. This was chosen, because in the smallest pixelscale one can see the variations of the line profile best. The different colored lines are the medians over the respective slitlets. So to speak it shows the variation over the pseudo slit. The upper images are from pre- upgrade data, while the lower ones are post- upgrade. The variation here has at least partly to do with the different beam footprints of the individual slitlets on the collimator mirrors and diffraction grating. The footprints of the slitlets are completely separated on M1, overlap slightly on M2, overlap significantly on M3 which is close to the pupil position of the spectrometer, and overlap nearly completely on the gratings. Since the grating is tilted with respect to the pupil, it is not exactly at pupil position.

Figure \ref{fig:slitlet_vatiation} however shows the variation of the hypersampled line profiles within slitlet 15, which was chosen, because it's close to the center of the detector an thus the corresponding mirror on the small image slicer is nearly not tilted, meaning that the defocus problem is minimized. The detector columns corresponding to slitlet 15 are around 783-846 before the upgrade and around 764-827 after the upgrade (depending on band and pixelscale). The columns changed because the detector was shifted in the upgrade in order to get the outermost slitlet 24, which was falling of the detector, aligned and further to move the dead spot near the center of the detector from slitlet 16 into slitlet 1 (see section \ref{sec:detector_adjustment}). In a comparison between pre- and post- upgrade data it can bee seen that the variation of the LSF across a single slitlet did not change by much, while the variation of the mean slitlet line profile across the detector was diminished by the upgrade. The variation within the slitlet is after the upgrade larger than the median variation of the LSF across the detector. The upgrade did not change the line profile variation across a single slitlet significantly in general.

Since the variation in the median line profile of each slitlet over the full length of the pseudo-slit was reduced by the upgrade, the most probable explanation for this improvement is the replacement of specifically the M2 collimator mirror, which is now much more uniform (compare with section \ref{sec:M2} on the single mirror measurements of M2.)

Further plots of the variation of the LSF are in the appendix in figure \ref{fig:slitlet_j_pre} to \ref{fig:slitlet_hk_post} for the variation within a slitlet and figure \ref{fig:median_j} to \ref{fig:median_hk} for the variation of the median line profile across the detector. 

One question that stays open is, why there is still such a strong variation of the line profiles within a single slitlet (see figure \ref{fig:slitlet_j_post} for J-band and \ref{fig:slitlet_h_post} for H-band). They are especially present in the J- and H-band in the smallest pixelscale. Assuming the spectral line profiles are caused by the diffraction gratings, as it is discussed in section \ref{sec:discussion_slitlet}, there are several possibilities for the strong variation of the line profiles within a single slitlet. Since they all build on the assumption that the gratings cause the spectral line profiles they are discussed after the next section which describes the possible deformations of the gratings

\subsection{Discussion about the Origin of the Line Profiles }\label{sec:discussion_lineprofiles}
\begin{figure}[htbp!]
	\begin{center}
		\resizebox{1.0\textwidth}{!}{
			\includegraphics[width=1.0\textwidth, trim={0.8cm 0 1.0cm 0cm}, clip=true]{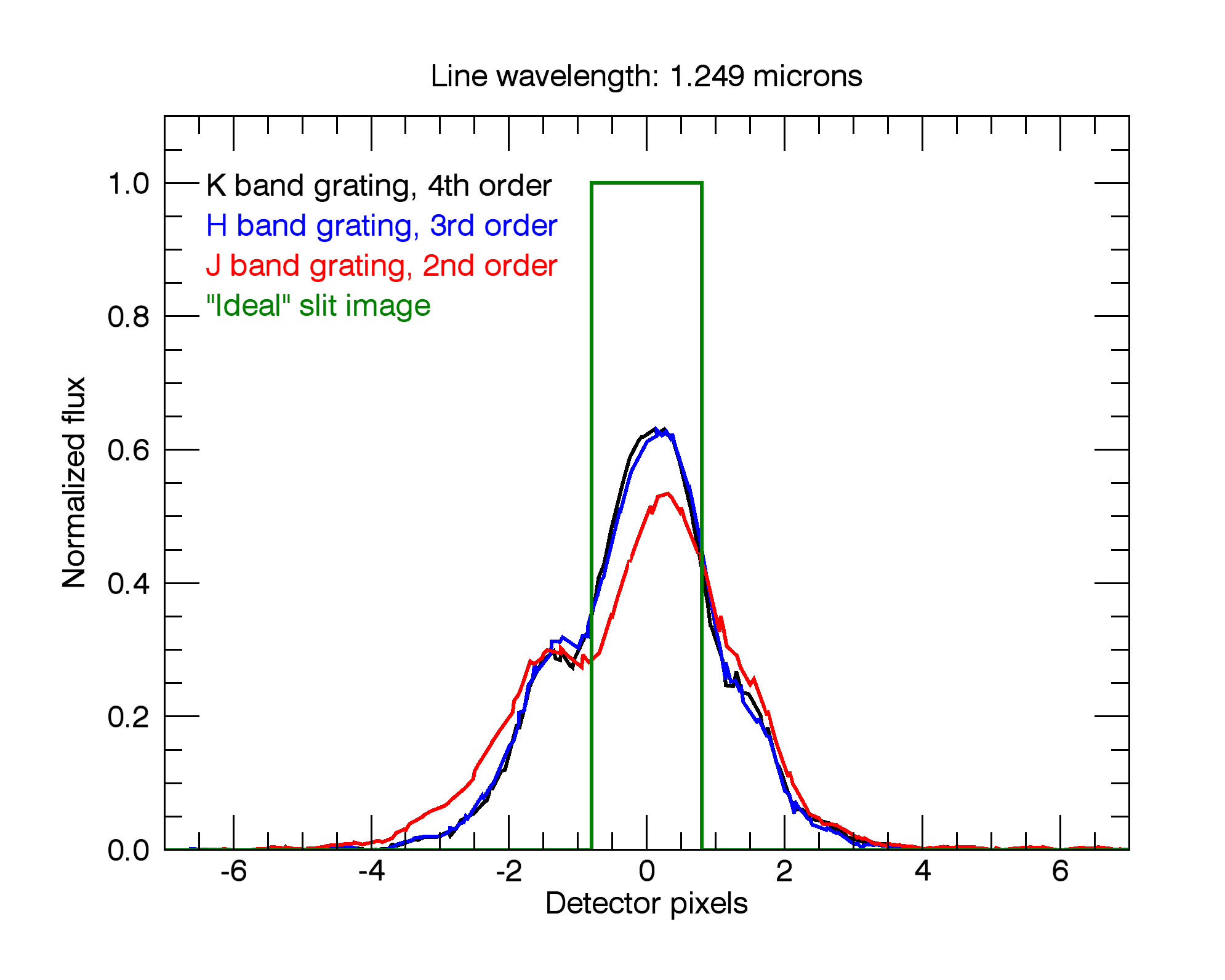}
			\includegraphics[width=1.0\textwidth, trim={0.8cm 0 1.0cm 0cm}, clip=true]{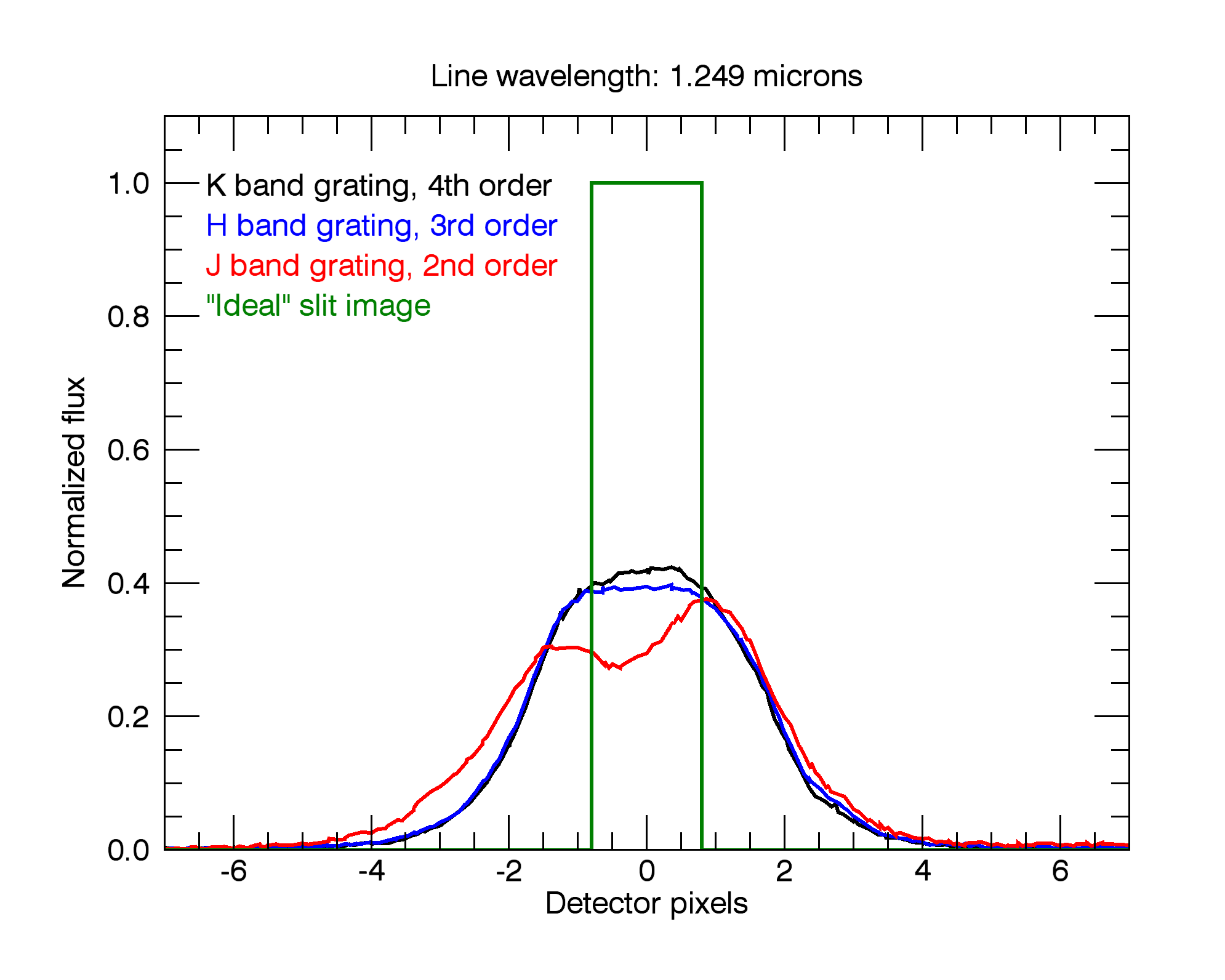}
			\includegraphics[width=1.0\textwidth, trim={0.8cm 0 1.0cm 0cm}, clip=true]{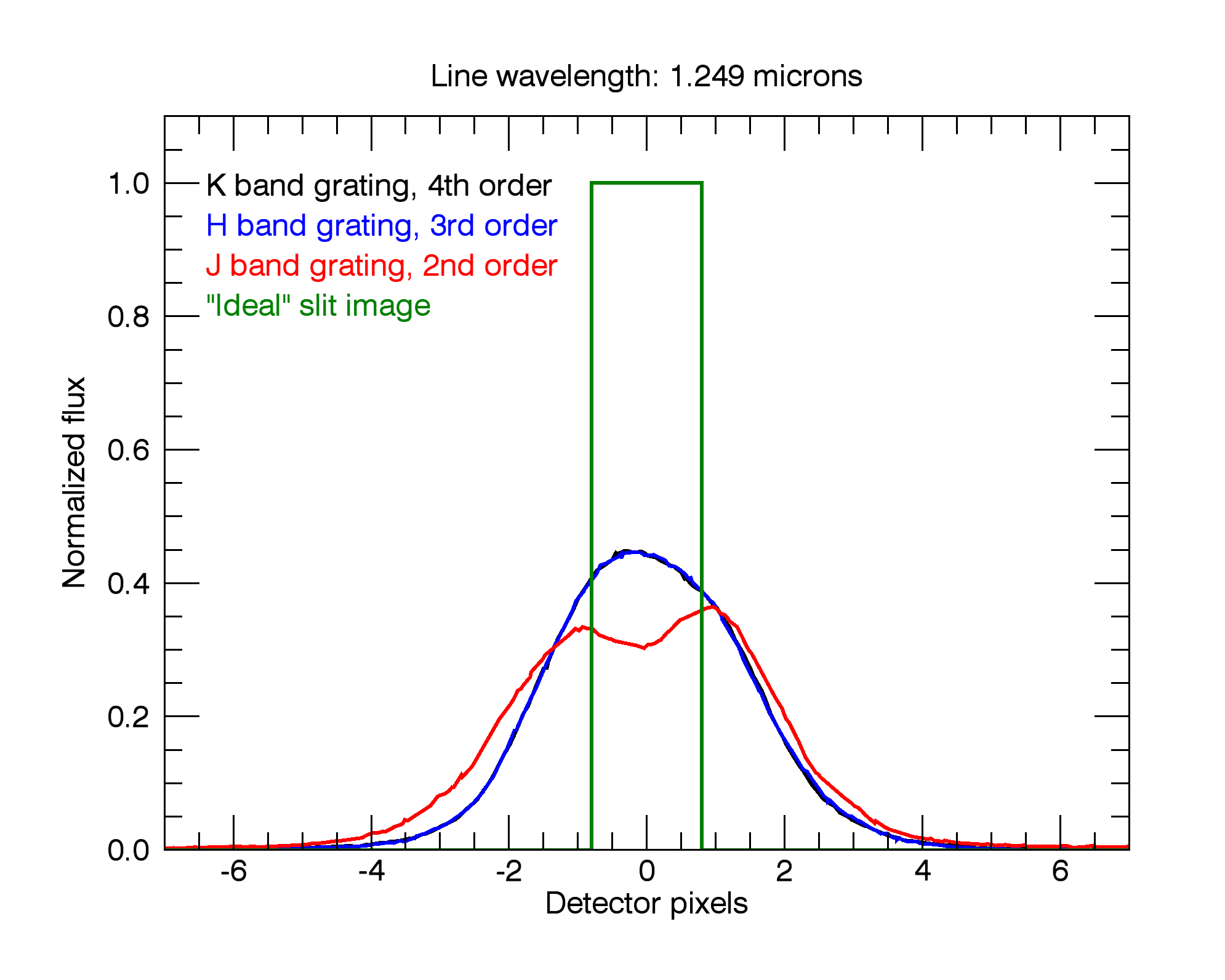}	
		}
		\caption[Line profiles for different diffraction gratings at one wavelengths]{Plots of single super-sampled lines on different diffraction gratings in all pixel pixelscale (left: 25 mas, center: 100 mas, right: 250 mas), normalized to the total integrated power in the line. For all pixelscales the same line at $\mathrm{\sim 1.25}$ micron is used together with the J-, H-, and K-band diffraction gratings. The line morphology is remarkably similar between the gratings, as expected by our optical model, though not identical. \cite{george16}}
		\label{fig:differentorders}
	\end{center}
\end{figure}

After the old collimator mirrors were replaced by new ones in this upgrade it was possible to measure the wavefronts of the old collimator mirrors that were in SPIFFI until 2016 (see chapter \ref{ch:chapter_mirrors}). From these measurements and from the only small increase in the resolution (see section \ref{sec:resolution}) of the spectrograph in comparison to the pre- upgrade resolution, it became clear that the diamond turning marks on the old mirrors and their surface deviation were only a minor contributing factor to the asymmetric and multi-peaked shape of the SPIFFI line profiles. Even with the third of a micron coma in the wavefront of the new spectrometer collimator, the surface form of the collimator mirrors can be ruled out for causing the complex spectral line shapes. The coma effects the line profiles by making them slightly unsymmetrical, and of course broadens the lines slightly, but the complex structure of the lines remaining is not an effect of the collimator mirrors. The only thing left in the spectrometer and could distort the wavefront enough to result in the observed line profiles are the diffraction gratings. 
To determine how the line profiles are affected by the gratings, measurements were taken with the different gratings in different orders for a given wavelength (shown in figure \ref{fig:differentorders} by \cite{george16}).

\begin{table}[htbp!]
	\begin{center}
		\begin{tabular}{c|c}
			\hline
			Band&Dispersion [$nm / px$]\\
			\hline
			\hline
			J	&	0.15\\
			H 	&	0.2\\
			K	&	0.25\\
			H+K	&	0.5\\
			\hline
		\end{tabular}
	\end{center}
	\caption[Dispersion on the detector]{Dispersion on the detector.}
	\label{tab:dispersion}
\end{table}

The gratings were operated in this measurement in different orders for a certain wavelength to have the incidence angle between the beam and the grating always as close as possible to the default angle. The scale on the x-axis is here in detector pixels. A conversion is given in table \ref{tab:dispersion} showing the dispersion on the detector for each band. The image of the one slitlet on the detector is $\mathrm{27 \mu m}$ wide due to anamorphic distortion of the grating. Since the size of one slitlet on the detector is $\mathrm{27 \mu m}$, this corresponds to 1.5 pixels indicated by the normalized box in the plots.

A difference between the J-band grating and H-band and K-band grating respectively can be seen. Also the bump in J-band in both larger pixelscales does only exist in the J-band grating. From the plots, especially in the larger pixelscales, one can tell, that the J-band grating behaves differently from the H- and K-band gratings producing this bump. The different behavior cannot be an effect of the collimator wavefront, since for all three plots the same wavelength is used. Fact is that the bumps for the line shapes with the J-band grating could only be seen as distinct, because the new collimator does not wash out this feature anymore as the old collimator mirrors did. 

The question that remains is, why in the small pixelscale the gratings behave more identically, and all show side peaks and shoulders independent of the grating? An answer to this could lie in the non-sinusoidal large scale structure of the grating resulting from bi-metallic bending stress. The small pixelscale would be affected different because only one or two periods of this semi-sinusoidal perturbation are illuminated, while in the large pixelscales, different frequencies and amplitudes of this perturbation would smear out small features as described in the next paragraphs.

\begin{figure}[htbp!]
	\begin{center}
		\includegraphics[width=12cm]{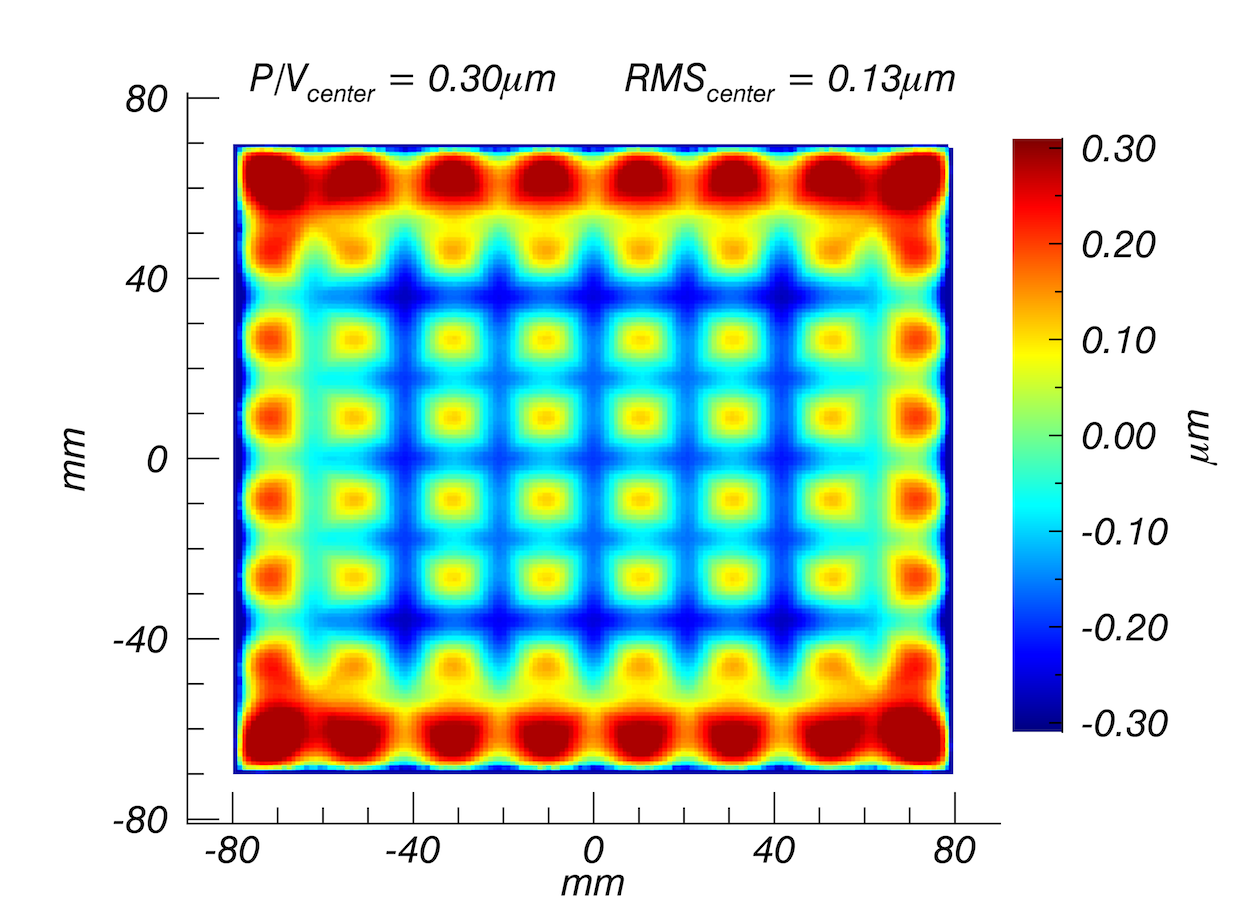}
		\caption[Finite-Element-Analysis of the gratings]{The surface deformation of the gratings caused by bimetallic bending stress between the nickel layer and the aluminum blank. The surface deformation follows the periodicity of the light-weighting holes in the aluminum blank.\cite{george16}}
		\label{fig:FEM}
	\end{center}
\end{figure}

One spare J-band grating is under investigation at the moment. The overall surface form will be measured interferometrically in a Cryostat at 80K. The goal is to find out how much deformation of the grating surface caused by bimetallic bending stresses one expects in SPIFFI. This bending stress is caused by the $\mathrm{\sim 100 \ \mu m }$ thick remaining nickel layer on the large surfaces and $\mathrm{\sim 125 \ \mu m }$ nickel layer on all other surfaces and in the light-weighting holes. (For more details about the grating see section \ref{sec:spiffi}.)
From a finite element analysis (FEA) of the grating one expects a deformation of the surface which is structured like the pattern of the light-weighting holes in the aluminum grating blank (see figure \ref{fig:FEM} from \cite{george16}). The distance of the light-weighting holes is 21 mm in the vertical direction and 18 mm in the horizontal direction for all gratings, since they were manufactured on identical blanks. From this, the surface deviation for the individual gratings should be quite similar. The expected FEM peak-valley deviation from a plane surface at 77 K is in the center region $\mathrm{P/V_{center} = 0.3 \ \mu m}$ with an RMS in the center of $\mathrm{RMS_{center} = 0.13 \ \mu m}$ \cite{george16}. However, right at the edges of the grating the deformation is higher, but this is masked in SPIFFI with an elliptical stop directly in front of the grating. Each grating was polished separately resulting in different nickel layer thicknesses and thus an uncertainty on the deformation values and the overall wavefront.

Because the surface deviation should be quite similar for the individual gratings, the wavefronts after the gratings should be also similar and thus have similar effects on the LSF. Nevertheless each grating has a unique grating constant and blaze angle, is further operated in different orders resulting in slightly different incident angles for different bands and thus different distances to the focal plane.

In a simple optical model of the spectrograph to test the hypothesis that the grating might cause the line profiles in SPIFFI, \cite{george16} applied the FEA model of the grating together with the measured collimator wavefront from figure \ref{fig:oldnewwavefront} to an optical model of the spectrograph. It was found, that by scaling the amplitude of the FEA model surface deviation by a factor of 1.25-1.5 it is possible to reproduce the shape of the LSF as observed. The model showed that the location of the shoulders in the line profiles is only caused by the frequency of the surface deviation, which is determined by the light weighting holes. The relative height of the shoulders between different wavelengths corresponds directly to the wavelength of the incoming light (larger wavelengths correspond to smaller shoulders), while the total height of the shoulders is correlated to the amplitude of the grating surface deviation. Additionally, the effect of the smearing out of the shoulders in the larger pixelscales is reproduced in this model. The non-sinusoidal shape of the surface deformation (the frequency varies slightly and also the amplitude varies at the edges of the grating) explains the washed out lines, as a larger fraction of the grating surface is illuminated at larger pixelscales. As a result the line profiles do not show as much substructure as in the AO pixelscale.

The hypothesis that the diffraction gratings may cause the non Gaussian line profiles in SPIFFI may be supported by the interferometric and cryogenic measurement that will be carried out in the laboratory. But ultimately these measurements can only show the size of the grating deformation on the spare grating, (not the ones in SPIFFI), and deliver new interferograms for simulations. If the measured deformations match the amplitude required in the simulation to reproduce the observed line profiles, then it builds a strong case that the diffraction gratings are to blame for the line profiles in the instrument. Though ultimately, only by replacing the gratings in SPIFFI with new ones a definitive answer will be reached.

\subsection{Discussion about the Origin of the LSF Variation within Single Slitlets}\label{sec:discussion_slitlet}

The strong variation of the spectral line profiles within a slitlet are still an open question. In J- and H-band in the smallest pixelscale the variation is the largest, while in the 100 mas and 250 mas pixelscale the LSF nearly does not vary across a single slitlet. If one assumes that the spectral line profiles are caused by the diffraction gratings as discussed above, there are several possibilities for the strong variation of the line profiles within a single slitlet.

The assumption is that different illumination spots on the grating affect the spectral line profiles in different ways due to the proposed surface deformations from section \ref{sec:discussion_lineprofiles}. Since the median line profile variation of the individual slitlets across the detector is smaller or quite similar (see J-band figure \ref{fig:slitlet_j_post} and \ref{fig:median_j} in the appendix) to the variation within a slitlet, it can be concluded that both are likely to have a similar or even the identical origin. In the two larger pixelscales the variations of the line profiles within a slitlet are much smaller than in the 25 mas pixelscale. For them larger parts of the gratings are illuminated, while for the 25 mas pixelscale the central maximum of the illumination spot is only affected by one or two bumps on the grating surface originating from the proposed bimetallic bending effects (see section \ref{sec:discussion_lineprofiles}). The result is that in the larger pixelscales the periodic surface deviation structure effects the wavefront, while for the smallest pixelscale the periodicity of the deviation does play a minor role for the variation of the line profiles. The periodicity mainly affects the spacing of the shoulders in the spectral line profiles. The amplitude of the LSF peak depends on where the pupil is located on the gratings. If the pupil is now shifted on the grating especially the smallest pixelscale will be affected by it.  This is because the central spot of the diffraction pattern would be disturbed by totally different surface deviations (peaks or valleys), while in the larger pixelscales still the periodic structure will have the main influence independent of the concrete position of the pupil on this periodic grid structure because the \lq footprint\rq \ is large enough to fill significant parts of the grating surfaces.

From the plots \ref{fig:slitlet_j_pre} to \ref{fig:slitlet_h_post} in the appendix it can be seen that both the variation within a single slitlet as well as the symmetry of the evolution of the LSF along a slitlet changed due to the upgrade. This may be caused by the fact that the pupil position on the grating changed in the upgrade. This shift due to the upgrade can be caused by different factors. One could be vertical shift of the pupil due to the strong surface deviation of M3. On the detector the shift of the image due to M3 is 22 pixel corresponding to an offset of 0.4 mm in the illumination of the pupil. A second factor is the position of the grating, which had to be adapted to the rotation of M3 which was applied to correct for the aberrations induced by its surface deviation. By this the pupil position changed by about 2 mm with respect to the grating.

There are also different factors that can cause a shift of the pupil on the grating depending on the field, which should be considered when evaluating the variation of the spectral line profiles. The position of the gratings is such that they only intersects with the pupil plane in one vertical line (relative to the instrument plate) and thus the other parts of the grating are not exactly in pupil. However, this is not likely to cause the variation in the line profiles. The illumination of the grating in the smallest pixelscale results in the movement of \lq footprint\rq \ locations that are not exactly in the pupil plane by less than a millimeter for the 6 degree field of the spectrometer. This sub-millimeter shift is not expected to affect the variation of line profiles that much, as measured. Additionally, for the small field of a single slitlet, the movement of the illumination on the grating caused by the effect that the grating surface is not identical with the pupil plane is negligible.

\begin{figure}[htbp!]
	\center
	\subfloat[\label{fig:focus_slitlet}Defocus across the slitlets]
	{\includegraphics[width=7.5cm]{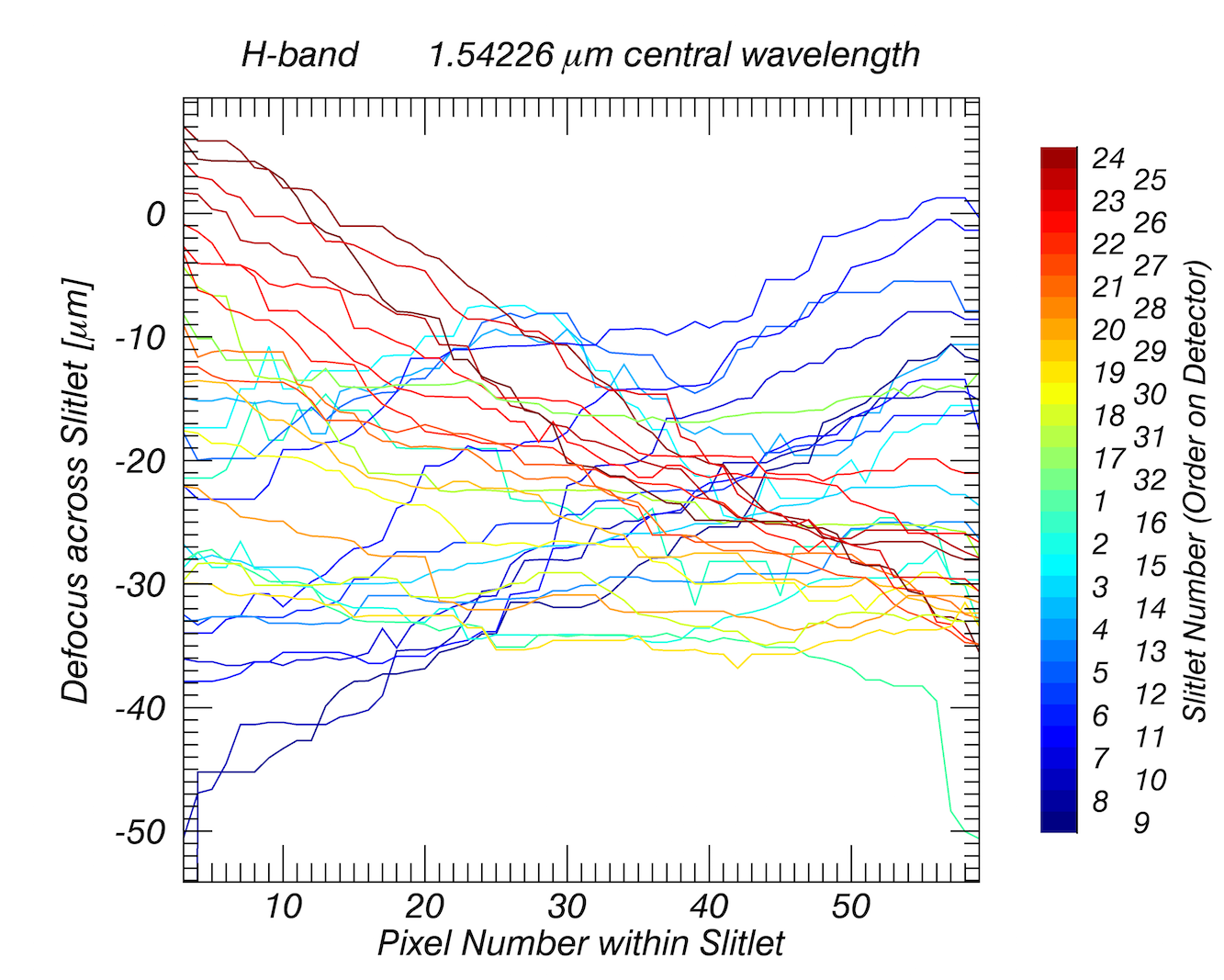}}
	\subfloat[\label{fig:focus_change}Maximal change of defocus for each slitlet]
	{\includegraphics[width=7.5cm]{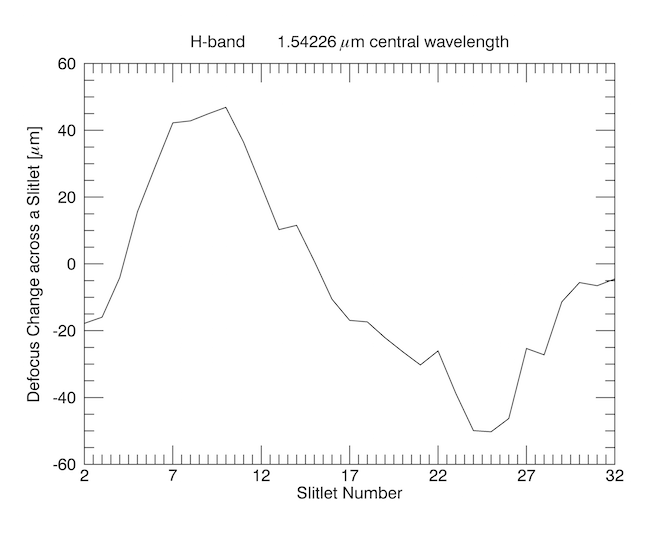}}
	\caption[Defocus of the individual slitlets]{Defocus of the individual slitlets. The left image shows for each slitlet how much out of focus the different parts of the slitlets are. The right plot shows the maximal change of defocus along a slitlet as a function of the slitlet position in the small image slicer.}
	\label{fig:focus}
\end{figure}

From the defocus measurements that were carried out during the upgrade intervention (see section \ref{sec:detector_adjustment}) one can not only find out the overall defocus of the detector, but also additionally the defocus as a function of position within a slitlet. Figure \ref{fig:focus_slitlet} shows the defocus as a function of the field position within a slitlet from these measurements. The different colored lines correspond to the individual slitlets. The order of the colors corresponds to the slitlet position of the detector, while this position directly corresponds to the tilt of a slitlet mirror in the small slicer. It can be seen that slitlets on the lower part of the detector (blue) show a positive slope, while those on the upper part of the detector (red) show a negative slope. In between the lines (green) are quite horizontal. The y-axis shows how far the corresponding pixel on the x-axis is out of focus. Positive numbers indicate that the position of the slitlet is behind the detector focus, while for negative numbers that part of the slitlet is in front of the detector focus. The fact that the centers of the individual slitlets are not exactly in focus (have the defocus value 0), but that there is a systematic offset, shows that the whole detector is in front of the focus by around 20 $\mathrm{\mu m}$ to 30 $\mathrm{\mu m}$. The plot shows that those slitlets with a high tilt in the small image slicer have also a high variation of the defocus dependent on the position on the slitlet. This is also shown clearly in figure \ref{fig:focus_change}, where the maximal change of the defocus along one slitlet is shown as a function of the slitlet number. It can be seen that the change in focus along the slitlet directly corresponds to the tilt of the slitlet mirror of the small slicer. The y-values in that plot correspond to the slope of a linear fit for each line in figure \ref{fig:focus_slitlet} multiplied by the length of one slitlet in pixels.

These defocus measurements were carried out only in the 250 mas pixelscale. There the defocus of the small slicer has the largest effect because the f-number of the beam from the pre-optics is the smallest compared to the other two pixelscales. As described above in this pixelscale there is nearly no variation of the line profiles within a single slitlet, because the size of the pupil is large enough, so that small shifts of the pupil depending on the field will not have an influence on the line profiles. For the smallest pixelscale the defocus effect of the small slicer is estimated to be smaller. The pupil size of the grating has only a tenth of the diameter compared to the 250 mas pixelscale, but the outer regions of the grating are illuminated by the diffraction pattern. However in the 25 mas pixelscale, effects of the exact location of the illumination of the grating deformation become relevant. In the optical model mentioned in section \ref{sec:discussion_lineprofiles}, a focus through-scan was applied with a magnitude similar to the measured defocus on the small slicer, and this showed that the defocus has a not negligible influence on the spectral line profiles in the 25mas pixel scale.

From figure \ref{fig:slitletvar} in the appendix, where for every slitlet the LSF variation in H-band in the 25 mas pixelscale is shown, it can be seen that also in the central slitlets on the detector (no tilt or small tilt on the small slicer) there is still a variation in the line profiles across a slitlet. For these slitlets the defocus on the small slicer does not play a role. The variation is slightly smaller (with exception of slitlet 16) than for the tilted slitlets. With respect to figure \ref{fig:focus_slitlet} for the slitlets with small slitlet numbers, the parts on the slitlet more to the right (larger detector column number) should be the ones with the highest amplitude. For the slitlets with large slitlet numbers, the left side of the slitlet (smaller detector column number) should have the the highest amplitude in the LSF. But for most slitlets the central slitlet parts have the highest amplitude, and for the slitlets with high slitlet numbers the behavior is contrary to what is expected from figure \ref{fig:focus_slitlet}. The conclusion is that the variance within a slitlet is not only caused by defocus effect, but the major component of the variance must be slitlet independent. This hypothesis could be confirmed with a defocus measurement that is carried out in the 25 mas pixelscale.

Another component that could cause a movement of the \lq footprint\rq on the gratings may be the individual slitlets of the small image slicer. For each slitlet the tolerance between input and output beam is 2' through the large slicer.\cite{tecza03} This corresponds to a possible offset of the pupil position between the individual slitlets of $\mathrm{\sim 1.5 \ mm}$. This is the case if the surfaces of the slicer mirrors are flat. If the slicer mirrors deviate from this, the pupil will move for the different field points on a slitlet. This surface deviation could for example be a convex or concave surface form of the small slicer mirrors or a torsion. For a concave or convex case, a sagitta of 1 micron leads to a 1 mm deplacement of the pupil in vertical direction. A torsion of one fringe in J-band however shifts the pupil for the outermost field points on a slitlet by around 1 cm, but in horizontal direction with respect to the instrument plate. When the image slicer was built, the measurements on the slitlets were done based on the interference pattern of a He-Ne laser and a reference glass plate. This method showed that the slitlet mirrors had a accuracy of a few fringes. Since this measurement method works in reflection, four fringes that were likely to be seen corresponds to $\mathrm{1.2 \ \mu m}$. A pupil shift of 1 mm vertically, produced by a concave or convex mirror surface deformation of the slitlet mirrors is not expected to cause such strong variations of the LSF within a slitlet as observed. However, a torsion of that amount and thus a pupil shift of about 1 cm could easily cause the variation measured. By the optical model from section \ref{sec:discussion_lineprofiles} it was confirmed that pupil shifts of several millimeters in the smallest pixelscale are likely to change the line profiles as measured in J- and H-band. 

The hypothesis described above builds on the assumption of the grating having a surface as shown in figure \ref{fig:FEM}. The conclusion of this hypothesis is that the grating causes the non ideal spectral line profiles as well as the strong variation of the line profiles within a slitlet. The latter may only be caused by the moving pupil for different fields within a slitlet. The variation of the line profiles across a slitlet is a combined result of the possible bumpy surface of the grating and the possible non-flatness of the slitlet mirrors of the small slicer. If the grating surface were not bumpy, the line profiles would not vary by such a large amount, even if the slitlet mirrors are not flat. However, what would remain is still a small variation in the line profiles caused by the defocus of the tilted slitlets.

Closely connected to the spectral line profiles is the spectral resolution that is in the end the relevant quantity for the performance of the spectrometer. To meet this demand, the next section investigates the resolution of the SPIFFI spectrometer.

\section{Spectral Resolution}\label{sec:resolution}
One of the most important characteristics of a spectrograph is its spectral resolution. It defines the smallest scale over which spectral features can be separated or distinguished. The design resolution of SPIFFI is $\mathrm{\sim 4000}$ for J-, H- and K-band as well as $\mathrm{\sim 2000}$ for H+K-band.\cite{eisenhauer00}

\subsection{Resolution Measurement}
In order to give numbers for the spectral resolution of SPIFFI which are relevant for the scientific users of the instrument, it is not helpful to give it a resolution measured on the hypersampled data. An additional complication with the hypersampled data is the question of how to define the resolution of a spectrograph where the spectral lines do not have the Gaussian shape, but have shoulders or double peaks. Since for most observers the resolution within a single exposure is relevant, here the resolution is calculated based on the individual single exposures used in the construction of the hypersampled line profiles. The benefit of the babysteps data in this case is, that the spectral lines are for each exposure shifted within a sub-pixel range, which leads to a slightly different shape of the undersampled LSF for each exposure, while the FWHM should stay roughly the same. In order to best match the actual scientific observation conditions, the convention chosen to define the resolution is to fit a Gaussian to the spectral lines in a single exposure and divide the wavelength of the spectral line by the FWHM of this Gaussian fit.

\begin{figure}[htbp!]
	\begin{center}
		\resizebox{1.0\textwidth}{!}{
			\includegraphics[width=7cm]{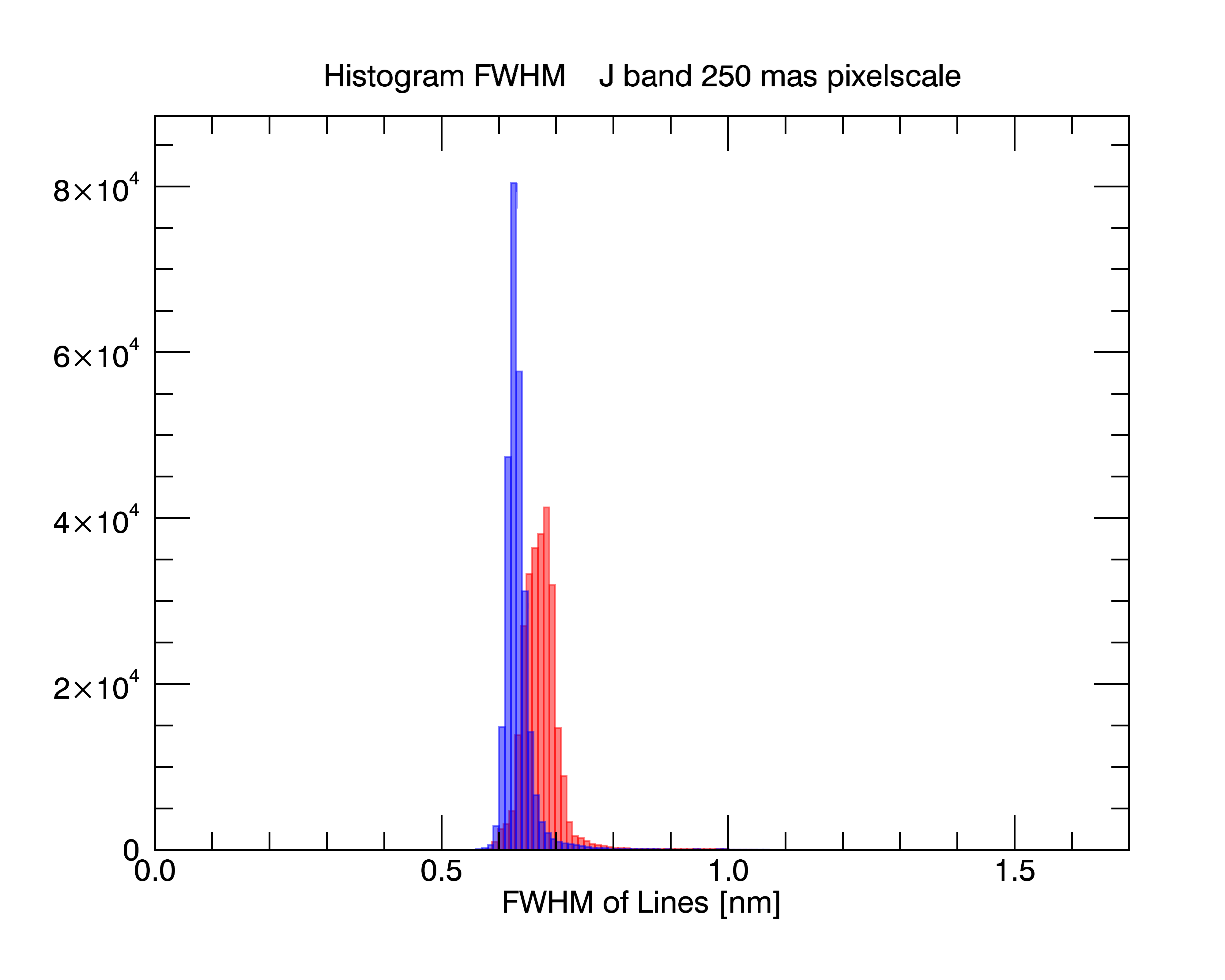}
			\includegraphics[width=7cm]{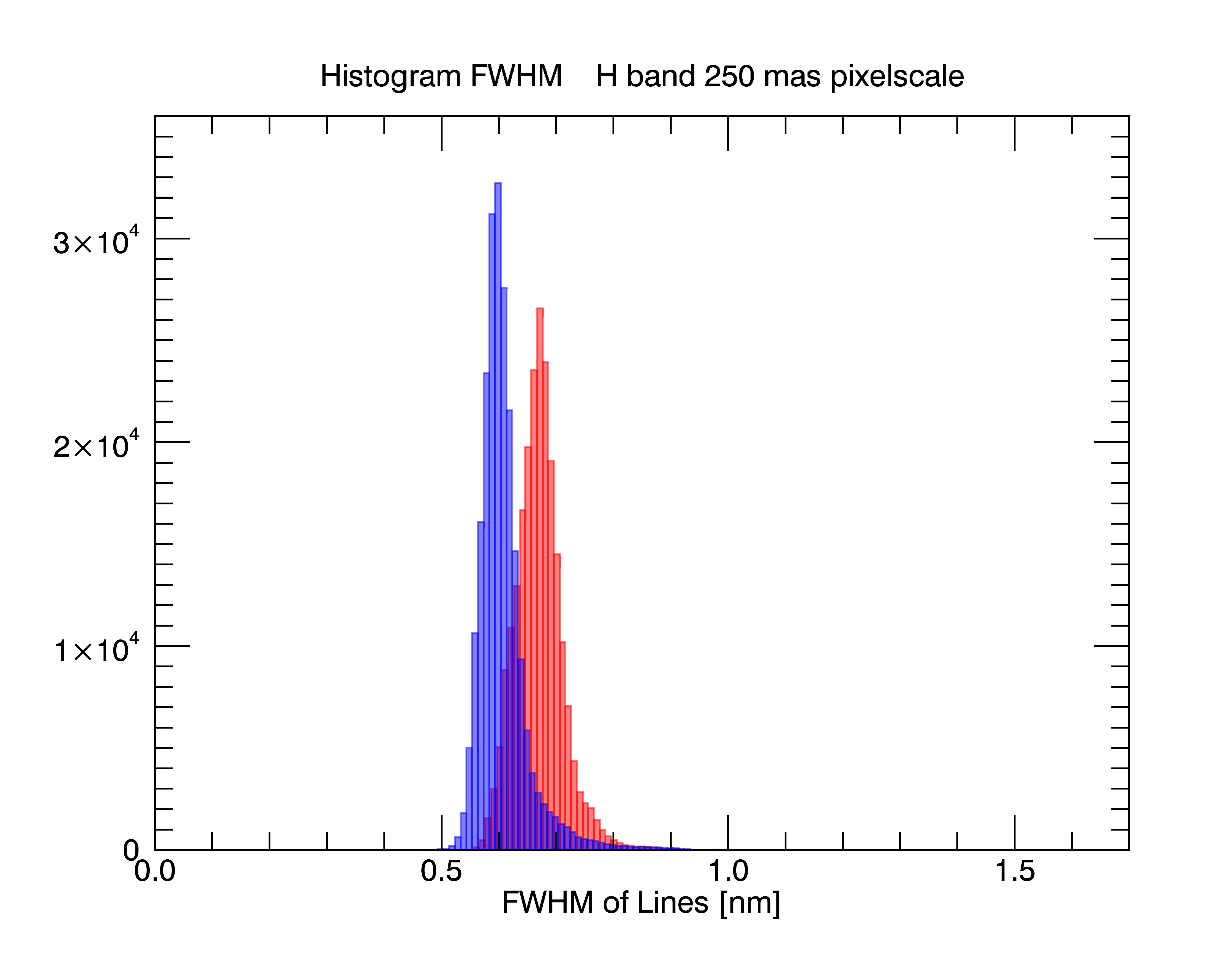}}
		\resizebox{1.0\textwidth}{!}{
			\includegraphics[width=7cm]{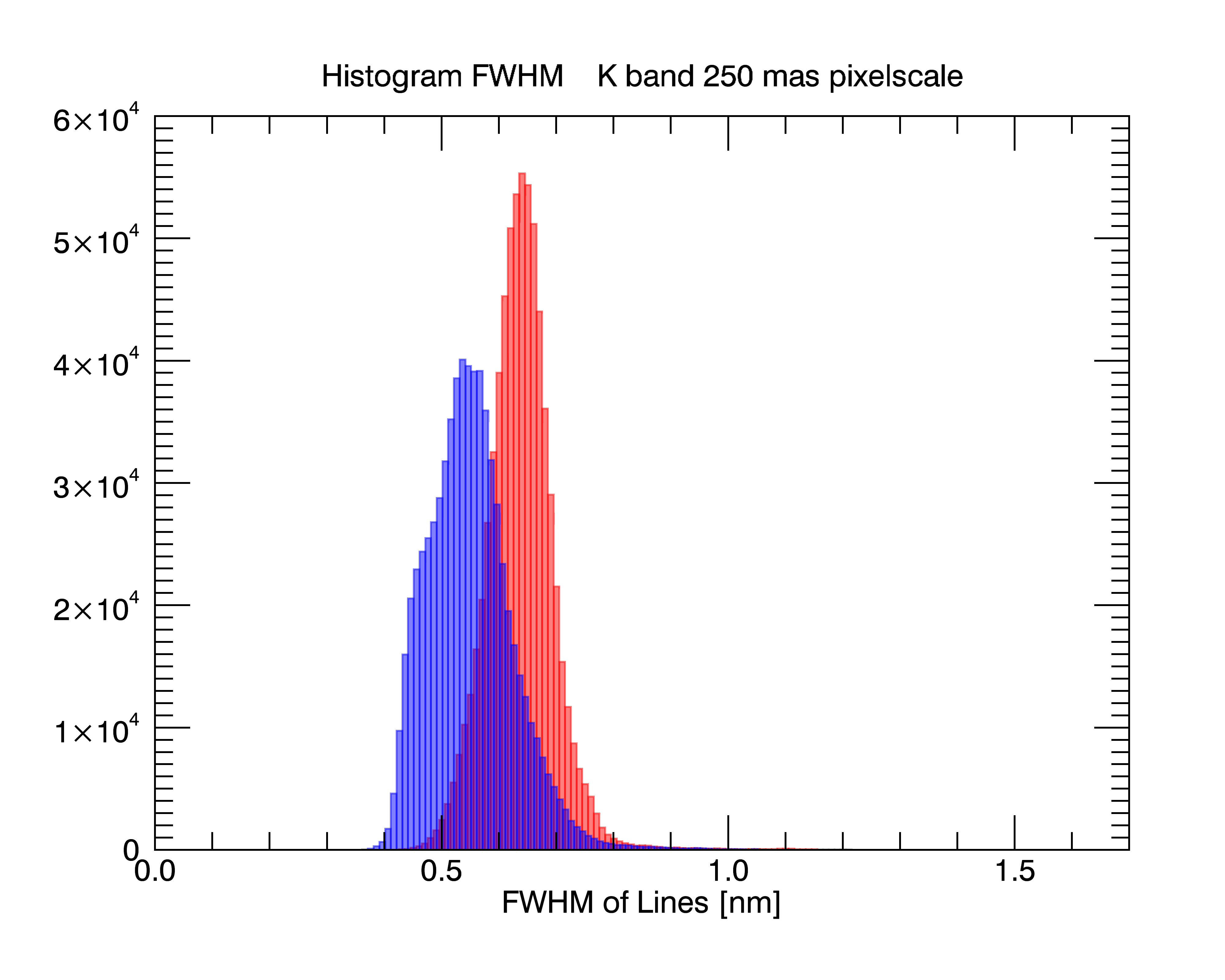}
			\includegraphics[width=7cm]{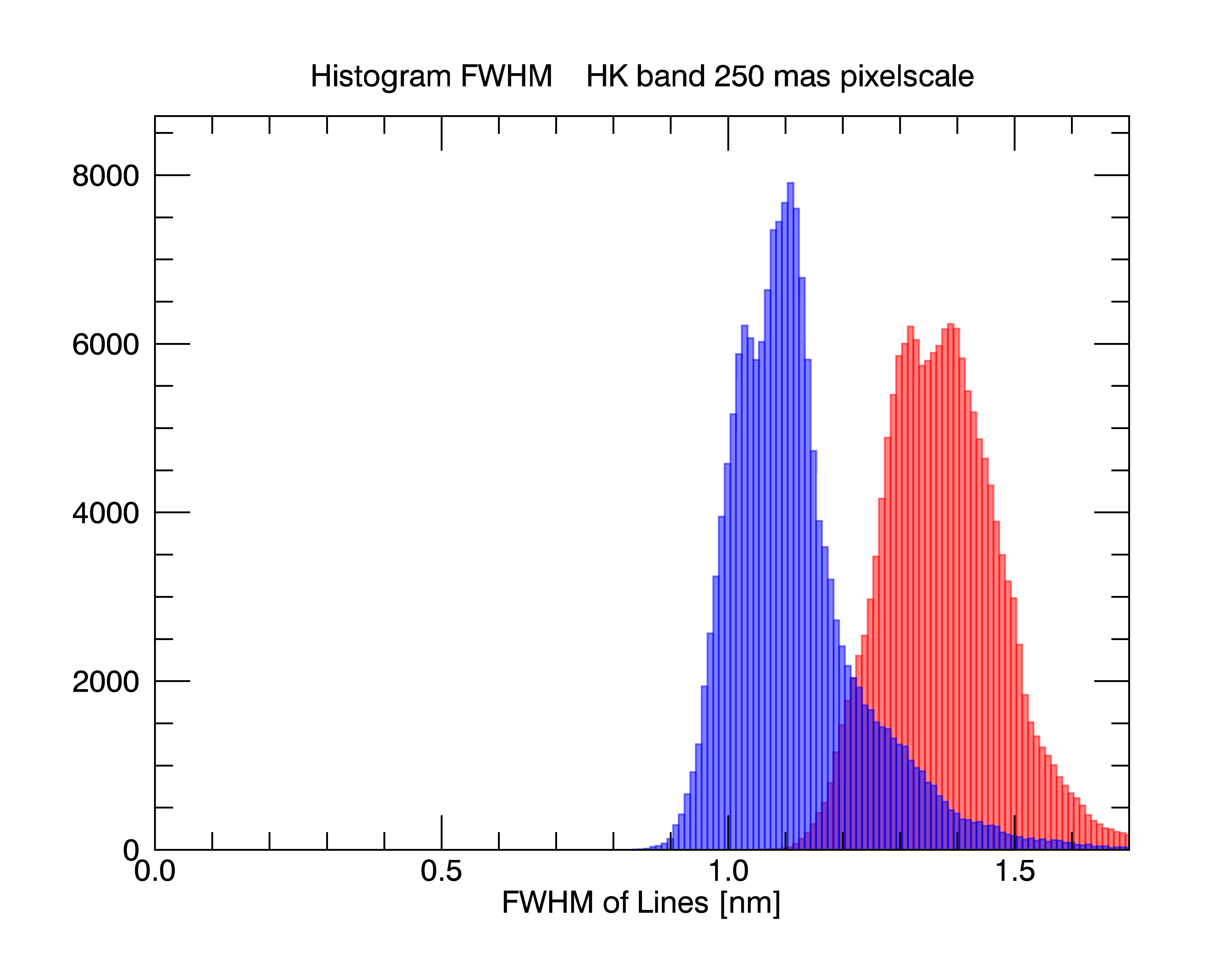}}
		\caption[Histograms of the FWHM of the LSF in the 250 mas pixelscale]{Histograms of the FWHM of the undersampled spectral lines with a binsize of 0.1 \AA. The red bars correspond to the pre - upgrade data, the blue bars to the post - upgrade data}
		\label{fig:histo}
	\end{center}
\end{figure}

That the upgrade of the SPIFFI instrument actually changed the resolution of the spectrograph can be seen best if one looks at histograms of the distribution the FWHMs of the LSFs. In figure \ref{fig:histo} this distribution is shown. The FWHM are measured on nearly all emission lines of the nobel-gas lamps that are also used for the wavelength calibration and across the detector, meaning in spatial direction. The red bars are from the pre- upgrade data, while the blue bars are from the post- upgrade data. As an example only the 250 mas pixelscale is shown here. For a complete set of plots in all pixelscales see \ref{fig:histograms_appendix}. The scale for the different bands is different, because the number of babysteps exposures taken for the respective bands differ.

The histograms of figure \ref{fig:histo} show, that in the 250 mas pixelscale the distribution of the FWHM of the LSFs was shifted due to the upgrade to the left, which means to smaller FWHMs. This results in a higher resolution of the spectrograph in the respective pixelscale. The most frequent value of the FWHMs corresponds to the x-value of the peak in the histogram. Since the distribution is nearly symmetrical, this is close to the median of the FWHMs. If the shape of the spectral lines were identical for each emission line and on all areas of the detector one would obtain a symmetrical Gaussian distribution due to errors in the fitting process. That the distribution is not a Gaussian curve shows that there are line profile features dependent on the detector position.

Further in the histograms the trend can be seen that the distribution is sharpest in J-band, meaning that there is the lowest variation of the LSF width. Additionally when looking on the different pixelscales in the histograms of figure \ref{fig:histograms_appendix} in the appendix, a behavior is visible that longer wavelength and smaller pixelscale the distribution is broadened resulting from more variations in the line profile width. K-band in the 25 mas pixelscale is here an exception and does not follow that behavior. This is, because in the AO pixelscale in K-band the amplitude of the side-peaks is so low that mainly the core of the LSF is fit, while for the other pixelscales the side-peaks are higher and thus contribute more to the Gauss-fit. Also in H+K-band this effect of a larger variation in smaller pixelscales cannot be seen. In general this comes from the variation of the line profiles which is the most variable in the AO pixelscale. The second influence here is the more complicated shape, especially in J-band that makes the Gaussian fit more imprecise in the smallest pixelscale. It shows that the fit works best for the 250 mas pixelscale which implies that the shape in this pixelscale is the closest to a Gauss curve and has the lowest variation. Since in H+K-band the line profiles look quite similar for all pixelscales this behavior cannot be seen in this band (compare also to the last row of figure \ref{fig:lineprofileshk} in the appendix). This shows that the spectral lines in H+K-band are close to the ideal shape and look quite similar for all pixelscales.

\begin{figure}[htbp!]
	\center
	\subfloat[\label{fig:fwhm_pre}pre - upgrade]
	{\includegraphics[width=7.5cm]{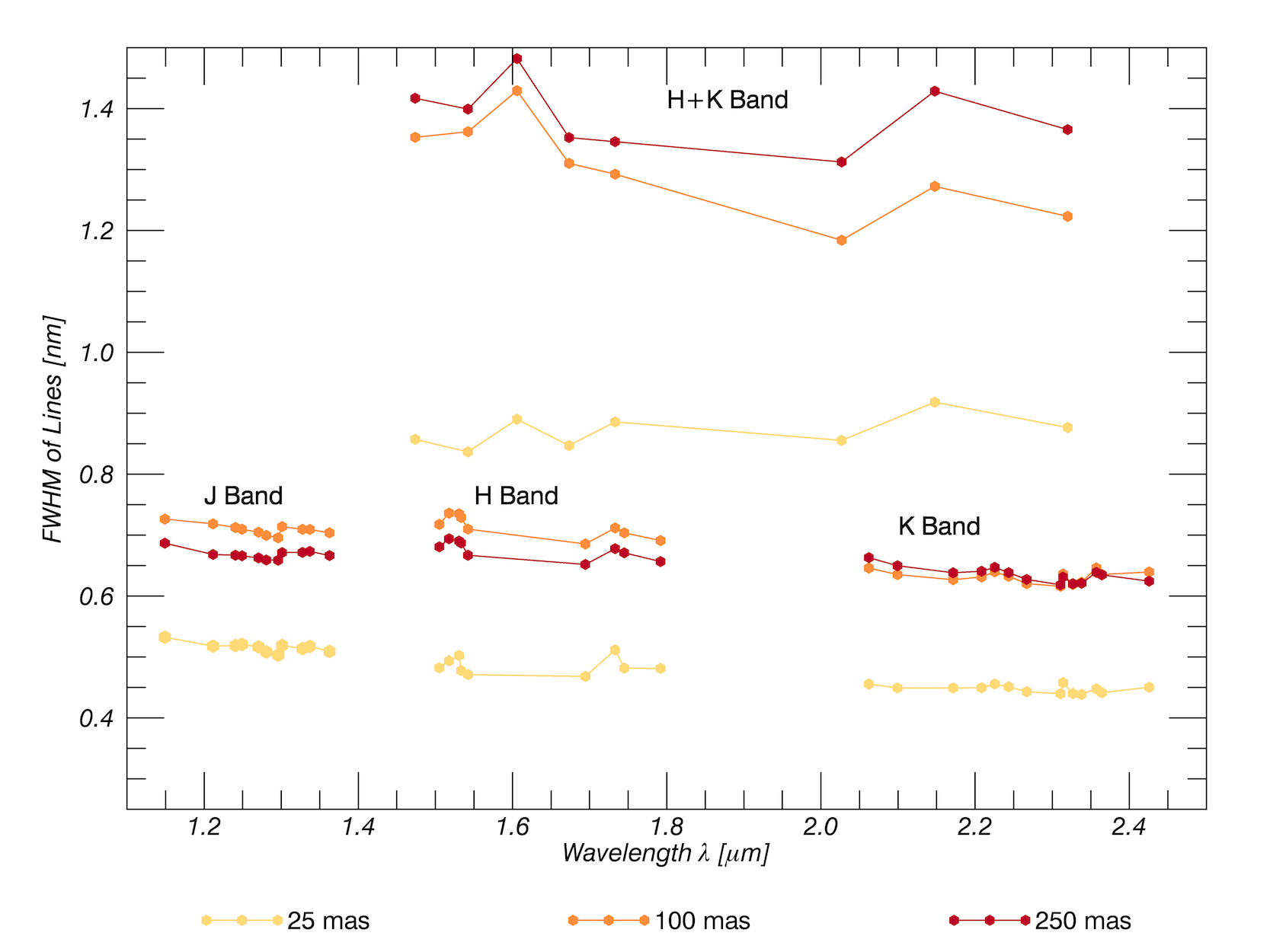}}
	\subfloat[\label{fig:fwhm_post}post - upgrade]
	{\includegraphics[width=7.5cm]{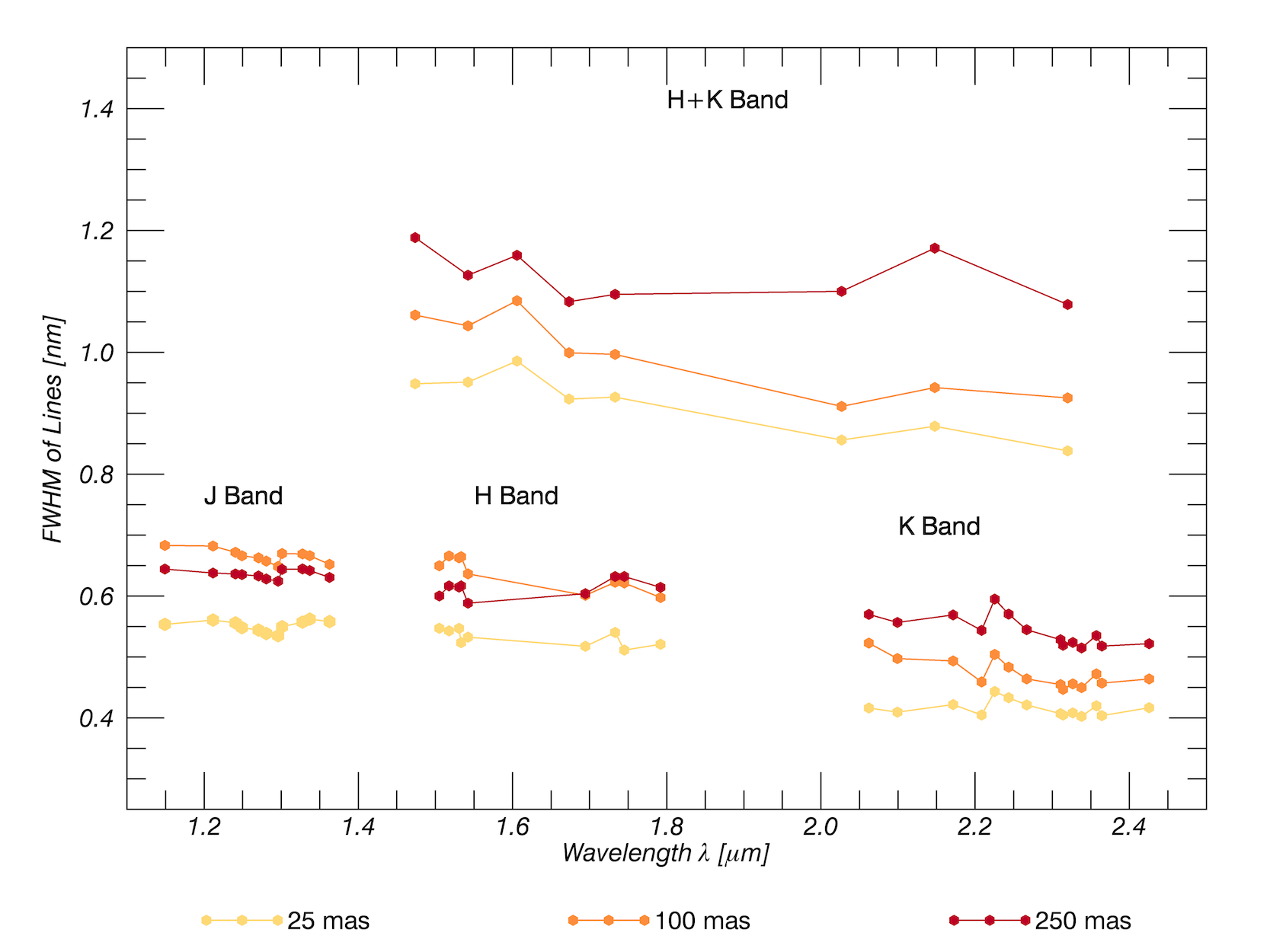}}
	\caption[FWHM of the spectral lines as a function of wavelength]{Plot of the FWHM of the spectral lines as a function of wavelength for the different gratings.}
	\label{fig:fwhm}
\end{figure}

In figure \ref{fig:fwhm} the FWHMsof the spectral lines are shown as a function of wavelength for the different gratings. The left plot shows the data taken before the upgrade of SPIFFI, while the right plot is made from data taken after the upgrade. Each FWHM data point is the weighted mean of the gaussian fits to the undersampled spectral lines over all detector columns and all babystep exposures at the wavelength of the given line.

One can recognize that the 25 mas pixelscale did not change a lot after the upgrade, while the larger pixelscales were shifted to smaller FWHMs. This behavior is described in more detail along with the plot of the resolution as a function of wavelength (which directly corresponds to the FWHM/wavelength), and also in section \ref{sec:interpretation}

The resolutions for each band and pixelscale with results from the FWHM measurements are listed in table \ref{tab:resolution}. They were calculated as $\lambda_c/\Delta\lambda$, where $\lambda_c$ are the central wavelengths of the bands ($\mathrm{J: \ \lambda_c=1.25 \mu m}$, $\mathrm{H: \ \lambda_c=1.65 \mu m}$, $\mathrm{K: \ \lambda_c=2.2 \mu m}$, $\mathrm{H+K: \ \lambda_c=1.95 \mu m}$). The FWHM is again the weighted average over the whole detector. The improvement in resolution is also shown in the last coulmn of the table. Other than the small pixelscale, the improvement of the resolution is around 10 \% to 30 \%. The tendency is that the upgrade affected mostly the longer wavelengths and larger pixelscales. In the larger pixelscales the out-smearing effect of the M2 and M3 mirror surfaces was strongest. The result of the upgrade is that there is now more power in the central line regions, respectively less in the shoulders of the line profiles leading to higher resolutions. That the improvement is higher in the longer wavelengths compared to the shorter ones is a relative effect. When the FWHM for all bands decreases by a equal amount, the fractional change will be largest for narrow line profiles like in K-band and only small for the wide profile of J-band. In J and H Band the improvement is within 10 \% while in the AO pixelscale the performance stayed equal or decreased. The negative numbers in the AO pixelscale in J-, H- and H+K-band come from the Gaussian-fit that fits a wider function to the line profiles, since the shoulders of the profiles are not as much washed out as before. So in general the central peak of the line profiles in J-, H-, and K-band became sharper due to the upgrade, while the outer parts are not as washed out as before, leading to fit that weights the outer parts more than before. In H+K-band the profile changed only very slightly. The change here is assumed to be dominated by the error of the fit. (See also the figures of the line profiles in section \ref{sec:lineprofiles}.) The error of this measurement is likely to be $\mathrm{\sim 5\%}$. 

\begin{table}[htbp!]
	\scalebox{0.7}{
		\begin{tabular}{cc|cccc|cccc|c}
			
			&& \multicolumn{4}{c|}{\textbf{pre - upgrade}} & \multicolumn{4}{c|}{\textbf{post - upgrade}} &\\
			\hline
			{\small Band}&{\small Scale}&{\small FWHM}&{\small $\sigma_{FWHM}$}&{\small Resolution}&{\small $\sigma_{Resolution}$}&{\small FWHM}&{\small $\sigma_{FWHM}$}&{\small Resolution}&{\small $\sigma_{Resolution}$}&{\small Improvement}\\
			&[mas/px]&[pixels]&[pixels]&$\lambda_c/\Delta\lambda$&$\lambda_c/\Delta\lambda$&[pixels]&[pixels]&$\lambda_c/\Delta\lambda$&$\lambda_c/\Delta\lambda$&[\%]\\
			\hline
			\hline
			J & 25				&3.4&0.6&2420&410&3.6&0.8&2270 &500			& -6 \\
			& 100				&4.7&0.4&1760&160&4.4&0.4&1880 &170			& 7\\
			& 250			 	&4.5&0.4&1870&160&4.2&0.4&1960 &200			& 5\\
			\hline
			H & 25				&2.4&0.5&3400&750&2.7&0.4&3090  &490			& -10\\  
			& 100				&3.6&0.3&2310&190&3.2&0.3&2570  &200			& 11\\
			& 250				&3.4&0.2&2440&150&3.0&0.2&2710  &200			& 11\\
			\hline
			K & 25				&1.8&0.5&4920&1350&1.6&0.4&5330  &1430			& 8\\ 
			& 100				&2.5&0.3&3480&430&1.9&0.3&4670  &870			& 34\\
			& 250				&2.5&0.3&3460&350&2.2&0.3&4070  &570			& 17\\
			\hline
			H+K&25				&1.7&0.3&2260&450&1.9&0.4&2120  &400			& -6\\
			& 100					&2.7&0.3&1460&150&2.0&0.3&1940 &300			& 33\\
			& 250					&2.8&0.3&1410&130&2.3&0.3&1730 &240			& 22\\
			\hline
		\end{tabular}
	}
	\caption[Instrument resolution]{Summary of the instrument resolution with standard detector sampling.}
	\label{tab:resolution}
\end{table}

The standard deviation of the single value in table \ref{tab:resolution} is around 10\% to 20 \%, while the standard deviation of the mean ist tiny and in the order of 1\permil. This is a result of the behavior of the LSF across the detector, especially in spatial direction and also within a slitlet. From the data one can see, that the standard deviation did not change by a lot due to the upgrade, which means that the FWHMs (and line profiles) still vary in similar amounts to before the upgrade. An example of this variation over the spatial dimension of the detector is shown in Figure 4.11 for the post-upgrade measurement in J band in the 25 mas pixelscale, because here its the most obvious.

\begin{figure}[htbp!]
	\center
	\subfloat[\label{fig:fwhm_var}The variance of the FWHM from the gaussian fit along one spectral line on the detector.]
	{\includegraphics[height=3.5cm, trim={0.8cm 0cm 0.8cm 0.2cm}, clip=true]{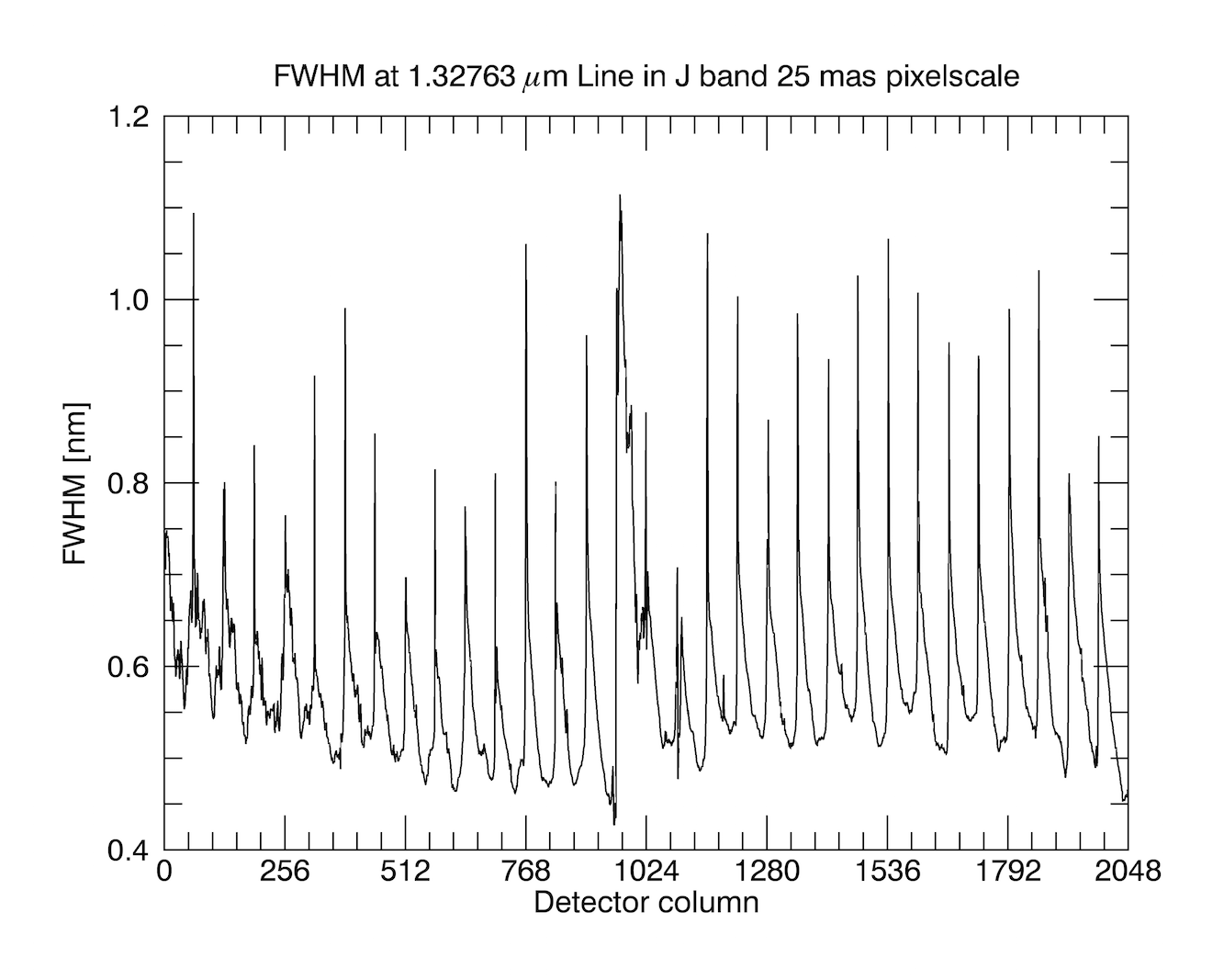}}
	\quad
	\subfloat[\label{fig:profile_slitlet}Spectral line profile across slitlet 15.]
	{\includegraphics[height=3.7cm, trim={1.5cm 0.5cm 2.0cm 0}, clip=true]{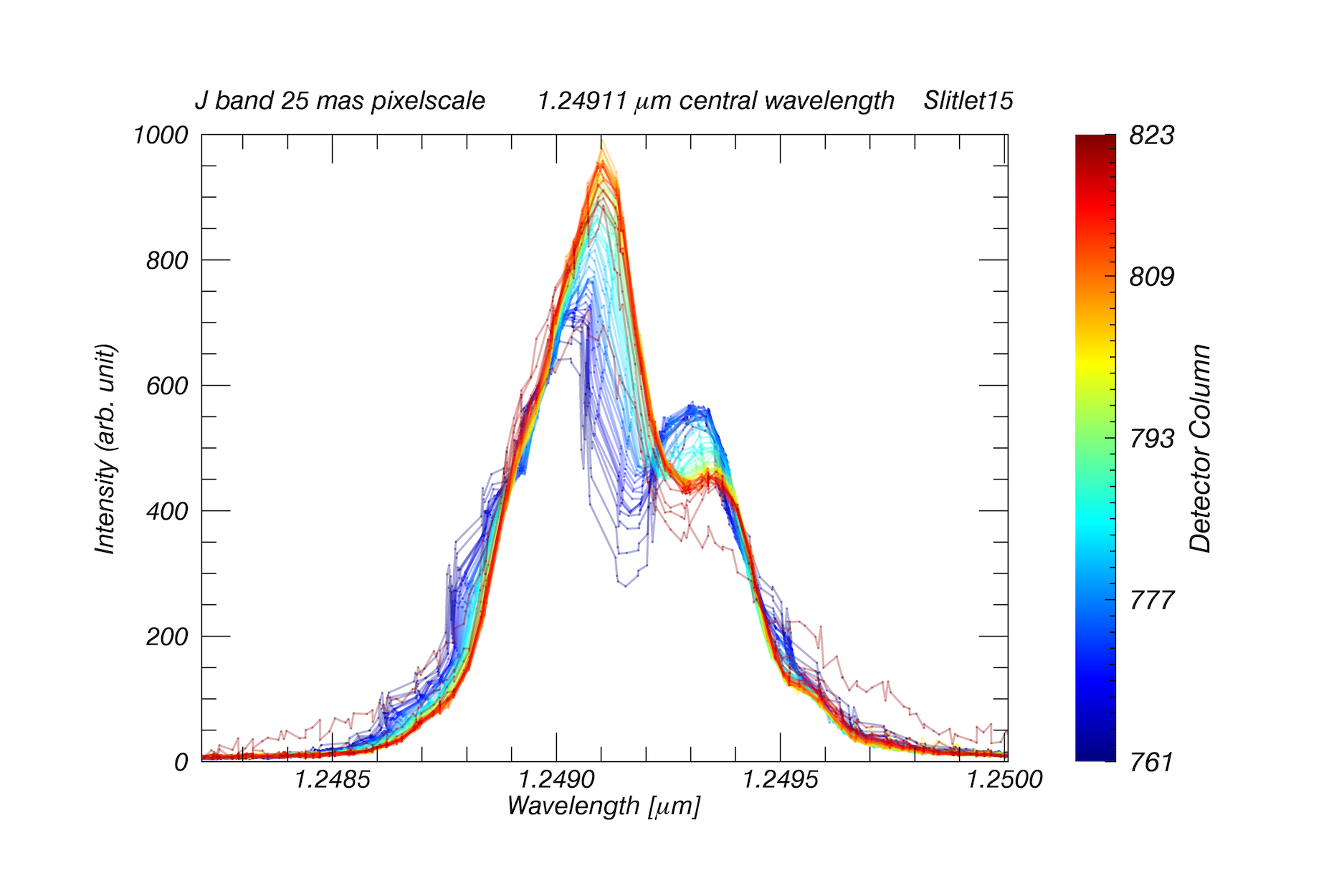}}
	\quad
	\subfloat[\label{fig:profile_detector}Median of the spectral line profiles of each slitlet across the detector]
	{\includegraphics[height=3.5cm, trim={1.1cm -0.5cm 1.0cm 0}, clip=true]{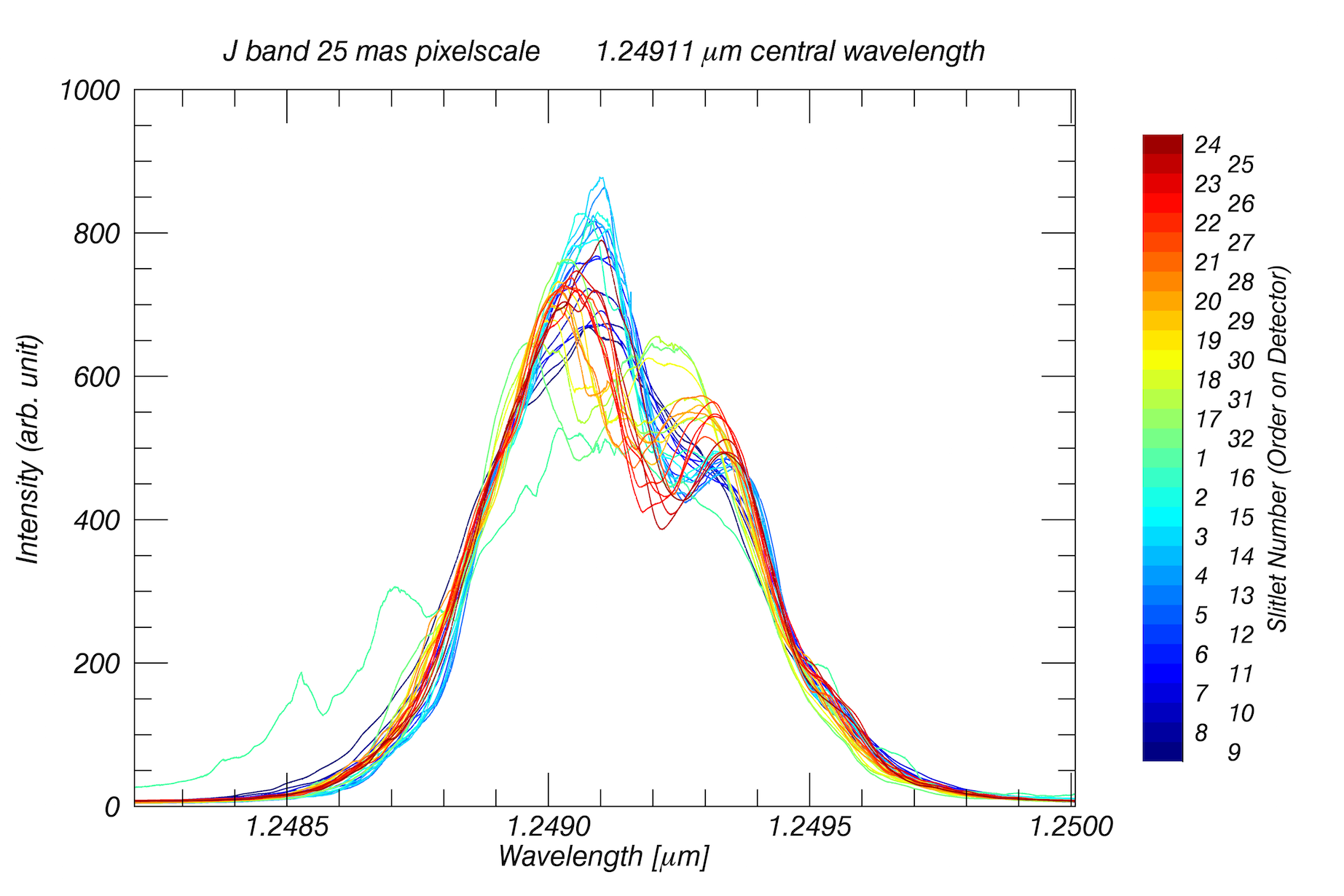}}
	\caption{Explanation of the standard deviation of the emission line FWHM.}
	\label{fig:explanation_resolution}
\end{figure}

Figure \ref{fig:fwhm_var} shows clearly, that after the upgrade the FWHM varies mostly over each slitlet. This is a result of the varying spectral line profiles within a slitlet (see figure \ref{fig:profile_slitlet}). Slitlet 15 is chosen for display, because it is close to the middle of the de- tector, which means that the small mirror corresponding to this slitlet on the small slicer is nearly not tilted, so defocus effects should be minimal. The conclusion must be, that this variation is not an effect of defocus of the small slicer, because in figure \ref{fig:fwhm_var} the variation within the slitlets is quite similar for all slitlets. One cause could be diffraction effects at the edges of the slitlets on the small slicer. These diffraction effects could widen the LSF. Figure \ref{fig:profile_detector} however shows the median of the the line profiles in each slitlet across the detector. It can be seen, that the variation of the median over the detector is less with agreement to \ref{fig:fwhm_var}. Also from the two right figures it is clear, that in the special case of the J-band a Gaussian-fit for the determination of the FWHM, respectively the resolution is not optimal because of the double-peaks. The less the side-peak is washed out - as it is the case for the post- upgrade data, the less applicable and reliable is the Gaussian-fit for the non-Gaussian line profiles.

\begin{figure}[htbp!]
	\center
	\subfloat[\label{fig:resolution_pre}pre - upgrade]
	{\includegraphics[width=7.5cm]{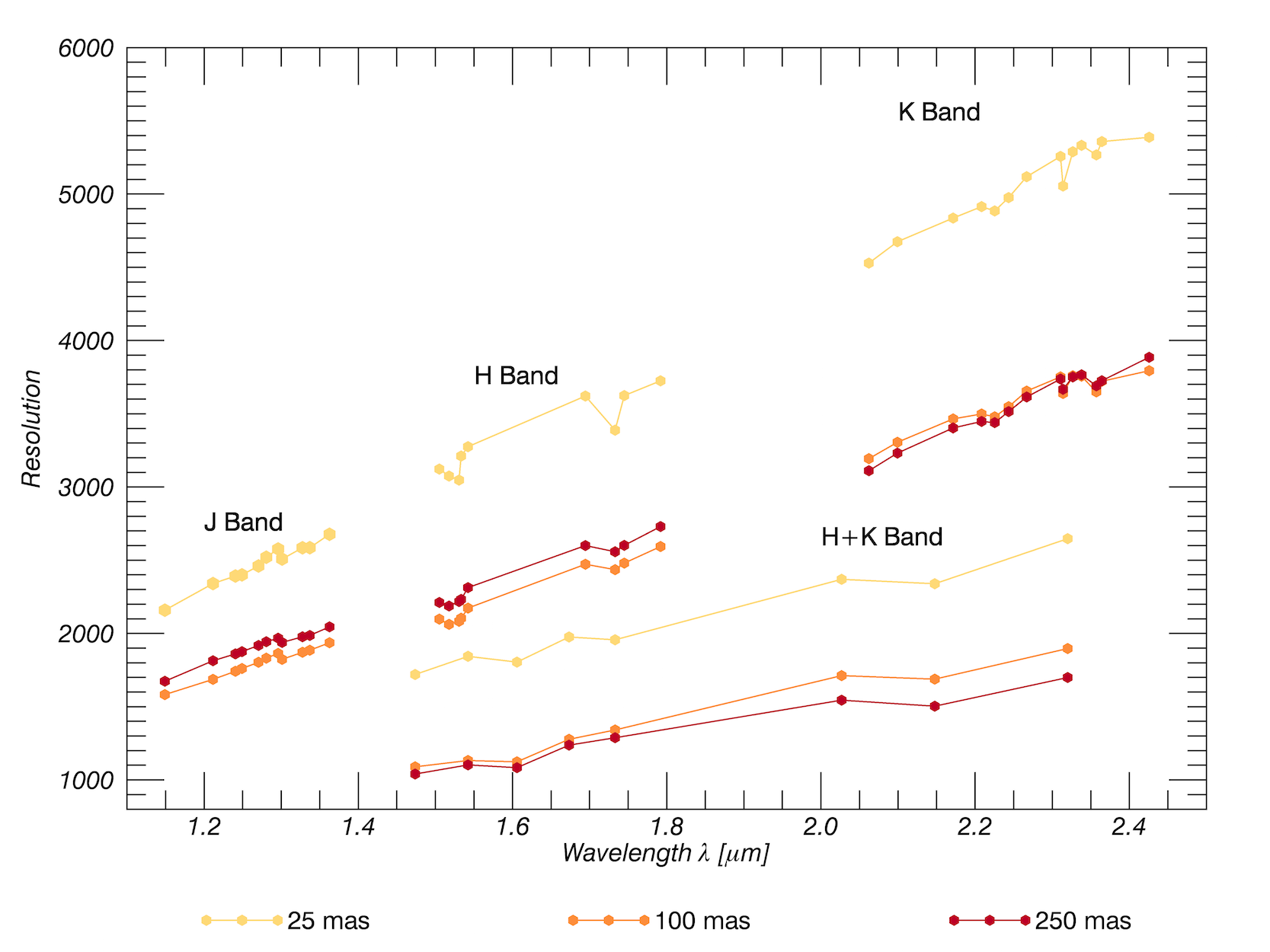}}
	\subfloat[\label{fig:resolution_post}post - upgrade]
	{\includegraphics[width=7.5cm]{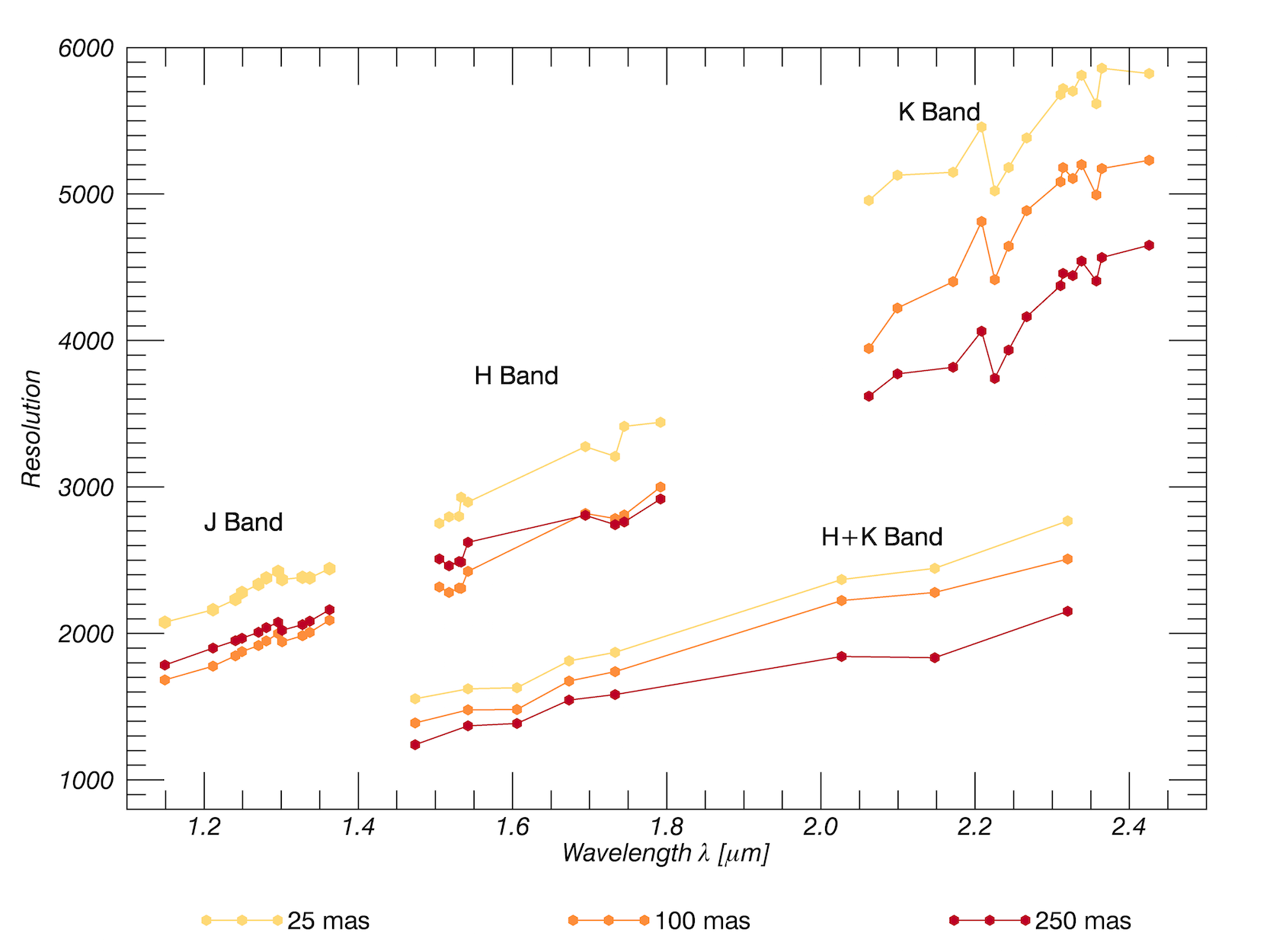}}
	\caption[Spectral resolution as a function of wavelength]{Plot of the spectral resolution $\lambda / \delta \lambda$ as a function of wavelength for the different gratings.}
	\label{fig:resolution}
\end{figure}

Figure \ref{fig:resolution} shows the same data as figure \ref{fig:fwhm}, but calculated and plotted as resolution as a function of wavelength. Resolution here is defined as $\lambda / \Delta\lambda$ where $\lambda$ is not the central wavelength as in table \ref{tab:resolution}, but the wavelength of each data point. $\Delta\lambda$ is the same as in figure \ref{fig:fwhm} and calculated as described before. Naturally the resolution is increasing with the wavelength, since the FWHM stays nearly constant over wavelength (see figure \ref{fig:fwhm}). When comparing the pre- and post-upgrade plot (figures \ref{fig:resolution_pre} and \ref{fig:resolution_post}), most obvious is that the larger pixelscales ()100 mas/px and 250 mas/px) changed most, while there is nearly no change in the 25 mas pixelscale. The larger pixelscales are so to speak shifted towards the AO pixelscale.

\subsection{Interpretation}\label{sec:interpretation}
While the process of fitting a Gaussian to the complex shaped line profiles of SPIFFI is not on optimal procedure to define resolution, it is a procedure that makes it possible to compare the performance of the SPIFFI spectrometer before and after the upgrade. As a result of the collimator exchange the shoulders and bumps in the line profiles do not smear out as much as before. Therefore in the large pixelscales there is more power in the peak, which makes the profile steeper an thus the location of the FWHM of the spectral lines is shifted upwards to higher amplitudes. This results then in a smaller width of the LSF at half maximum height even if the shoulders are more distinct. In general the center peak was made narrower by the upgrade and relative to it the shoulders do have less amplitude.

In the AO pixelscale the situation looks a bit different. There the effect of washing out of lineshapes was not so strong before the upgrade, because the footprint of the beam on the mirrors was smaller, so the wavefront was only affected by a smaller surface deviation of M2 and M3 and less diamond turning marks. Some of the left-side peaks of the LSFs in J- and H- band almost completely vanished after the upgrade. Despite from smoothing out these side peaks in the 25 mas pixelscale, the upgrade did not narrow the shape of the line profile. Thus the resolution in the AO pixelscale nearly did not change.

Due to the different behavior in the AO pixelscale and the larger pixelscales and also the fact, that the shape of the spectral line profiles in the 25 mas pixelscale changed significantly, it can be concluded that the increased resolution is indeed caused by the exchanged collimator mirrors and is not only affected by the fact that the detector is now better in focus. The change in focus of the detector of around 10 micron corresponds for the f/2.8 beam of the spectrometer camera to a change in FWHM of $\mathrm{\sim 0.2}$ pixels, which is also in the regime of the measured change for the resolution. However, the fact that the absolute focus position improved cannot explain the observed behavior of the distinct line profiles.

\chapter{Summary and Outlook}\label{ch:summary_and_outlook}
In this Master thesis the upgrade of the VLT integral filed spectrometer SPIFFI, which is part of the Cassegrain instrument SINFONI is described. The focus of this thesis is always on the resulting performance of the instrument, not on certain procedures applied during the upgrade or in the laboratory. The optical components of the instrument are introduced in the very first chapter, while in the next chapter the upgrade of the optical components is explained. Chapter 3 deals with the measurements of the collimator mirrors in the lab and their influence on the spectrometer performance. This instrument performance is then investigated in chapter 4 in order to determine the new instrument capabilities.

The new collimator mirrors exchanged in the upgrade have a much smoother surface in comparison to the pre- upgrade post- polished mirrors. There are nearly no diamond turning marks on the surfaces and overall the surface is smooth and uniform. Only for the last of the three collimator mirrors M3 there is a larger deviation in turning center and radius. This results mostly in one third micron coma in the collimator wavefront leading to an RMS wavefront deviation of 0.20 $\mathrm{\mu m}$ from a flat aberration-free wavefront. On the smaller scales the new collimator wavefront shows essentially no distortions that can be resolved properly with the used interferometer, while in the pre- upgrade wavefront a complex structure of residuals from diamond turning marks and post-polished surface deformations together with larger scale aberrations exists.

Mainly due to the exchange of the band-pass filters in J-, H- and K-bands. the throughput increased and especially the transmission became more constant as a function of wavelength. Because of the protection layer on the gold coating of the mirror surfaces, the through-put gained a very slight dependence on the wavelength within the individual bands. Overall, astronomy on faint objects will benefit from the higher throughput, especially in J-band studies using the H$\alpha$ line of red-shifted galaxies at $\mathrm{z \sim 1}$. The large throughput gain will make it easier to study galaxy formation and evolution.

The Strehl ratio was improved in H- and K-band in the upgrade by around 10\%, while in J- and H+K-band it stayed quite constant with no significant improvement. The effect is mostly caused the exchange of the pre-optics. With the improved Strehl ratio science done in fields of the sky that a crowded with many luminous objects like the Galactic center will benefit. Due to the upgrade the signal to noise of faint objects will be higher and the distinction between different objects will be easier.

With the new collimator mirrors the shapes of the spectral line profiles are now partly smoother and with less variation over the detector. In the seeing limited pixelscales the side peaks became more distinct, since they were washed out by the surface irregularities of the collimator mirrors before. By replacing the M3 mirror in the SPIFFIER upgrade in 2019 with a mirror that has a surface form within the tolerances, the RMS wavefront error of the collimator will be much smaller and thus the LSF will be more symmetrical. With the upgrade and the investigations done on the collimator mirrors, it is now clear that the unsymmetrical shape of the spectral line profiles with distinct shoulders is not mainly caused by the collimator mirrors, and was only to a smaller degree affected by the residual diamond turning marks and the polishing surface structure of the pre- upgrade collimator mirrors. In the analysis of the individual gratings there are hints showing up that bimetallic bending effects between the nickel layer and the aluminum blanks of the gratings could cause line profiles as they exist in SPIFFI. To verify this expected behavior a cryogenic interferometric surface form measurement of one J-band spare grating will be implemented in the MPE laboratory soon. The origin of the variation of the line profiles within a slitlet is still an open question.

The resolution of the spectrograph is strongly coupled to the line profile issue. Since the pre- upgrade collimator mirrors had only a minor influence on the shape of the spectral lines the resolution improved mainly in the seeing limited pixelscales, where the line profiles suffered in the past from the collimator wavefront that washed out and broadened the line profiles. With a new M3 installed in 2019, the resolution, especially in the large pixelscale, will benefit from the better wavefront. Due to the increase in resolution in the large pixelscales the velocity dispersion of galaxies can be measured more accurate. Furthermore it is possible to measure metallicities of red supergiants making them ideal targets for in the near infrared for measuring the metallicity of star forming galaxies up to 10 Mpc.\cite{gazak14}

The first half of the instrument upgrade of SPIFFI for its re-use in ERIS was completed in the upgrade in January 2016 described in this thesis. The remaining items before SPIFFI will become a part of ERIS will be installed in the SPIFFIER upgrade in 2019 and will further improve the performance of the instrument. The main optical improvement will be in the spectral line profiles, and respectively the resolution of the spectrometer, as well as the detector performance relating to quantum efficiency, persistence, and cosmetics. Also SPIFFIER will take advantage of the adaptive secondary mirror of VLT UT4 and the four-laser guide star facility. The science done with the SPIFFI, and then SPIFFIER instrument as a part of ERIS, will benefit from these upgrades, leading to a higher scientific output of the instrument.

\let\cleardoublepage\clearpage
\appendix
\chapter{Optics of MACAO}
This chapter of the appendix refers to section \ref{sec:discussion_lineprofiles}. Here, a schematic overview of the SINFONI adaptive optics unit MACAO shown. The light enters the instrument from the top, is then folded by M1 and by M2 the pupil is imaged on the deformable mirror M3. M4 creates the output focus and M5 folds the beam into the focal plane of SPIFFI. The M5 mirror is used for the pupil alignment between MACAO and SPIFFI. Before the beam reaches the focal plane of SPIFFI it is divided by the dichroic into a science beam directly entering SPIFFI and the AO beam. The calibration unit directly acts at the telescope focus at the entrance of MACAO. 
\begin{figure}[htbp!]
	\begin{center}
		\includegraphics[width=0.4\textwidth, trim={0.4cm 3cm 0.4cm 3.5cm},clip=true]{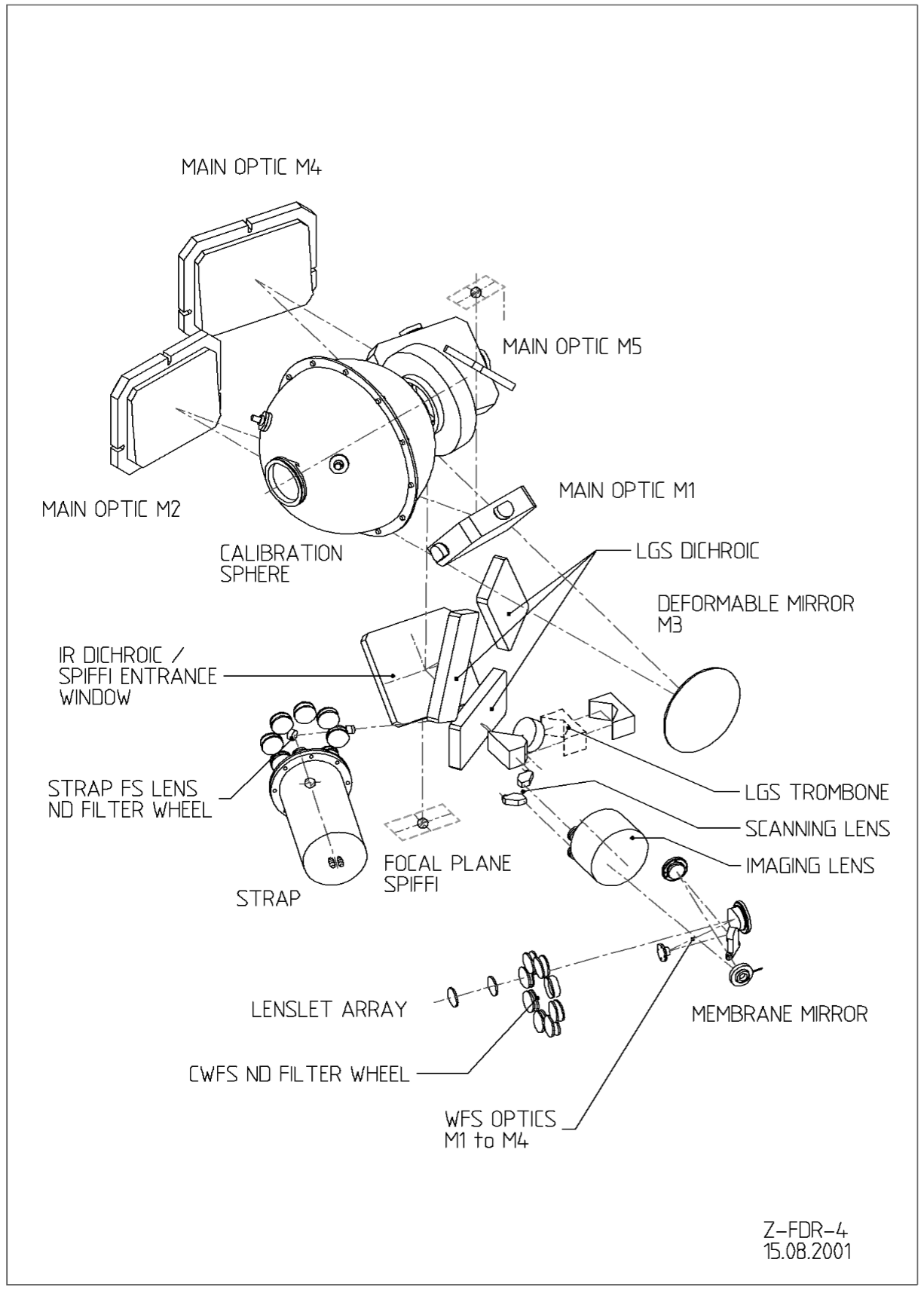}
		\caption[Optical elements of MACAO]{Optical elements of the MACAO adaptive optics module. \cite{bonnet03}}
		\label{fig:scheme_macao}
	\end{center}
\end{figure}

\chapter{Mirror Measurements}
\section{Cusp on M1}
Referring to section \ref{sec:M1}, a zoom-in to the turning center on the M1 interferograms from section \ref{sec:M1} is shown here. The cusp from the manufacturing procedure is obvious in the series 02. The concentric rings on series 01 are internal reflection in the interferometer.
\begin{figure}[htbp!]
	\begin{center}
		\subfloat[new M1 S01]
		{\includegraphics[width=0.4\textwidth, trim={0.cm 0 0cm 0.1cm}, clip=true]{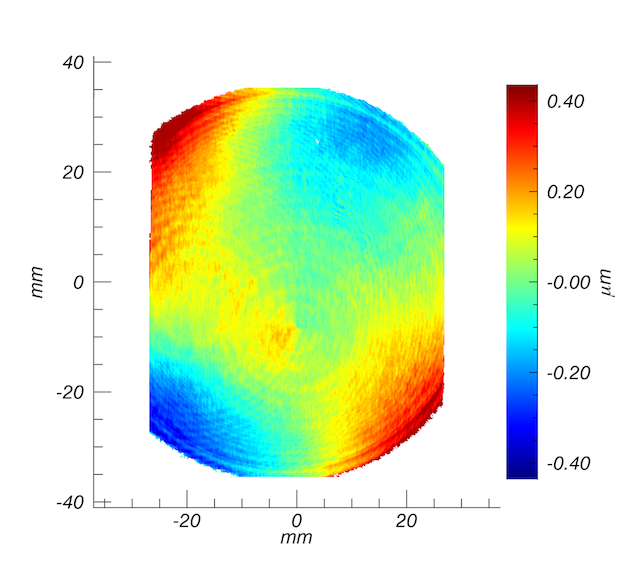}}
		\subfloat[new M1 S02]
		{\includegraphics[width=0.4\textwidth, trim={0.cm 0 0cm 0.1cm}, clip=true]{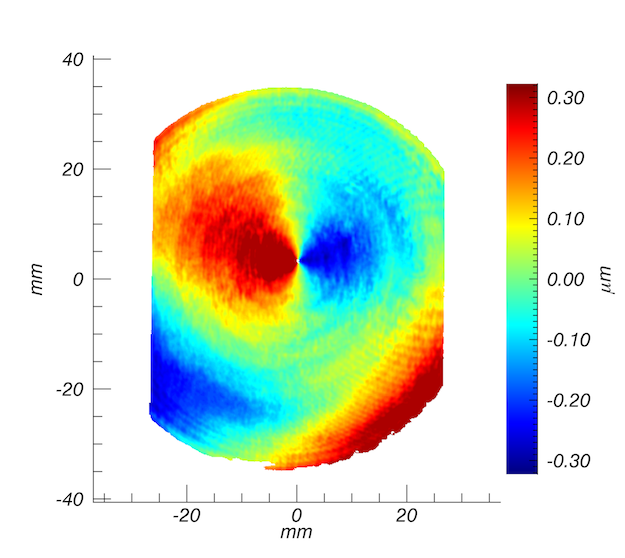}}
		\caption[M1 small center footprint]{The centers of the spherical M1 mirrors. In the right surface plot one clearly sees a cusp which comes from the turning process. The right plot is also slightly out of focus. The footprints shown are slightly shifted against each other. On the left image the turning center is a bit lower than in the right image. Note the different surface scales!}
		\label{fig:cusp}
	\end{center}
\end{figure}
\clearpage

\section{Alignment Procedure for the M2 Single Mirror Measurement}{\label{sec:M2alignment}}
In order to provide detailed information about the M2 measurement from section \ref{sec:M2}, the alignment procedure of the M2 single mirror setup is described here. The alignment is done with a FARO Gage measurement arm. The interferometer is the FISBA Twyman-Green interferometer with a 6 mm collimated output beam at 632.8 nm.

\begin{figure}[htbp!]
	\begin{center}
		\includegraphics[width=1.\textwidth]{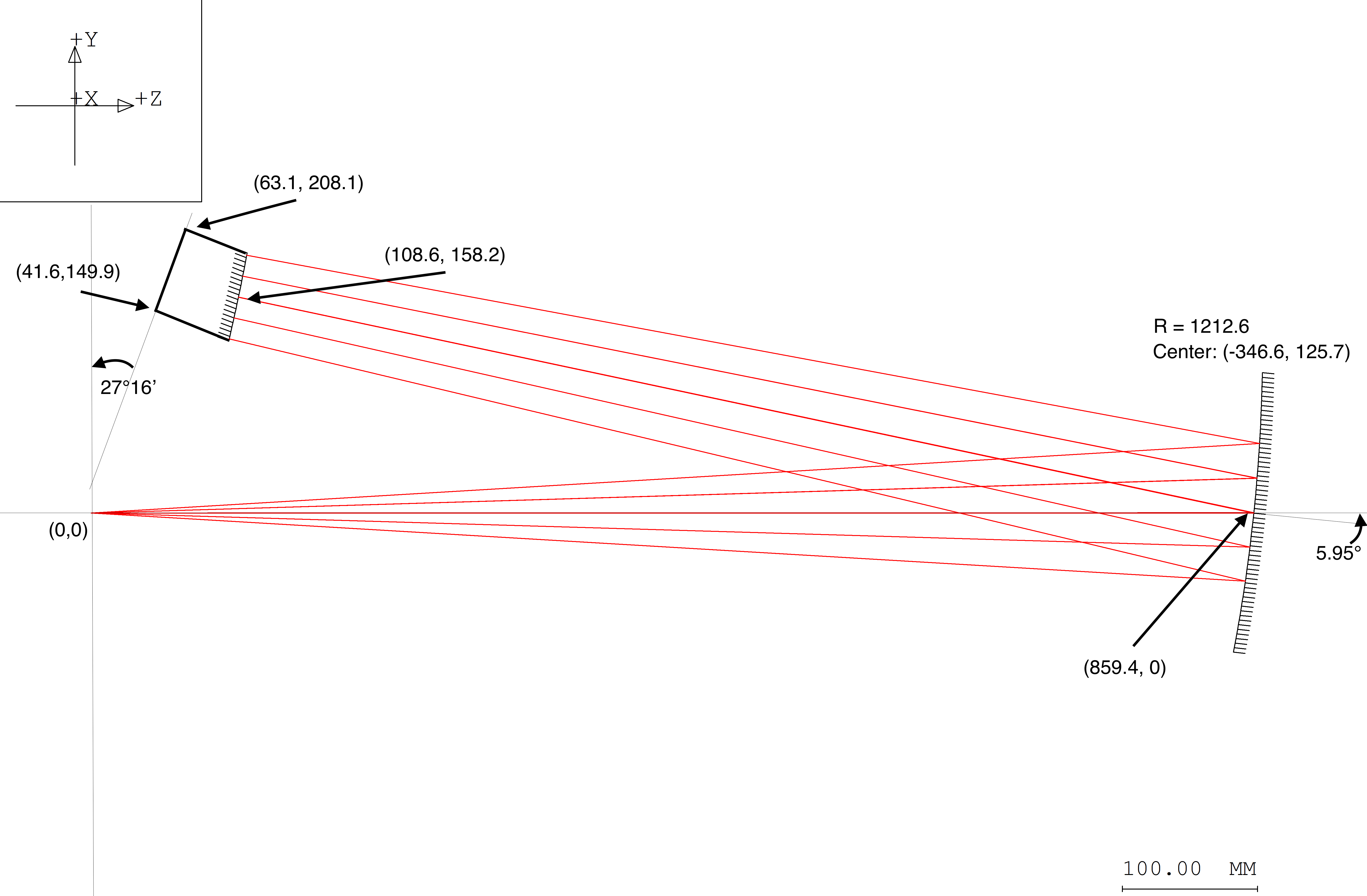}
		\caption[M2 test setup with coordinates]{M2 single mirror measurement setup with coordinates of the optical components. The origin of the coordinate system is at the focus position.}
		\label{fig:M2setupnumbers}
	\end{center}
\end{figure}

Preparations for M2 FISBA-interferometer test:
\begin{itemize}
	\item First place the FARO arm on a place of the optics table where all optical components can be reached. Then make a first coordinate system with the FARO arm computer program so that the x-axis is orthogonal to the table surface and the y- and z-axis is parallel to the holes in the table. Use a long metal ruler (1m) for this and push it against screws in the table. The origin of the coordinate system does not matter.
	\item Mount the interferometer on a scissors stage at a hight that the beam is roughly retro reflected by M2 back into the interferometer.
	\item Align the FISBA beam parallel to the optical table. Do this by aligning two pinholes with the FARO arm at an identical height that is roughly the beam height. Place one pinhole therefor close to the interferometer and one as far as possible away, but still reachable with the FARO arm. Tilt the FISBA so that the beam is parallel to the optics table. 
	\item Align the FISBA beam parallel to the grid of the optics table. Do this again with the two pinholes by measuring their position with the FARO arm and adjusting the position until they are aligned to the grid of the optics table. Rotate the FISBA so that the beam is aligned to the grid of the optical table.
	\item Align the FISBA at the right height. To do so, retro reflect the beam with M2 on its xy-translation stage in a distance of roughly 1m.
	\item Adjust the large spherical mirror on its stage to the right height. For this put the mirror in roughly the right position and retro reflect the collimated beam with the spherical mirror. Adjust the tilt and rotation of the mirror so that the reflected beam fills the whole FISBA aperture on the RTD. Leave the mirror there.
	\item Position the lens. Use a f = 25 mm lens to illuminate all relevant parts of M2. Mount it roughly 25 mm in front of the exit pupil of the interferometer. Align the lens laterally so that the reflected beam (small dot on the RTD) is centered on the RTD.
	\item Place a pinhole at the focus of the lens. This position will be the origin of the new coordinate system.
	\item Make a new coordinate system with the FARO arm:
	\begin{itemize}
		\item For the measurement of the first plane surface take the optics table. The x-vector will be the normal vector of this surface.
		\item To get the x-axis measure a line from the pinhole in the focus to the second pinhole that you already aligned.
		\item For the determination of the origin of your coordinate system first measure a second plane surface. To do so, take the aperture of the pinhole. Second, measure a circle around the pinhole and give the FARO program as reference the surface plane of the pinhole.
		\item $\rightarrow$ Chose for the coordinate system: plane, circle, line in the FARO program.
		This is the new coordinate system.
	\end{itemize}
	\item Put the mirrors roughly in the correct positions with the help of the FARO arm as shown in figure \ref{fig:M2setupnumbers}. The first coordinate is to the right and the second upwards.
	\item For the large spherical mirror find the right angle by measuring a sphere. Calculate the angle from the output coordinates of the center of the sphere. Another possibility is simply to rotate it slowly, after M2 is in its correct position, until an interferogram is visible on the RTD.
	\item Check if the pupil is in focus. This means, that the image of M2 on the FISBA RTD is sharp. One can see this best by looking at the edges of the mirror on the RTD. If they are sharp, the pupil is in focus. If not, move the lens along the beam line to get a sharp image. For this the positions of M2 and the spherical mirror have to be roughly in the right spot, so that fringes can be seen.) After moving the lens you have to find the new focus point, place the pinhole there and make a new coordinate system with Faro arm. This is an iterative process.
	\item When the image of the mirror surface is sharp the mirrors has to positioned very precisely. Therefor make sure that the coordinate system and the origin of it is correct.
	\item Place M2 at nominal position. For this the angle of M2 with respect to the beam line and the position with respect to the focus has to be correct. To get the right angle measure the backside of M2 with the Faro arm and get the angle with the axis along the beam line. For the positions of the edges of M2 intersect respectively the three corresponding surfaces of M2 in the FARO program and get from this with the coordinates of the edges.
	\item Optimize interferogram by adjusting spherical mirror for rotation (not for tilt!) and the lens for the hight. 
	\item The M2 footprint should be centered, otherwise the laser beam is not parallel to the coordinate system (because of non linear effects of a lens with small focal lengths). Move the lens slightly perpendicular to the beam line. Make your coordinate system again, positioning of M2 and spherical mirror again.
	\item Calibrate the lens by placing a return sphere so that the focus of it is identical with the focus of the lens. Remove the return sphere afterwards
	\item Measure surface of M2 with FISBA.
\end{itemize}

\clearpage
\chapter{Plots of Line Profiles and Resolution}
The plots shown in this appendix extend the plots from section \ref{sec:lineprofiles} for all bands and pixelscales of SPIFFI. They show nearly entirely the behavior of the LSF.
\section{Comparison of the LSF Pre- and Post-Upgrade}\label{compare_lsf}
The accumulated line profiles of three spectral lines on the lower, central and upper part of the detector are shown. The line profiles of the 2048 columns of the detector are over-plotted resulting in a data point cloud representing the variation of the line profiles along one spectral line across the detector. The red data is pre- upgrade, while the blue is post- upgrade.

\begin{figure}[htbp!]
	\begin{center}
		\resizebox{1.0\textwidth}{!}{
			\includegraphics[width=1.0\textwidth, trim={1.0cm 0 1.0cm 0cm}, clip=true]{j25_wlength_1_24911_compare.png}
			\quad
			\includegraphics[width=1.0\textwidth, trim={1.3cm 0 0.7cm 0cm}, clip=true]{j100_wlength_1_24911_compare.png}
			\includegraphics[width=1.0\textwidth, trim={1.3cm 0 0.7cm 0cm}, clip=true]{j250_wlength_1_24911_compare.png}
		}
		\resizebox{1.0\textwidth}{!}{
			\includegraphics[width=1.0\textwidth, trim={1.0cm 0 1.0cm 0cm}, clip=true]{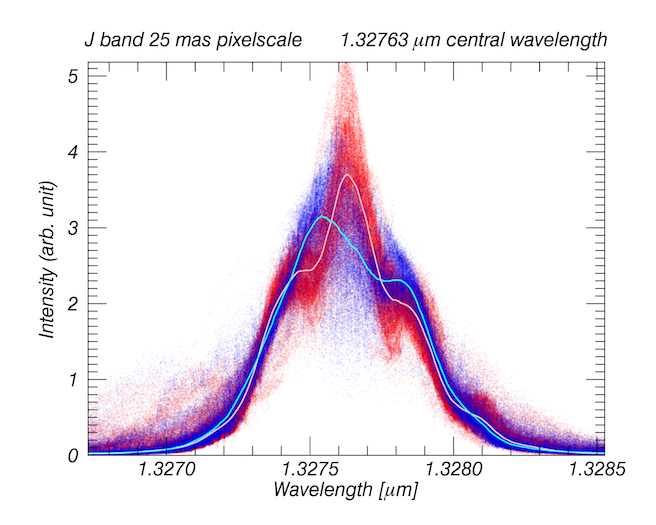}
			\quad
			\includegraphics[width=1.0\textwidth, trim={1.3cm 0 0.7cm 0cm}, clip=true]{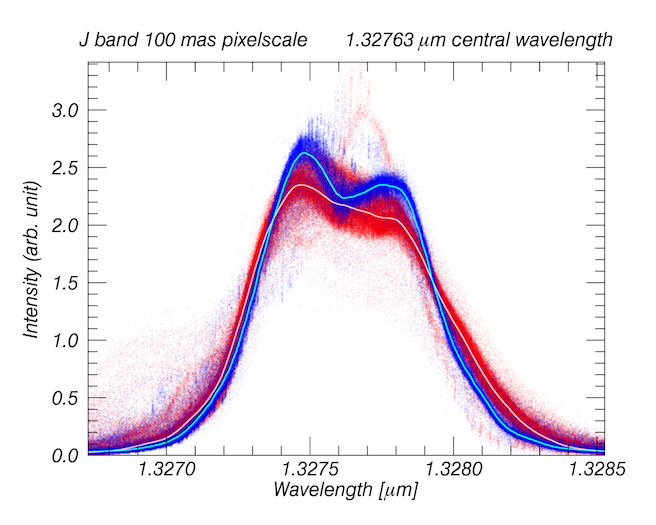}
			\includegraphics[width=1.0\textwidth, trim={1.3cm 0 0.7cm 0cm}, clip=true]{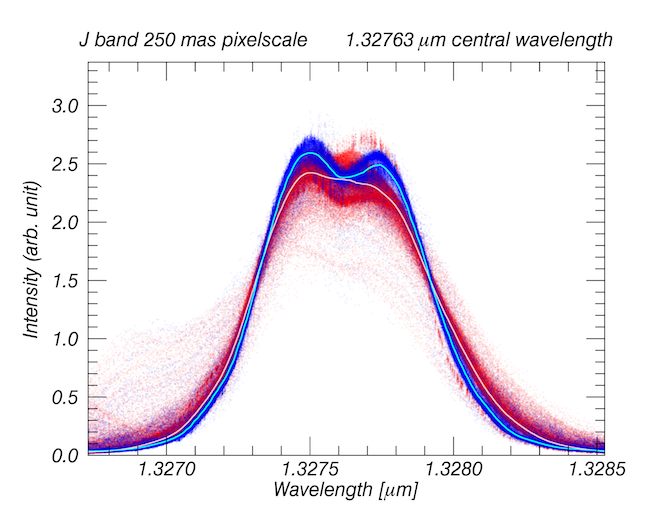}
		}
		\resizebox{1.0\textwidth}{!}{
			\includegraphics[width=1.0\textwidth, trim={1.0cm 0 1.0cm 0cm}, clip=true]{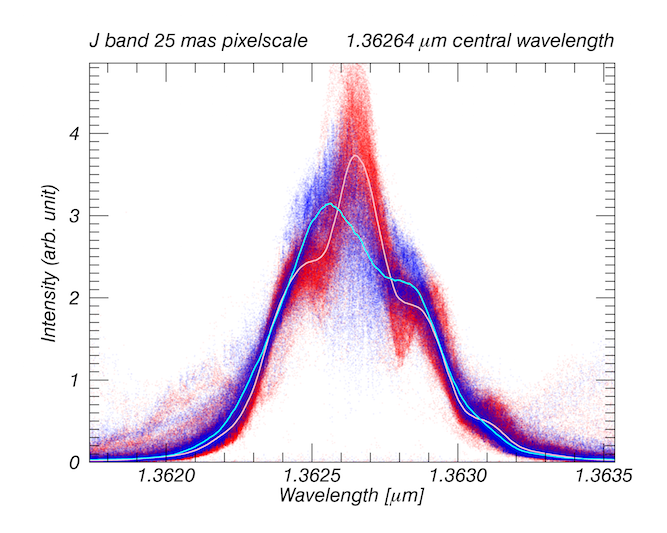}	
			\quad
			\includegraphics[width=1.0\textwidth, trim={1.3cm 0 0.7cm 0cm}, clip=true]{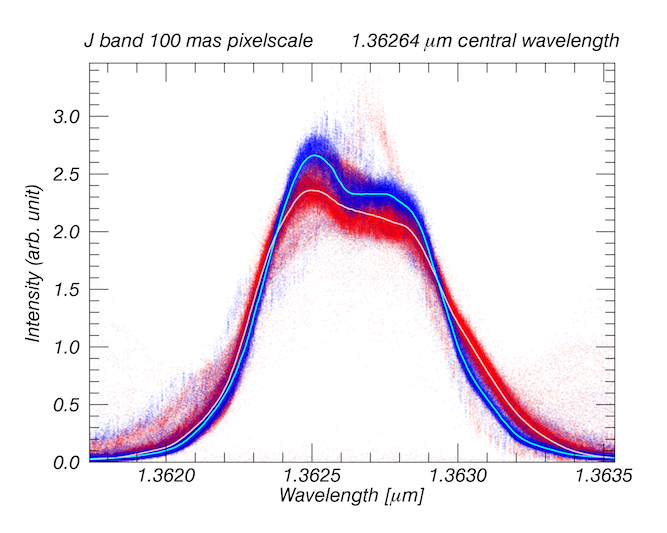}
			\includegraphics[width=1.0\textwidth, trim={1.3cm 0 0.7cm 0cm}, clip=true]{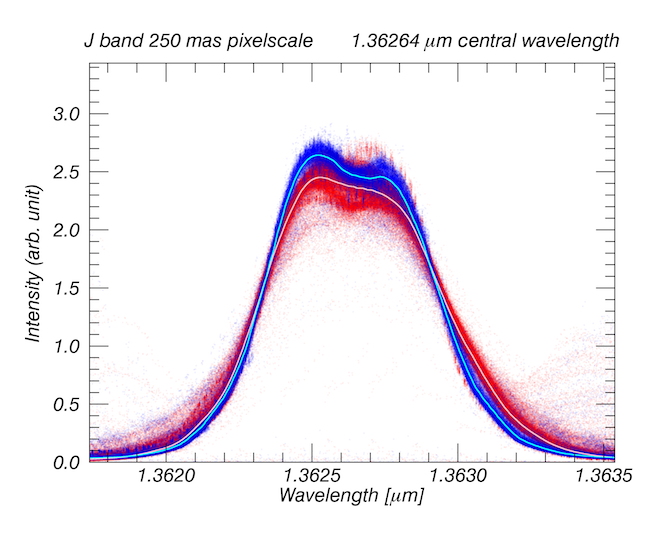}
		}
		\caption[Accumulated line profile in J-band]{Accumulated line profile along different wavelengths and in each pixelscale in J-band.}
		\label{fig:lineprofilesj}
	\end{center}
\end{figure}

\begin{figure}[htbp!]
	\begin{center}
		\resizebox{1.0\textwidth}{!}{
			\includegraphics[width=1.0\textwidth, trim={1.0cm 0 1.0cm 0cm}, clip=true]{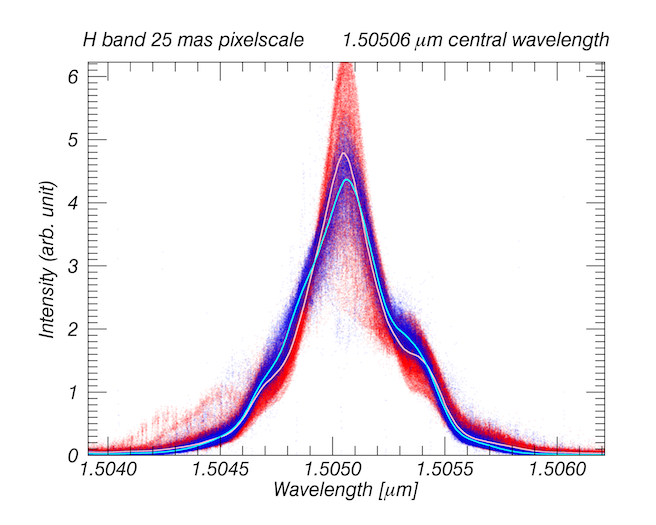}
			\quad
			\includegraphics[width=1.0\textwidth, trim={1.5cm 0 0.5cm 0cm}, clip=true]{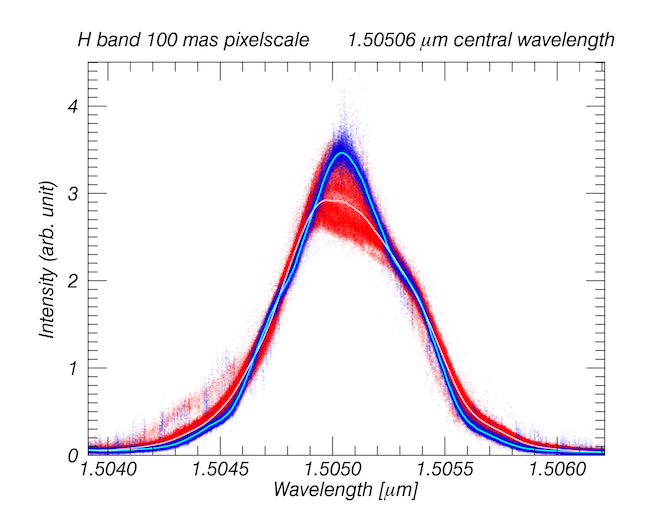}
			\includegraphics[width=1.0\textwidth, trim={1.5cm 0 0.5cm 0cm}, clip=true]{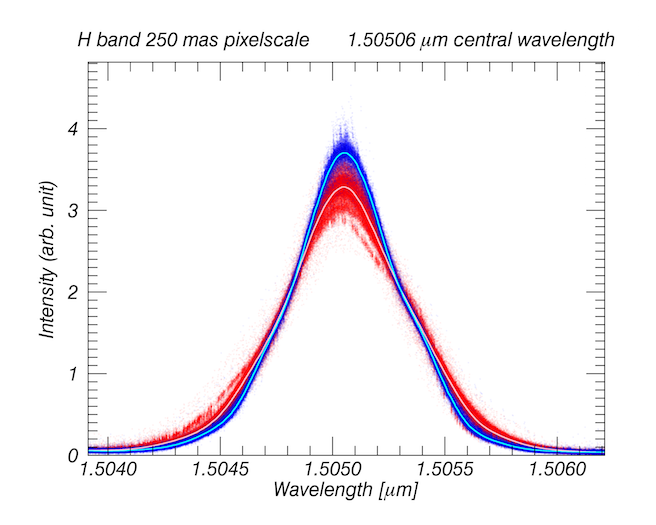}
		}
		\resizebox{1.0\textwidth}{!}{
			\includegraphics[width=1.0\textwidth, trim={1.0cm 0 1.0cm 0cm}, clip=true]{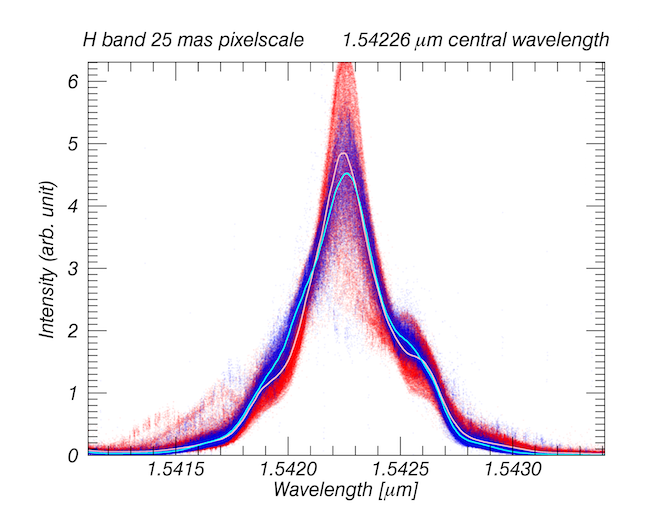}
			\quad
			\includegraphics[width=1.0\textwidth, trim={1.5cm 0 0.5cm 0cm}, clip=true]{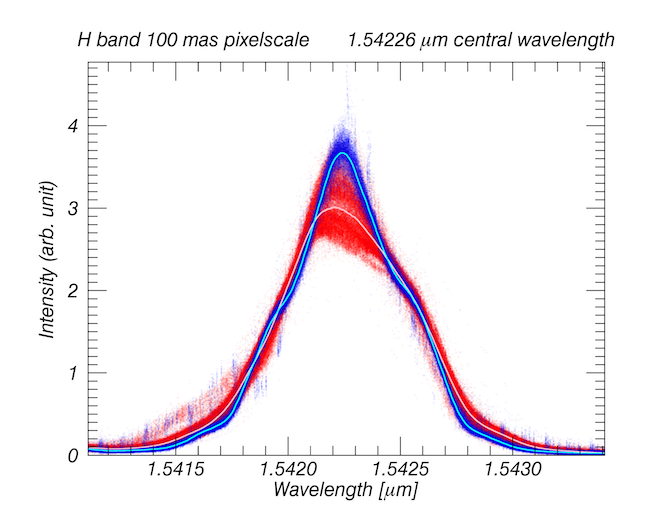}
			\includegraphics[width=1.0\textwidth, trim={1.5cm 0 0.5cm 0cm}, clip=true]{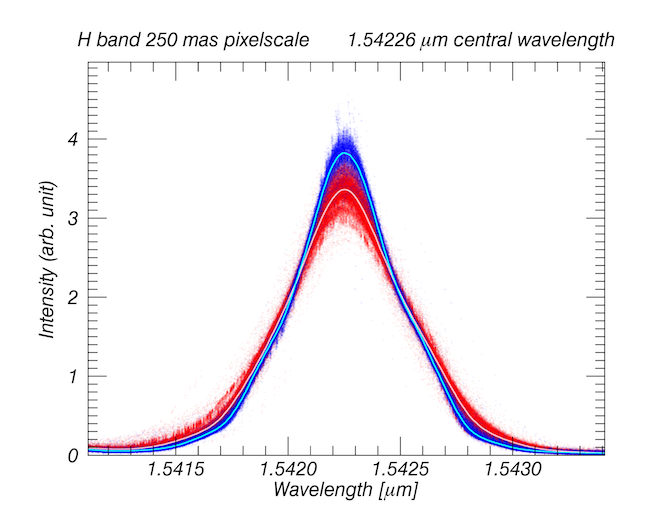}
		}
		\resizebox{1.0\textwidth}{!}{
			\includegraphics[width=1.0\textwidth, trim={1.0cm 0 1.0cm 0cm}, clip=true]{h25_wlength_1_79196_compare.png}
			\quad
			\includegraphics[width=1.0\textwidth, trim={1.5cm 0 0.5cm 0cm}, clip=true]{h100_wlength_1_79196_compare.png}
			\includegraphics[width=1.0\textwidth, trim={1.5cm 0 0.5cm 0cm}, clip=true]{h250_wlength_1_79196_compare.png}
		}
		\caption[Accumulated line profile in H-band]{Accumulated line profile along different wavelengths and in each pixelscale in H-band.}
		\label{fig:lineprofilesh}
	\end{center}
\end{figure}

\begin{figure}[htbp!]
	\begin{center}
		\resizebox{1.0\textwidth}{!}{
			\includegraphics[width=1.0\textwidth, trim={1.0cm 0 1.0cm 0cm}, clip=true]{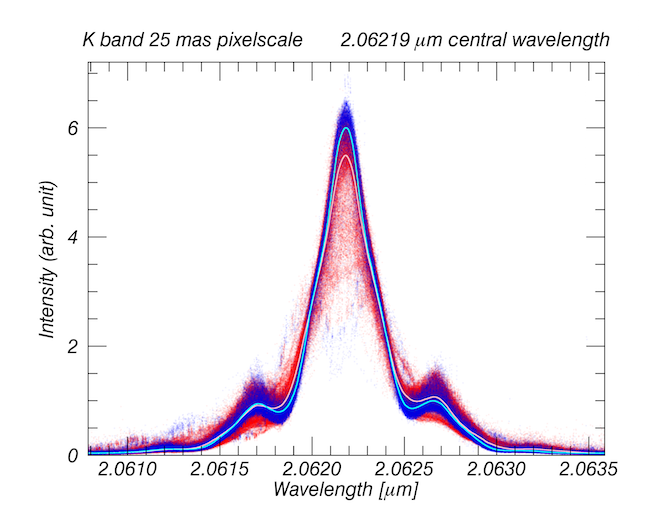}
			\quad
			\includegraphics[width=1.0\textwidth, trim={1.5cm 0 0.5cm 0cm}, clip=true]{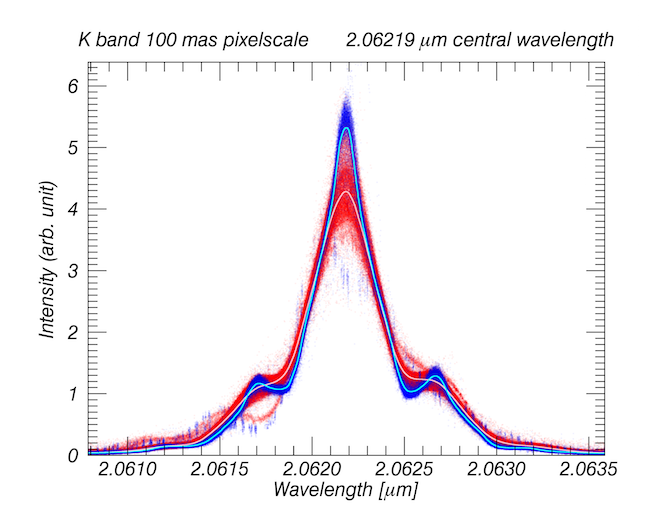}
			\includegraphics[width=1.0\textwidth, trim={1.5cm 0 0.5cm 0cm}, clip=true]{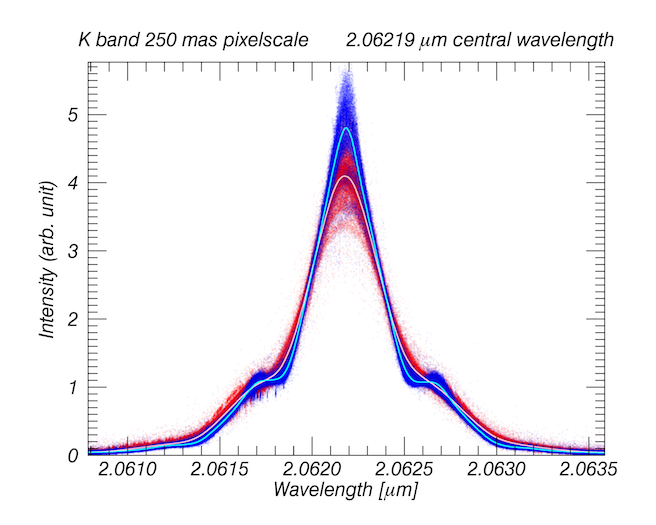}
		}
		\resizebox{1.0\textwidth}{!}{
			\includegraphics[width=1.0\textwidth, trim={1.0cm 0 1.0cm 0cm}, clip=true]{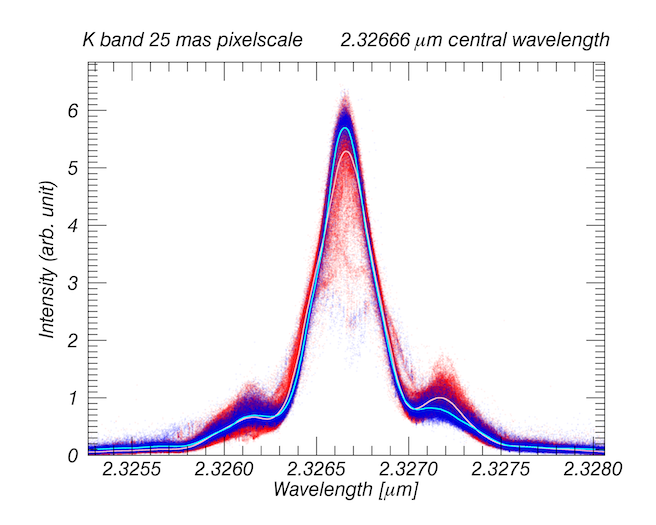}
			\quad
			\includegraphics[width=1.0\textwidth, trim={1.5cm 0 0.5cm 0cm}, clip=true]{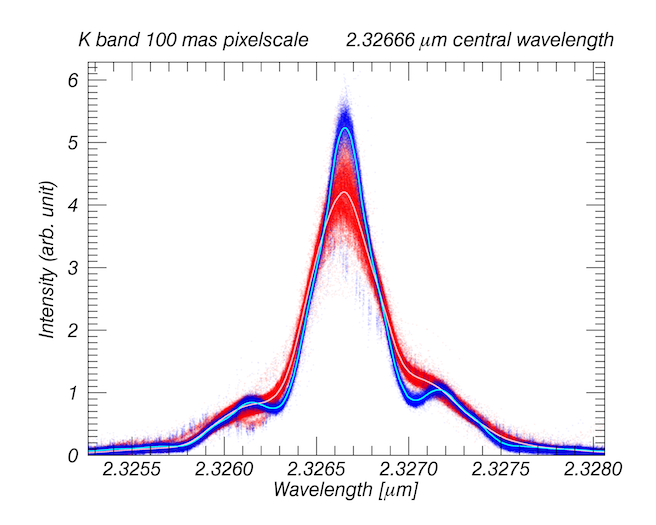}
			\includegraphics[width=1.0\textwidth, trim={1.5cm 0 0.5cm 0cm}, clip=true]{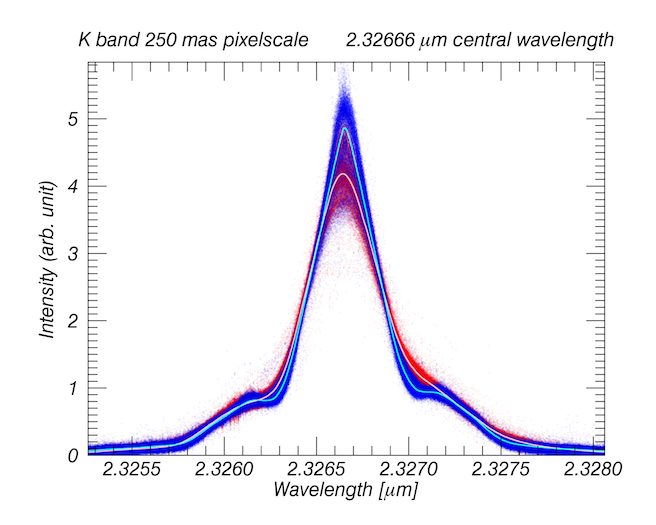}
		}
		\resizebox{1.0\textwidth}{!}{
			\includegraphics[width=1.0\textwidth, trim={1.0cm 0 1.0cm 0cm}, clip=true]{k25_wlength_2_36429_compare.png}
			\quad
			\includegraphics[width=1.0\textwidth, trim={1.5cm 0 0.5cm 0cm}, clip=true]{k100_wlength_2_36429_compare.png}
			\includegraphics[width=1.0\textwidth, trim={1.5cm 0 0.5cm 0cm}, clip=true]{k250_wlength_2_36429_compare.png}
		}
		\caption[Accumulated line profile in K-band]{Accumulated line profile along different wavelengths and in each pixelscale in K-band.}
		\label{fig:lineprofilesk}
	\end{center}
\end{figure}

\begin{figure}[htbp!]
	\begin{center}
		\resizebox{1.0\textwidth}{!}{
			\includegraphics[width=1.0\textwidth, trim={1.0cm 0 1.0cm 0cm}, clip=true]{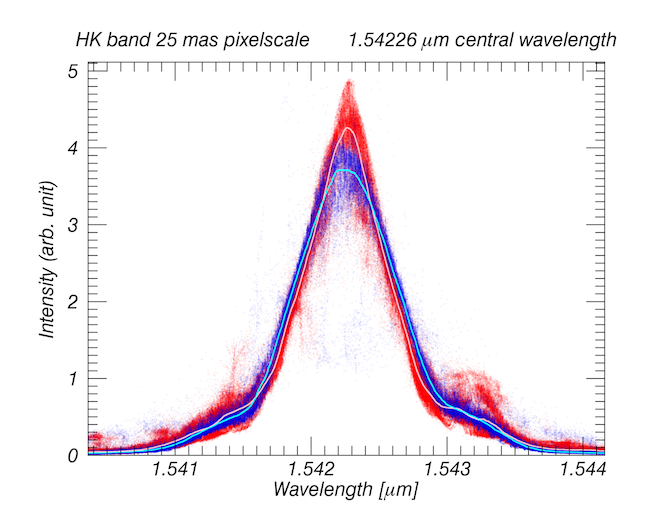}
			\quad
			\includegraphics[width=1.0\textwidth, trim={1.5cm 0 0.5cm 0cm}, clip=true]{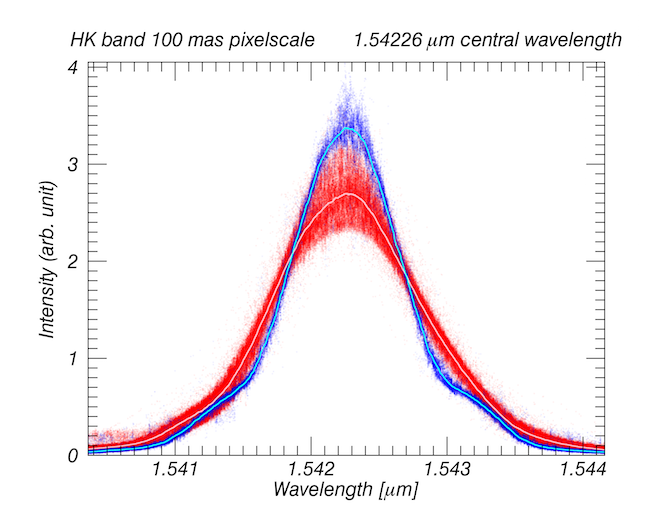}
			\includegraphics[width=1.0\textwidth, trim={1.5cm 0 0.5cm 0cm}, clip=true]{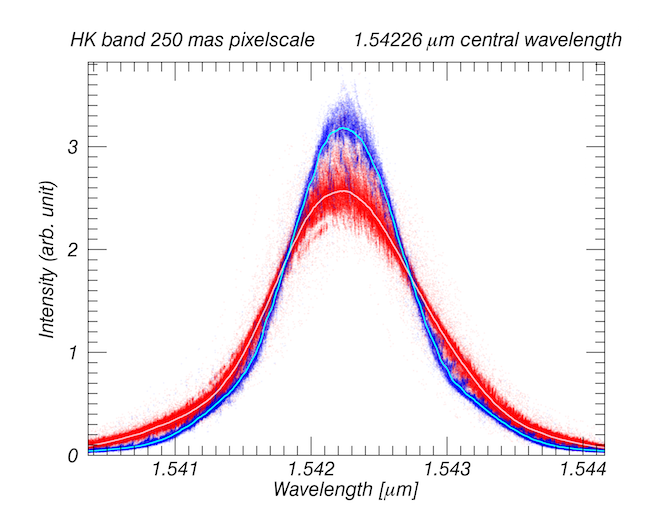}
		}
		\resizebox{1.0\textwidth}{!}{
			\includegraphics[width=1.0\textwidth, trim={1.0cm 0 1.0cm 0cm}, clip=true]{hk25_wlength_1_67327_compare.png}
			\quad
			\includegraphics[width=1.0\textwidth, trim={1.5cm 0 0.5cm 0cm}, clip=true]{hk25_wlength_1_67327_compare.png}
			\includegraphics[width=1.0\textwidth, trim={1.5cm 0 0.5cm 0cm}, clip=true]{hk25_wlength_1_67327_compare.png}
		}
		\resizebox{1.0\textwidth}{!}{
			\includegraphics[width=1.0\textwidth, trim={1.0cm 0 1.0cm 0cm}, clip=true]{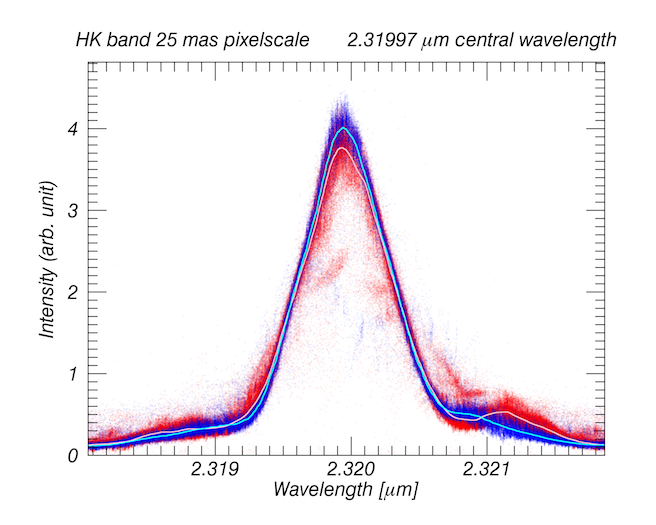}
			\quad
			\includegraphics[width=1.0\textwidth, trim={1.5cm 0 0.5cm 0cm}, clip=true]{hk25_wlength_2_31997_compare.png}
			\includegraphics[width=1.0\textwidth, trim={1.5cm 0 0.5cm 0cm}, clip=true]{hk25_wlength_2_31997_compare.png}
		}
		\caption[Accumulated line profile in H+K-band]{Accumulated line profile along different wavelengths and in each pixelscale in H+K-band.}
		\label{fig:lineprofileshk}
	\end{center}
\end{figure}

\clearpage
\section{Variations of the Line Profiles Within a Slitlet}
This appendix section refers to section \ref{sec:var_line} and the discussion about the variation of the line profiles within a slitlet from section \ref{sec:discussion_slitlet}. In order to show further the variations of the LSF within a slitlet, plots for slitlet 15 (close to the center of the detector) and slitlet 25 (close to the edge of the detector and also one of the most tilted slitlets on the small slicer) are shown. The columns are the pixelscales 25 mas, 100 mas and 250 mas, while the first row of each figure is always slitlet 15 and the second row slitlet 25. The upper figure on each page are the pre- upgrade plots, while the lower figure are the post- upgrade plots.
\newpage
\begin{figure}[htbp!]
	\begin{center}
		\resizebox{1.0\textwidth}{!}{
			\includegraphics[height=6.5cm, trim={1.5cm 0 5.5cm 0cm}, clip=true]{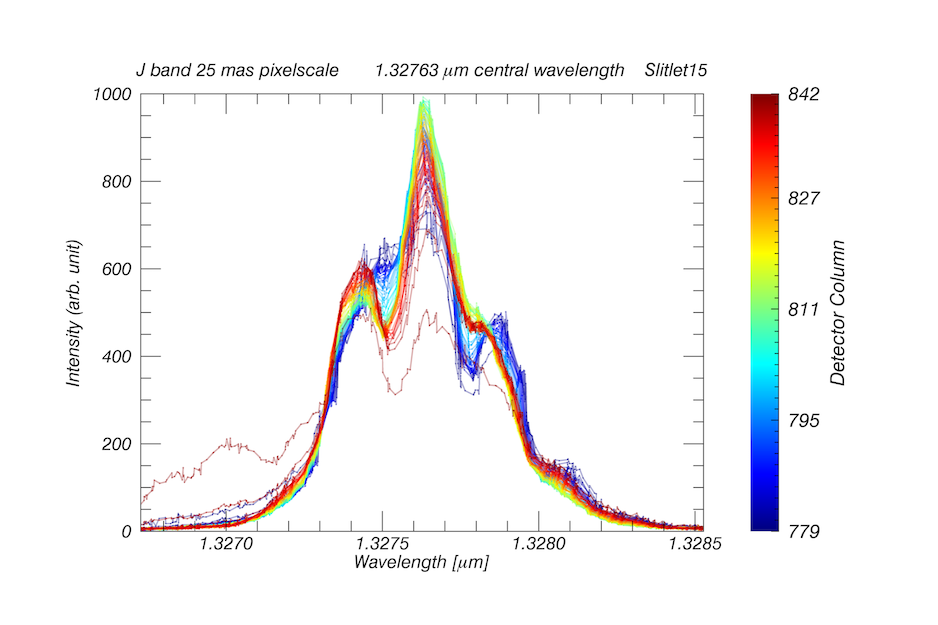}
			\includegraphics[height=6.5cm, trim={2.4cm 0 5.5cm 0cm}, clip=true]{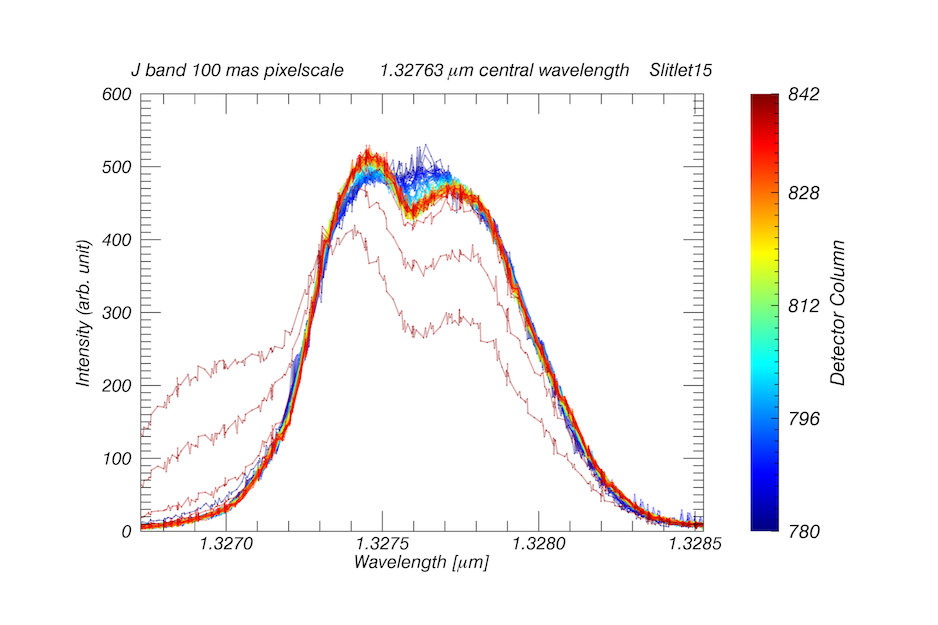}
			\includegraphics[height=6.5cm, trim={2.4cm 0 2.0cm 0cm}, clip=true]{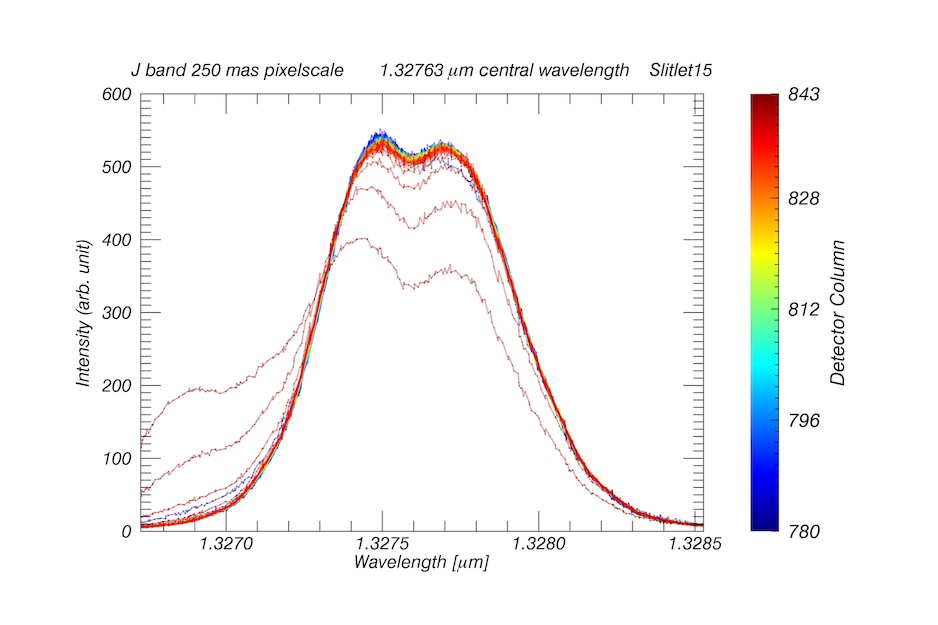}
		}
		\resizebox{1.0\textwidth}{!}{
			\includegraphics[height=6.5cm, trim={1.5cm 0 5.5cm 0cm}, clip=true]{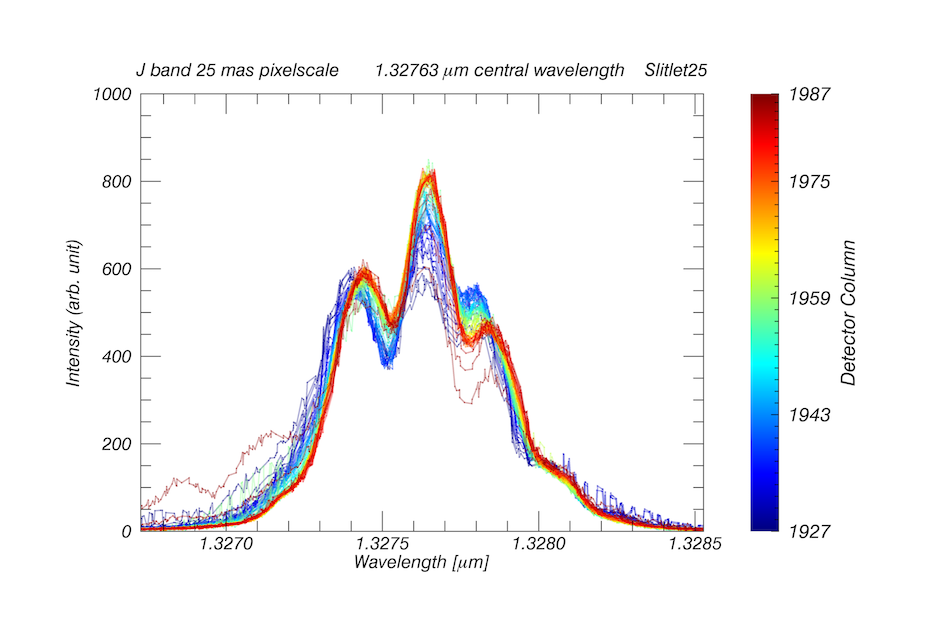}
			\includegraphics[height=6.5cm, trim={2.4cm 0 5.5cm 0cm}, clip=true]{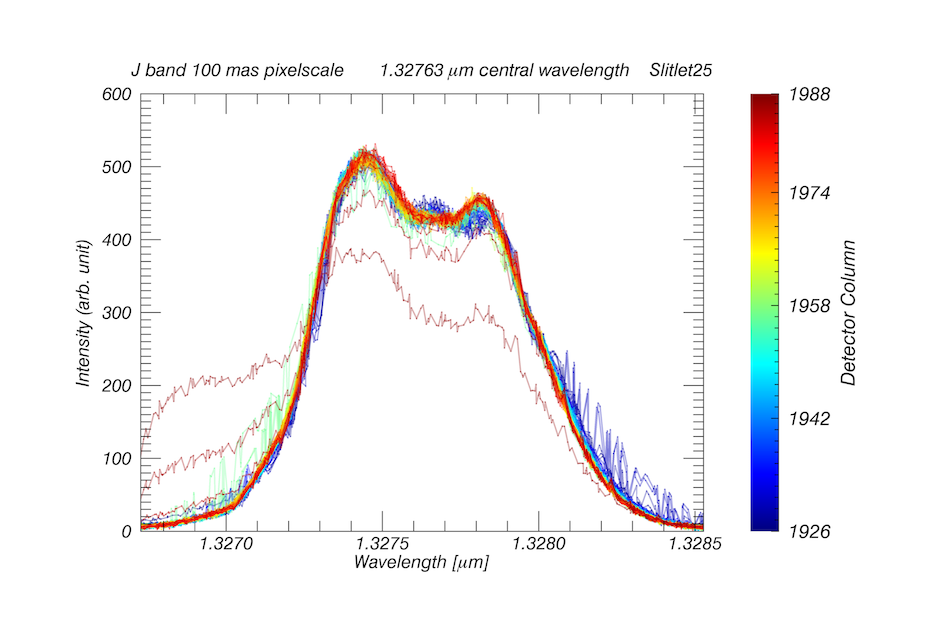}
			\includegraphics[height=6.5cm, trim={2.4cm 0 2.0cm 0cm}, clip=true]{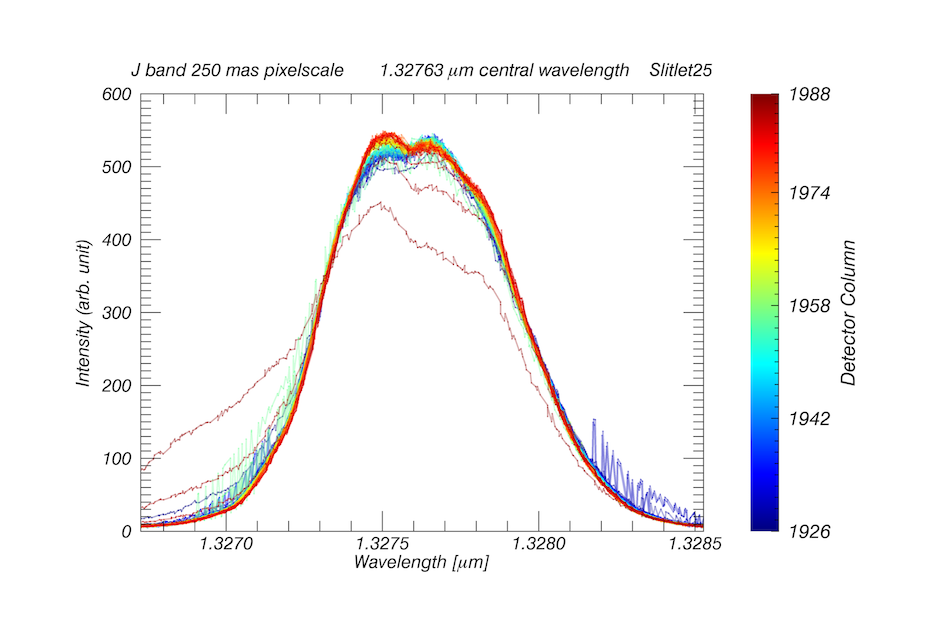}
		}	
		\caption[]{Variation of the line profiles within a slitlet in J-band pre- upgrade.}
		\label{fig:slitlet_j_pre}
	\end{center}
\end{figure}

\begin{figure}[htbp!]
	\begin{center}
		\resizebox{1.0\textwidth}{!}{
			\includegraphics[height=6.5cm, trim={1.5cm 0 5.5cm 0cm}, clip=true]{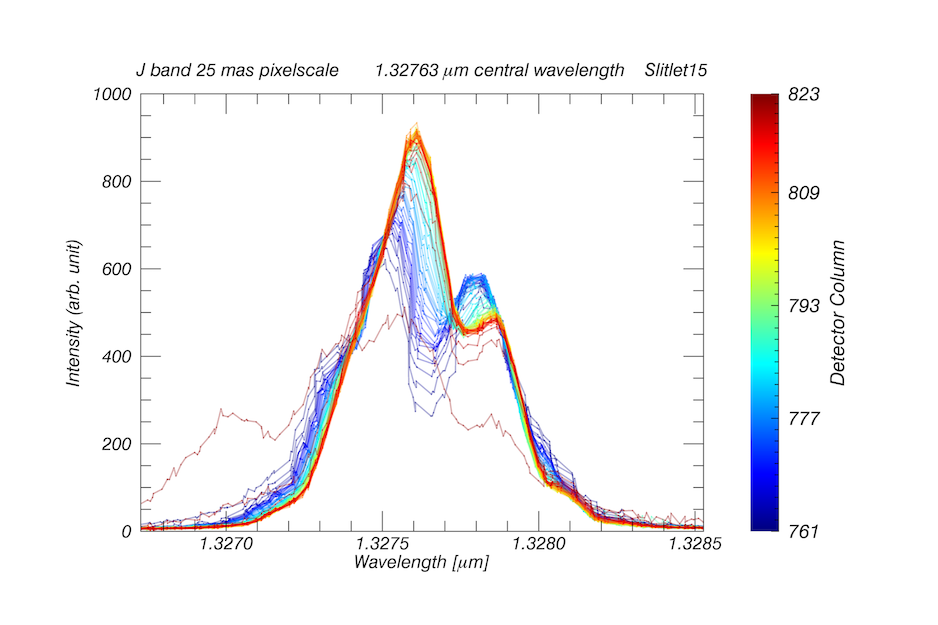}
			\includegraphics[height=6.5cm, trim={2.4cm 0 5.5cm 0cm}, clip=true]{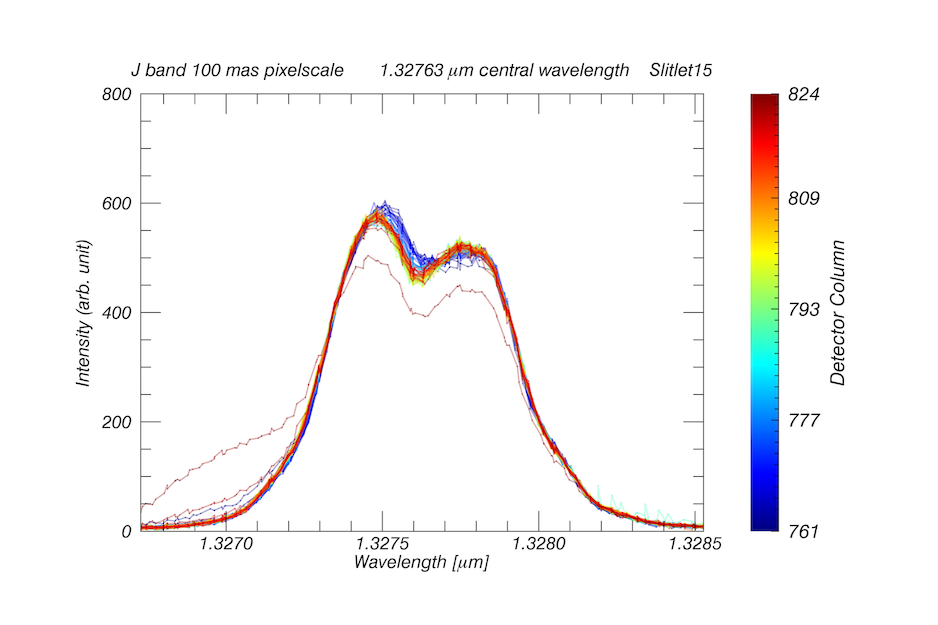}
			\includegraphics[height=6.5cm, trim={2.4cm 0 2.0cm 0cm}, clip=true]{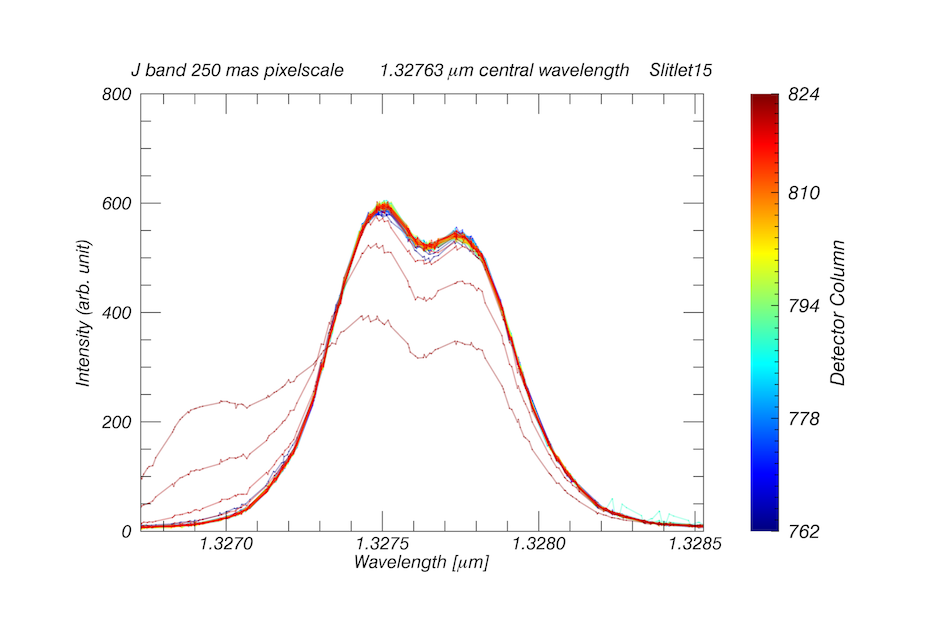}
		}
		\resizebox{1.0\textwidth}{!}{
			\includegraphics[height=6.5cm, trim={1.5cm 0 5.5cm 0cm}, clip=true]{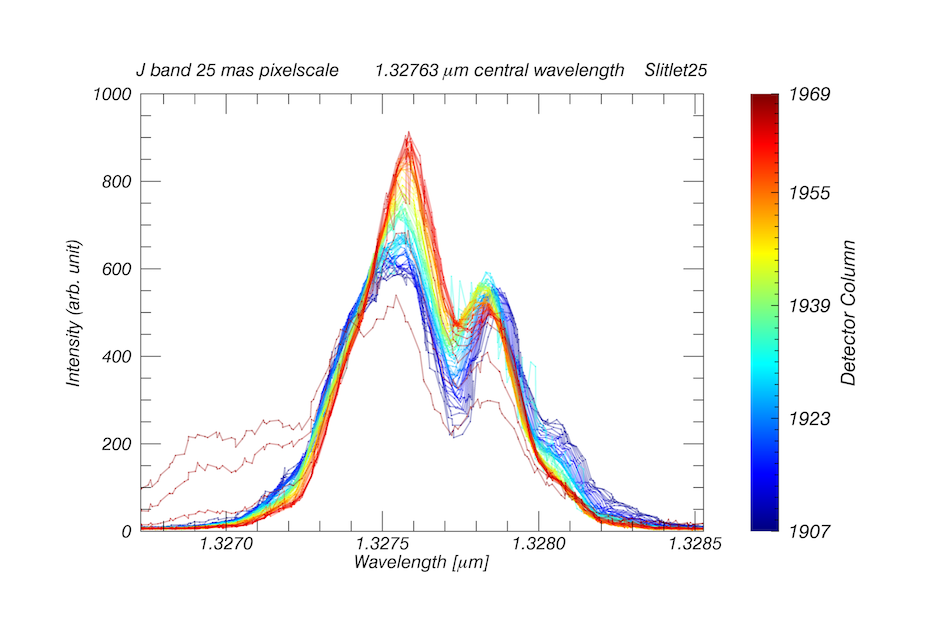}
			\includegraphics[height=6.5cm, trim={2.4cm 0 5.5cm 0cm}, clip=true]{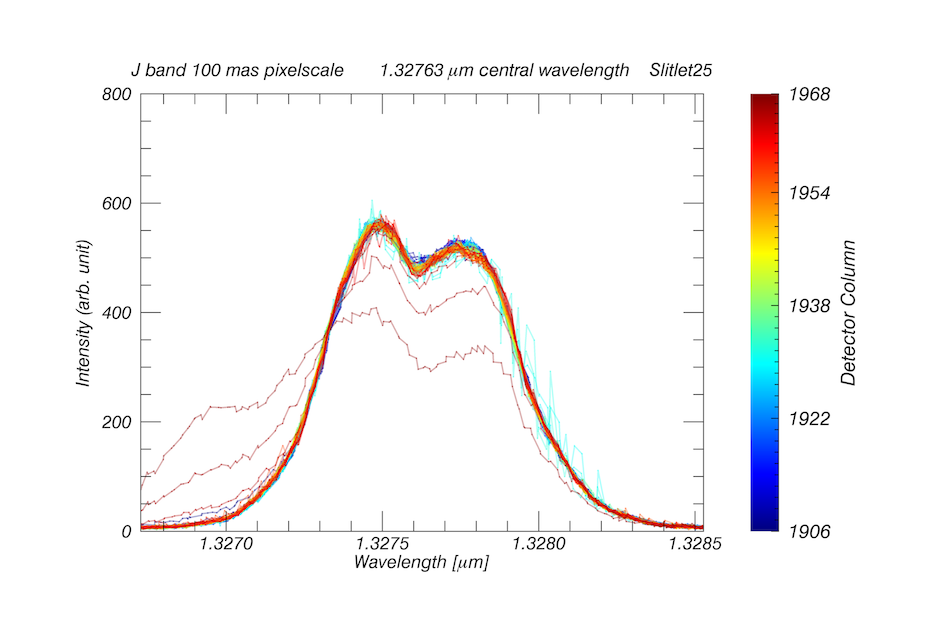}
			\includegraphics[height=6.5cm, trim={2.4cm 0 2.0cm 0cm}, clip=true]{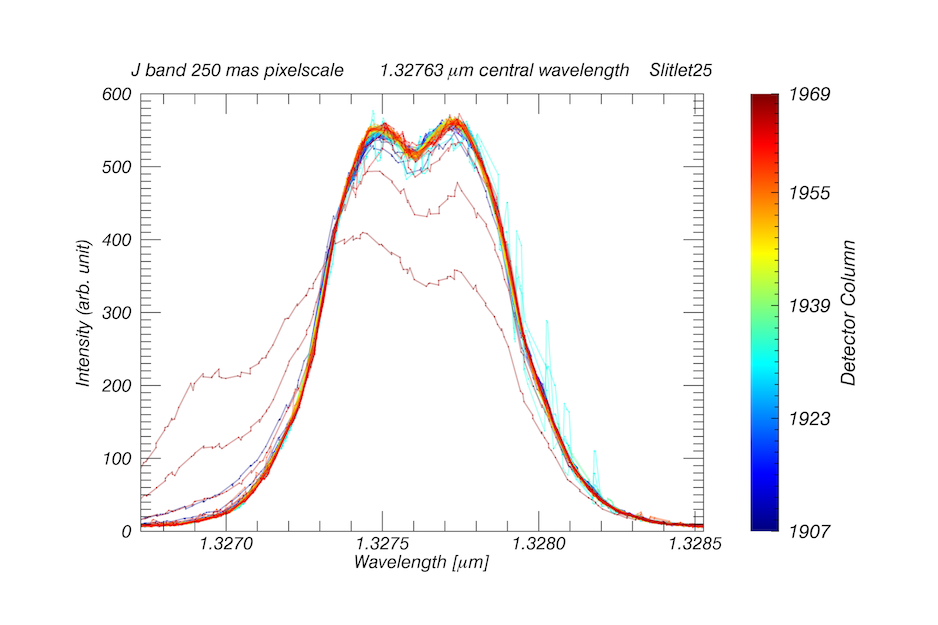}
		}	
		\caption{Variation of the line profiles within a slitlet in J-band post- upgrade.}
		\label{fig:slitlet_j_post}
	\end{center}
\end{figure}

\begin{figure}[htbp!]
	\begin{center}
		\resizebox{1.0\textwidth}{!}{
			\includegraphics[height=6.5cm, trim={1.5cm 0 5.5cm 0cm}, clip=true]{{h25old_wlength_1.54226_slitlet15}.png}
			\includegraphics[height=6.5cm, trim={2.4cm 0 5.5cm 0cm}, clip=true]{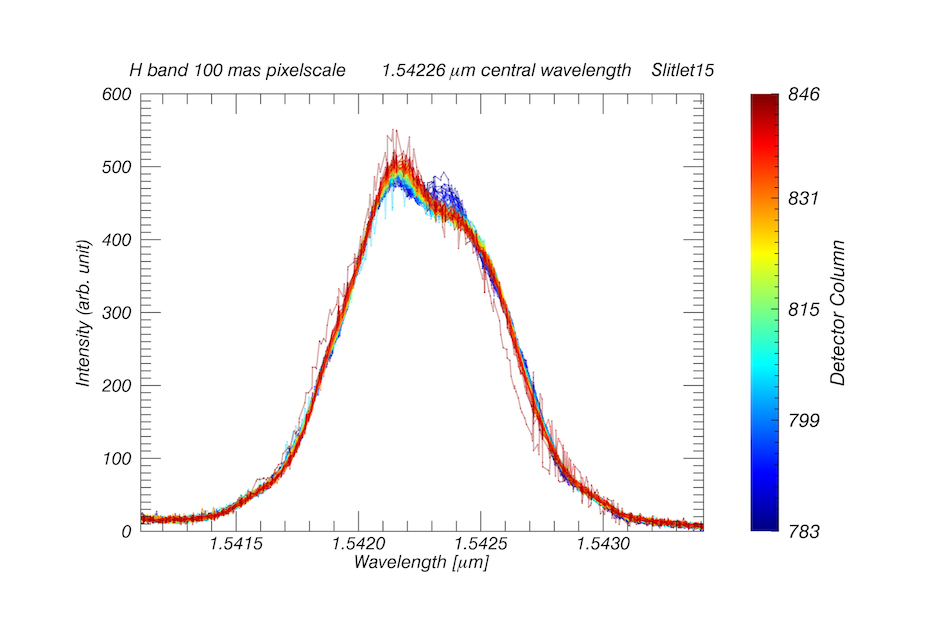}
			\includegraphics[height=6.5cm, trim={2.4cm 0 2.0cm 0cm}, clip=true]{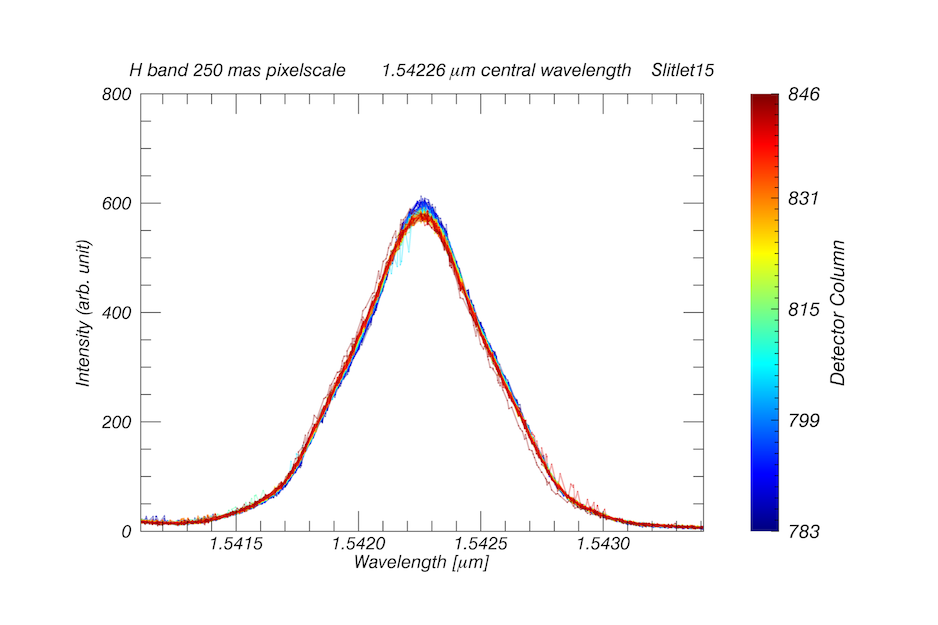}
		}
		\resizebox{1.0\textwidth}{!}{
			\includegraphics[height=6.5cm, trim={1.5cm 0 5.5cm 0cm}, clip=true]{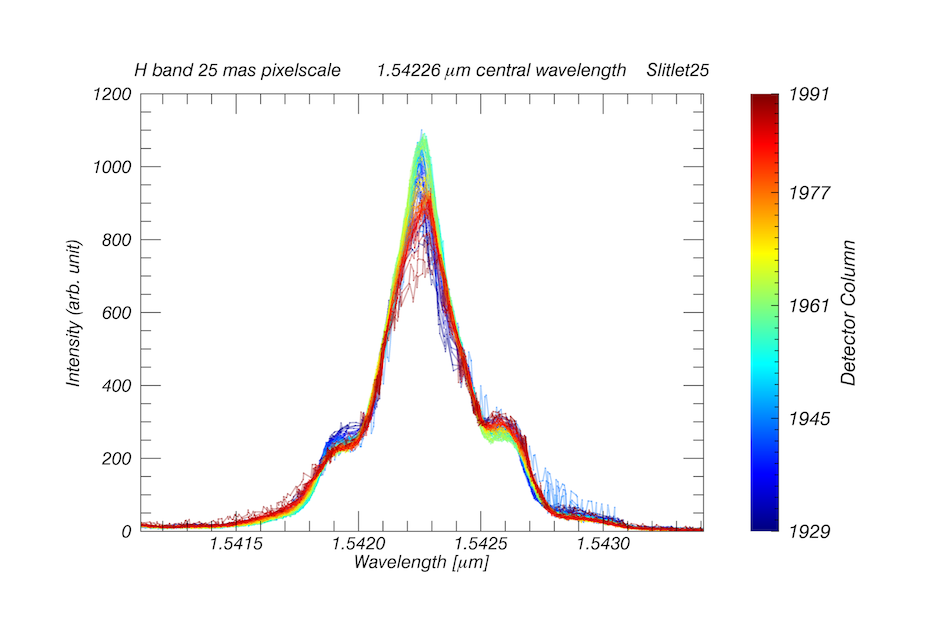}
			\includegraphics[height=6.5cm, trim={2.4cm 0 5.5cm 0cm}, clip=true]{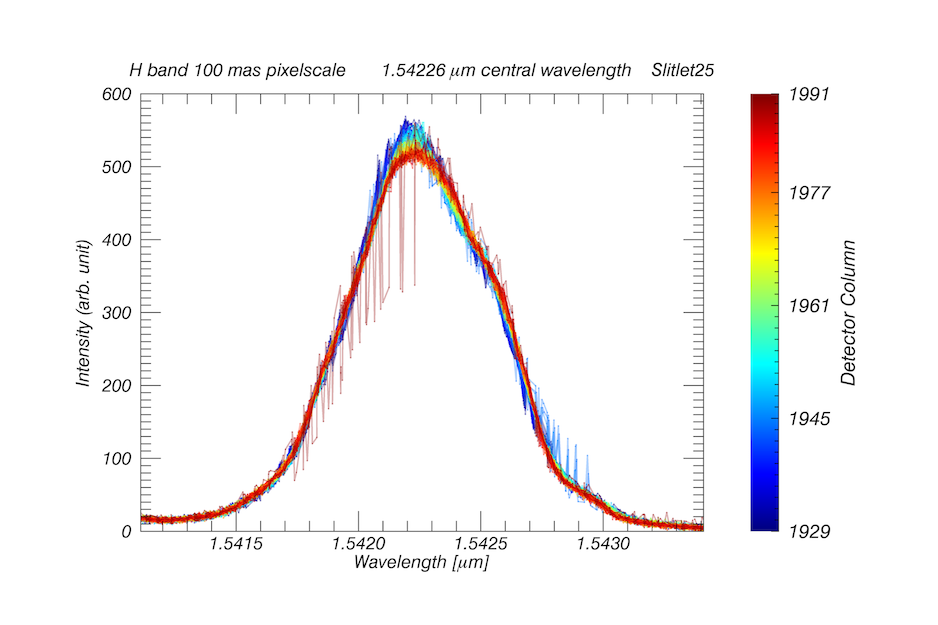}
			\includegraphics[height=6.5cm, trim={2.4cm 0 2.0cm 0cm}, clip=true]{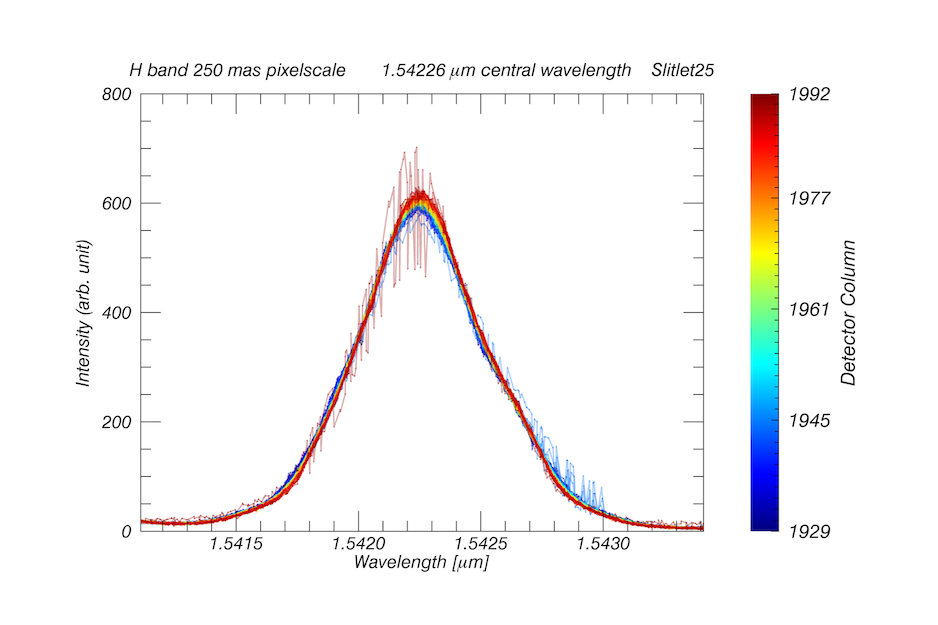}
		}	
		\caption{Variation of the line profiles within a slitlet in H-band pre- upgrade.}
		\label{fig:slitlet_h_pre}
	\end{center}
\end{figure}

\begin{figure}[htbp!]
	\begin{center}
		\resizebox{1.0\textwidth}{!}{
			\includegraphics[height=6.5cm, trim={1.5cm 0 5.5cm 0cm}, clip=true]{{h25new_wlength_1.54226_slitlet15}.png}
			\includegraphics[height=6.5cm, trim={2.4cm 0 5.5cm 0cm}, clip=true]{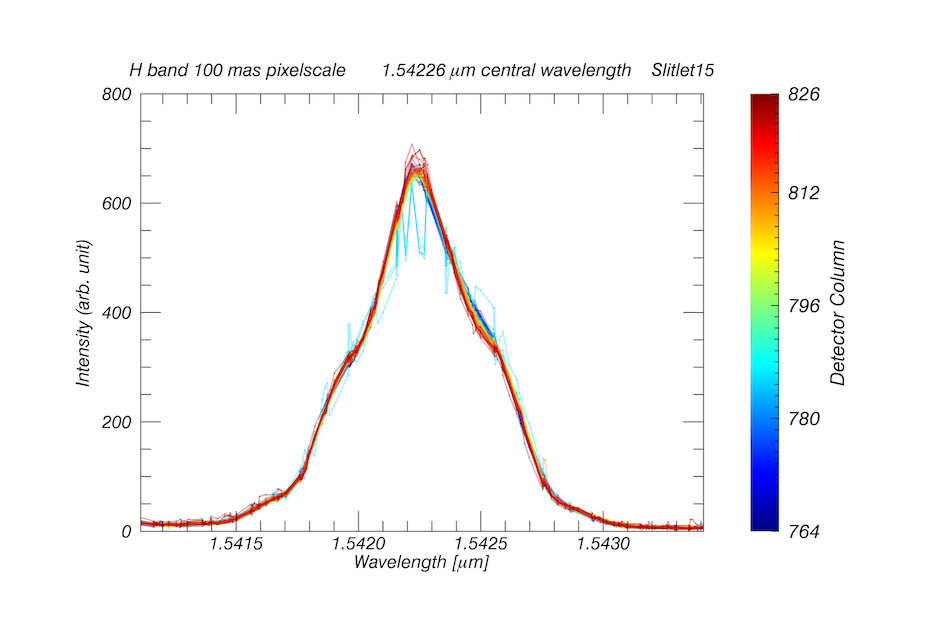}
			\includegraphics[height=6.5cm, trim={2.4cm 0 2.0cm 0cm}, clip=true]{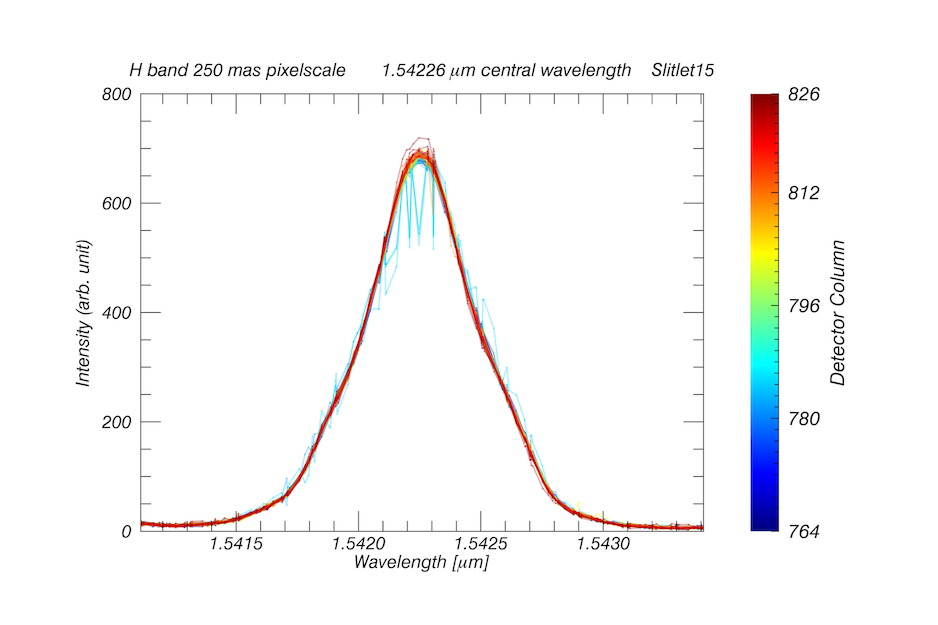}
		}
		\resizebox{1.0\textwidth}{!}{
			\includegraphics[height=6.5cm, trim={1.5cm 0 5.5cm 0cm}, clip=true]{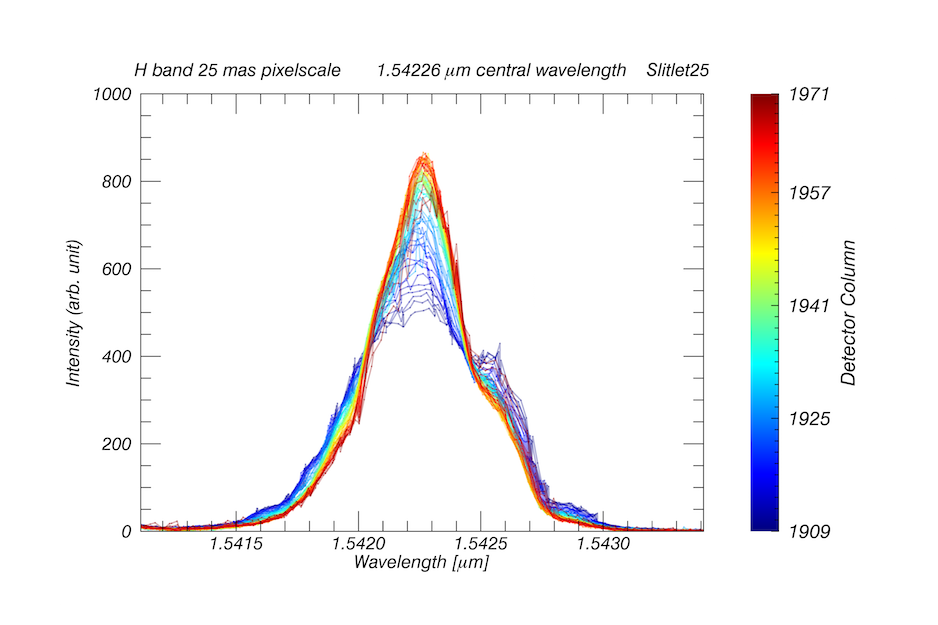}
			\includegraphics[height=6.5cm, trim={2.4cm 0 5.5cm 0cm}, clip=true]{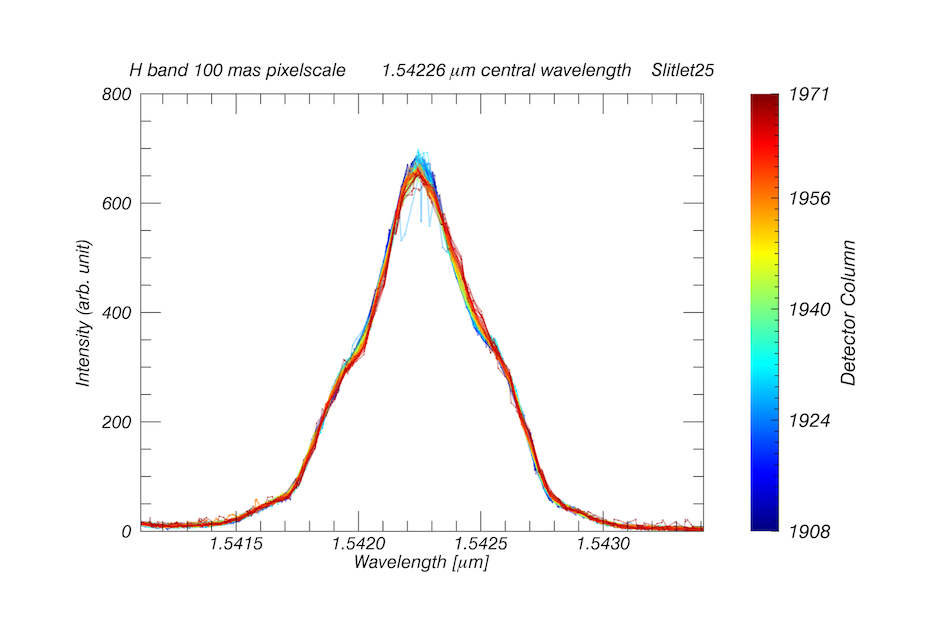}
			\includegraphics[height=6.5cm, trim={2.4cm 0 2.0cm 0cm}, clip=true]{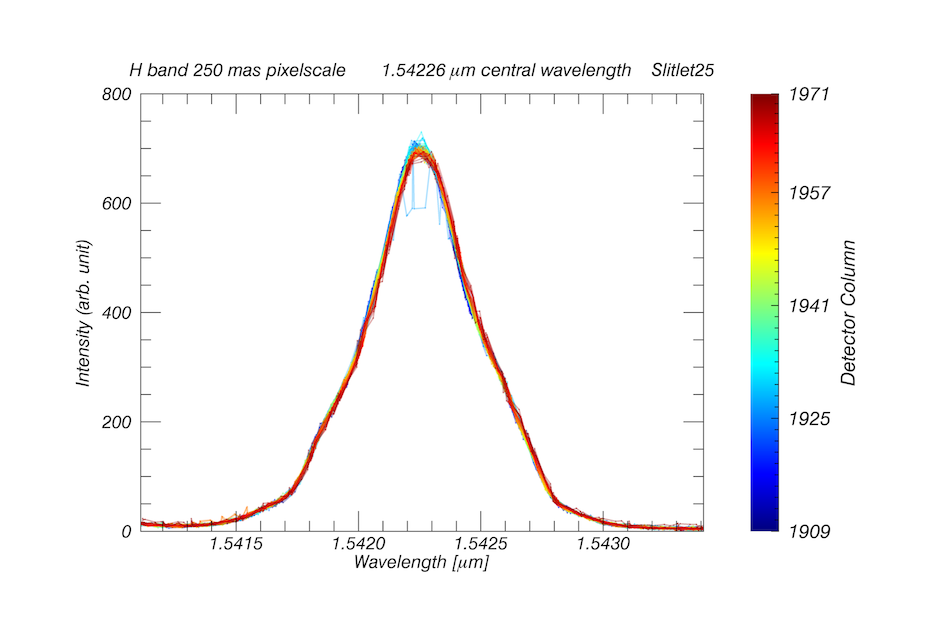}
		}	
		\caption{Variation of the line profiles within a slitlet in H-band post- upgrade.}
		\label{fig:slitlet_h_post}
	\end{center}
\end{figure}

\begin{figure}[htbp!]
	\begin{center}
		\resizebox{1.0\textwidth}{!}{
			\includegraphics[height=6.5cm, trim={1.5cm 0 5.5cm 0cm}, clip=true]{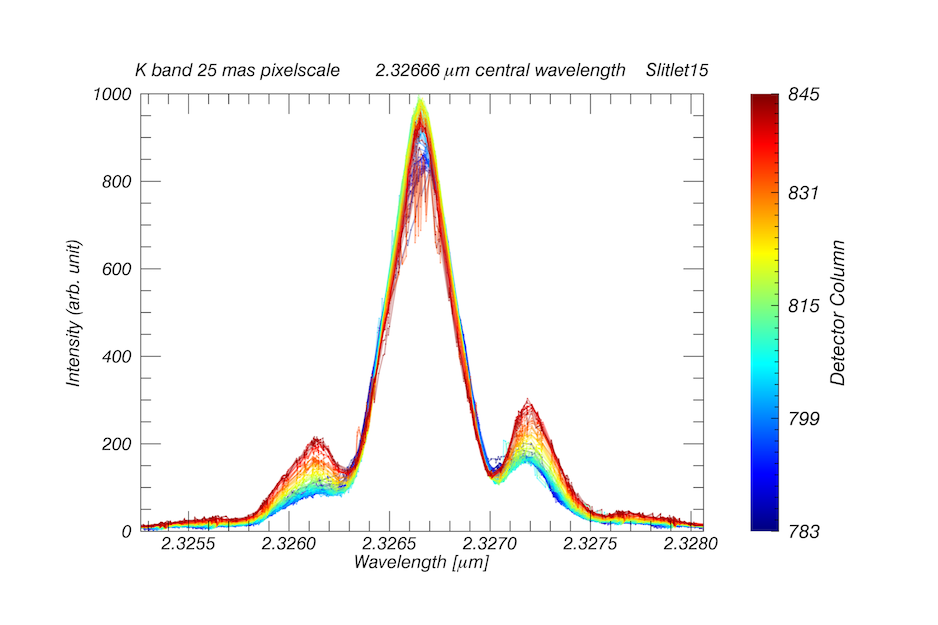}
			\includegraphics[height=6.5cm, trim={2.4cm 0 5.5cm 0cm}, clip=true]{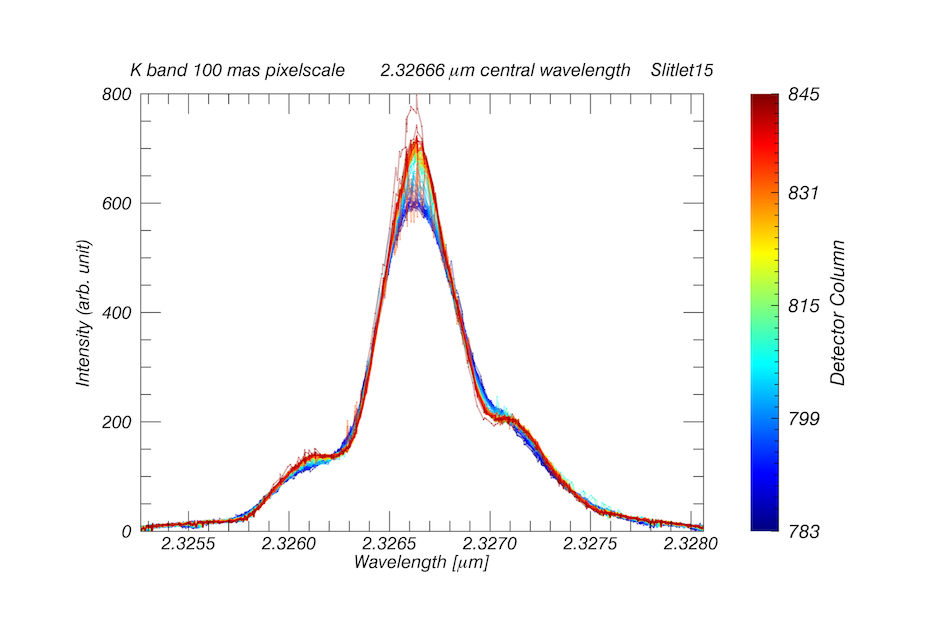}
			\includegraphics[height=6.5cm, trim={2.4cm 0 2.0cm 0cm}, clip=true]{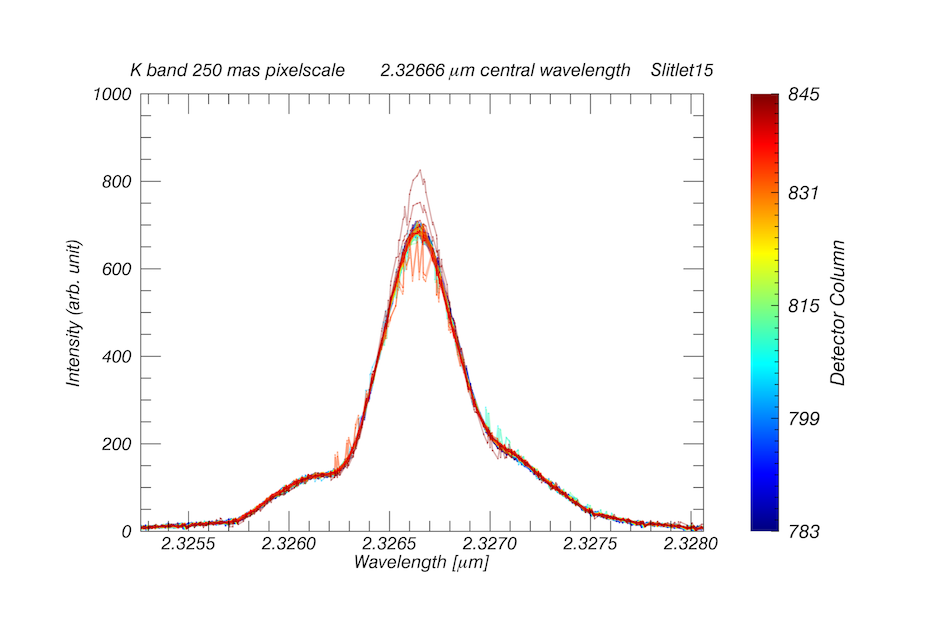}
		}
		\resizebox{1.0\textwidth}{!}{
			\includegraphics[height=6.5cm, trim={1.5cm 0 5.5cm 0cm}, clip=true]{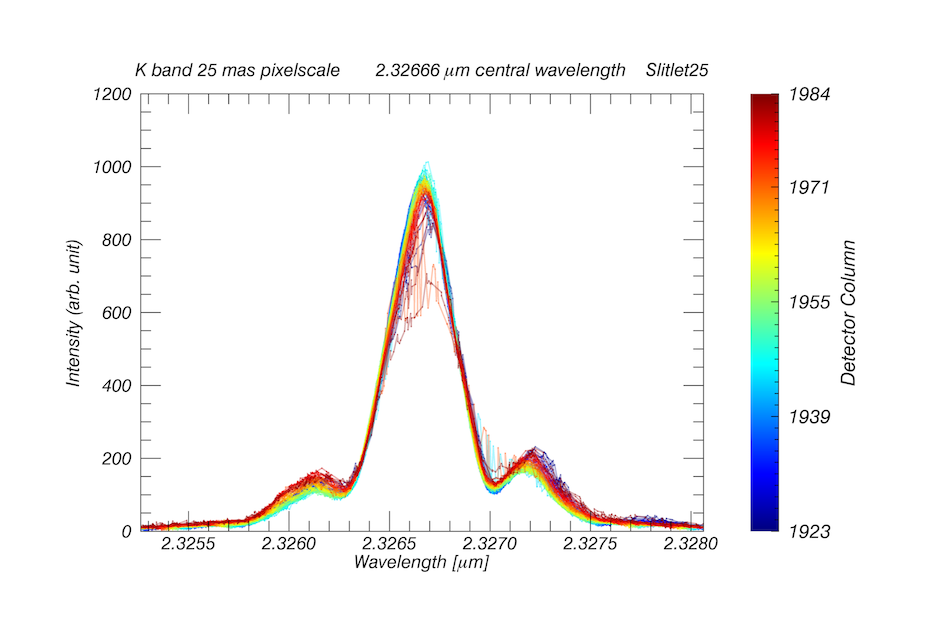}
			\includegraphics[height=6.5cm, trim={2.4cm 0 5.5cm 0cm}, clip=true]{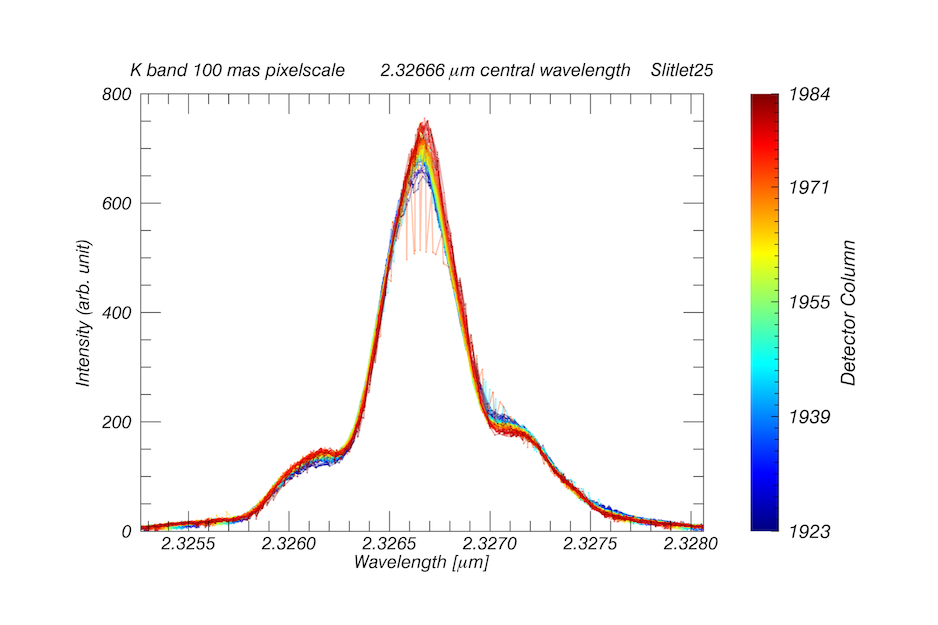}
			\includegraphics[height=6.5cm, trim={2.4cm 0 2.0cm 0cm}, clip=true]{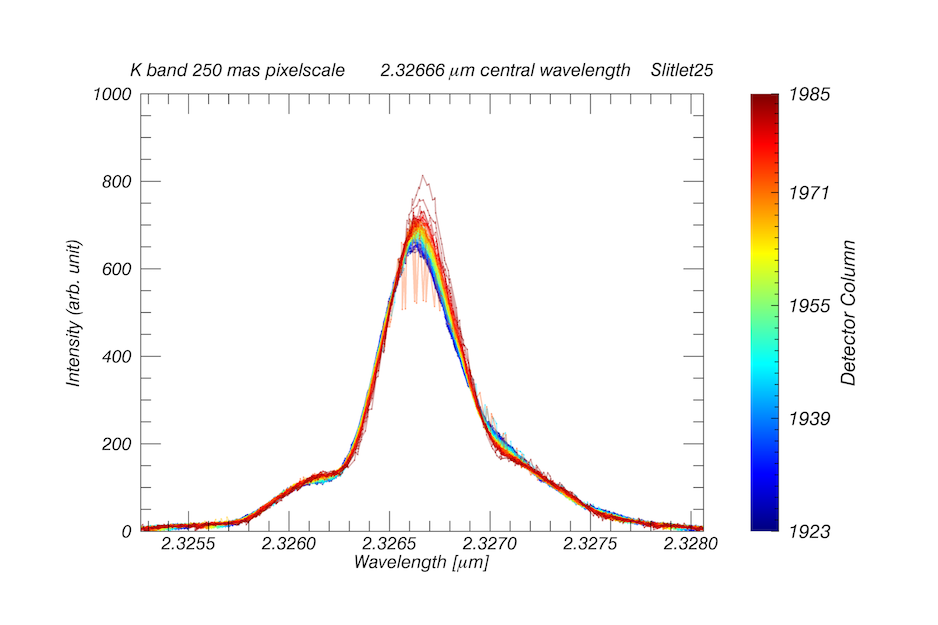}
		}	
		\caption{Variation of the line profiles within a slitlet in K-band pre- upgrade.}
		\label{fig:slitlet_k_pre}
	\end{center}
\end{figure}

\begin{figure}[htbp!]
	\begin{center}
		\resizebox{1.0\textwidth}{!}{
			\includegraphics[height=6.5cm, trim={1.5cm 0 5.5cm 0cm}, clip=true]{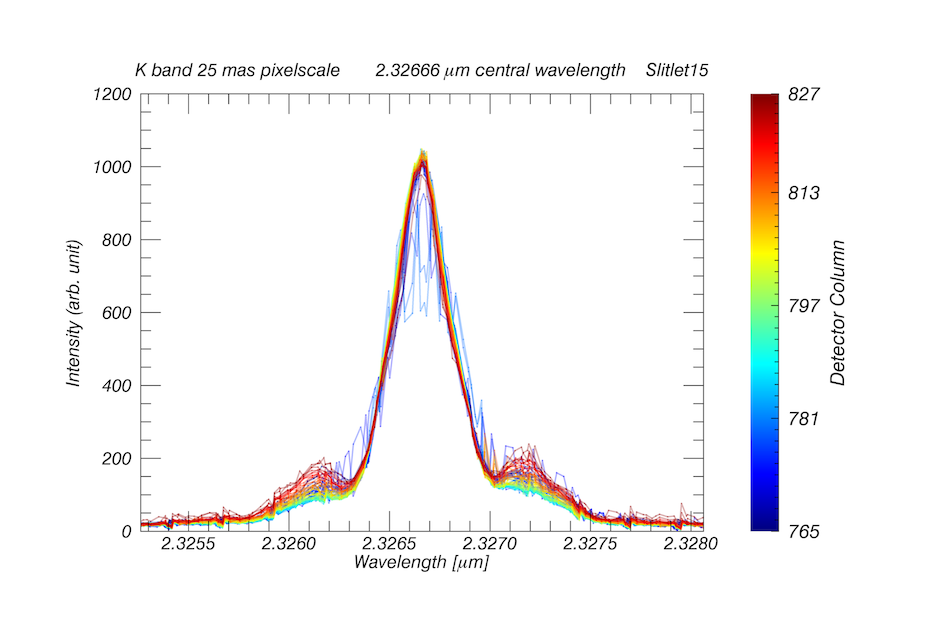}
			\includegraphics[height=6.5cm, trim={2.4cm 0 5.5cm 0cm}, clip=true]{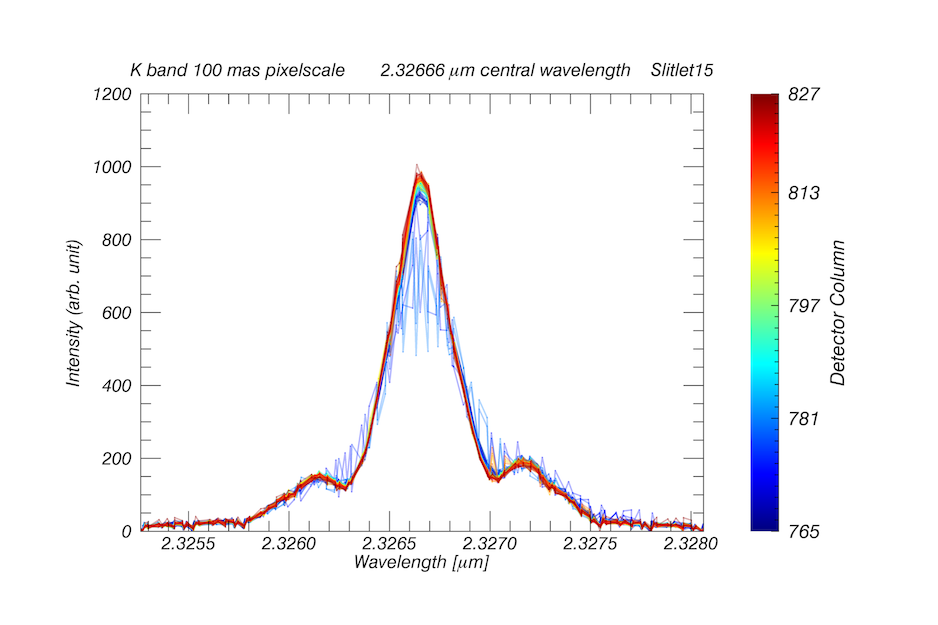}
			\includegraphics[height=6.5cm, trim={2.4cm 0 2.0cm 0cm}, clip=true]{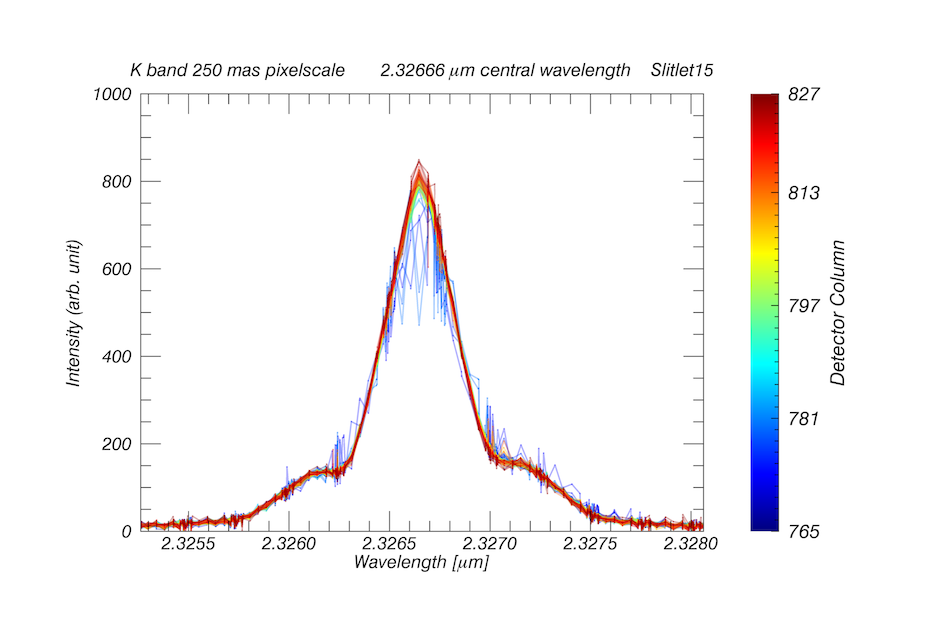}
		}
		\resizebox{1.0\textwidth}{!}{
			\includegraphics[height=6.5cm, trim={1.5cm 0 5.5cm 0cm}, clip=true]{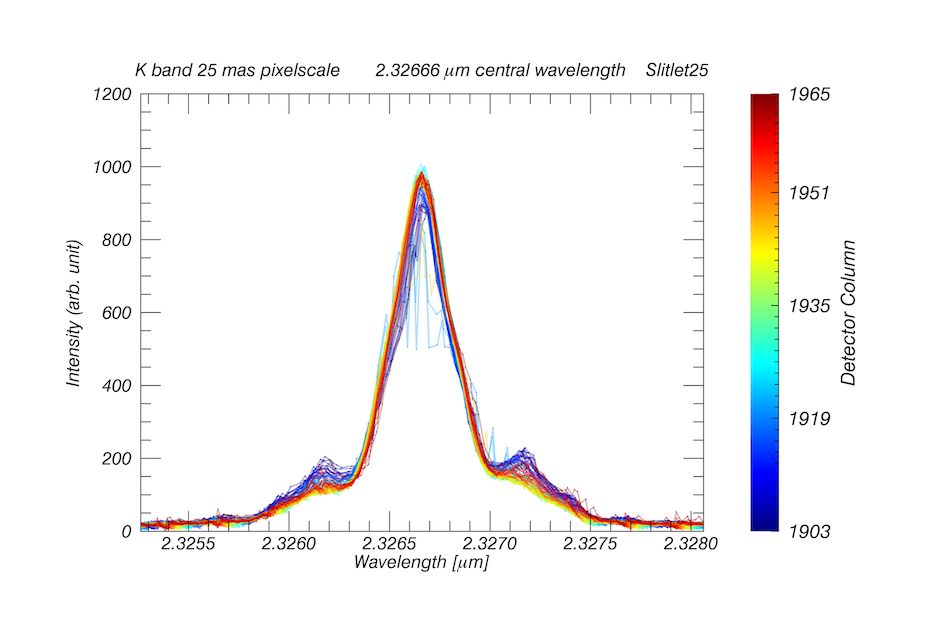}
			\includegraphics[height=6.5cm, trim={2.4cm 0 5.5cm 0cm}, clip=true]{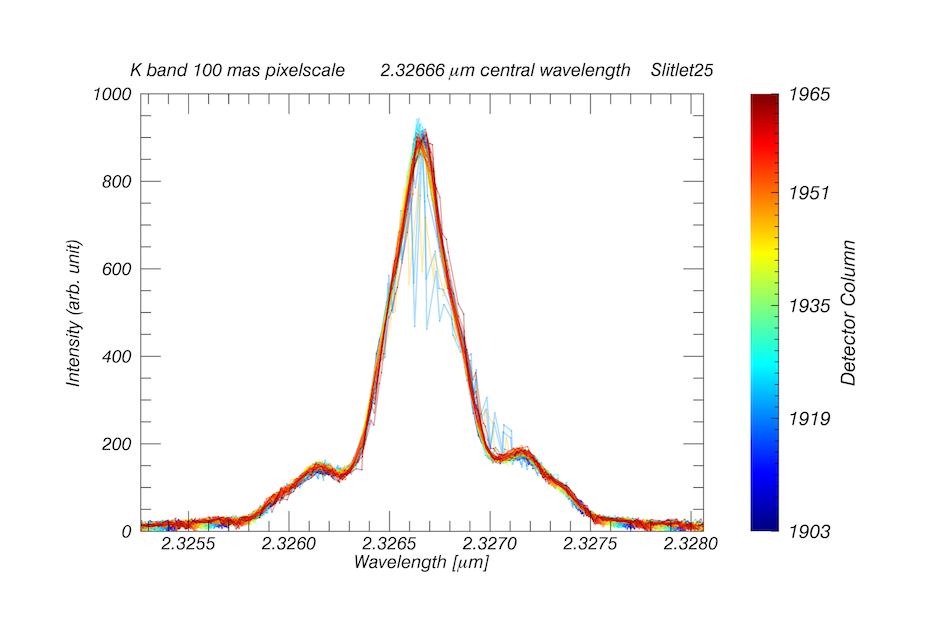}
			\includegraphics[height=6.5cm, trim={2.4cm 0 2.0cm 0cm}, clip=true]{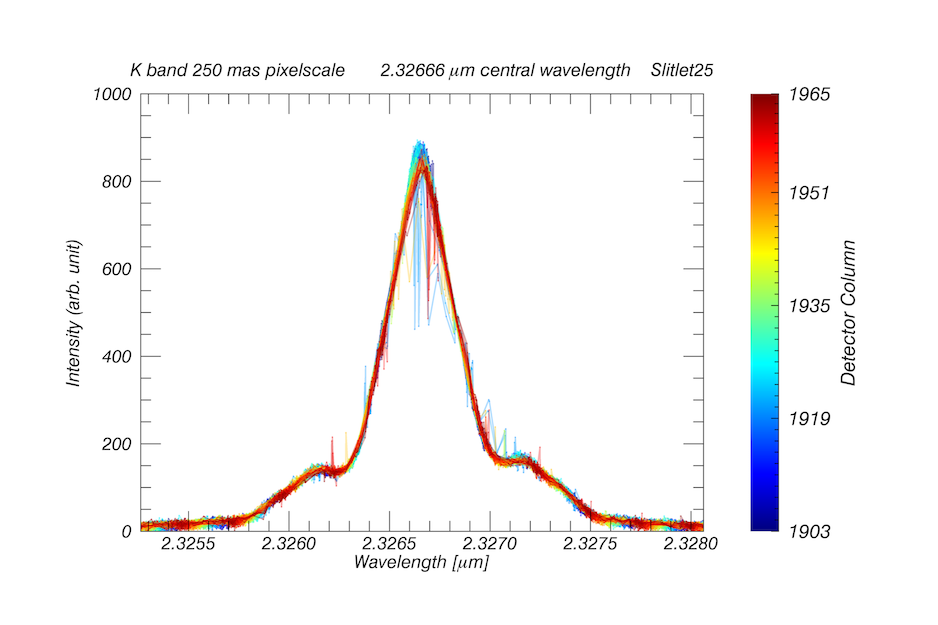}
		}	
		\caption{Variation of the line profiles within a slitlet in K-band post- upgrade.}
		\label{fig:slitlet_k_post}
	\end{center}
\end{figure}

\begin{figure}[htbp!]
	\begin{center}
		\resizebox{1.0\textwidth}{!}{
			\includegraphics[height=6.5cm, trim={1.5cm 0 5.5cm 0cm}, clip=true]{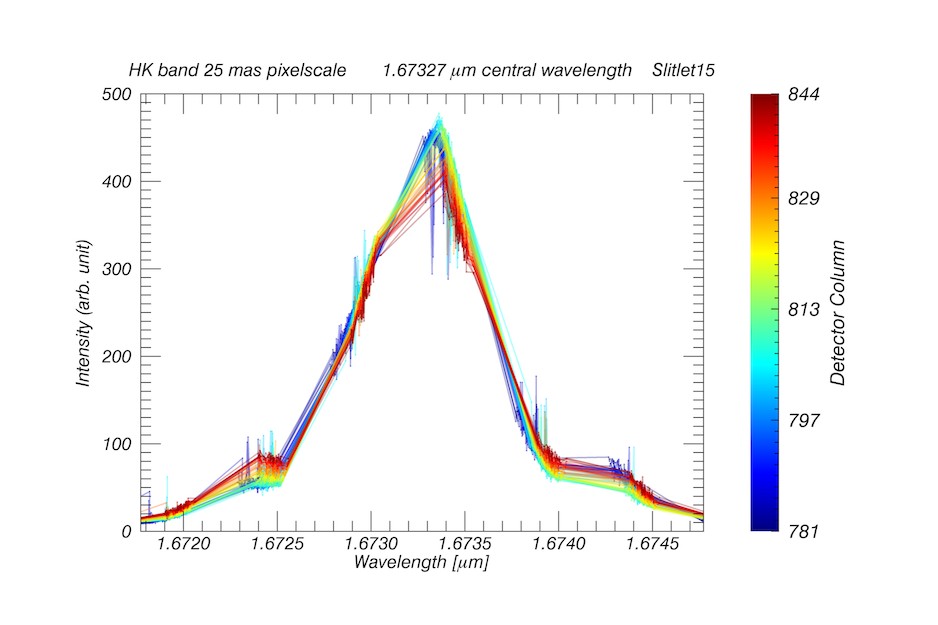}
			\includegraphics[height=6.5cm, trim={2.4cm 0 5.5cm 0cm}, clip=true]{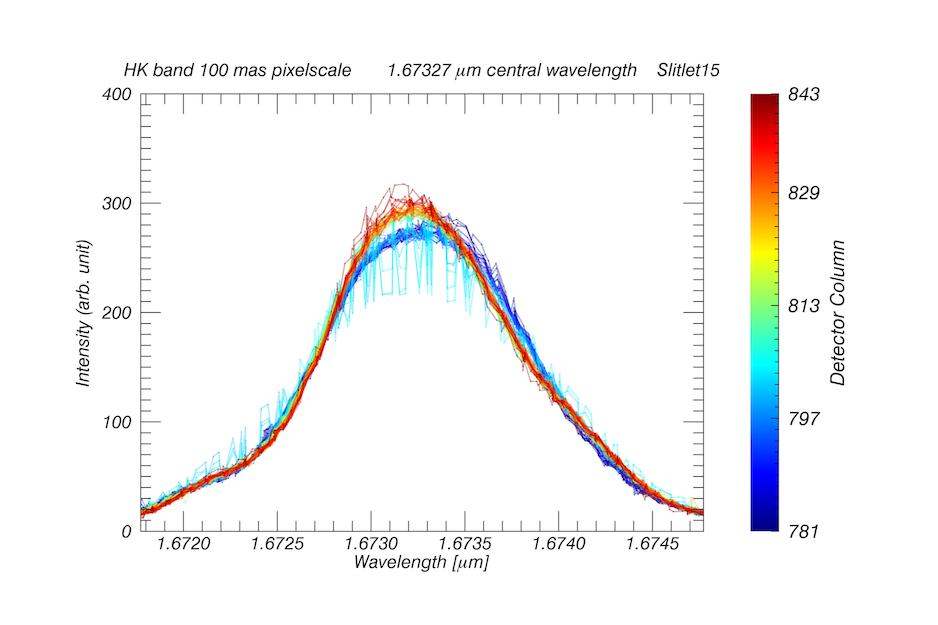}
			\includegraphics[height=6.5cm, trim={2.4cm 0 2.0cm 0cm}, clip=true]{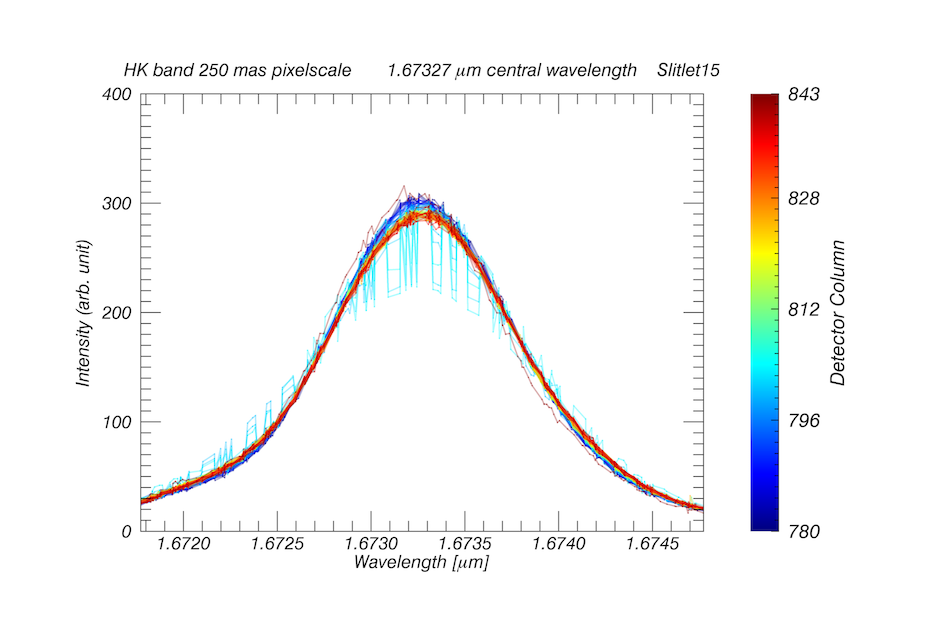}
		}
		\resizebox{1.0\textwidth}{!}{
			\includegraphics[height=6.5cm, trim={1.5cm 0 5.5cm 0cm}, clip=true]{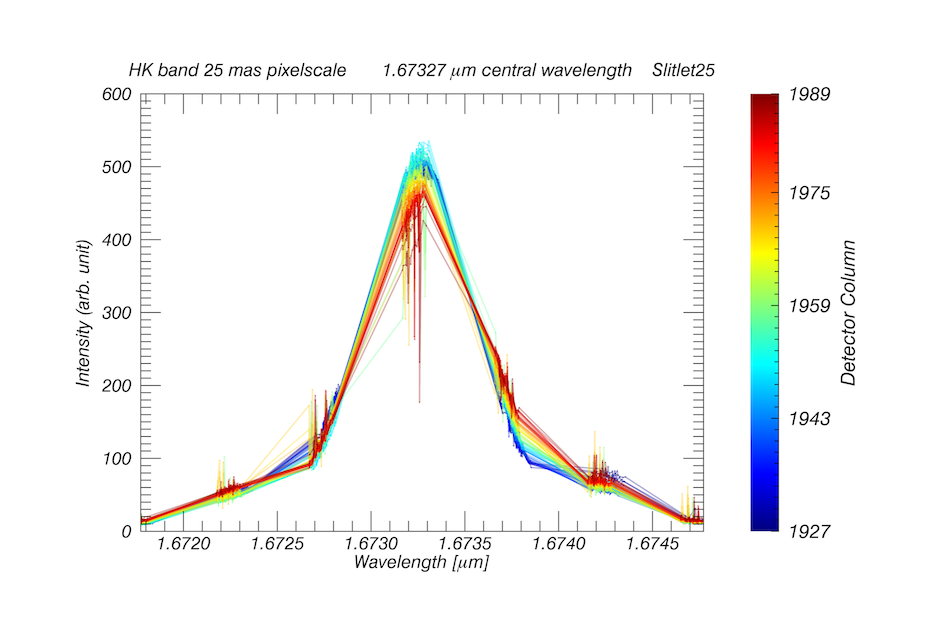}
			\includegraphics[height=6.5cm, trim={2.4cm 0 5.5cm 0cm}, clip=true]{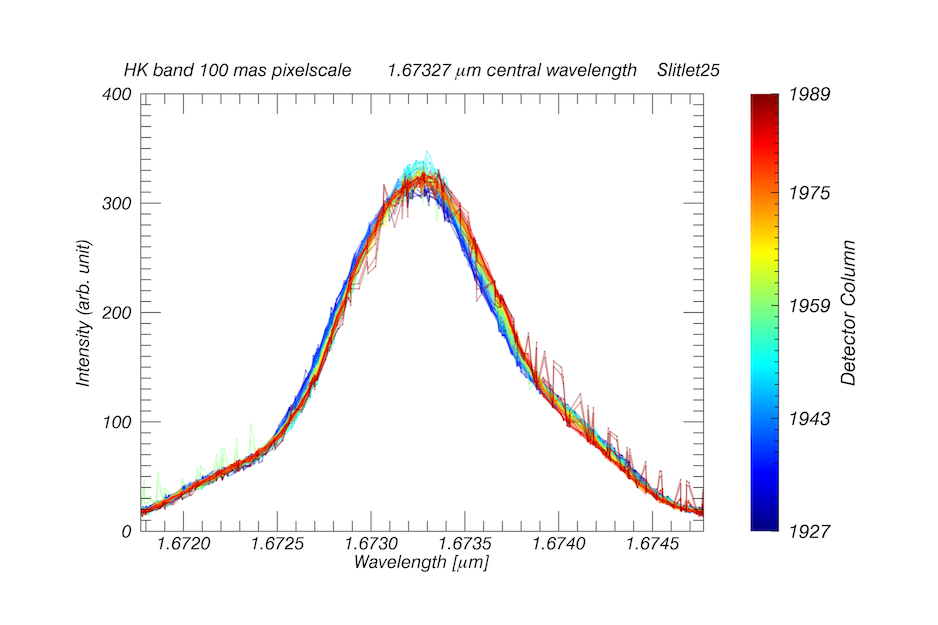}
			\includegraphics[height=6.5cm, trim={2.4cm 0 2.0cm 0cm}, clip=true]{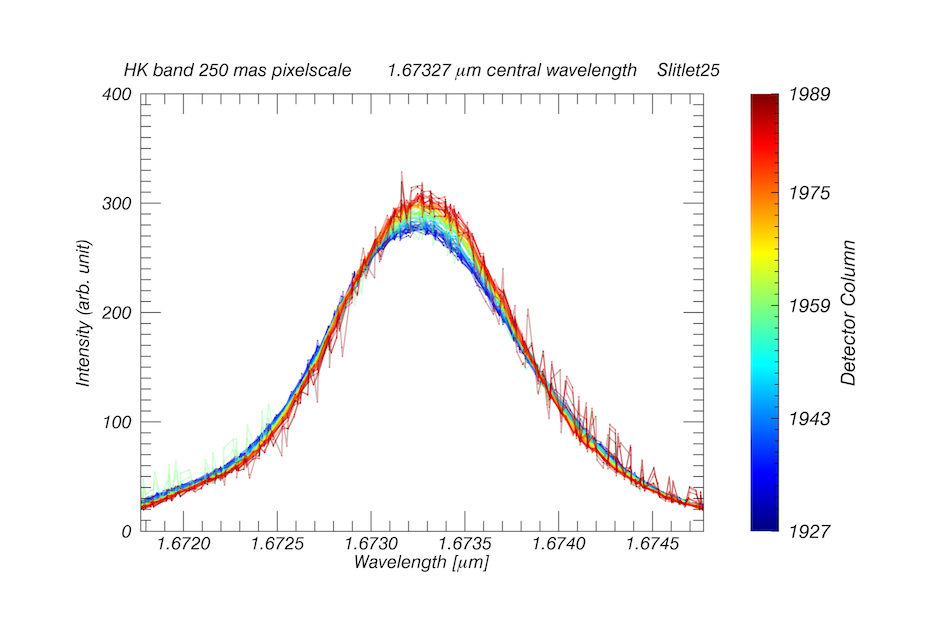}
		}	
		\caption[Variation of the line profiles within a slitlet in H+K-band pre- upgrade.]{Variation of the line profiles within a slitlet in H+K-band pre- upgrade. Remark: in the 25 mas pixelscale the sampling of the datapoints was not well distributed over the line profile.}
		\label{fig:slitlet_hk_pre}
	\end{center}
\end{figure}

\begin{figure}[htbp!]
	\begin{center}
		\resizebox{1.0\textwidth}{!}{
			\includegraphics[height=6.5cm, trim={1.5cm 0 5.5cm 0cm}, clip=true]{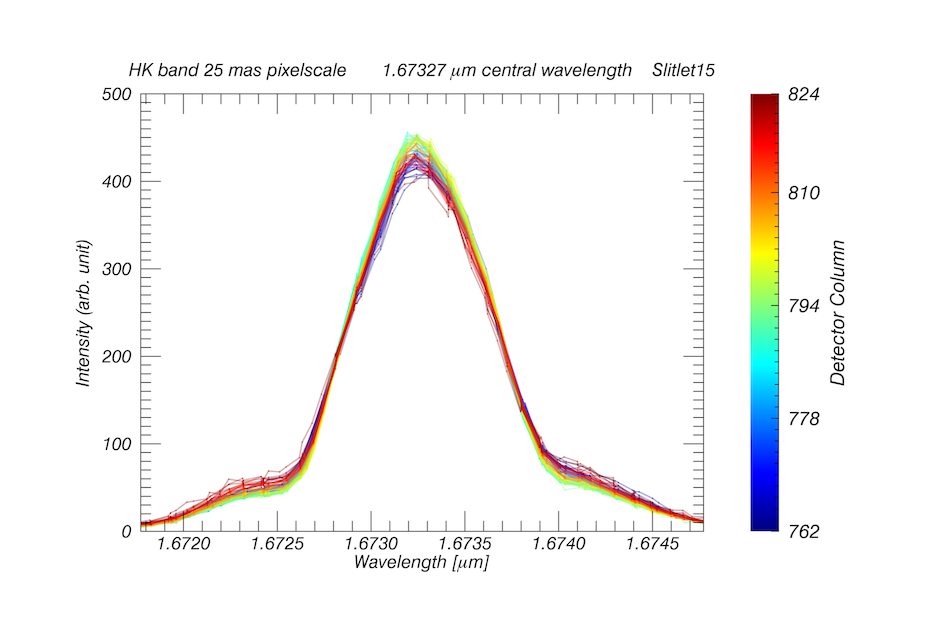}
			\includegraphics[height=6.5cm, trim={2.4cm 0 5.5cm 0cm}, clip=true]{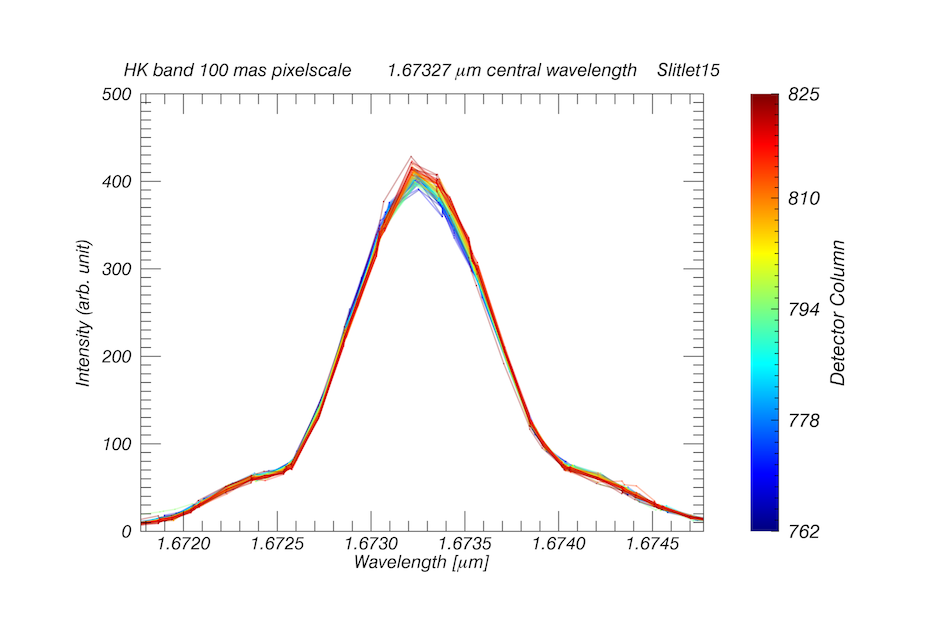}
			\includegraphics[height=6.5cm, trim={2.4cm 0 2.0cm 0cm}, clip=true]{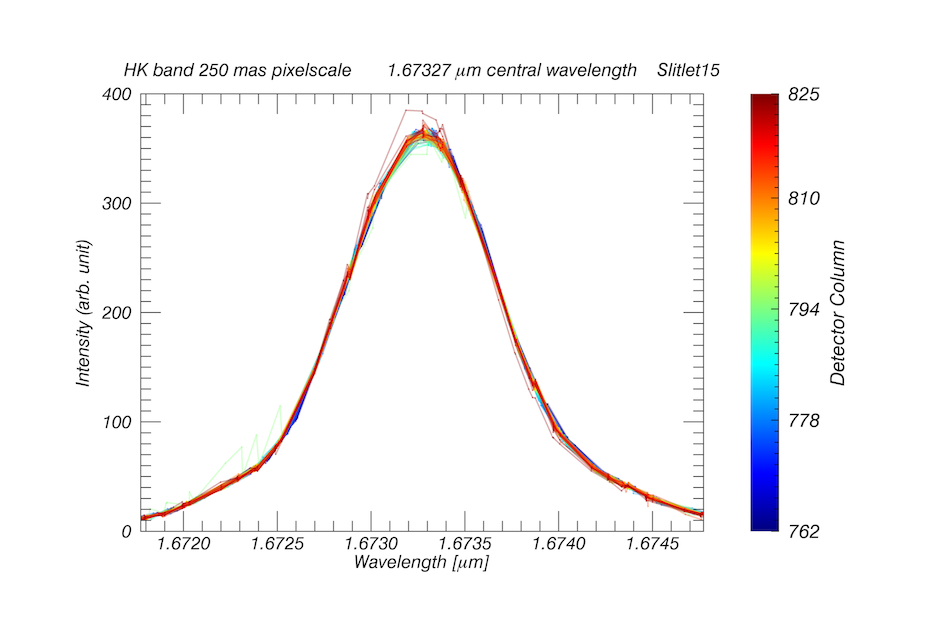}
		}
		\resizebox{1.0\textwidth}{!}{
			\includegraphics[height=6.5cm, trim={1.5cm 0 5.5cm 0cm}, clip=true]{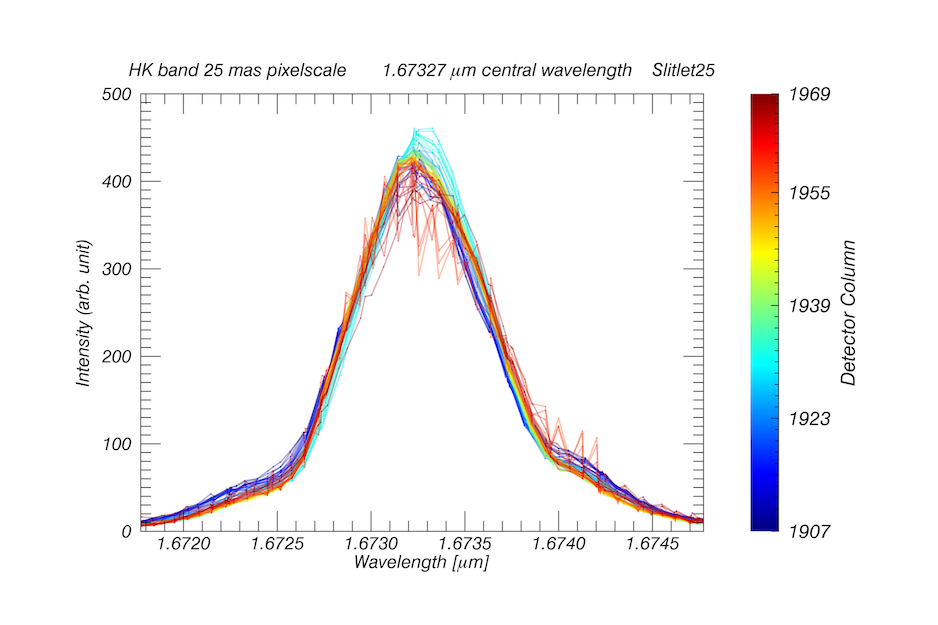}
			\includegraphics[height=6.5cm, trim={2.4cm 0 5.5cm 0cm}, clip=true]{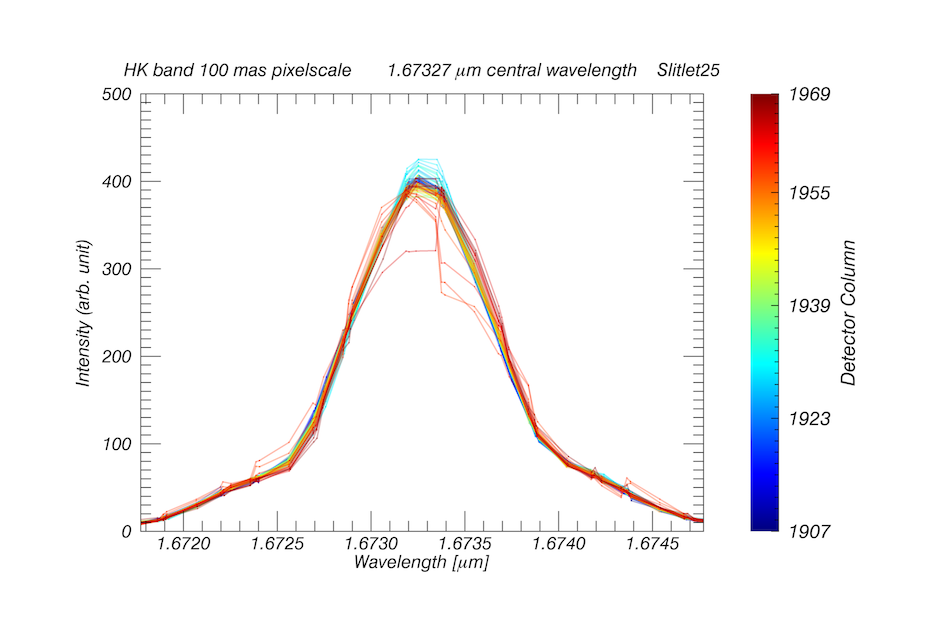}
			\includegraphics[height=6.5cm, trim={2.4cm 0 2.0cm 0cm}, clip=true]{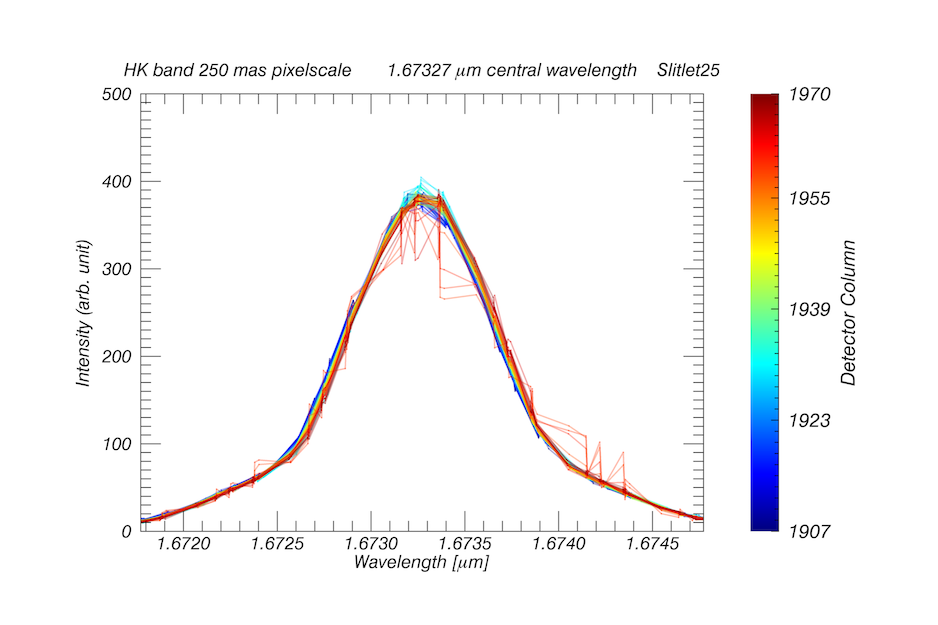}
		}	
		\caption{Variation of the line profiles within a slitlet in H+K-band post- upgrade.}
		\label{fig:slitlet_hk_post}
	\end{center}
\end{figure}
\clearpage

\section{Variation of the Line Profiles for each Slitlet}
Referring to section \ref{sec:discussion_slitlet}, in the following plots the variation of the line profiles are shown for all slitlets (beside slitlet 1) in H-band in the 25 mas pixelscale. The order of the plots is the same as on the detector. It can be seen that slitlets on the right side o the detector behave different from slitlets on the left side. The variation is smallest in the central parts of the detector. For most slitlets the most peaky LSF is the one of the central parts of the slitlets.

\begin{figure}[htbp!]
	\begin{center}
		\resizebox{1.0\textwidth}{!}{
			\includegraphics[height=6.5cm, trim={1.5cm 0 2.0cm 0cm}, clip=true]{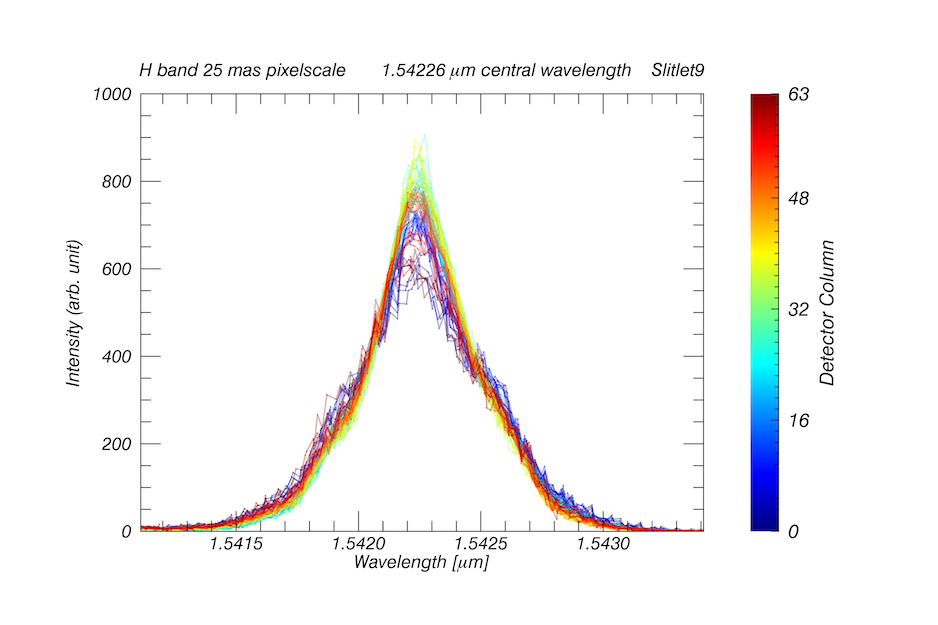}
			\includegraphics[height=6.5cm, trim={1.5cm 0 2.0cm 0cm}, clip=true]{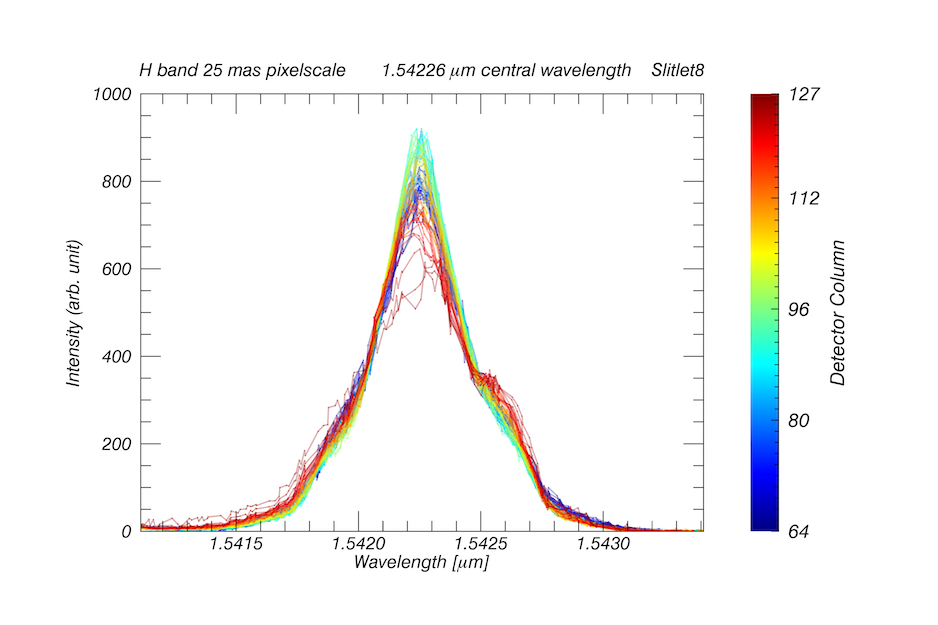}
			\includegraphics[height=6.5cm, trim={1.5cm 0 2.0cm 0cm}, clip=true]{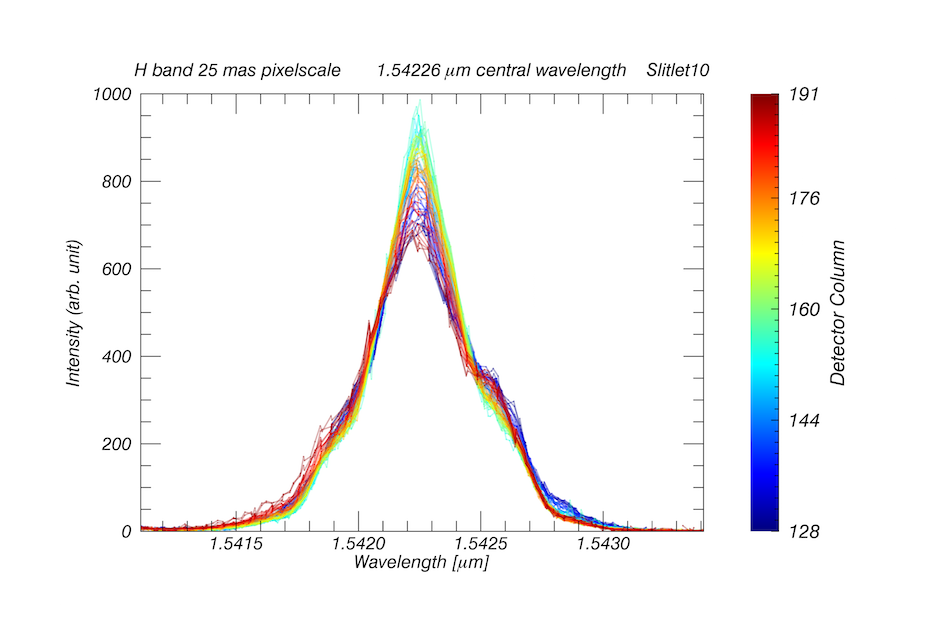}
		}
		\resizebox{1.0\textwidth}{!}{
			\includegraphics[height=6.5cm, trim={1.5cm 0 2.0cm 0cm}, clip=true]{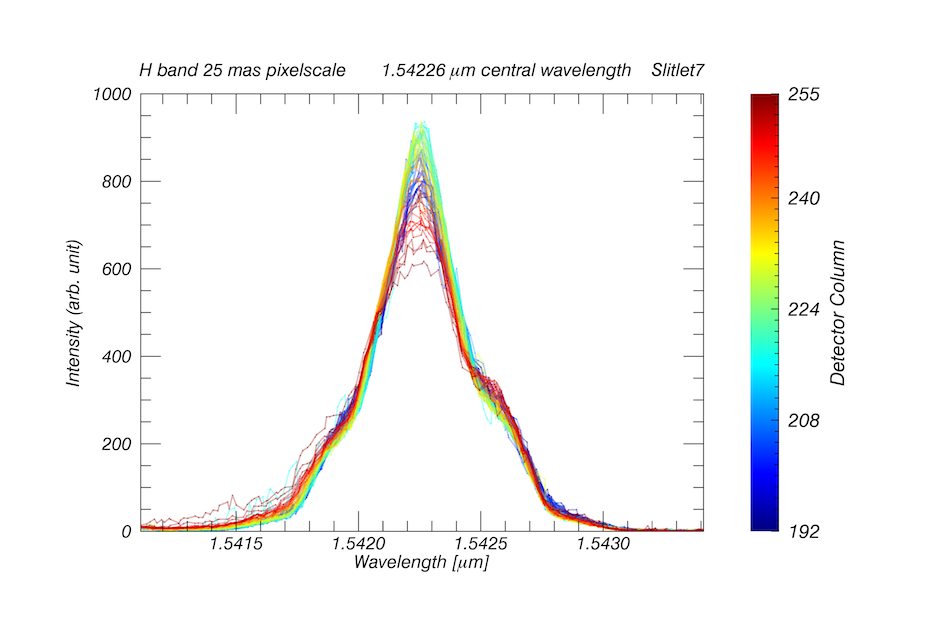}
			\includegraphics[height=6.5cm, trim={1.5cm 0 2.0cm 0cm}, clip=true]{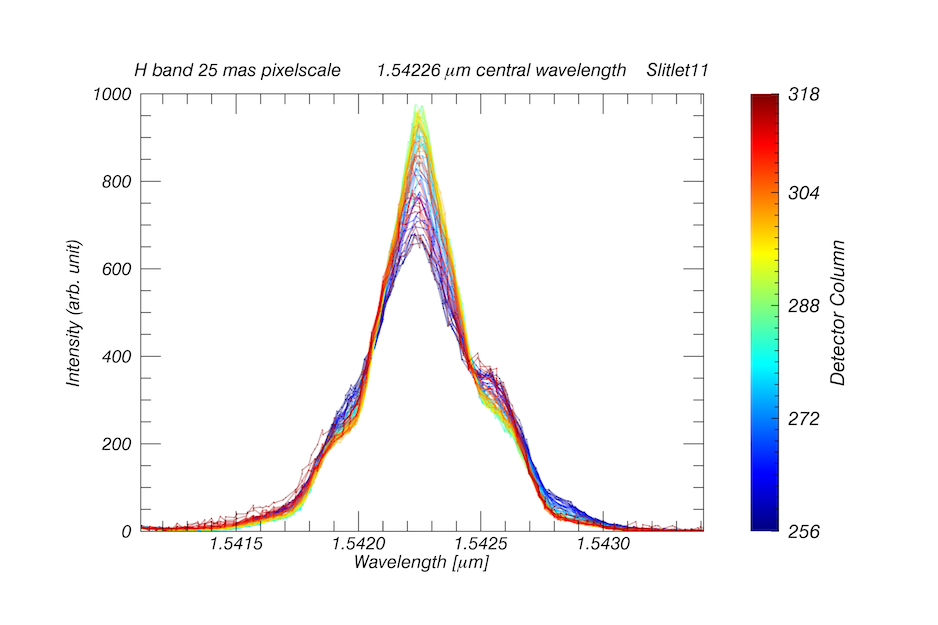}
			\includegraphics[height=6.5cm, trim={1.5cm 0 2.0cm 0cm}, clip=true]{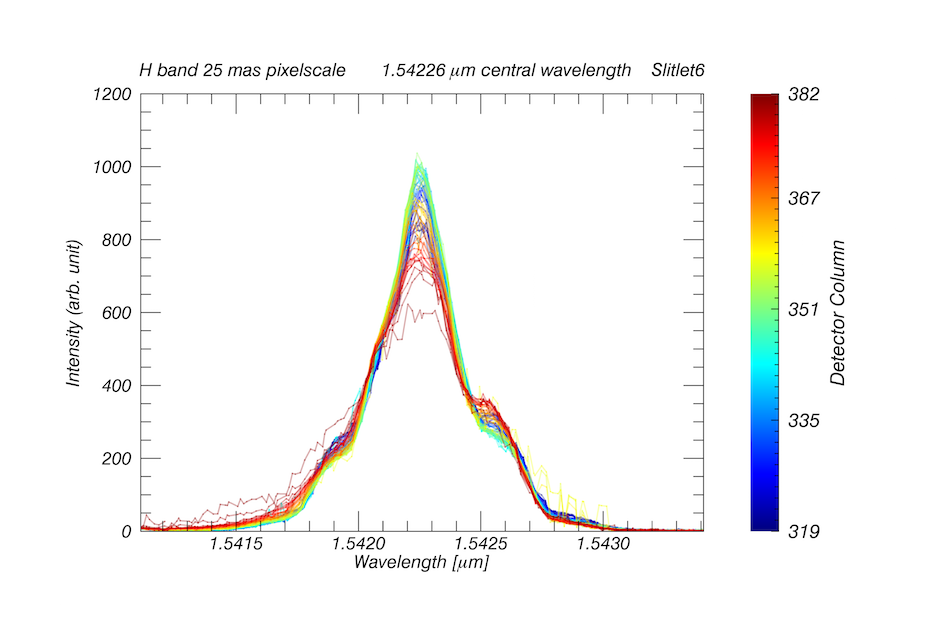}
		}
		\resizebox{1.0\textwidth}{!}{
			\includegraphics[height=6.5cm, trim={1.5cm 0 2.0cm 0cm}, clip=true]{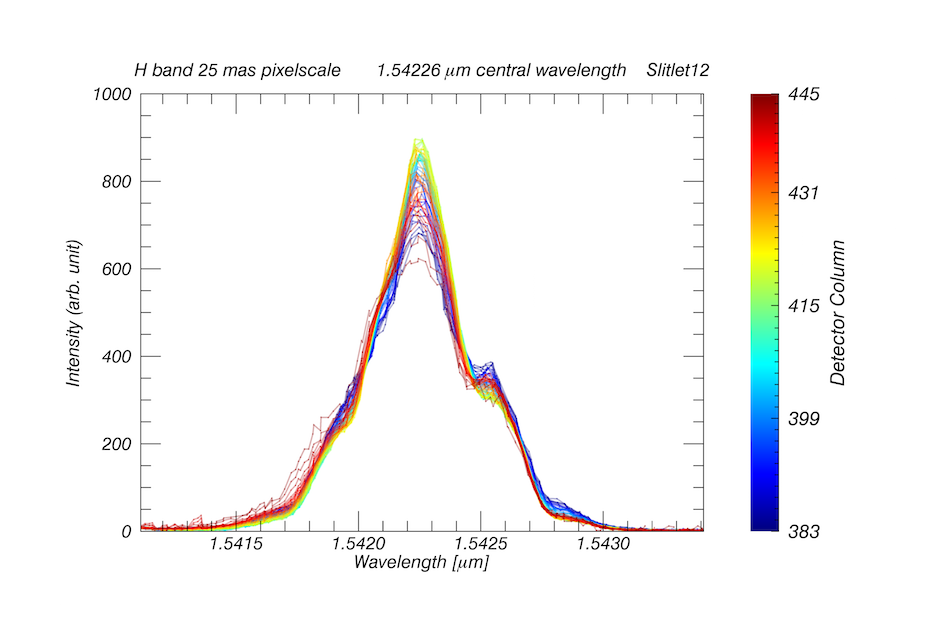}
			\includegraphics[height=6.5cm, trim={1.5cm 0 2.0cm 0cm}, clip=true]{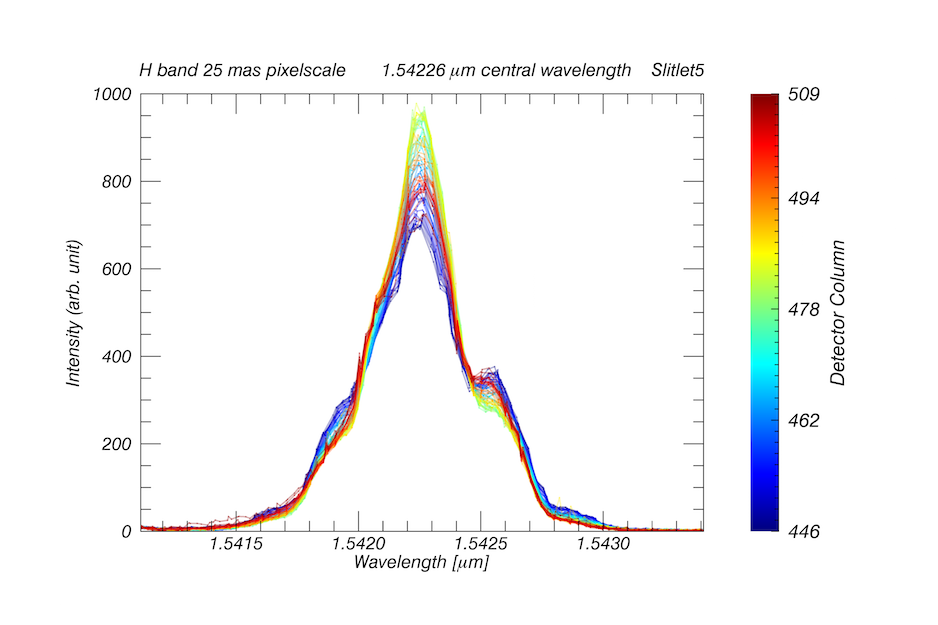}
			\includegraphics[height=6.5cm, trim={1.5cm 0 2.0cm 0cm}, clip=true]{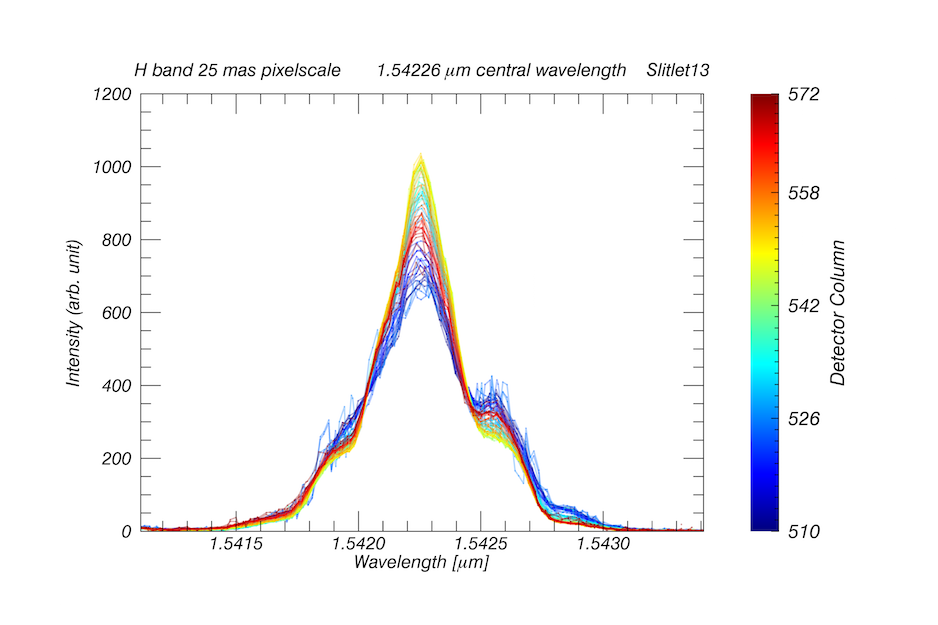}
		}
		\resizebox{1.0\textwidth}{!}{
			\includegraphics[height=6.5cm, trim={1.5cm 0 2.0cm 0cm}, clip=true]{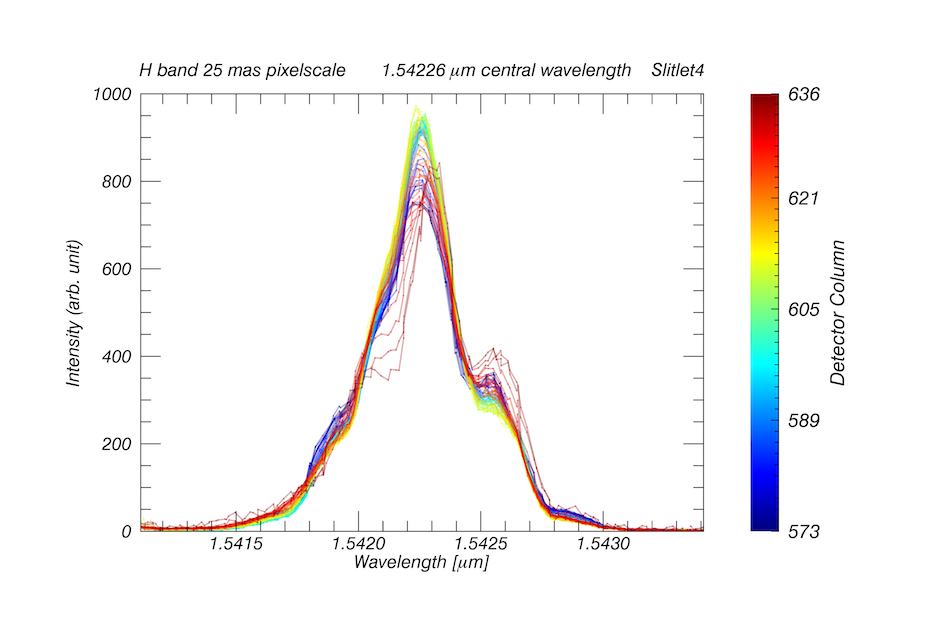}
			\includegraphics[height=6.5cm, trim={1.5cm 0 2.0cm 0cm}, clip=true]{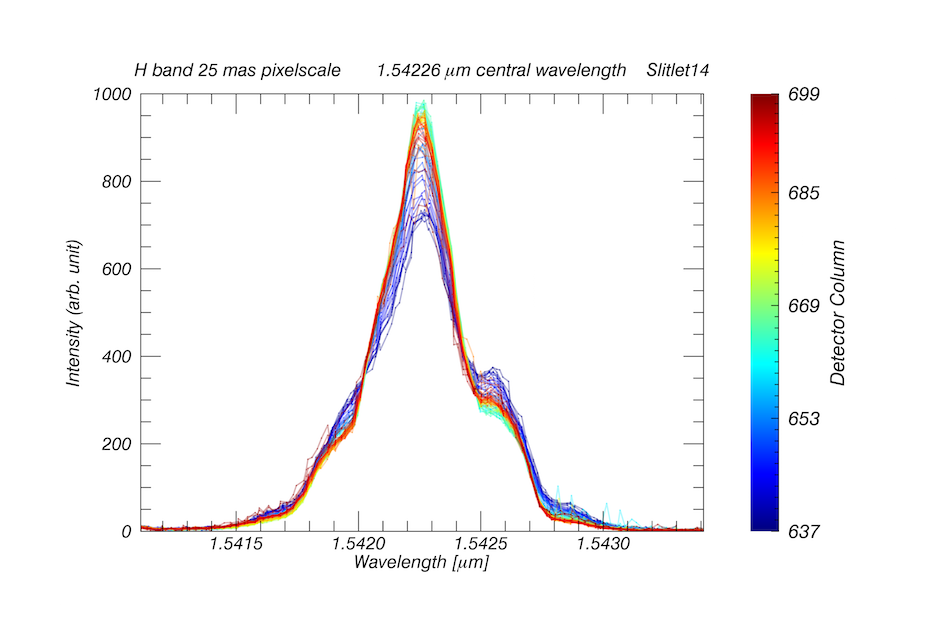}
			\includegraphics[height=6.5cm, trim={1.5cm 0 2.0cm 0cm}, clip=true]{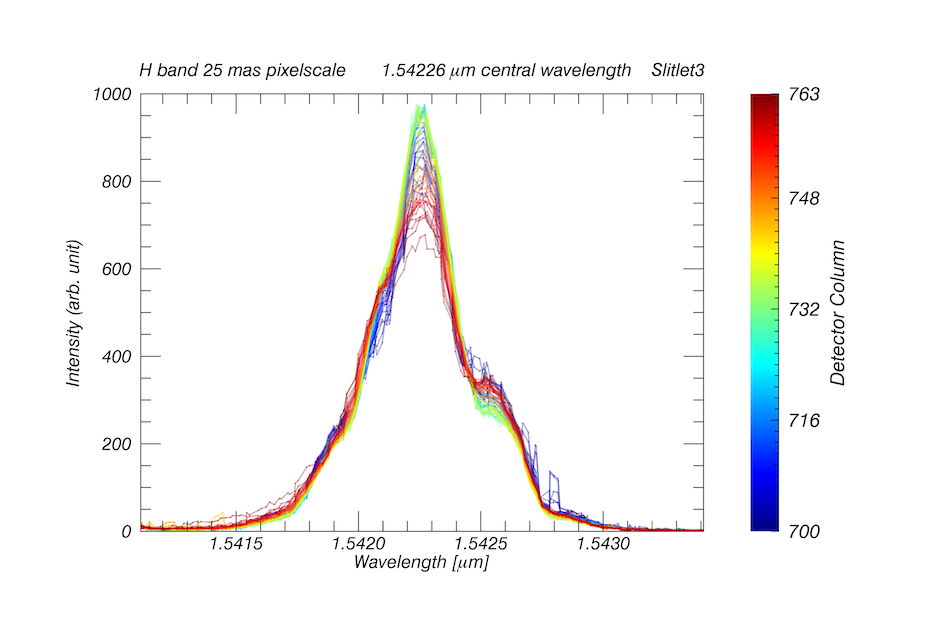}
		}

	\end{center}
\end{figure}

\begin{figure}[htbp!]
	\begin{center}
		\resizebox{1.0\textwidth}{!}{
			\includegraphics[height=6.5cm, trim={1.5cm 0 2.0cm 0cm}, clip=true]{{h25new_wlength_1.54226_slitlet15}.png}
			\includegraphics[height=6.5cm, trim={1.5cm 0 2.0cm 0cm}, clip=true]{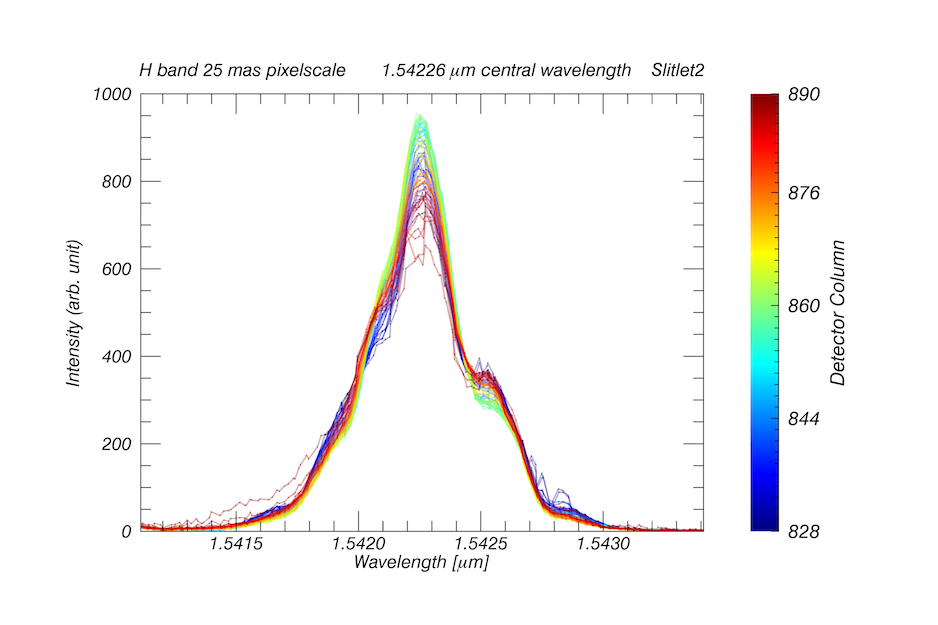}
			\includegraphics[height=6.5cm, trim={1.5cm 0 2.0cm 0cm}, clip=true]{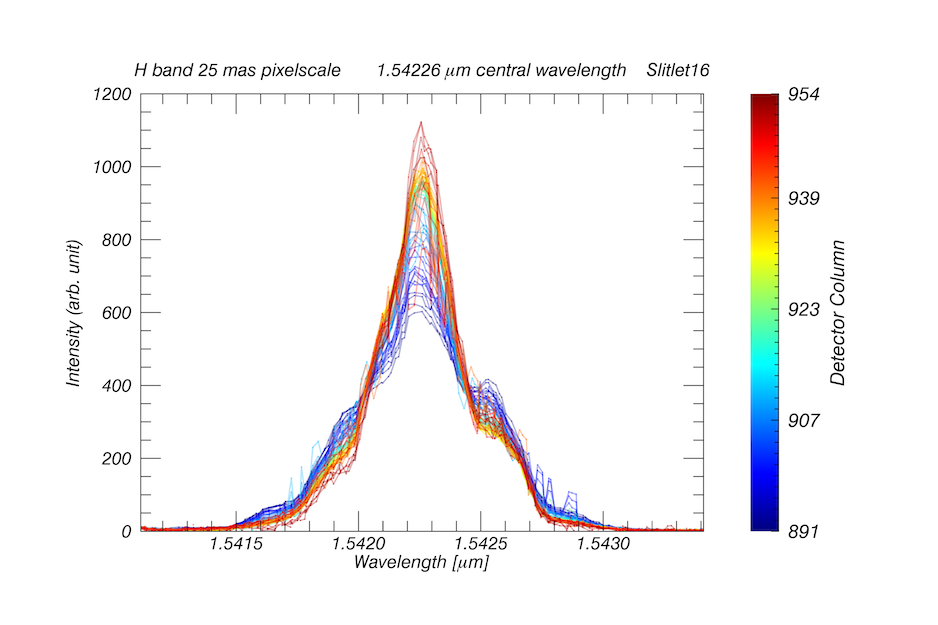}
		}
		\resizebox{1.0\textwidth}{!}{
			\includegraphics[height=6.5cm, trim={1.5cm 0 2.0cm 0cm}, clip=true]{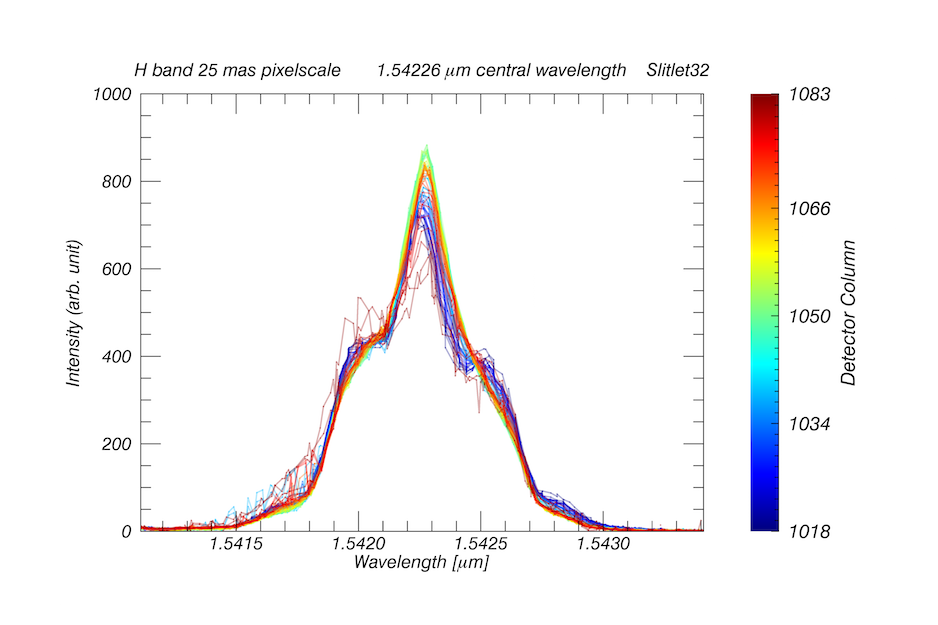}
			\includegraphics[height=6.5cm, trim={1.5cm 0 2.0cm 0cm}, clip=true]{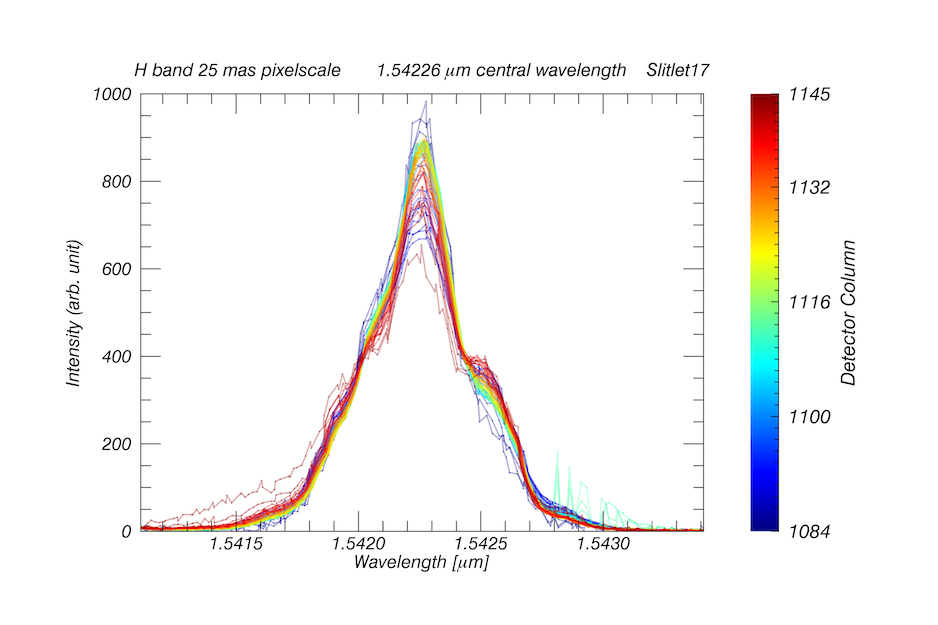}
			\includegraphics[height=6.5cm, trim={1.5cm 0 2.0cm 0cm}, clip=true]{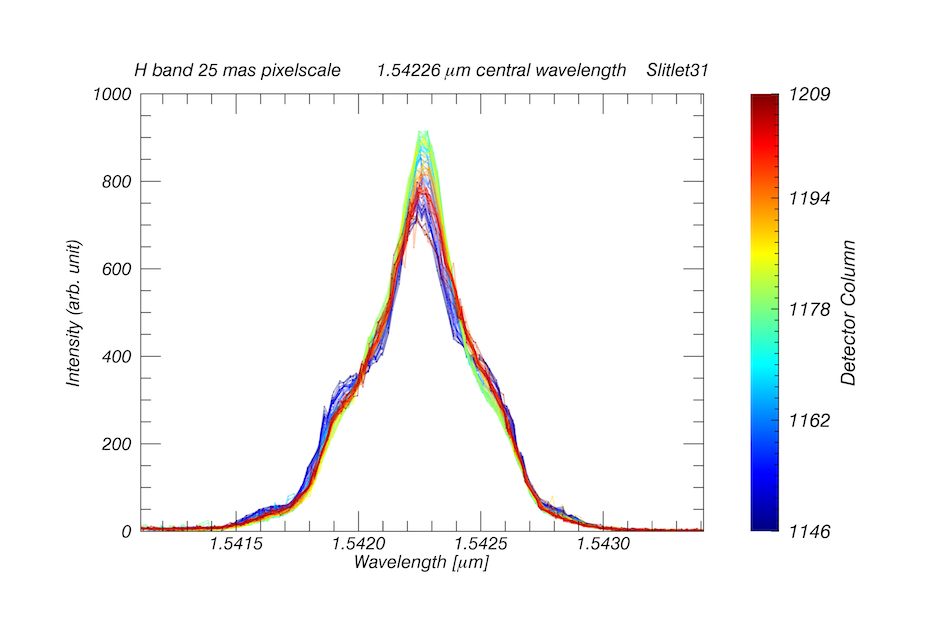}
		}
		\resizebox{1.0\textwidth}{!}{
			\includegraphics[height=6.5cm, trim={1.5cm 0 2.0cm 0cm}, clip=true]{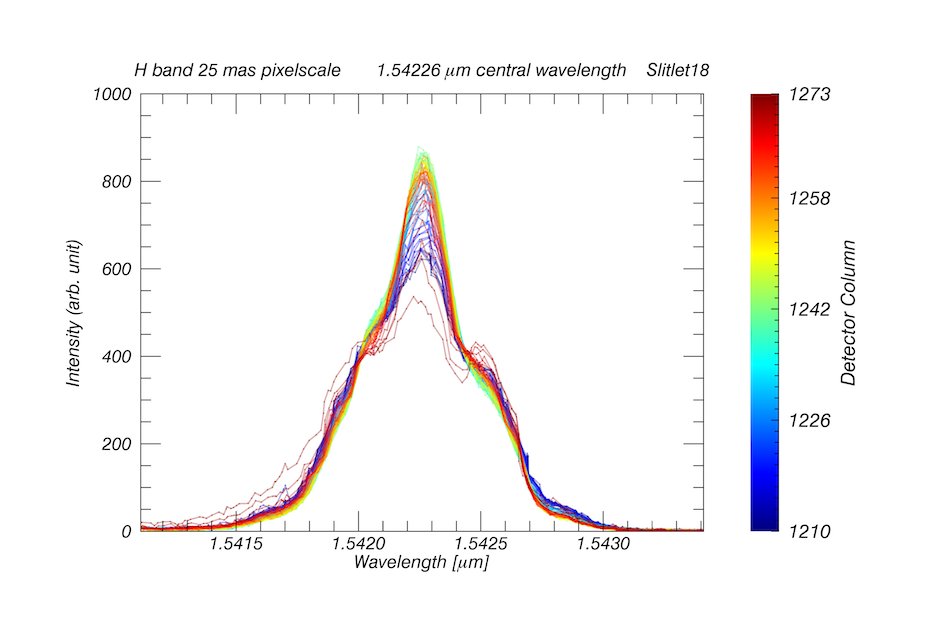}
			\includegraphics[height=6.5cm, trim={1.5cm 0 2.0cm 0cm}, clip=true]{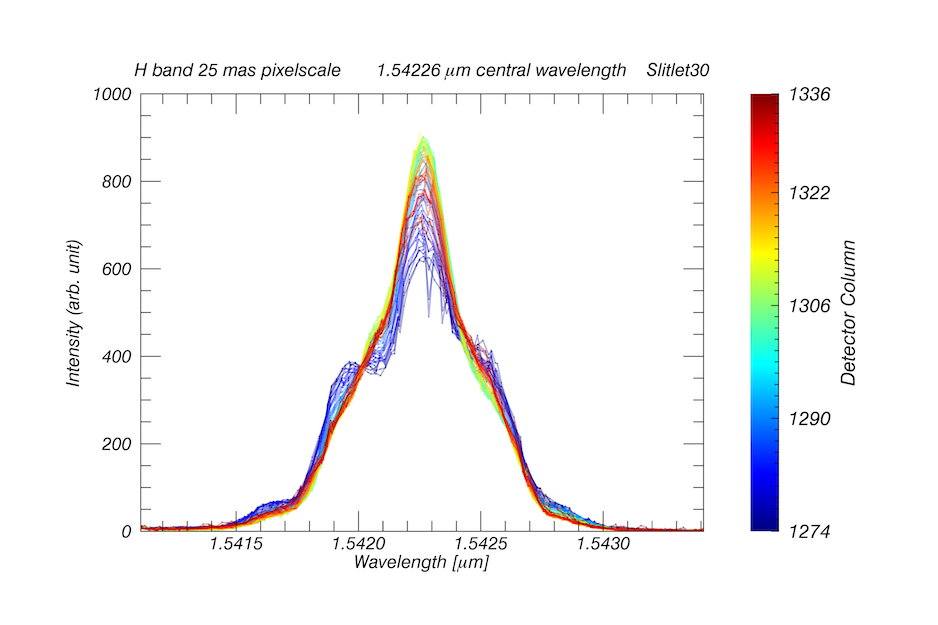}
			\includegraphics[height=6.5cm, trim={1.5cm 0 2.0cm 0cm}, clip=true]{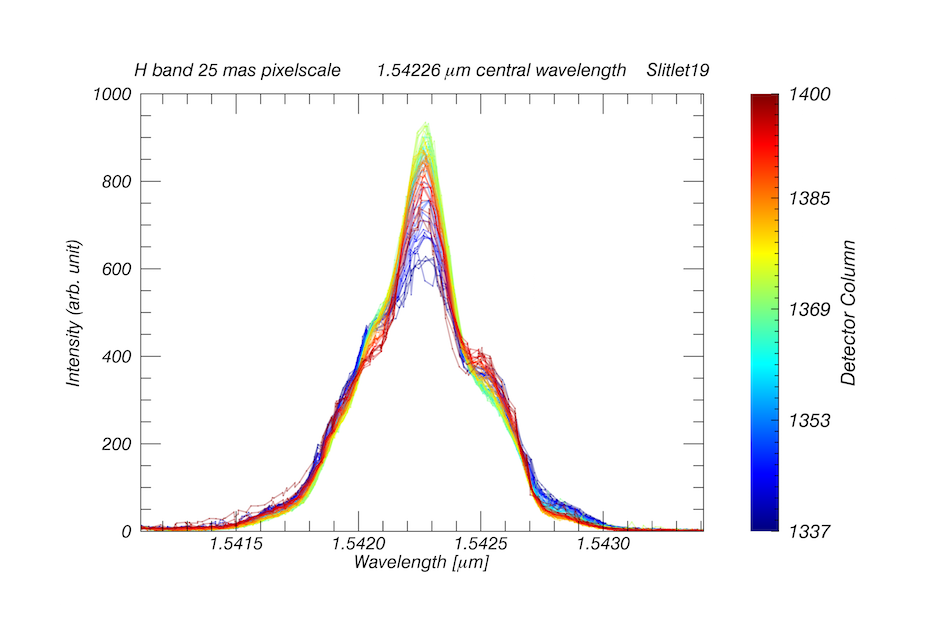}
		}
		\resizebox{1.0\textwidth}{!}{
			\includegraphics[height=6.5cm, trim={1.5cm 0 2.0cm 0cm}, clip=true]{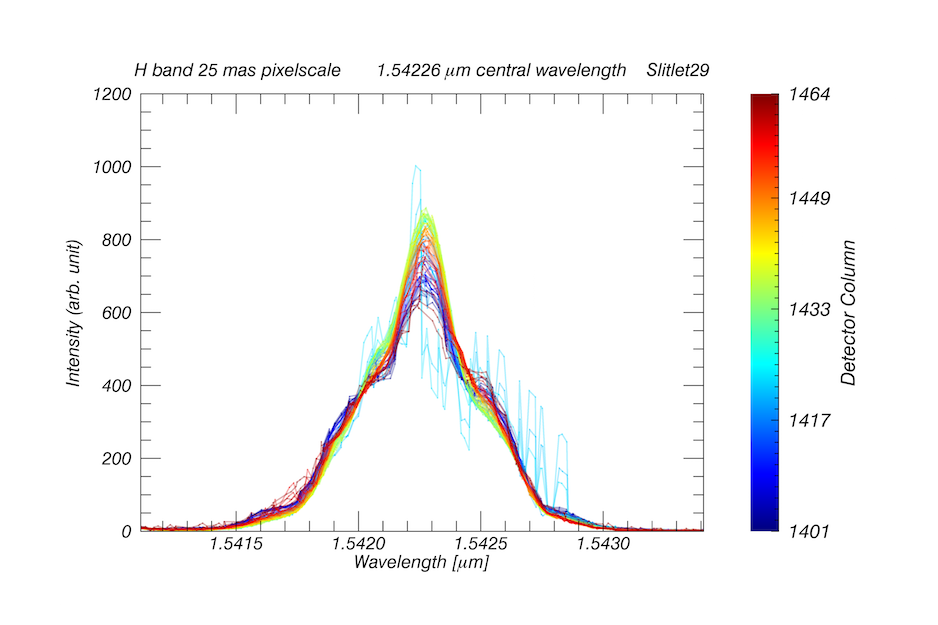}
			\includegraphics[height=6.5cm, trim={1.5cm 0 2.0cm 0cm}, clip=true]{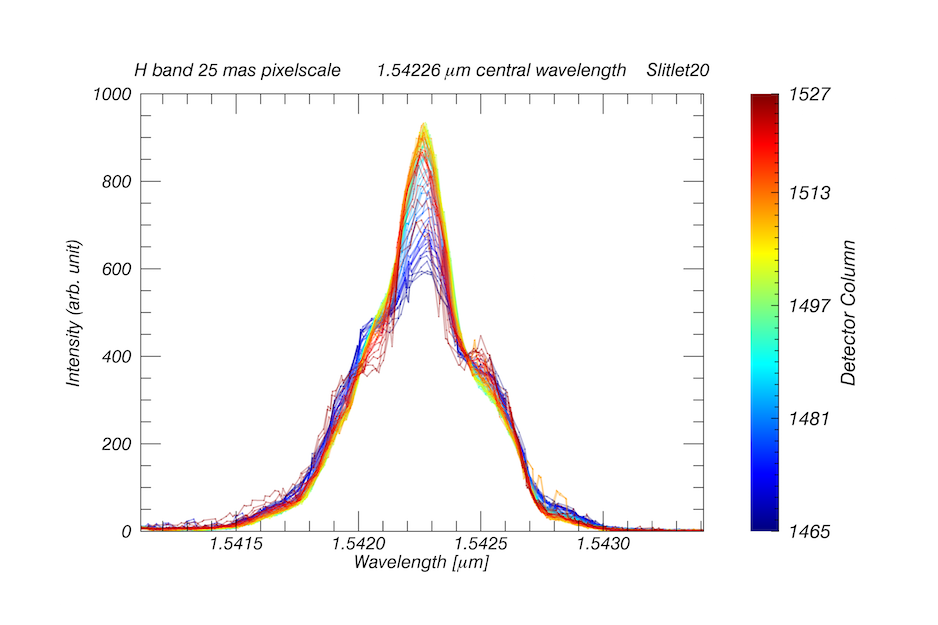}
			\includegraphics[height=6.5cm, trim={1.5cm 0 2.0cm 0cm}, clip=true]{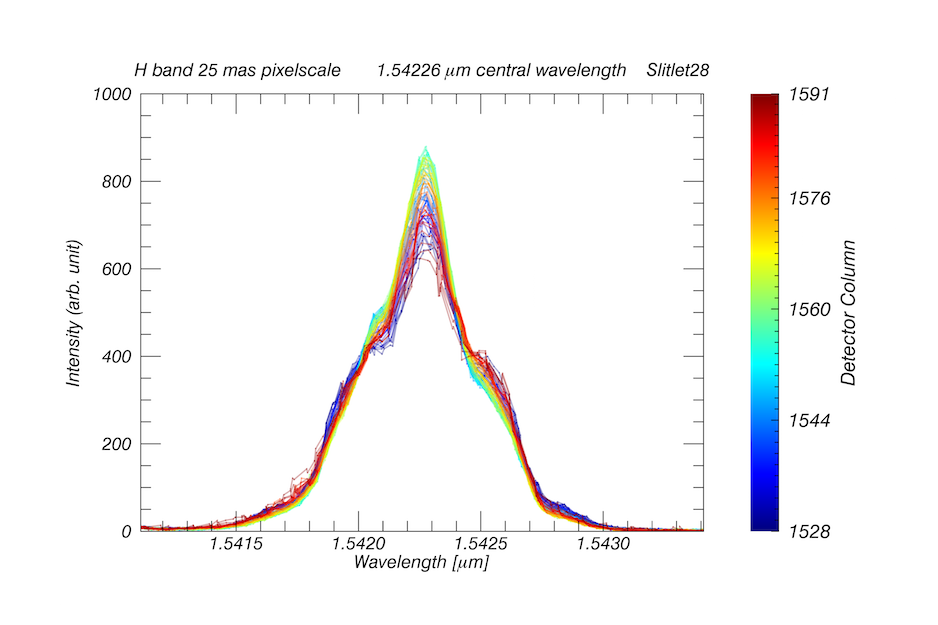}
		}
		\resizebox{1.0\textwidth}{!}{
			\includegraphics[height=6.5cm, trim={1.5cm 0 2.0cm 0cm}, clip=true]{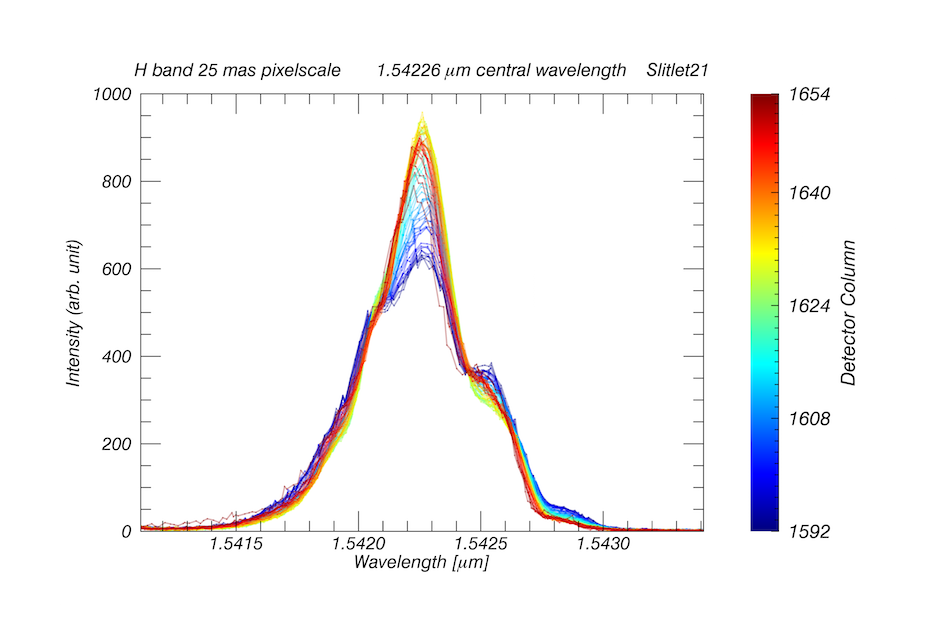}
			\includegraphics[height=6.5cm, trim={1.5cm 0 2.0cm 0cm}, clip=true]{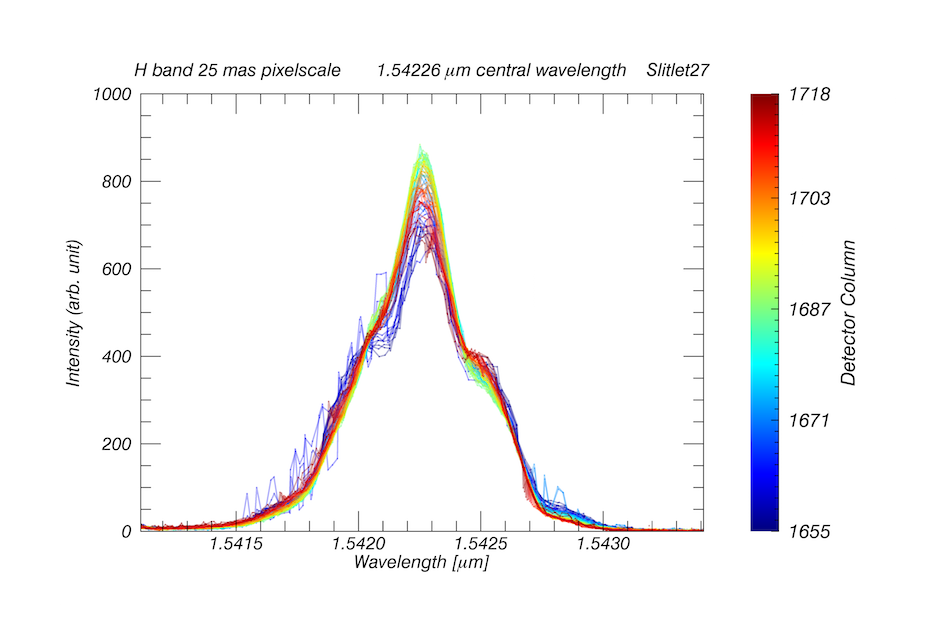}
			\includegraphics[height=6.5cm, trim={1.5cm 0 2.0cm 0cm}, clip=true]{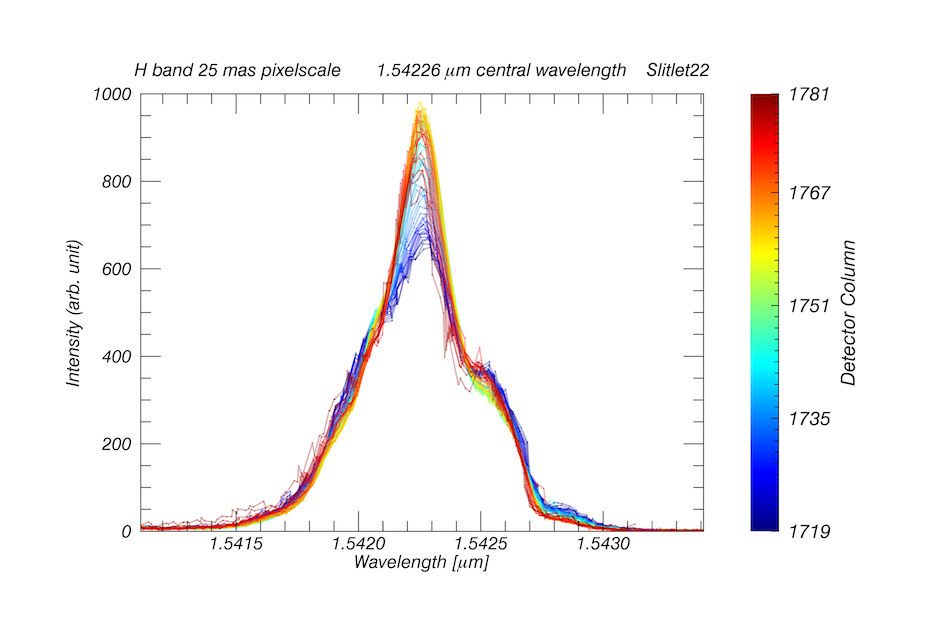}
		}
	\end{center}
\end{figure}

\begin{figure}
	\begin{center}
		\resizebox{1.0\textwidth}{!}{
			\includegraphics[height=6.5cm, trim={1.5cm 0 2.0cm 0cm}, clip=true]{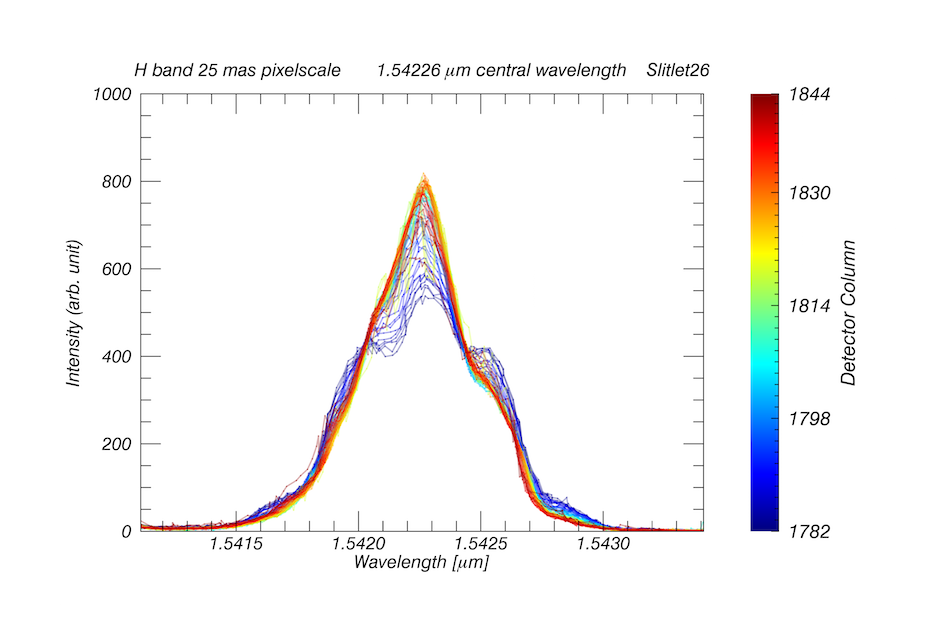}
			\includegraphics[height=6.5cm, trim={1.5cm 0 2.0cm 0cm}, clip=true]{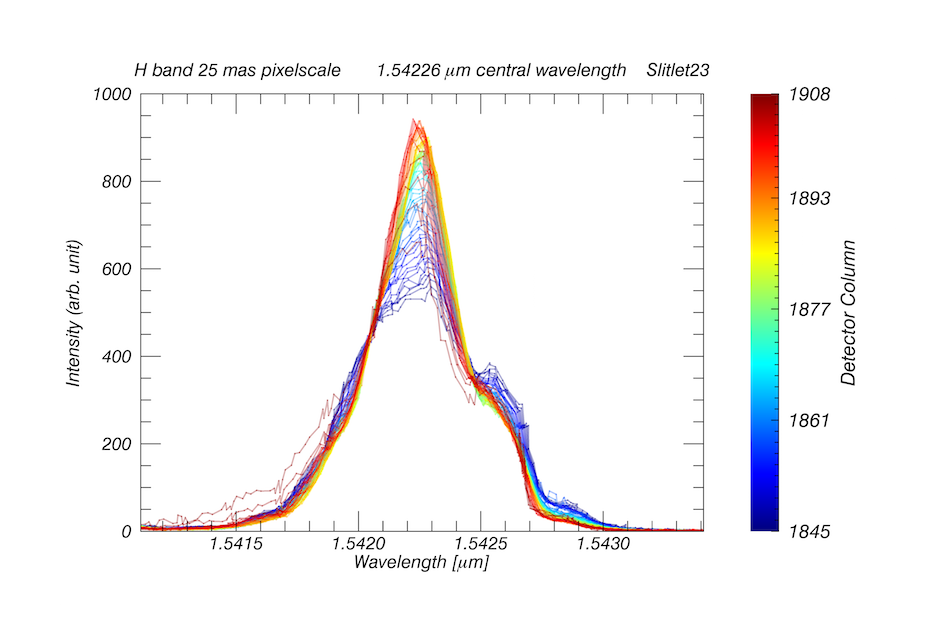}
			\includegraphics[height=6.5cm, trim={1.5cm 0 2.0cm 0cm}, clip=true]{{h25new_wlength_1.54226_slitlet25}.png}
		}
		\resizebox{0.33\textwidth}{!}{
			\includegraphics[height=6.5cm, trim={1.5cm 0 2.0cm 0cm}, clip=true]{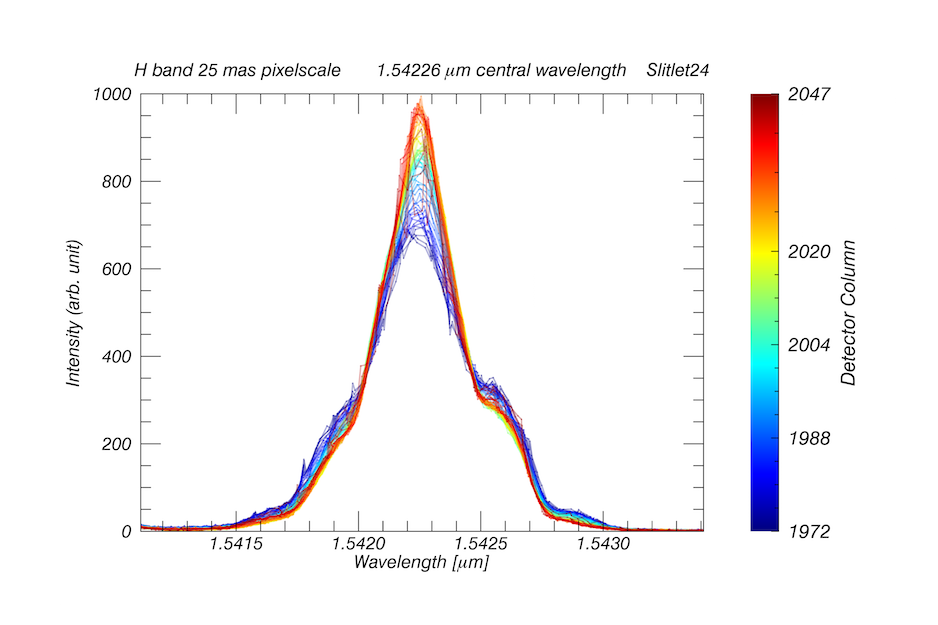}
		}
		
		\caption{Variation of the line profiles within a slitlet in H-band post- upgrade for all slitlets. The order of the plots is like on the detector}
		\label{fig:slitletvar}
	\end{center}
\end{figure}

\clearpage

\section{Variation of the Median Slitlet LSF Across the Detector}
This appendix section shows further plots of the median slitlet variation across the detector referring to section \ref{sec:var_line}. The upper row of each figure is from pre- upgrade data, the lower one from post- upgrade data.
\begin{figure}[htbp!]
	\begin{center}
		\resizebox{1.0\textwidth}{!}{
			\includegraphics[height=6.5cm, trim={0.3cm 0 3.7cm 0cm}, clip=true]{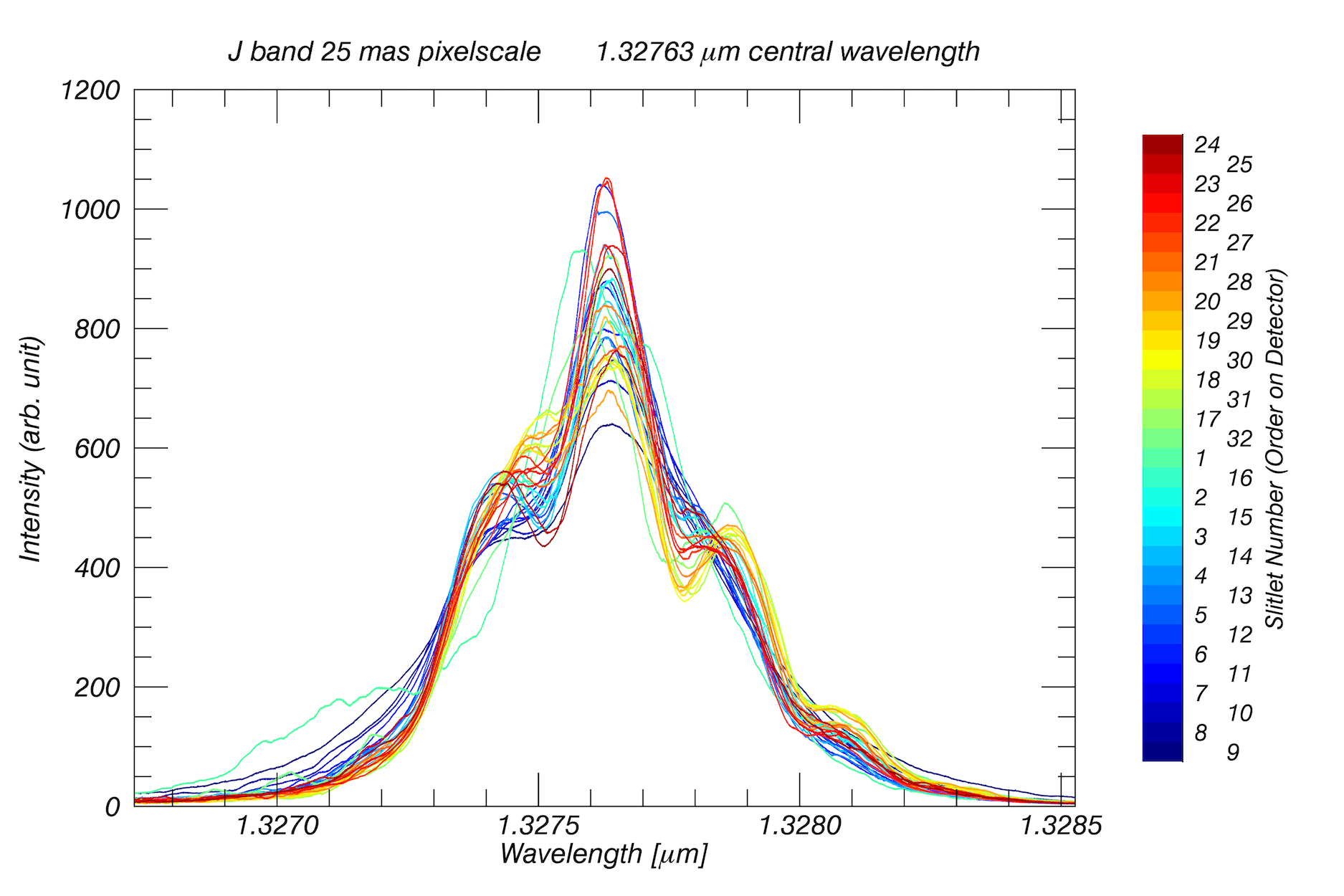}
			\includegraphics[height=6.5cm, trim={1.1cm 0 3.7cm 0cm}, clip=true]{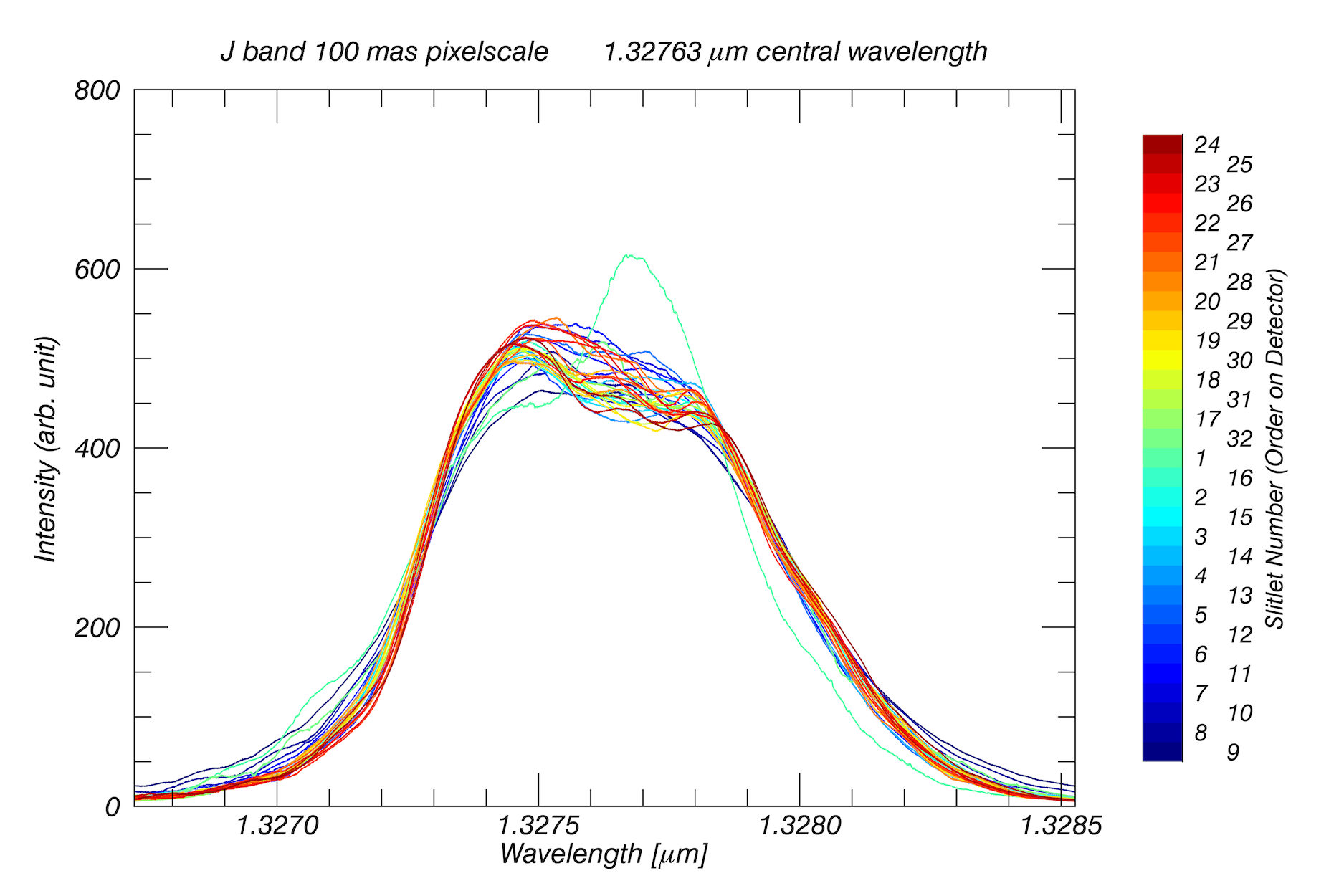}
			\includegraphics[height=6.5cm, trim={1.1cm 0 1.0cm 0cm}, clip=true]{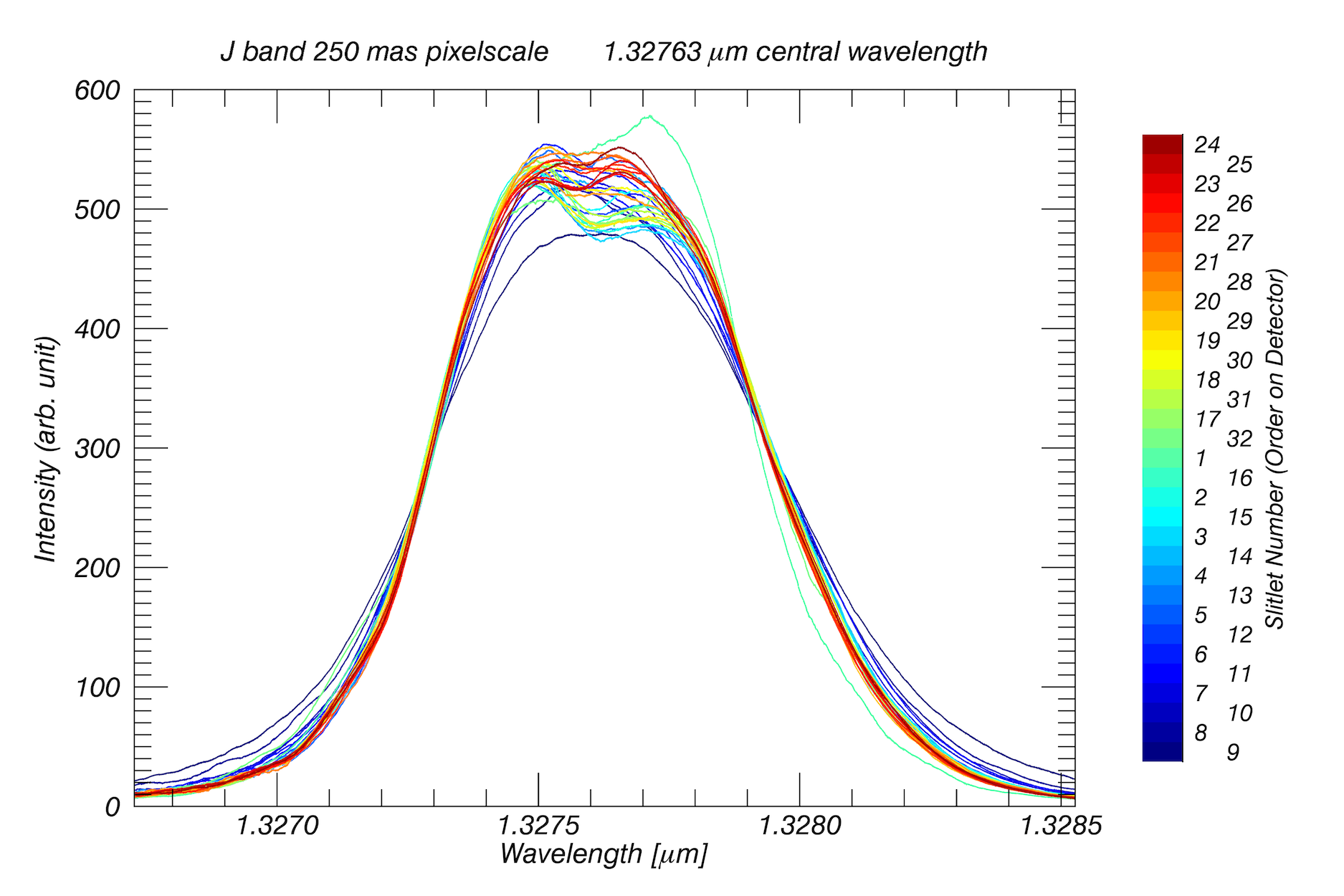}
		}
		\resizebox{1.0\textwidth}{!}{
			\includegraphics[height=6.5cm, trim={0.3cm 0 3.7cm 0cm}, clip=true]{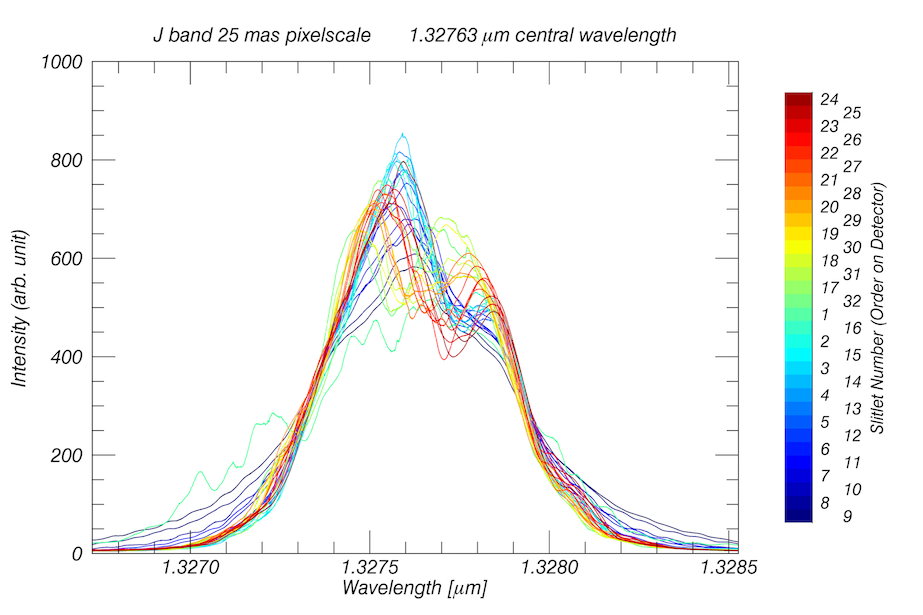}
			\includegraphics[height=6.5cm, trim={1.1cm 0 3.7cm 0cm}, clip=true]{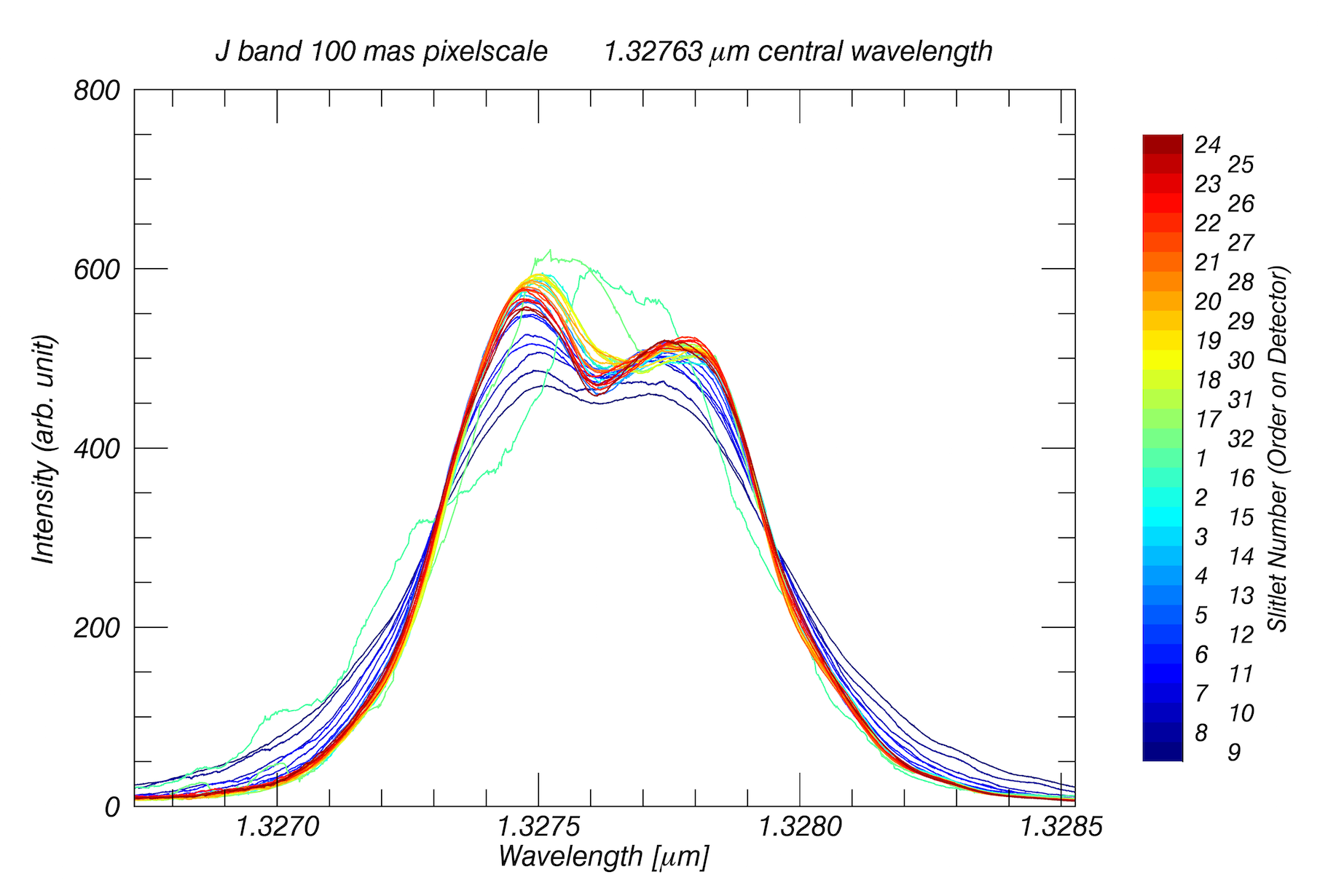}
			\includegraphics[height=6.5cm, trim={1.1cm 0 1.0cm 0cm}, clip=true]{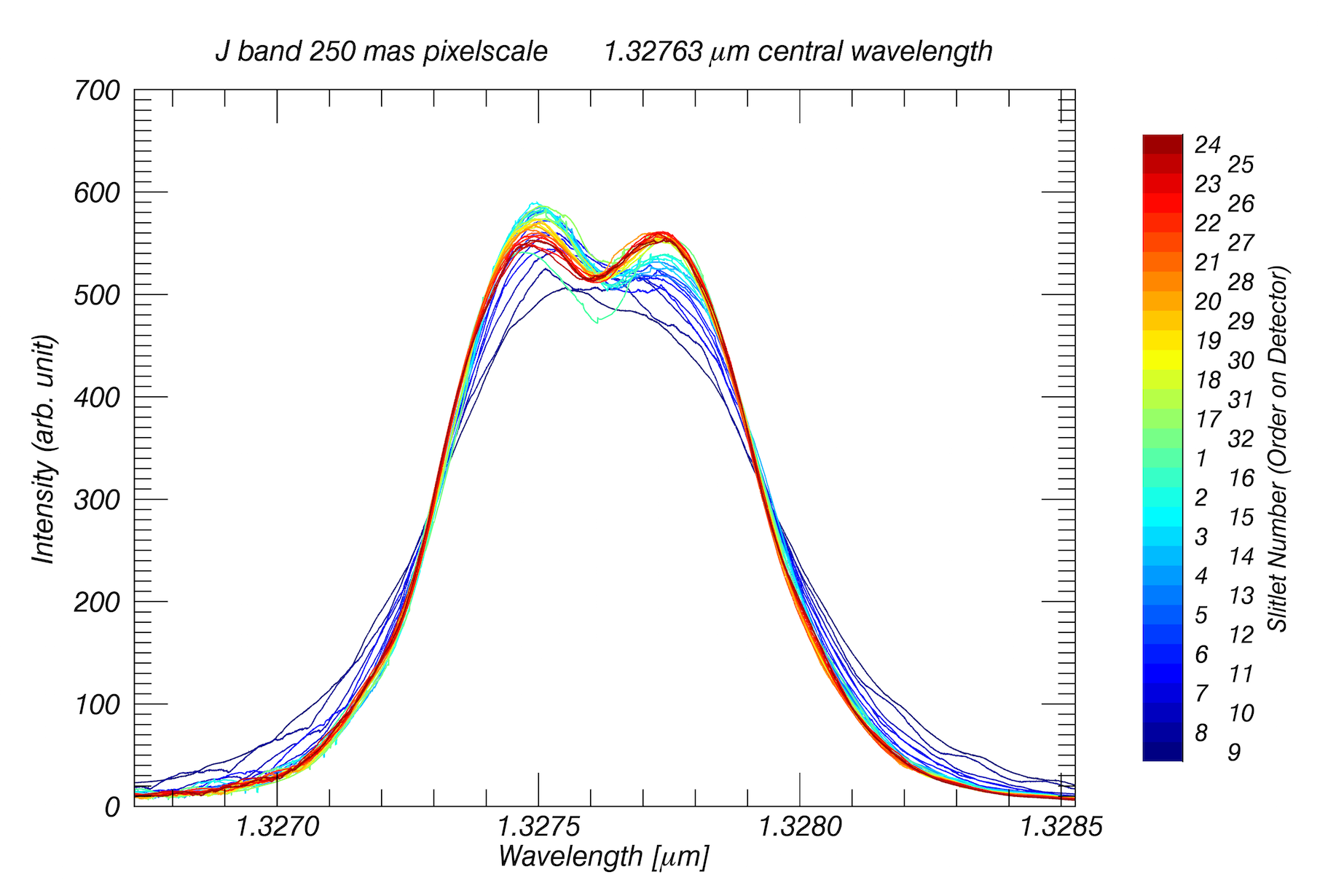}
		}
		\caption[Median LSF of individual slitlets in J-band]{J-band median LSF of the individual slitlets across the detector. Upper row: pre- upgrade, lower row: post- upgrade}
		\label{fig:median_j}
	\end{center}
\end{figure}

\begin{figure}[htbp!]
	\begin{center}
		\resizebox{1.0\textwidth}{!}{
			\includegraphics[height=6.5cm, trim={0.3cm 0 3.7cm 0cm}, clip=true]{{h25old_wlength_1.54226_slitlet_color}.png}
			\includegraphics[height=6.5cm, trim={1.1cm 0 3.7cm 0cm}, clip=true]{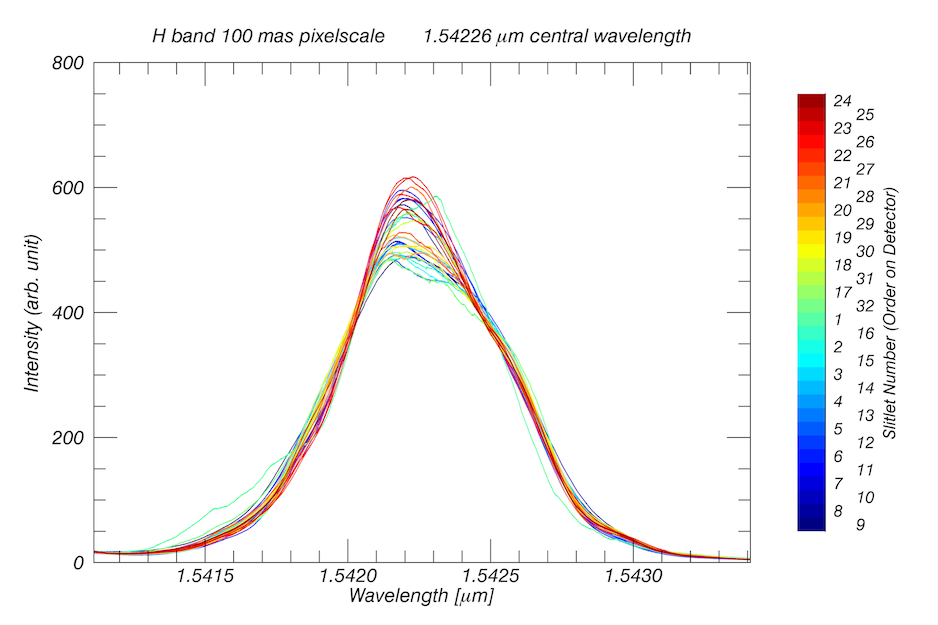}
			\includegraphics[height=6.5cm, trim={1.1cm 0 1.0cm 0cm}, clip=true]{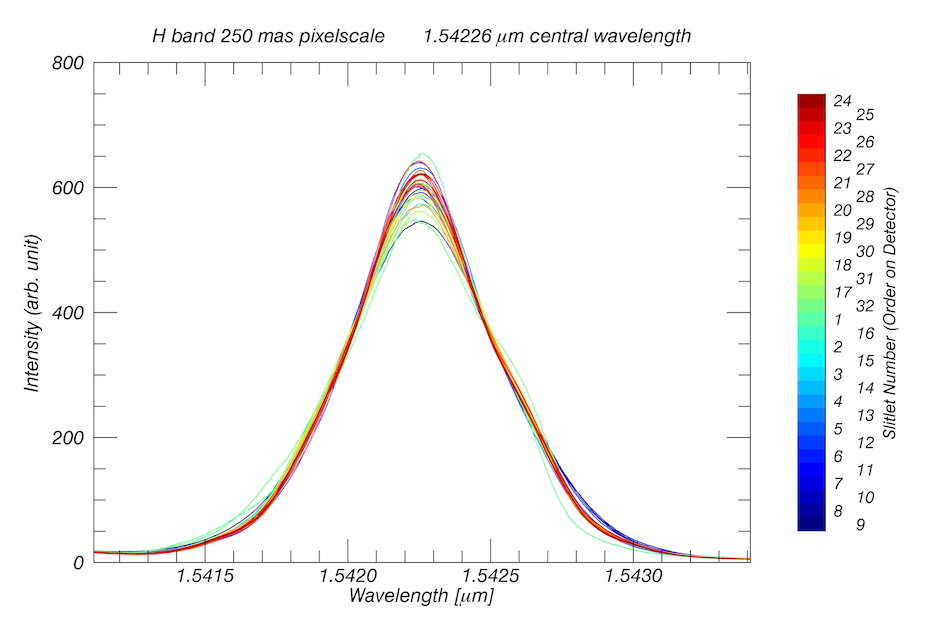}
		}
		\resizebox{1.0\textwidth}{!}{
			\includegraphics[height=6.5cm, trim={0.3cm 0 3.7cm 0cm}, clip=true]{{h25new_wlength_1.54226_slitlet_color}.png}
			\includegraphics[height=6.5cm, trim={1.1cm 0 3.7cm 0cm}, clip=true]{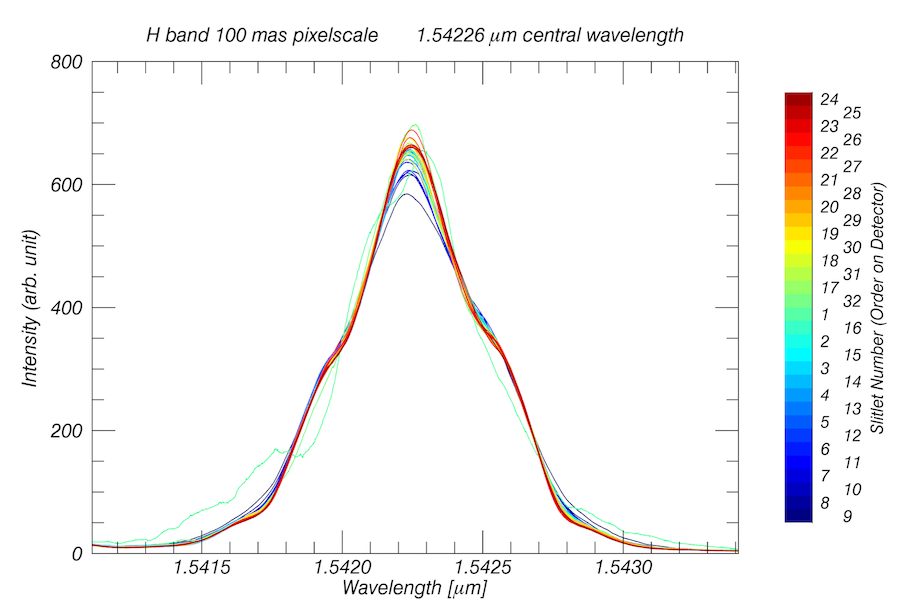}
			\includegraphics[height=6.5cm, trim={1.1cm 0 1.0cm 0cm}, clip=true]{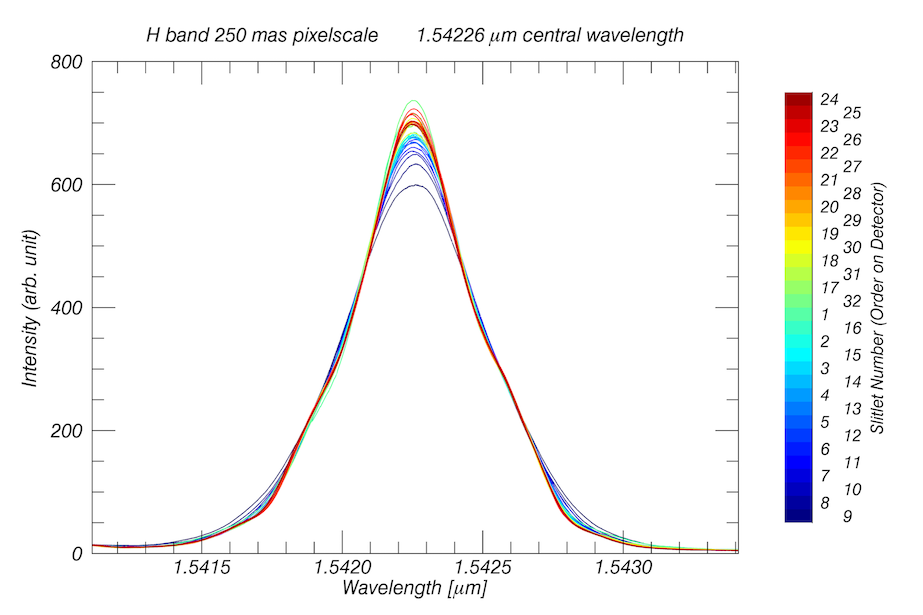}
		}
		\caption[Median LSF of individual slitlets in H-band]{H-band median LSF of the individual slitlets across the detector. Upper row: pre- upgrade, lower row: post- upgrade}
		\label{fig:median_h}
	\end{center}
\end{figure}

\begin{figure}[htbp!]
	\begin{center}
		\resizebox{1.0\textwidth}{!}{
			\includegraphics[height=6.5cm, trim={0.3cm 0 3.7cm 0cm}, clip=true]{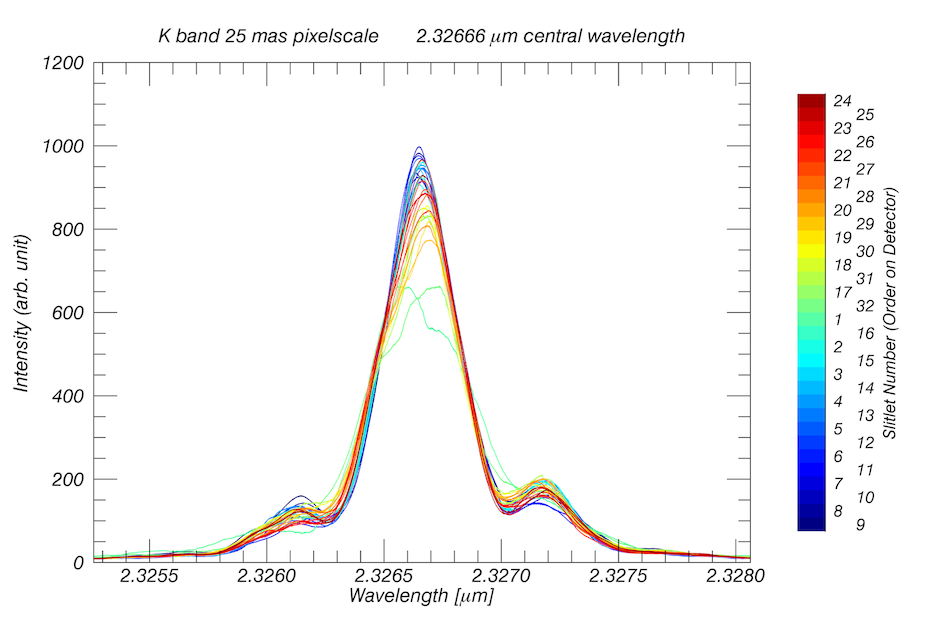}
			\includegraphics[height=6.5cm, trim={1.1cm 0 3.7cm 0cm}, clip=true]{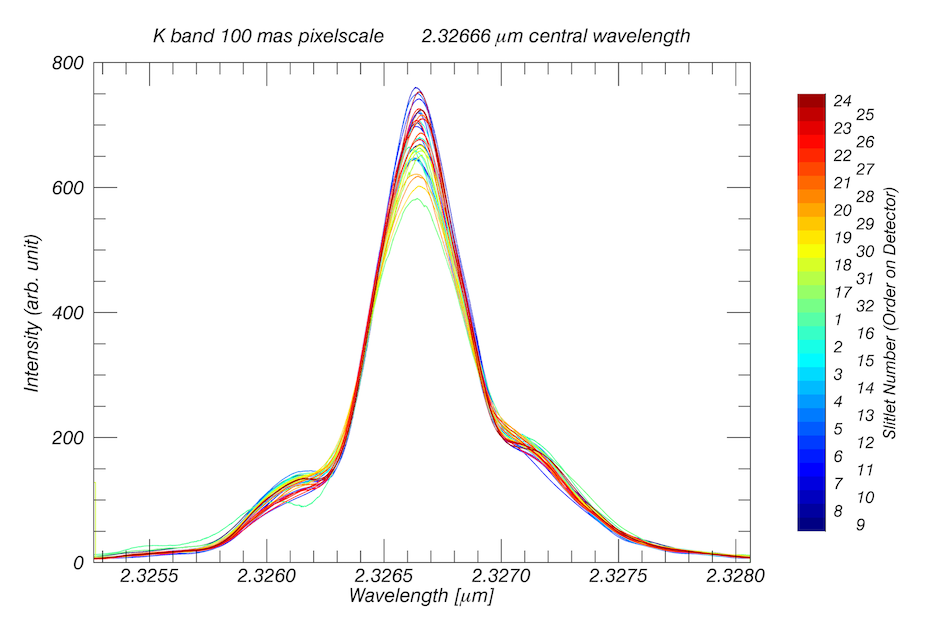}
			\includegraphics[height=6.5cm, trim={1.1cm 0 1.0cm 0cm}, clip=true]{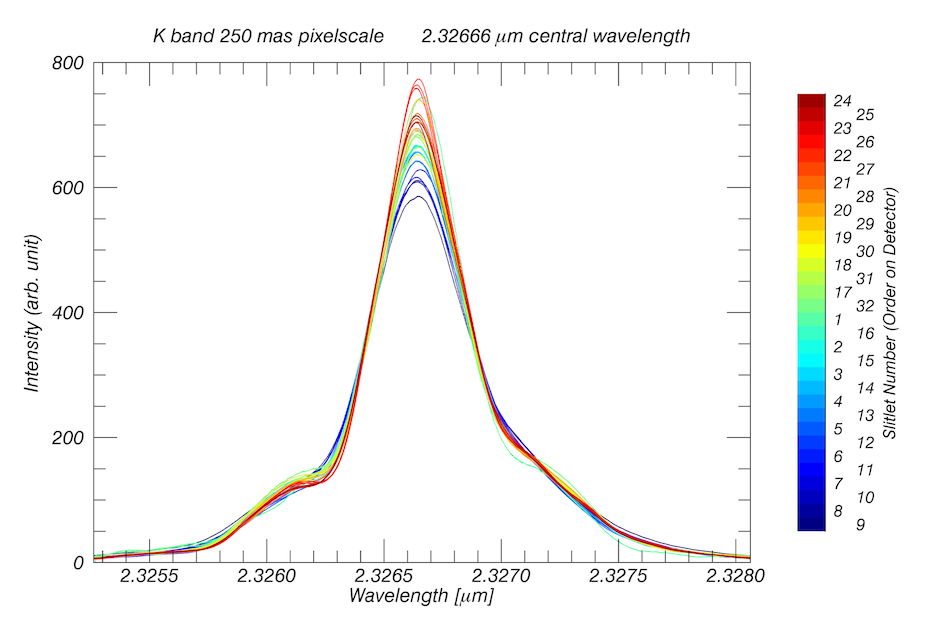}
		}
		\resizebox{1.0\textwidth}{!}{
			\includegraphics[height=6.5cm, trim={0.3cm 0 3.7cm 0cm}, clip=true]{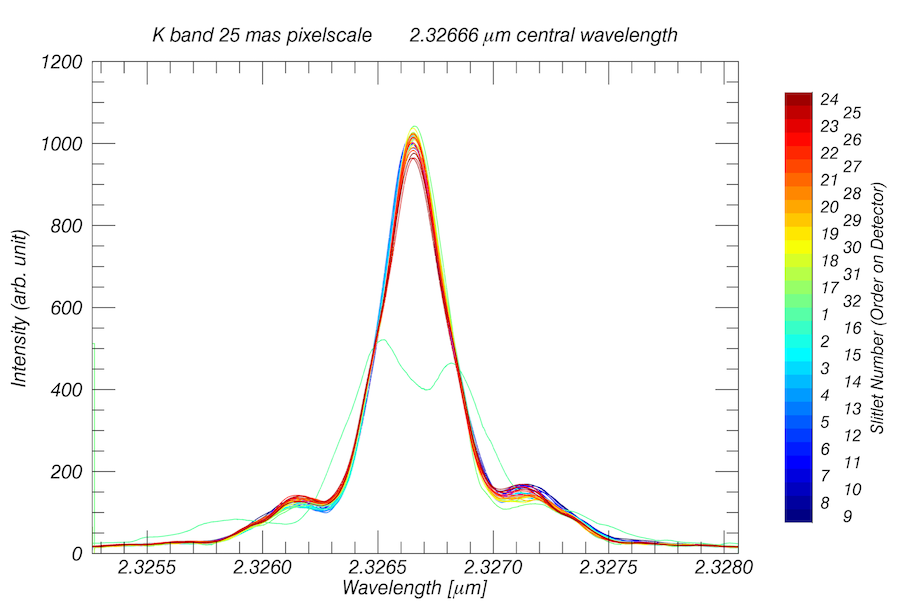}
			\includegraphics[height=6.5cm, trim={1.1cm 0 3.7cm 0cm}, clip=true]{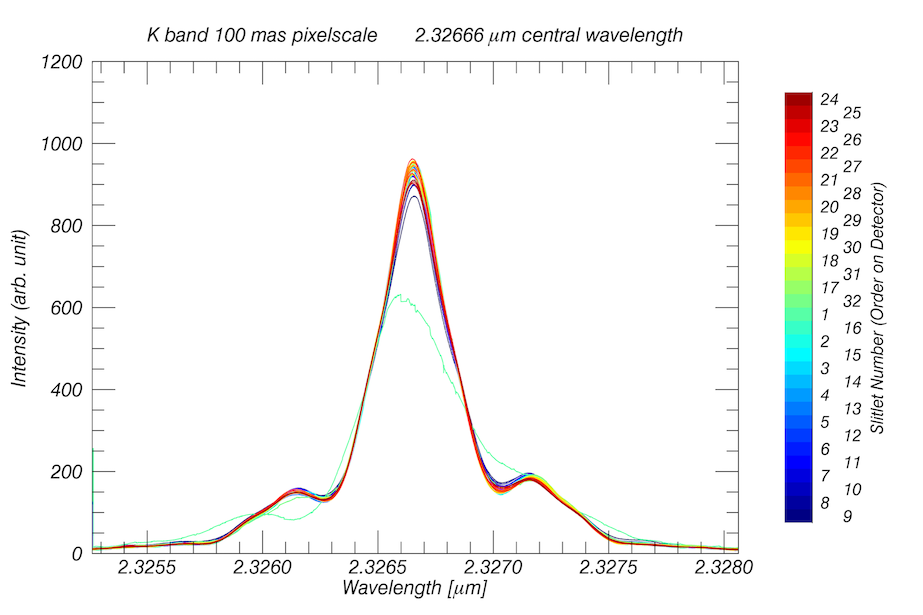}
			\includegraphics[height=6.5cm, trim={1.1cm 0 1.0cm 0cm}, clip=true]{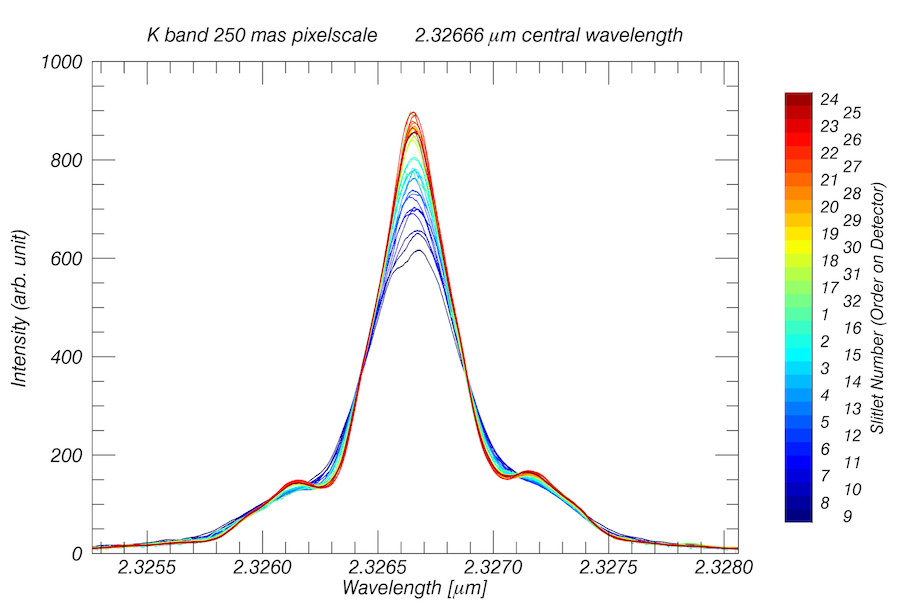}
		}
		\caption[Median LSF of individual slitlets in K-band]{K-band median LSF of the individual slitlets across the detector. Upper row: pre- upgrade, lower row: post- upgrade}
		\label{fig:median_k}
	\end{center}
\end{figure}		

\begin{figure}[htbp!]
	\begin{center}
		\resizebox{1.0\textwidth}{!}{
			\includegraphics[height=6.5cm, trim={0.3cm 0 3.7cm 0cm}, clip=true]{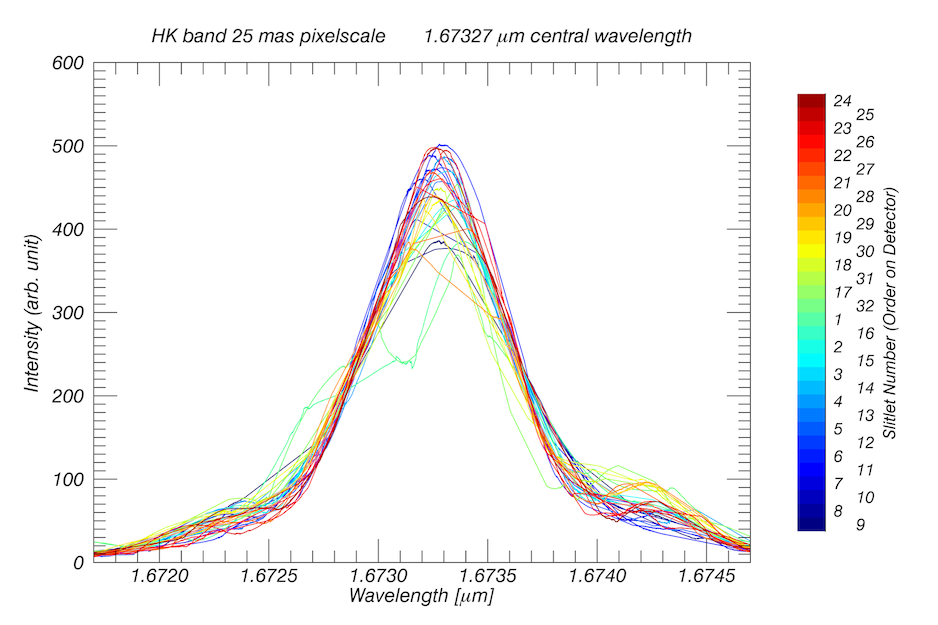}
			\includegraphics[height=6.5cm, trim={1.1cm 0 3.7cm 0cm}, clip=true]{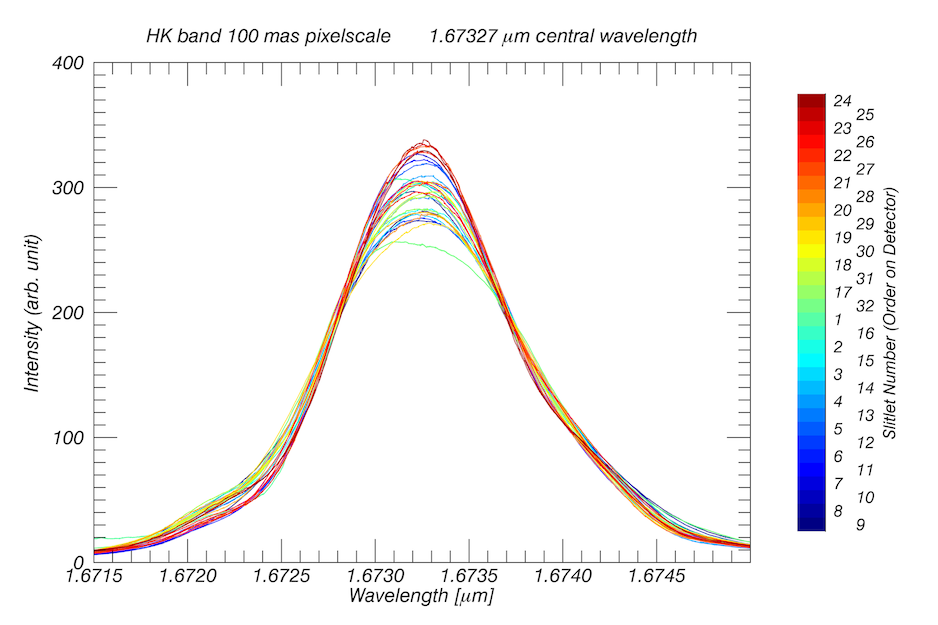}
			\includegraphics[height=6.5cm, trim={1.1cm 0 1.0cm 0cm}, clip=true]{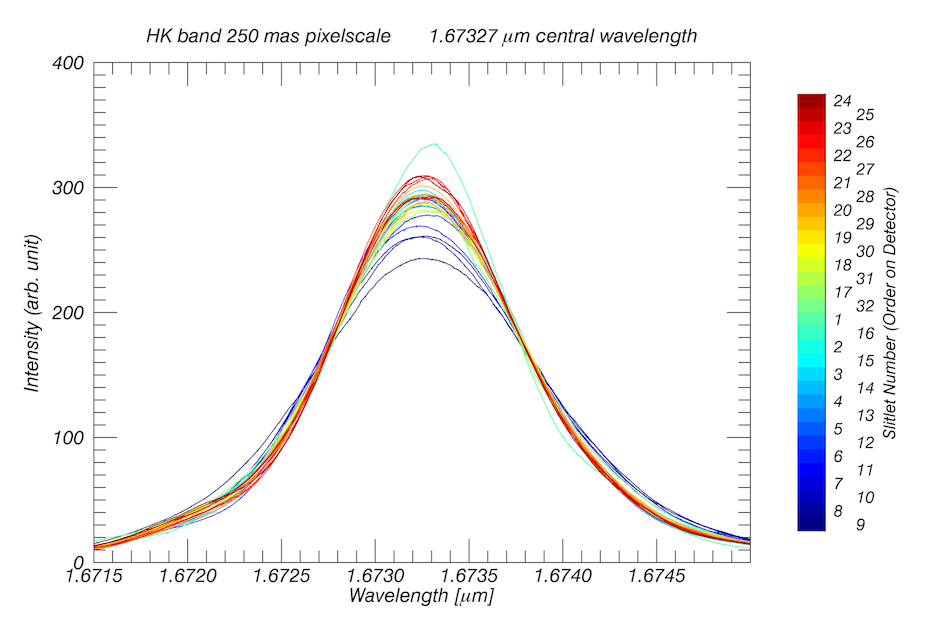}
		}
		\resizebox{1.0\textwidth}{!}{
			\includegraphics[height=6.5cm, trim={0.3cm 0 3.7cm 0cm}, clip=true]{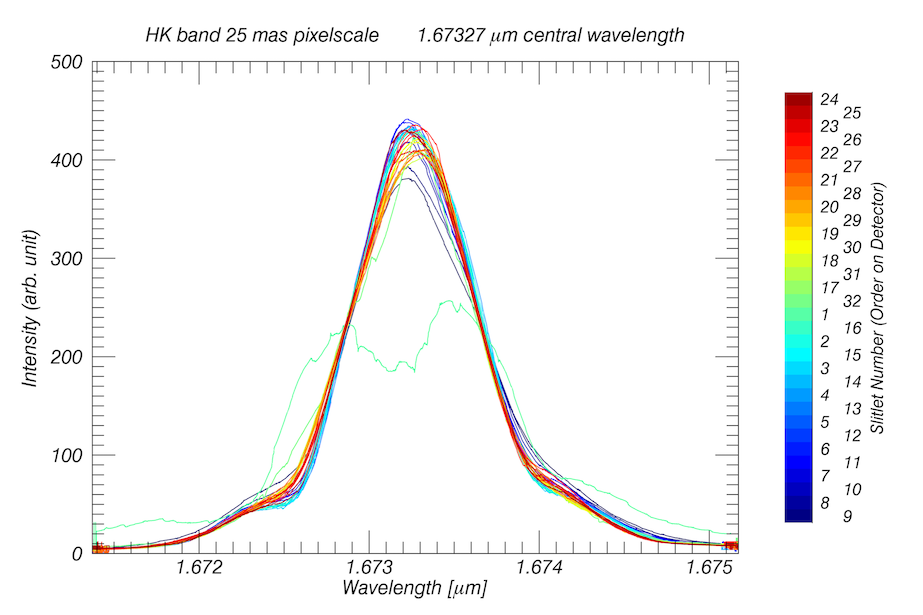}
			\includegraphics[height=6.5cm, trim={1.1cm 0 3.7cm 0cm}, clip=true]{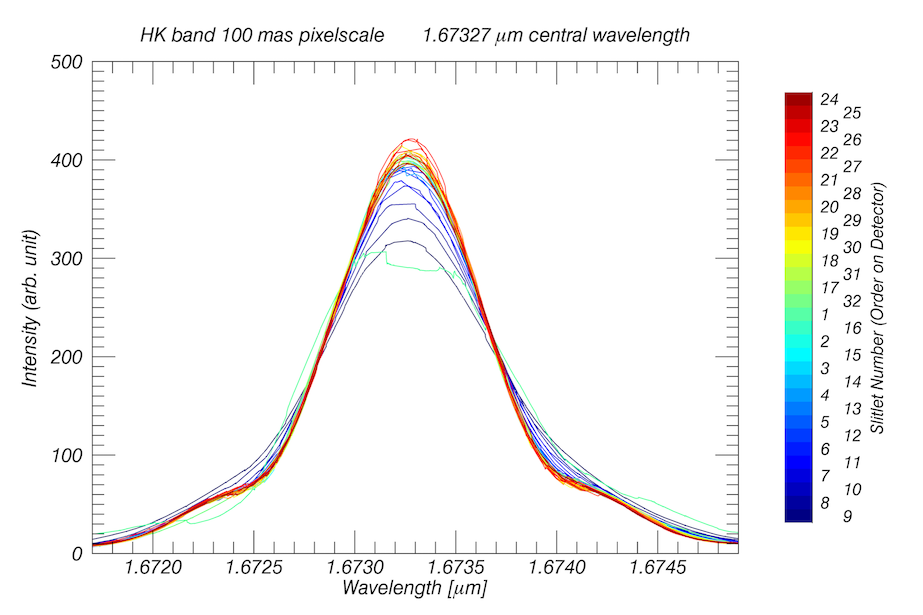}
			\includegraphics[height=6.5cm, trim={1.1cm 0 1.0cm 0cm}, clip=true]{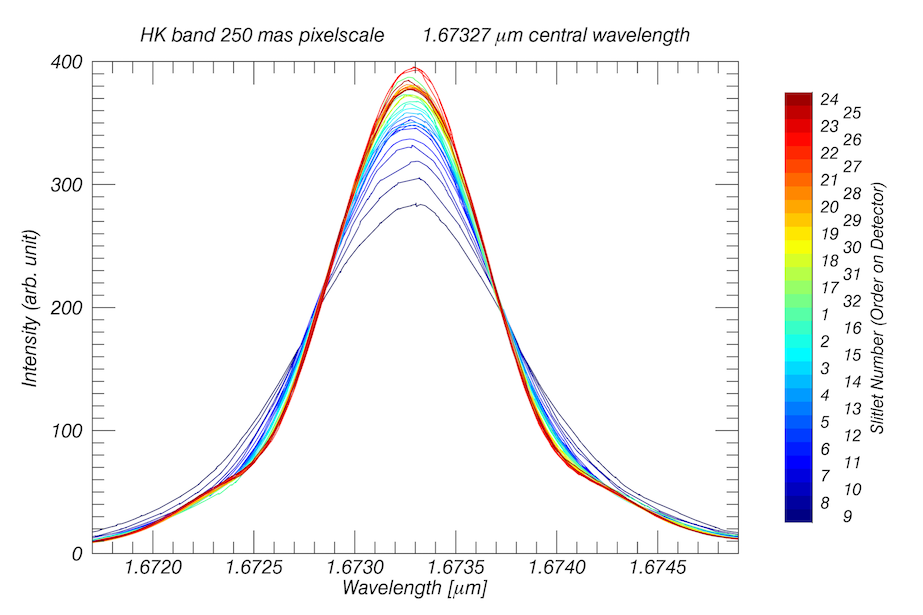}
		}
		\caption[Median LSF of individual slitlets in H+K-band]{H+K-band median LSF of the individual slitlets across the detector. Upper row: pre- upgrade, lower row: post- upgrade, Remark: in the 25 mas pixelscale of the pre- upgrade data the sampling of the datapoints was not well distributed over the line profile.}
		\label{fig:median_hk}
	\end{center}
\end{figure}
\clearpage
\section{FWHM Histograms}
This appendix section refers to section \ref{sec:resolution}. Here the complete resolution histograms are shown for all pixelscales and bands. The scale on the y-axis is again arbitrary. What can be seen additionally to the plots in the main chapter is that the distribution is sharpest in J-band, meaning that there is the lowest variation of the LSF width. With longer wavelength and smaller pixelscale the distribution is broadened resulting from more variations in the line profile width. K-band in the 25 mas pixelscale is here an exception and does not follow that behavior. This is, because in the AO pixelscale in K-band the amplitude of the side-peaks is so low that mainly the core of the LSF is fit, while for the other pixelscales the side-peaks are higher and thus contribute more to the Gauss-fit. Also in H+K-band this effect of a larger variation in smaller pixelscales cannot be seen.
\begin{figure}[htbp!]
	\begin{center}
		\resizebox{1.0\textwidth}{!}{
			\includegraphics[width=1.0\textwidth, trim={0.4cm 0 0.5cm 0cm}, clip=true]{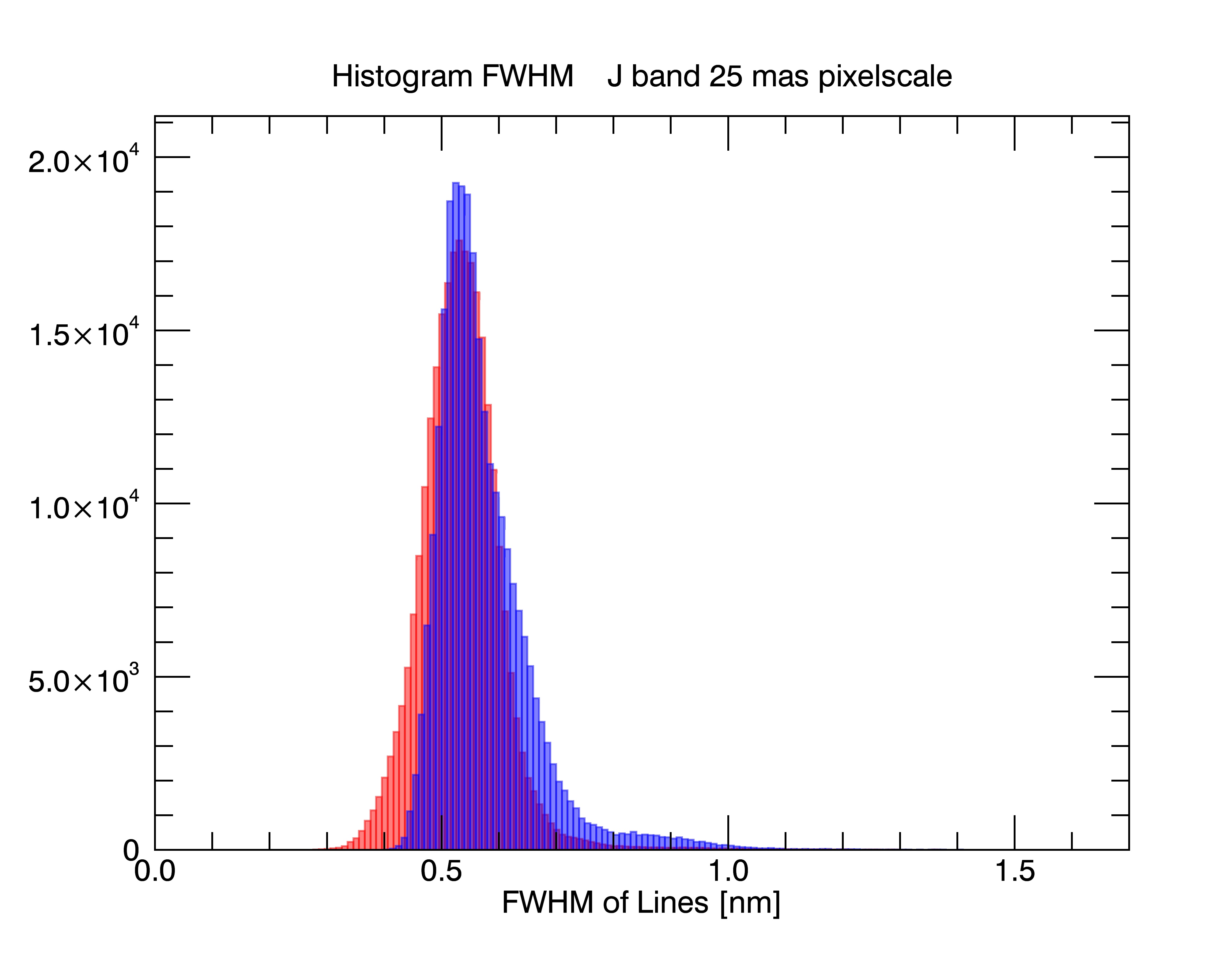}
			\includegraphics[width=1.0\textwidth, trim={0.4cm 0 0.5cm 0cm}, clip=true]{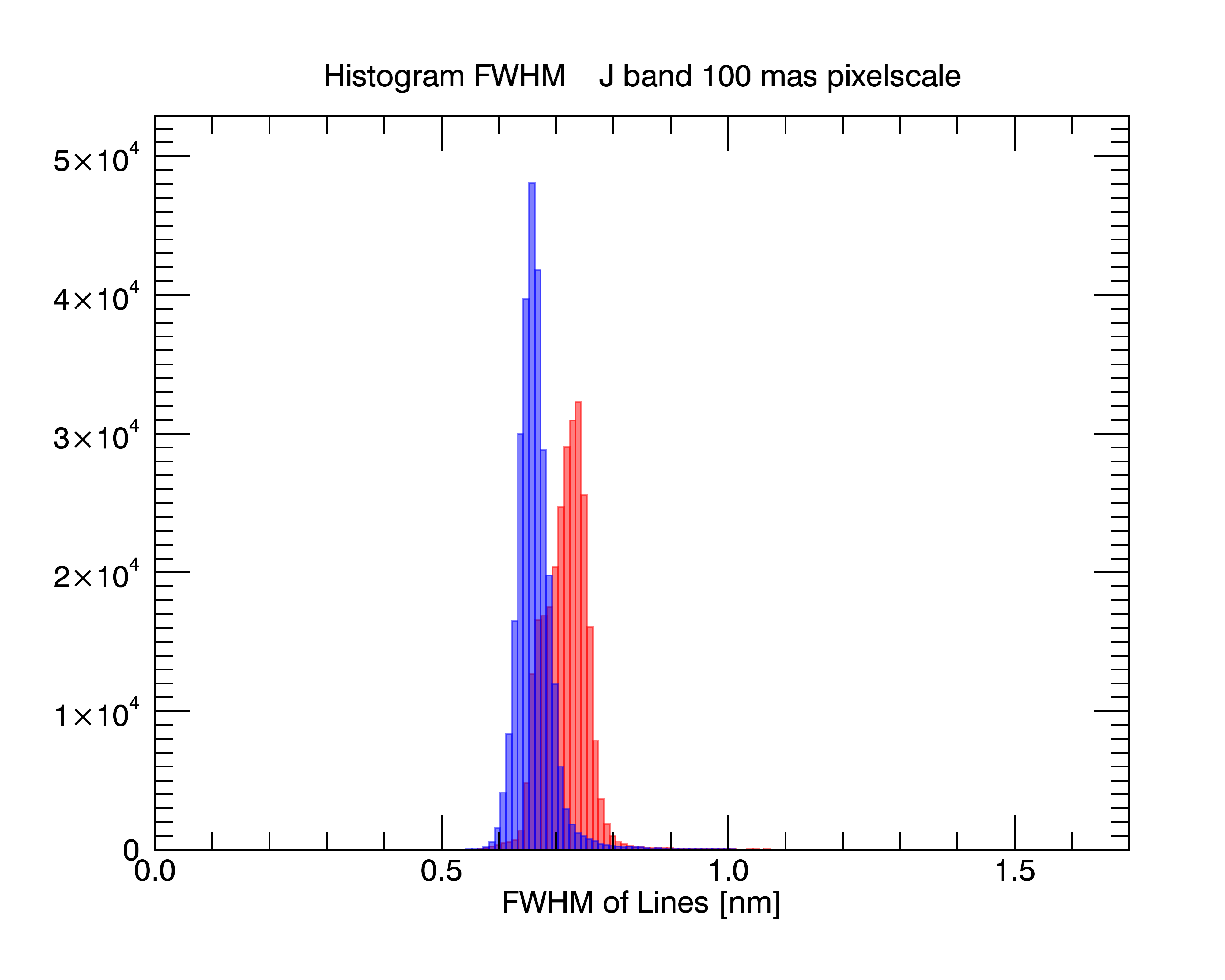}
			\includegraphics[width=1.0\textwidth, trim={0.4cm 0 0.5cm 0cm}, clip=true]{histogram_j250_nyquist_compare.png}
		}
		\resizebox{1.0\textwidth}{!}{
			\includegraphics[width=1.0\textwidth, trim={0.4cm 0 0.5cm 0cm}, clip=true]{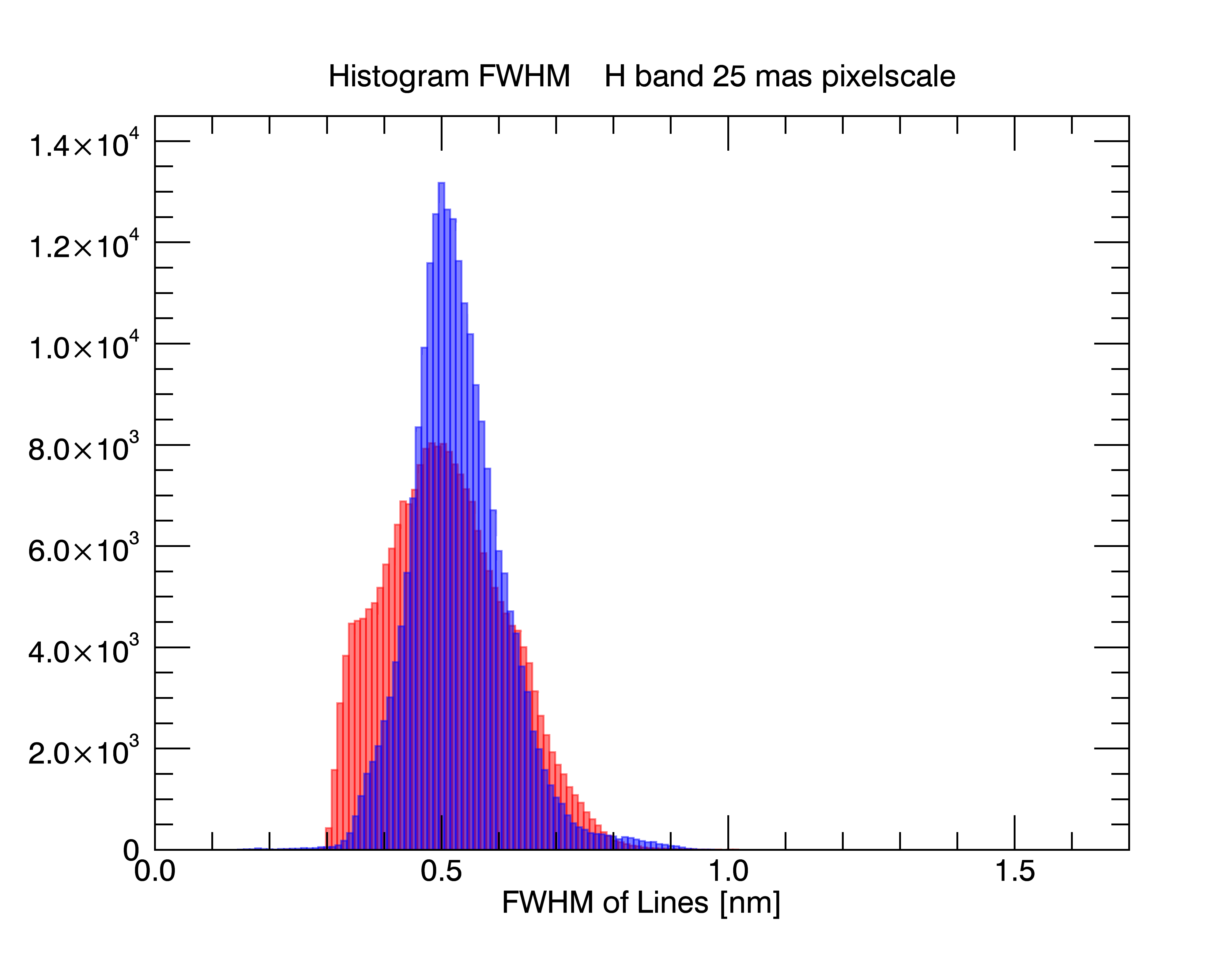}
			\includegraphics[width=1.0\textwidth, trim={0.4cm 0 0.5cm 0cm}, clip=true]{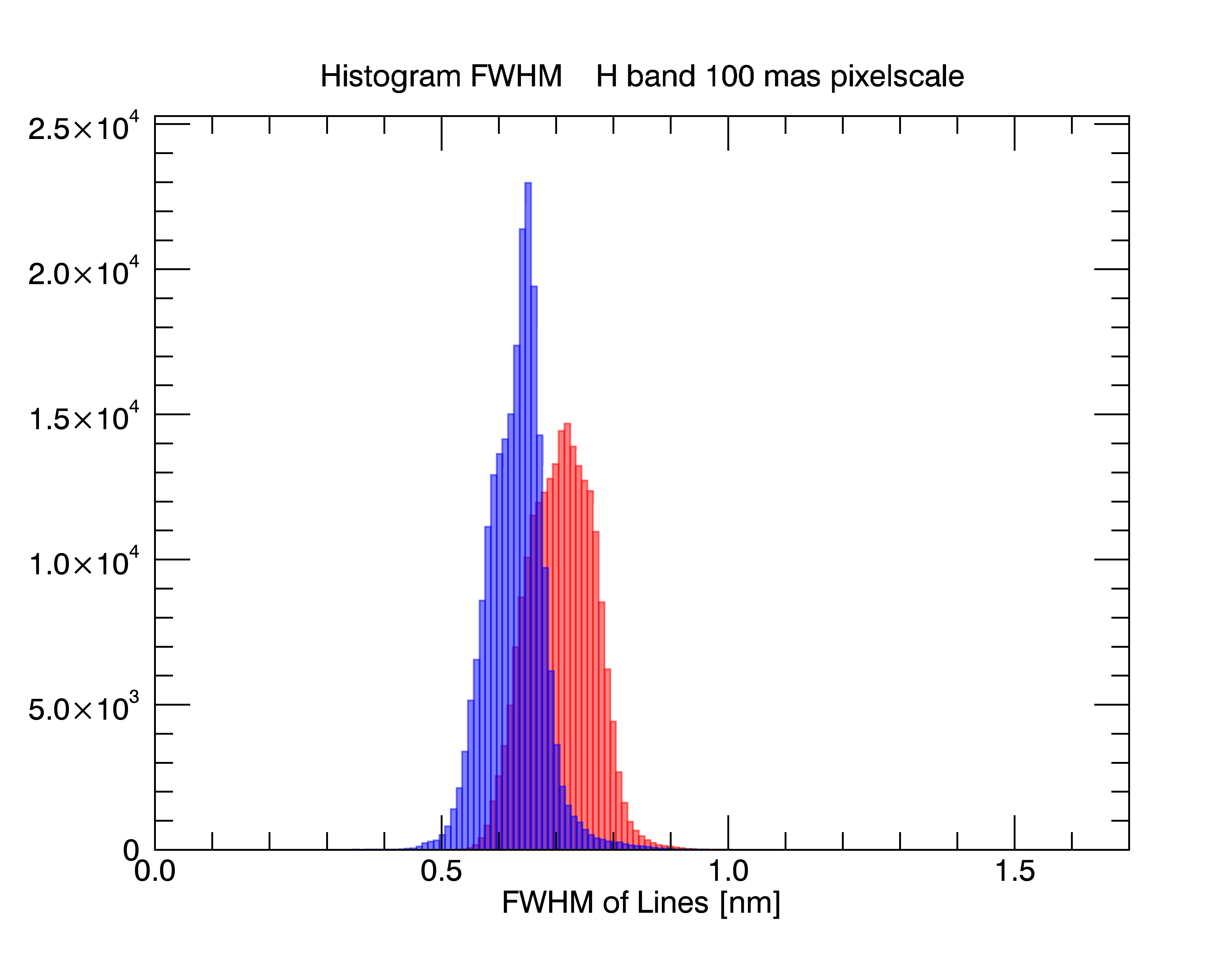}
			\includegraphics[width=1.0\textwidth, trim={0.4cm 0 0.5cm 0cm}, clip=true]{histogram_h250_nyquist_compare.png}
		}
		\resizebox{1.0\textwidth}{!}{
			\includegraphics[width=1.0\textwidth, trim={0.4cm 0 0.5cm 0cm}, clip=true]{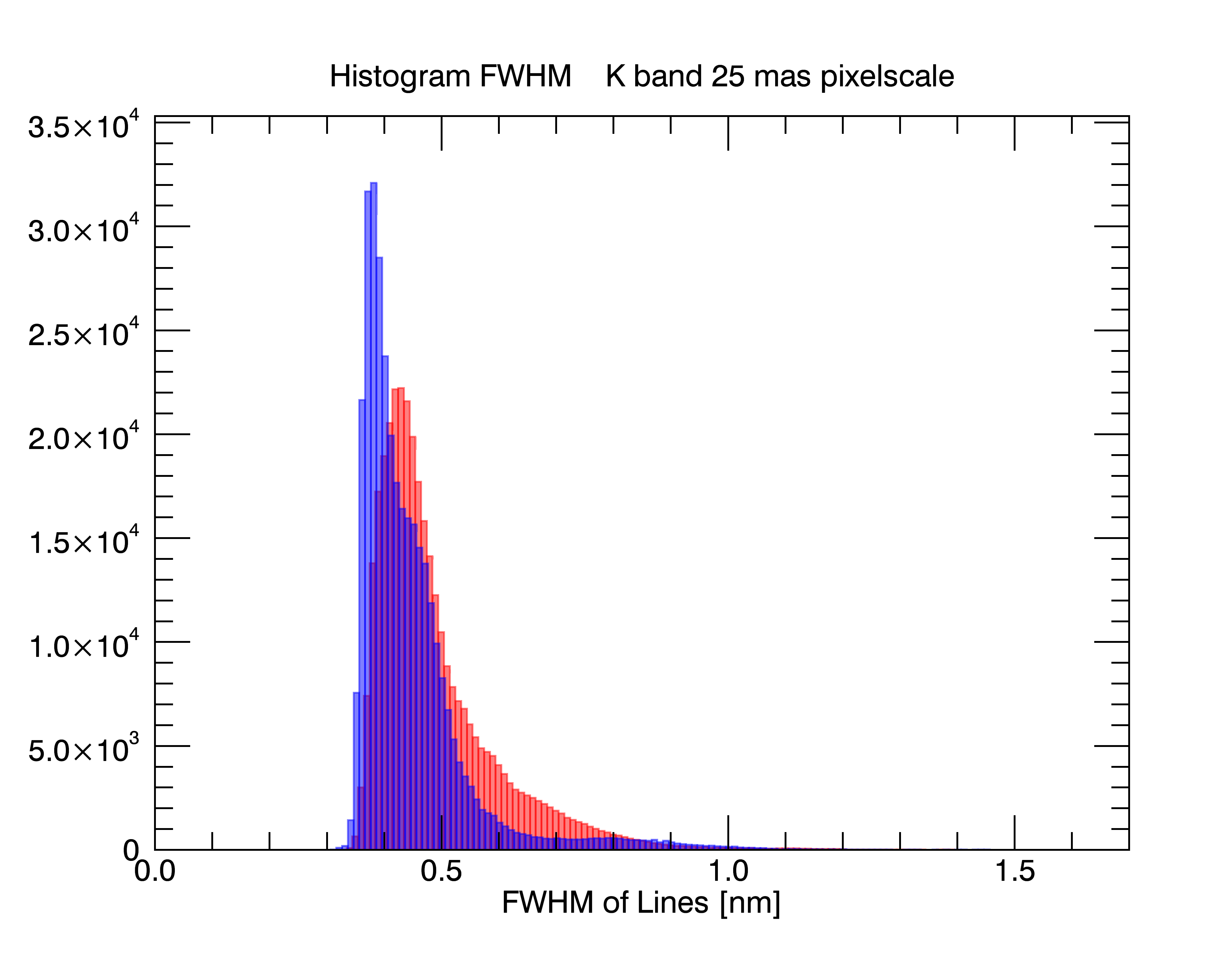}
			\includegraphics[width=1.0\textwidth, trim={0.4cm 0 0.5cm 0cm}, clip=true]{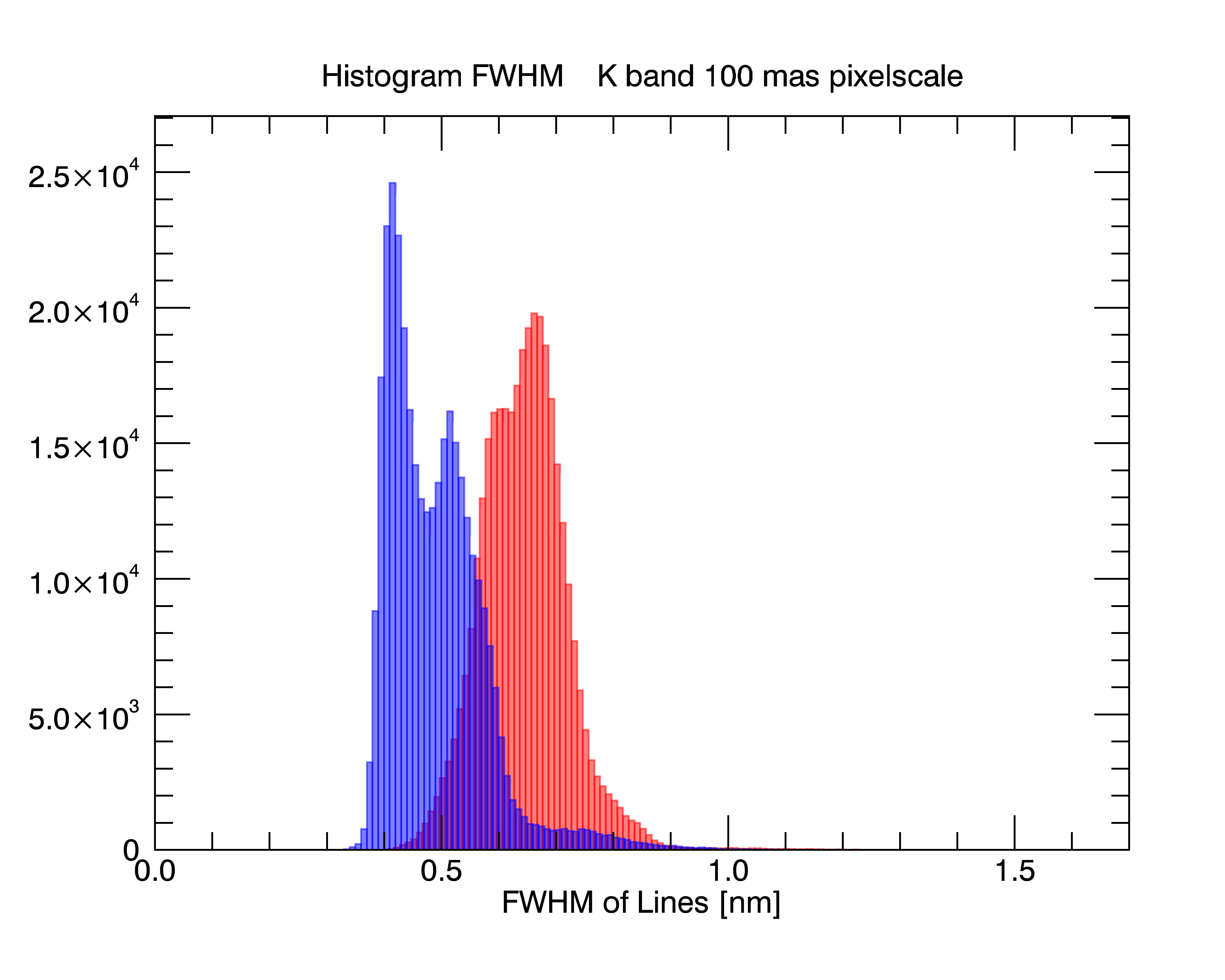}
			\includegraphics[width=1.0\textwidth, trim={0.4cm 0 0.5cm 0cm}, clip=true]{histogram_k250_nyquist_compare.png}
		}
		\resizebox{1.0\textwidth}{!}{
			\includegraphics[width=1.0\textwidth, trim={0.4cm 0 0.5cm 0cm}, clip=true]{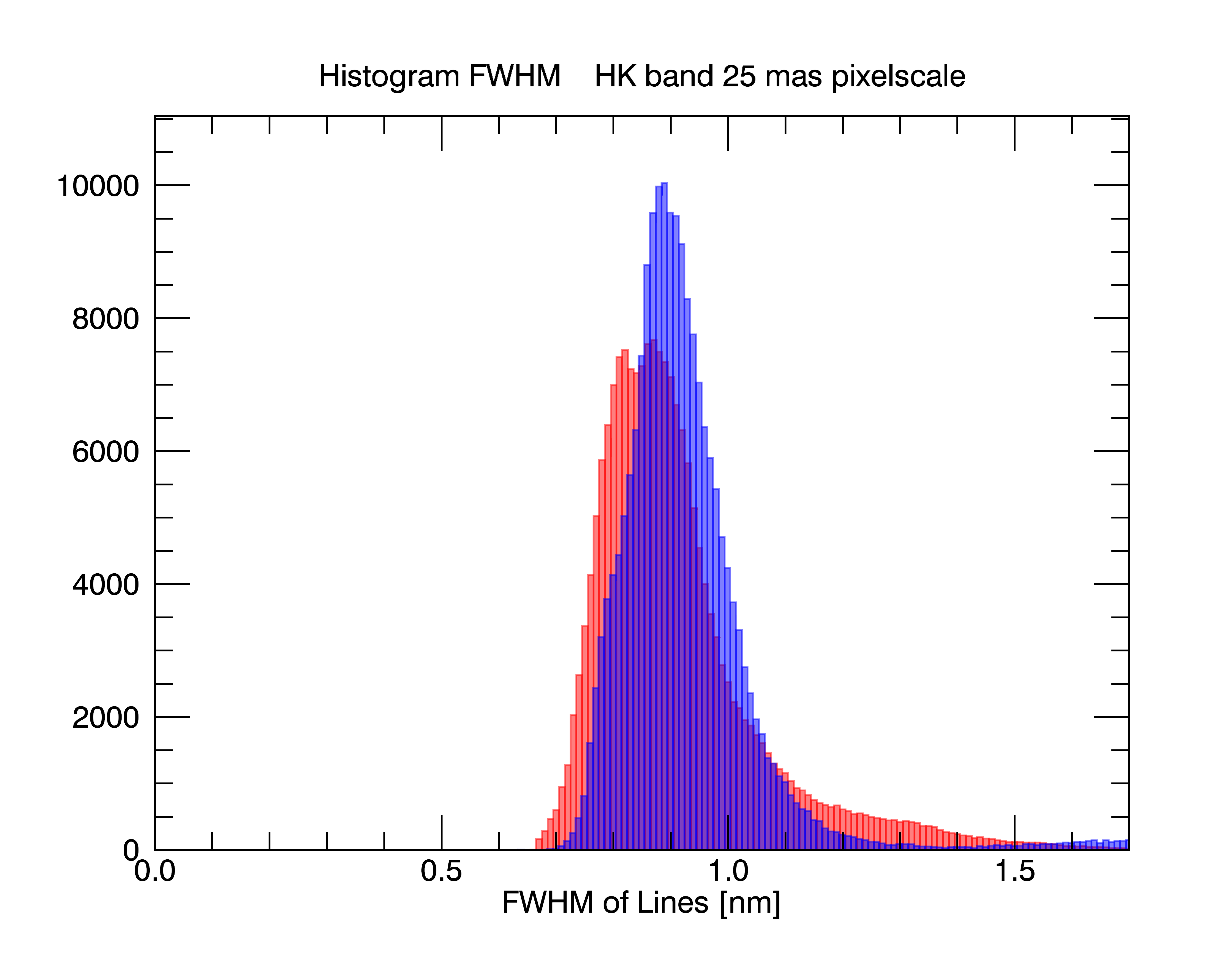}
			\includegraphics[width=1.0\textwidth, trim={0.4cm 0 0.5cm 0cm}, clip=true]{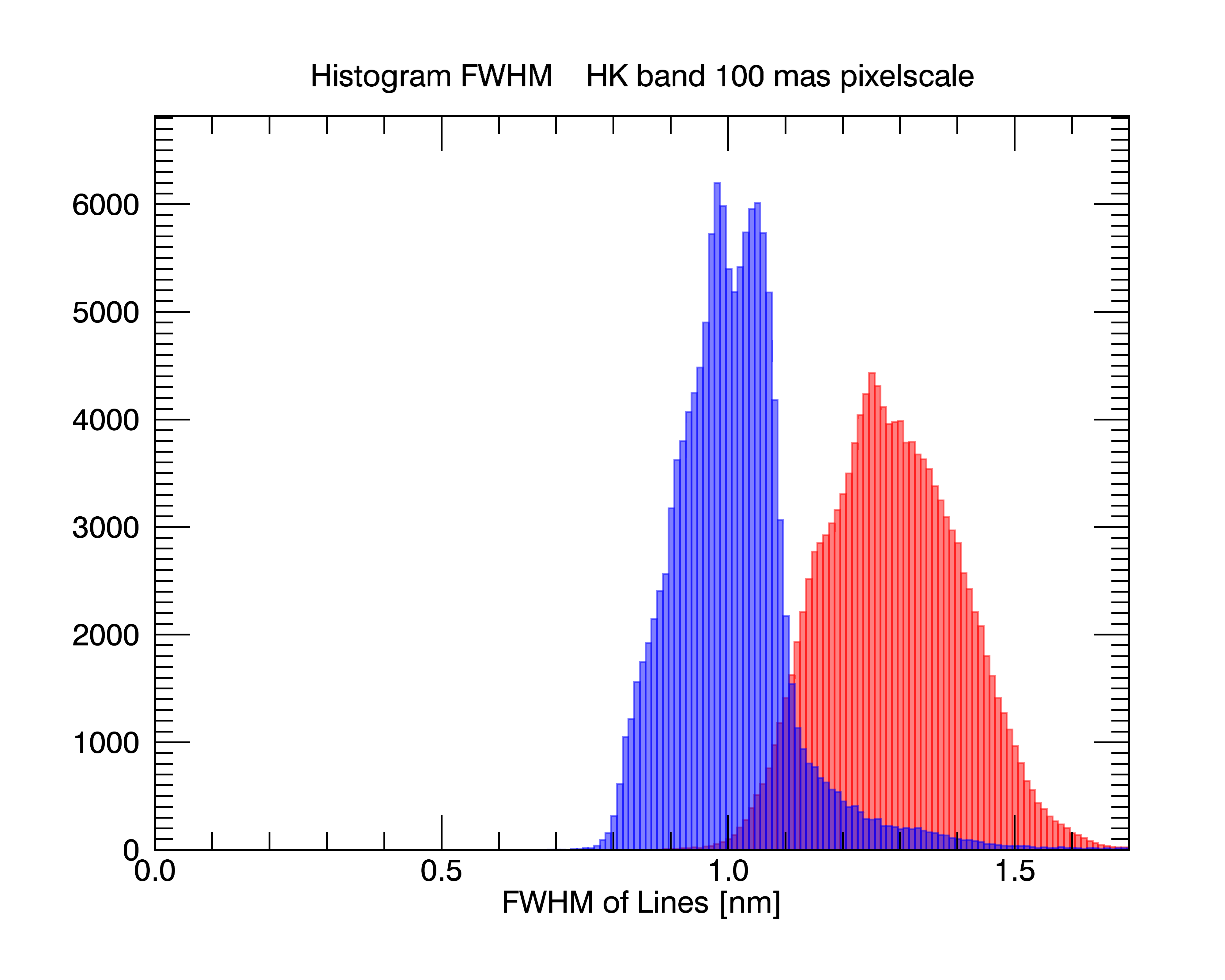}
			\includegraphics[width=1.0\textwidth, trim={0.4cm 0 0.5cm 0cm}, clip=true]{histogram_hk250_nyquist_compare.png}
		}
		\caption[Histograms of the FWHM of the spectral lines]{Histograms of the FWHM of the undersampled spectral lines with a binsize of 0.1 \AA. The red bars correspond to the pre- upgrade data, the blue bars to the post- upgrade data. Rows from top to bottom: J, H, K, H+K gratings. Columns from left to right: 25, 100, 250 mas pixels scales.}
		\label{fig:histograms_appendix}
	\end{center}
\end{figure}

\newpage
\addcontentsline{toc}{chapter}{Bibliography}
\bibliography{Thesis.bbl}{}
\bibliographystyle{apalike}

\addcontentsline{toc}{chapter}{List of Figures}
\listoffigures
\addcontentsline{toc}{chapter}{List of Tables}
\listoftables

\chapter*{Acknowledgments}

Finally I want to thank everybody who contributed to the success of this Master's thesis. In particular my thanks applies to:\\

\noindent \textbf{Reinhard Genzel} for giving me the possibility to work within one of the leading research groups in IR-astronomy.

\noindent \textbf{Frank Eisenhauer} for the possibility to work on this exciting instrument upgrade. I got to know a large variety of fields of work in instrumentation for astronomy.

\noindent \textbf{Liz George} for the excellent supervision during the last year. Thanks for being a great partner at work and for your friendship.

\noindent \textbf{Michael Hartl} for any help with optics whether in theory or in the lab.

\noindent \textbf{Helmut Feuchtgruber} for integrating me in the SPIFFI-upgrade and ERIS team. 

\noindent Our electronics and mechanics team on Paranal: \textbf{Alexander Buron}, \textbf{Johannes Hartwig}, \textbf{David Huber}, \textbf{Stefan Huber} for being a great team to work with.

\noindent All \textbf{people in the IR-group} for the nice and friendly atmosphere at work and in the breaks.\\

\noindent Mein ganz besonderer Dank gilt \textbf{meinen Eltern}, die mir mein Studium erm\"oglicht haben.

\end{document}